\documentclass[11pt,twoside]{mitthesis}
\usepackage{lgrind}
\pdfoutput=1

\usepackage{color}
\usepackage{graphicx}
\usepackage{amsmath}
\usepackage{amssymb}
\usepackage{xspace}
\usepackage[small]{subfigure}
\usepackage[hyperfootnotes=false]{hyperref}
\usepackage[numbers,sort&compress]{natbib}

\newcommand{\eq}[1]{Eq.~\eqref{eq:#1}}
\newcommand{\eqs}[2]{Eqs.~\eqref{eq:#1} and \eqref{eq:#2}}
\newcommand{\ch}[1]{Ch.~\ref{ch:#1}}
\newcommand{\app}[1]{App.~\ref{app:#1}}
\renewcommand{\sec}[1]{Sec.~\ref{sec:#1}}

\newcommand{\subsec}[1]{Sec.~\ref{subsec:#1}}
\newcommand{\fig}[1]{Fig.~\ref{fig:#1}}
\newcommand{\figs}[2]{Figs.~\ref{fig:#1} and \ref{fig:#2}}

\newcommand{\abs}[1]{\lvert#1\rvert}
\newcommand{\Abs}[1]{\bigl\lvert#1\bigr\rvert}
\newcommand{\ord}[1]{\mathcal{O}(#1)}
\newcommand{\ORd}[1]{\mathcal{O}\Bigl(#1\Bigr)}
\newcommand{\ORD}[1]{\mathcal{O}\biggl(#1\biggr)}

\newcommand{\mae}[3]{\langle#1\rvert#2\rvert#3\rangle}
\newcommand{\Mae}[3]{\bigl\langle#1\bigr\rvert#2\bigr\rvert#3\bigr\rangle}
\newcommand{\MAe}[3]{\Bigl\langle#1\Bigr\rvert#2\Bigr\rvert#3\Bigr\rangle}
\newcommand{\bra}[1]{\langle#1\rvert}
\newcommand{\ket}[1]{\lvert#1\rangle}

\newcommand{\intlim}[3]{\int_{#1}^{#2}\! \df #3 \,}

\newcommand{\df}{\mathrm{d}}

\newcommand{\img}{\mathrm{i}}

\renewcommand{\Im}{\mathrm{Im}}

\newcommand{\sdt}{\!\cdot\!}
\newcommand{\tr}{\textrm{tr}}

\newcommand{\lra}{\leftrightarrow}

\newcommand{\al}{\alpha}
\newcommand{\bt}{\beta}
\newcommand{\ga}{\gamma}
\newcommand{\de}{\delta}
\newcommand{\eps}{\epsilon}
\newcommand{\ve}{\varepsilon}
\newcommand{\la}{\lambda}
\newcommand{\si}{\sigma}
\newcommand{\om}{\omega}
\newcommand{\w}{\omega}

\newcommand{\Tau}{\mathcal{T}}

\newcommand{\Ga}{\Gamma}
\newcommand{\De}{\Delta}

\newcommand{\cB}{{\mathcal B}}

\newcommand{\cI}{{\mathcal I}}
\newcommand{\cL}{{\mathcal L}}

\newcommand{\cP}{{\mathcal P}}

\newcommand{\cY}{{\mathcal Y}}

\newcommand{\tB}{\widetilde{B}}
\newcommand{\tM}{\widetilde{M}}
\newcommand{\tS}{\widetilde{S}}
\newcommand{\tga}{ {\tilde{\gamma}} }
\newcommand{\tZ}{\widetilde{Z}}

\newcommand{\hp}{\hat{p}}
\newcommand{\hJ}{\widehat{J}}

\newcommand{\bn}{\bar{n}}
\newcommand{\bnP}{\overline {\mathcal P}}
\newcommand{\bC}{\bar C}
\newcommand{\bT}{\overline{T}}

\newcommand{\Dslash}{D\!\!\!\!\slash}

\newcommand{\nslash}{n\!\!\!\slash}
\newcommand{\bnslash}{\bar{n}\!\!\!\slash}
\newcommand{\pslash}{p\!\!\!\slash}

\newcommand{\qslash}{q\!\!\!\slash}

\newcommand{\ellslash}{\ell\!\!\!\slash}

\newcommand{\GeV}{\,\mathrm{GeV}}
\newcommand{\TeV}{\,\mathrm{TeV}}
\newcommand{\MSbar}{$\overline{\text{MS}}$ }

\newcommand{\nn}{\nonumber}

\newcommand{\lqcd}{\Lambda_\mathrm{QCD}}
\newcommand{\alem}{\alpha_\mathrm{em}}

\newcommand{\lb}{ {\tilde{b}} }        
        
\newcommand{\lp}{ {\tilde{p}} }        
\newcommand{\lw}{ {\tilde{\omega}} } 

\newcommand{\cm}{\mathrm{cm}} 

\newcommand{\soft}{{s}}        
\newcommand{\parton}{\mathrm{part}} 
\newcommand{\thr}{\mathrm{thr}}   
  
\newcommand{\hemiin}{\mathrm{ihemi}}

\newcommand{\hemiout}{\mathrm{hemi}}

\newcommand{\hH}{\widehat{H}}
\newcommand{\hS}{\widehat{S}}

\newcommand{\Obs}{O}

\newcommand{\Ecm}{E_\mathrm{cm}}
\newcommand{\ECM}{E_\mathrm{cm}}

\newcommand{\Disc}{\mathrm{Disc}}

\newcommand{\cut}{\mathrm{cut}}
\renewcommand{\max}{\mathrm{max}}
\newcommand{\tree}{\mathrm{tree}}
\newcommand{\oneloop}{\mathrm{1loop}}
\newcommand{\cusp}{\mathrm{cusp}}
\newcommand{\bare}{\mathrm{bare}}

\newcommand{\incl}{\mathrm{incl}}
\renewcommand{\det}{\mathrm{det}}

\newcommand{\zero}{{(0)}}
\newcommand{\one}{{(1)}}

\newcommand{\SCETa}{\ensuremath{{\rm SCET}_{\rm I}}\xspace}
\newcommand{\SCETb}{\ensuremath{{\rm SCET}_{\rm II}}\xspace}

\newcommand{\op}{{\mathcal{O}}}
\newcommand{\tiop}{\widetilde{\mathcal{O}}}
\newcommand{\oq}{{\mathcal{Q}}}

\allowdisplaybreaks[2]

\definecolor{mygreen}{rgb}{0,0.65,0}
\definecolor{myblue}{rgb}{0,0,1}
\definecolor{myorange}{rgb}{1,0.5,0}
\definecolor{mydarkblue}{rgb}{0,0,0.6}
\definecolor{myred}{rgb}{1,0,0}

\newcommand{\beamc}[1]{\textcolor{mygreen}{#1}}

\newcommand{\softc}[1]{\textcolor{myorange}{#1}}
\newcommand{\hardc}[1]{\textcolor{mydarkblue}{#1}}

\begin{document}

\title{Factorization at the LHC: From PDFs to Initial State Jets}

\author{Wouter J.~Waalewijn}

\department{Department of Physics}

\degree{Doctor of Philosophy in Physics}

\degreemonth{June}
\degreeyear{2010}
\thesisdate{May 12, 2010}

\supervisor{Iain W.~Stewart}{Associate Professor}

\chairman{Krishna Rajagopal}{Associate Department Head for Education}

\maketitle

\cleardoublepage

 \pagestyle{empty}
\setcounter{savepage}{\thepage}
\begin{abstractpage}

New physics searches at the LHC or Tevatron typically look for a certain number of hard jets, leptons and photons. The constraints on the hadronic final state lead to large logarithms, which need to be summed to obtain reliable theory predictions. These constraints are sensitive to the strong initial-state radiation and resolve the colliding partons inside initial-state jets. We find that the initial state is properly described by ``beam functions".

We introduce an observable called ``beam thrust" $\tau_B$, which measures the hadronic radiation relative to the beam axis. By requiring $\tau_B \ll 1$, beam thrust can be used to impose a central jet veto, which is needed to reduce the large background in $H \to WW \to \ell\nu\ell\bar\nu$ from $t\bar t \to WWb\bar b$. We prove a factorization theorem for ``isolated'' processes, $pp \to XL$ where the hadronic final state $X$ is restricted by $\tau_B \ll 1$ and $L$ is non-hadronic. This factorization theorem enables us to sum large logarithms $\al_s^n \ln^m \tau_B$ in the cross section and involves beam functions. The beam thrust spectrum allows us to study initial-state radiation in perturbation theory, which can be compared with experiment and Monte Carlo. We present results for the beam thrust cross section for Drell-Yan and Higgs production through gluon fusion at next-to-next-to-leading-logarithmic order.

The beam functions depend on the momentum fraction $x$ and the (transverse) virtuality $t$ of the colliding partons. The $t$ dependence can be calculated in perturbation theory by matching beam functions onto the parton distribution functions (PDFs) at an intermediate scale $\mu_B \sim \sqrt{t}$. We calculate all the one-loop matching coefficients. Below $\mu_B$ we have the usual DGLAP evolution for the PDFs. Above $\mu_B$, the evolution of the beam function is in $t$ and does not change $x$ or the parton type. 

We introduce the event shape ``N-jettiness" $\tau_N$, which generalizes beam thrust to events with $N$ signal jets. Requiring $\tau_N \ll 1$ restricts radiation between the signal jets and vetoes additional undesired jets. This yields a factorization formula with inclusive beam and jet functions and allows us to sum the large logarithms $\al_s^n \ln^m \tau_N$ from the phase space restriction.

\end{abstractpage}

\cleardoublepage

\section*{Acknowledgments}

It has been a pleasure to work under the supervision of Prof.~Iain Stewart, and I want to thank him for all the advice and time he has given me. The work in this thesis has been done in collaboration with Dr.~Frank Tackmann, who I would like to thank for many enjoyable discussions and for sharing some of his physical intuition and passion for physics. I also want to thank Dr.~Carola Berger and Claudio Marcantonini with whom I collaborated on some of the research presented in this thesis. Thanks go to my thesis committee members Prof.~Robert Jaffe and Prof.~Allan Adams. I have enjoyed and benefitted from being part of a research group that has included Dr.~Keith Lee, Dr.~Chris Arnesen, Dr.~Ambar Jain, Dr.~Massimiliano Procura, Riccardo Abbate, Teppo Jouttenus and Roberto Sanchis. 

I am very grateful for the love and support of my parents Caspar and Heleen and my family. I especially want to thank my wife Eileen. Her love, care and encouragement have been a joy and help throughout my PhD. The MIT Graduate Christian Fellowship has been a great place of friendship, fun and growth, in particular the Sidney-Pacific bible study group. I am thankful for the many friends, both in an outside of the Center of Theoretical Physics that have been part of my life during my time at MIT. I realize that all my gifts and talents are not something I deserve, and I am grateful to God for this opportunity as well as the results.

This work was supported by the Office of Nuclear Physics of the U.S.\ Department
of Energy, under the grant DE-FG02-94ER40818.

\pagestyle{plain}

\tableofcontents
\newpage
\listoffigures
\newpage
\listoftables


\chapter{Introduction}
\label{ch:intro}

\section{Effective Physics}

Over the centuries we have been studying the world around us, discovering new physics at new length scales. Starting from our everyday experience which ranges from scales of about $10^3$ to $10^{-3}$ meters (m), we encounter phenomena like gravity, contact forces, fluid dynamics and light. When we move on to the smaller scales of cells at $10^{-5}$ m or molecules and atoms around $10^{-10}$ m, we pass entire fields of science such as biology and chemistry. It is surprising that the underlying physics that leads to this rich variety of phenomena is simply (quantum) electrodynamics. The nucleus of the atom at scales of $10^{-15}$ m is made of protons and neutrons and is held together by the nuclear force. This is a residual force coming from the strong force, also known as Quantum Chromodynamics (QCD). QCD is responsible for confining the quarks to protons and neutrons. The remaining force that is part of the well established Standard Model of particle physics is the weak force, which for example plays a role in some hadron decays. 

With the Large Hadron Collider (LHC) at CERN starting to take data, we will be probing nature at smaller length scales than before and hope to discover new physics. In order to reconstruct new short distance physics from the measured leptons and QCD radiation, one typically looks for a signal with a certain number of jets of energetic hadrons. This thesis explores the effect of such restrictions on the hadronic final state in theoretical calculations, with an emphasis on the 0-jet case which allows us to study initial-state radiation. An introduction to the work in this thesis can be found in the next section.

Our current understanding of nature as the strong, weak and electromagnetic force, together with gravity, is beautiful in its simplicity as a theory. Yet its computational complexity makes it ill-suited to describe most phenomena in nature. In practice one adopts an effective description appropriate for the length scale one is looking at. We are already used to doing this in much simpler theories such as Newtonian gravity, where for our everyday purposes a constant gravitational acceleration is an appropriate description. Effectively we are treating the earth as an infinite plane, which is a reasonable approximation given that we are small compared to the size of the earth. If we instead would try to describe the motion of the planets in our solar system, the size of the planets and the sun becomes negligibly small and we can treat them as points. 

We encounter a similar situation in electrostatics, when we want to determine the electric field due to an object at a distance $r$ that is large compared to the size $s$ of the object. We can start by approximating the object as a point with a total charge $Q$ and systematically improve this result by using the so-called multipole expansion. This is an expansion in $s/r \ll 1$, where each next order contains more details of the charge distribution. The first correction comes from the dipole moment $\vec p$ of the configuration and is suppressed by a factor of $s/r$. The next correction comes from the quadrupole moment $Q_{ij}$ and is suppressed by $s^2/r^2$ etc. We have thus found an effective description of the charge configuration, where the details of the full charge configuration are captured by only a few constants: $Q$, $\vec p$, etc. 

This simplification to an effective description, where the relevant details of the complete physical picture are condensed into just a few constants, will be a theme that continues as we study effective descriptions of quantum field theories. In this thesis we will use Soft-Collinear Effective Theory (SCET) , which describes initial and final state radiation and provides us with a powerful tool for studying processes at the LHC. In the next section we will present an introduction to the work in this thesis. In \sec{EFT_weak} we will look at an effective description for a decay involving the weak force, which will allow us to discuss general features of effective field theories. We present an intuitive introduction to Soft-Collinear Effective Theory in \sec{SCET_intro}.

\section{Predicting Collisions at the LHC}
\label{sec:pred_LHC}

In this section we introduce the work described in this thesis. Starting with a schematic picture of proton-proton collisions, we review the topic of factorization and discuss the relevance of experimental restrictions on the hadronic final state. Our work focusses on how such constraints can be incorporated into factorization theorems and we give a preview of our results.

\begin{figure}
\centering
\includegraphics[width=0.6\textwidth]{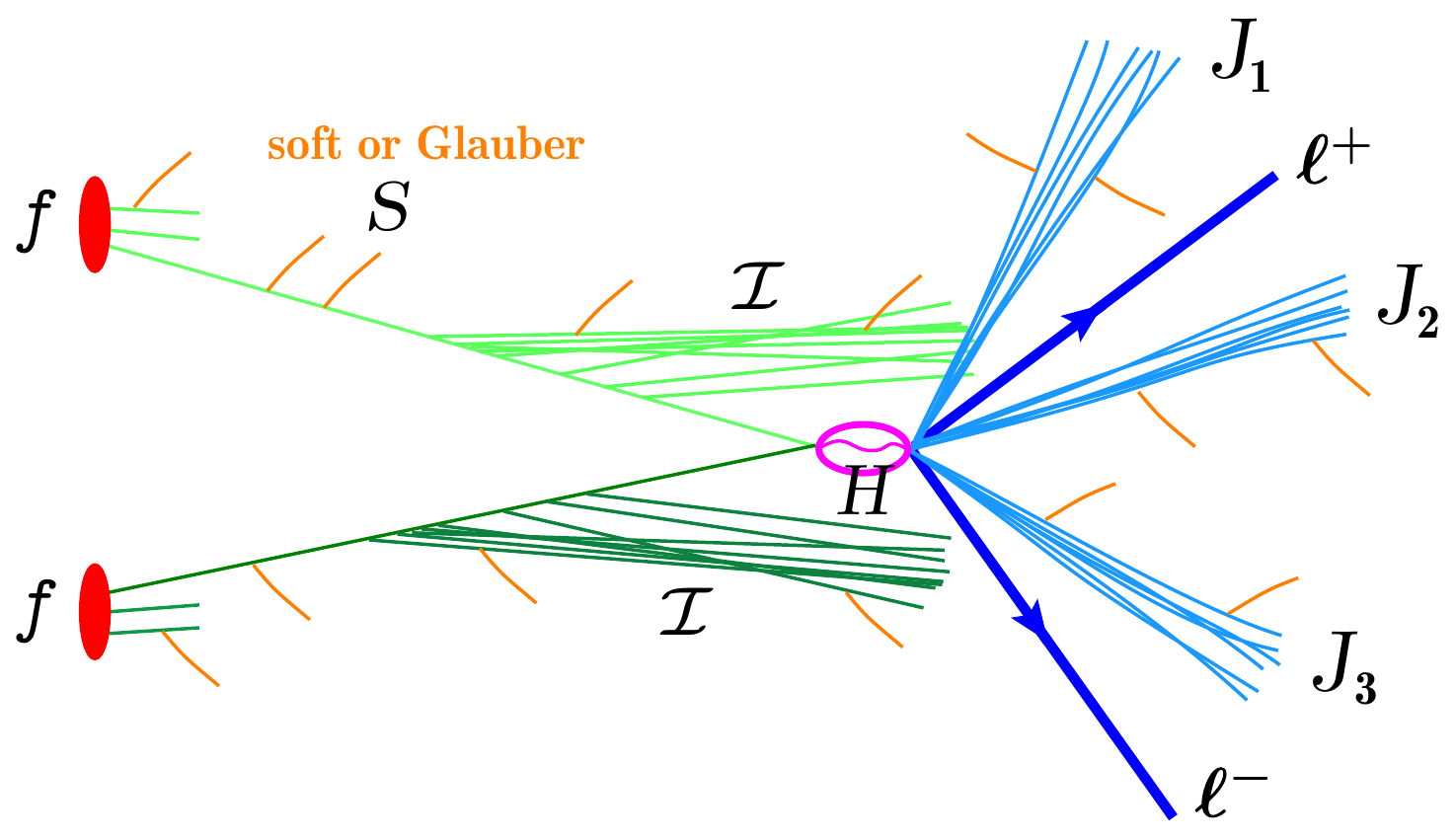}%
\caption[Schematic picture of a proton-(anti)proton collision at the LHC or Tevatron.]{Schematic picture of a collision at the LHC or Tevatron. It shows the evolution in time from left to right, starting with the proton and (anti)proton in the initial state and ending with jets, leptons and soft radiation in the final state.}
\label{fig:LHC}
\end{figure}

The primary goal of the experiments at the LHC and Tevatron is to search for the Higgs particle and physics beyond the Standard Model through collisions at the energy frontier. The fact that the short-distance processes of interest are interlaced with QCD interactions complicates the search.  A schematic picture of a proton-proton collision is displayed in \fig{LHC}. A quark or gluon is extracted from each proton (the red circles labeled $f$), and emits strong initial-state radiation ($\cI$) prior to the hard short-distance collision (at $H$). The hard collision produces strongly interacting partons which hadronize into collimated jets of hadrons ($J_{1,2,3}$), as well as non-strongly interacting particles (represented in the figure by the $\ell^+\ell^-$). Finally, all the strongly interacting particles, including the spectators in the proton, interact with low-momentum soft gluons and can exchange perpendicular momentum by Glauber gluons (both indicated by the short orange lines labeled $S$).

Factorization is one of the most fundamental concepts in the theoretical description of these collisions. It is the statement that the cross section can be computed as a product of probability functions, each describing a part of \fig{LHC}. For a review of factorization see Ref.~\cite{Collins:1989gx}, for examples of factorization theorems see \sec{DYfact}. Factorization allows one to separate the new hard physics from the array of QCD interactions in the initial and final states and is therefore key in the search for new physics. It is also necessary for controlling QCD effects. For example, the momentum distributions of the colliding partons in the protons are nonperturbative, but factorization can imply that these are described by universal distributions which have been measured in earlier experiments.

Monte Carlo programs are a widely used numerical method and employ notions from factorization and properties of QCD in the soft and collinear limits to model the ingredients in the full cross section, $\df\sigma = H\otimes f\otimes f \otimes \cI\otimes \cI \otimes \prod_i J_i \otimes S$. They have the virtue of providing a general tool for any observable, but have the disadvantage of making model-dependent assumptions to combine all the ingredients and to calculate some of them. When a factorization theorem for a specific observable is available this provides a better approach, since the various ingredients are defined and combined in a rigorous way.

In order to probe the physics of the hard interaction, measurements often impose restrictions on the hadronic final state, requiring a certain number of hard leptons or jets~\cite{:1999fr,:1999fq, Bayatian:2006zz,Ball:2007zza}. For example, a typical new physics search looking for missing transverse energy may also require a minimum number of jets with $p_T$ above some threshold.  To identify the new physics and determine the masses of new-physics particles, one has to reconstruct decay chains with a certain number of jets and leptons in the final state.

However, for the majority of processes of interest at hadron colliders where one distinguishes properties of the hadronic final state, so far no rigorous field-theoretic derivation of a factorization theorem to all orders in perturbation theory exists. The most well-known factorization theorem is
\begin{align} \label{eq:sigff}
\df\sigma = \sum_{i,j}\df\sigma^\parton_{ij}
  \otimes f_i(\xi_a)\otimes f_j(\xi_b)
\,.\end{align}
Here $f_i$ and $f_j$ are the standard parton distribution functions (PDFs) that describe the probability to extract a parton $i,j=\{g,u,\bar u,d,\ldots\}$ out of the proton, with a fraction $\xi_{a,b}$ of the proton momentum. The hard scattering of $i$ and $j$ is described by the partonic cross section $\df\sigma^\parton_{ij}$ calculated in fixed-order perturbation theory. In \eq{sigff}, the hadronic final state is treated as fully inclusive. Hence, in the presence of experimental restrictions that make a process less inclusive, \eq{sigff} is a priori not applicable. 

Factorization theorems for processes near threshold are a well-studied case where \eq{sigff} can be extended to sum large phase-space logarithms~\cite{Sterman:1986aj, Catani:1989ne, Kidonakis:1998bk, Bonciani:1998vc, Laenen:1998qw, Catani:2003zt, Idilbi:2006dg, Becher:2007ty, Chiu:2008vv}. (We discuss large logarithms in \sec{EFT_weak}.) However, threshold production requires the limit $\xi_{a,b} \to 1$. The PDFs fall off steeply in this limit, so this is not as relevant for the LHC~\cite{Campbell:2006wx}.

Our goal is to study factorization for a situation where the hard interaction occurs between partons with momentum fractions away from the limit $\xi_{a,b} \to 1$, and where the hadronic final state is measured and restricted by constraints on certain event shapes.  These restrictions allow one to probe more details about the final state and may be used experimentally to control backgrounds.

\begin{figure}[t!]
\subfigure[\hspace{1ex}3 central jets.]{%
\includegraphics[width=0.48\textwidth]{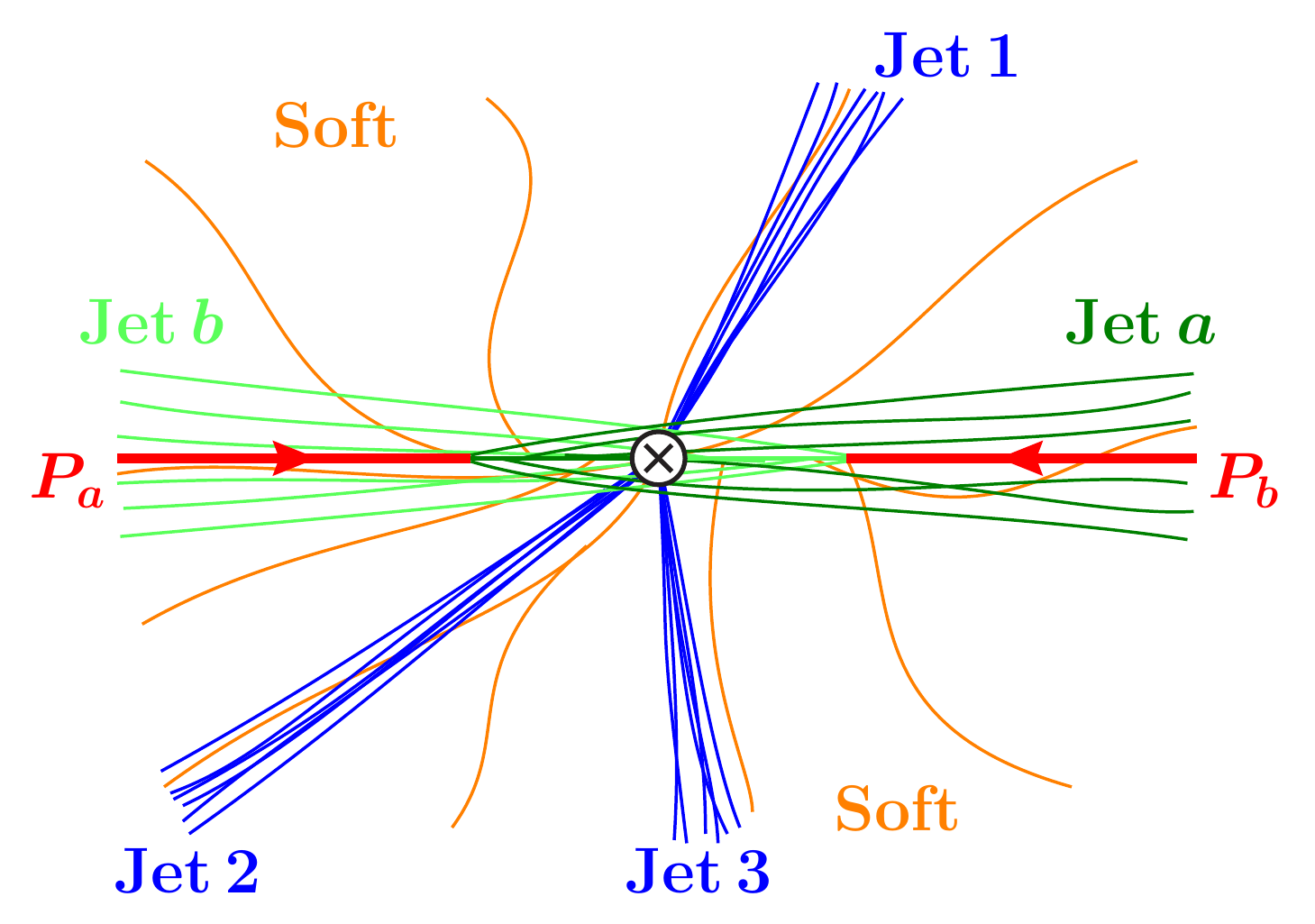}%
\label{fig:3jetLHCa}%
}\qquad
\subfigure[\hspace{1ex}$\ell^+ \ell^-$ + 0 central jets.]{%
\includegraphics[width=0.48\textwidth]{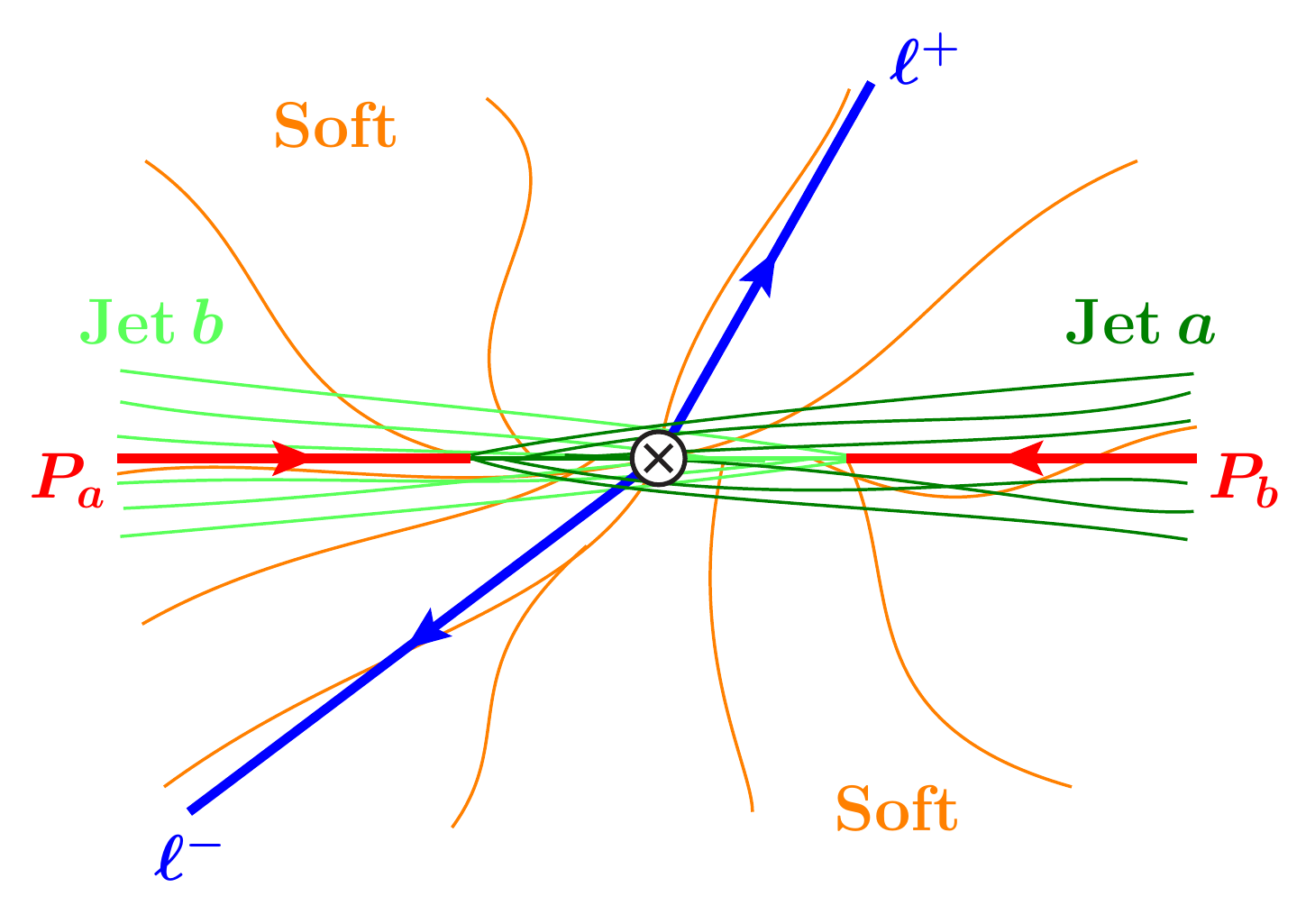}%
\label{fig:3jetLHCb}%
}
\caption[Examples of events at the LHC.]{Examples of events at the LHC. The $P_{a,b}$ denote the incoming protons. Jet $a$ and $b$ are due to initial-state radiation.}
\label{fig:3jetLHC}%
\end{figure}

A typical event at the LHC with three high-$p_T$ jets is illustrated in \fig{3jetLHCa}.  There are several complications one has to face when trying to derive a factorization theorem in this situation. First, experimentally the number and properties of the final-state jets are determined with a jet algorithm. Second, to enhance the ratio of signal over background, the experimental analyses have to apply kinematic selection cuts.  Third, in addition to the jets produced by the hard interaction, there is soft radiation everywhere (which is part of what is sometimes called the ``underlying event''). Fourth, a (large) fraction of the total energy in the final state is deposited near the beam axes at high rapidities.  Some of this radiation can contribute to measurements, and when it does, it cannot be neglected in the factorization. In this thesis we focus on the last three items. In chapter \ref{ch:njet} we take some steps towards resolving the first item. Methods for including jet algorithms in factorization have been studied in Refs.~\cite{Kidonakis:1998bk, Trott:2006bk, Bauer:2008jx}

To explore the implications of restrictions on the hadronic final state, we consider the simpler situation shown in \fig{3jetLHCb} where there are no hard central jets. This is already interesting for $h \to WW$ decaying non-hadronically, since it could help reduce the background from top quarks decaying into a $W$ plus $b$-jet. We will now construct an observable that we use in our analysis to impose this central jet veto. We define two hemispheres, $a$ and $b$, orthogonal to the beam axis and two unit lightlike vectors $n_a$ and $n_b$ along the beam axis pointing into each hemisphere. Taking the beam axis along the $z$ direction, hemisphere $a$ is defined as $z > 0$ with $n_a^\mu = (1,0,0,1)$, and hemisphere $b$ as $z < 0$ with $n_b^\mu = (1,0,0,-1)$. We now divide the total momentum $p_X$ of the hadronic final state into the contributions from particles in each hemisphere, $p_X = p_{X_a} + p_{X_b}$.  We measure the components $B_a^+$ and $B_b^+$ of the hemisphere momenta defined by
\begin{equation} \label{eq:Badef}
  B_a^+ = n_a \cdot p_{X_a}\,,
  \qquad
  B_b^+ = n_b \cdot p_{X_b} 
\,.\end{equation}
Because of the dot product with $n_a$ or $n_b$, energetic particles near the beam axes only give small contributions to $B_a^+$ or $B_b^+$.  In particular, any contributions from particles at very large rapidities outside the detector reach, including the remnant of unscattered partons in the proton, are negligible. Demanding that $B_{a,b}^+$ are small only allows highly energetic particles inside jets along the beam directions labeled ``Jet a'' and ``Jet b'' in \fig{3jetLHCb}. Hence, measuring and constraining $B_{a,b}^+$ provides a theoretically clean method to control the remaining particles in the hadronic final state. 

In this thesis, we will prove a factorization theorem for Drell-Yan $pp\to X\ell^+\ell^-$ where $X$ is allowed to have hard jets close to the beam but no hard central jets, corresponding to \fig{3jetLHCb}. We call this ``isolated Drell-Yan''.  Our proof of factorization uses Soft-Collinear Effective Theory (SCET)~\cite{Bauer:2000ew, Bauer:2000yr, Bauer:2001ct, Bauer:2001yt} plus additional arguments to rule out effects of Glauber gluons based in part on Refs.~\cite{Collins:1988ig, Aybat:2008ct}. Our factorization theorem applies to processes $pp\to XL$, were the lepton pair is replaced by other non-strongly interacting particles, such as Higgs or $Z'$ decaying non-hadronically. This is particularly interesting for $H \to W^+W^- \to \ell^+ \nu \ell^- \bar \nu$ where a central jet veto is required to remove the large background from top quarks decaying into a $W$ plus a $b$-jet. We will also write down a factorization formula with beam functions for processes with hard central jets. This is not as rigorous as the 0-jet case, in particular we have not yet shown the cancellation of Glauber gluons. 

Our main result is to show that process-independent ``beam functions'', $B_i(t,x)$ with $i=\{g,u, \bar u,d, \ldots\}$, are required to properly describe the initial state. Generically, by restricting $X$ one performs an indirect measurement of the proton prior to the hard collision.  At this point, the proton is resolved into a colliding hard parton inside a cloud of collinear and soft radiation. The proper description of this initial-state jet is given by a beam function in conjunction with a soft function describing the soft radiation in the event. 

One might worry that the collision of partons inside initial-state jets rather than partons inside protons could drastically change the physical picture. The changes are not dramatic but there are important implications. The beam functions can be computed in an operator product expansion, giving
\begin{equation} \label{eq:Bisf}
  B_i(t,\xi,\mu_B) = \delta(t) \, f_i(\xi, \mu_B) + \mathcal{O}[\alpha_s(\mu_B)]
\,,\end{equation}
where $\mu_B$ is an intermediate perturbative scale and $t$ is an invariant-mass variable closely related to the off-shellness of the colliding parton (and the Mandelstam variable $t$). Thus, the beam functions reduce to standard PDFs at leading order. For what we call the gluon beam function, this was already found in Ref.~\cite{Fleming:2006cd}, where the same matrix element of gluon fields appeared in their computation of $\gamma\,p \to J/\psi X$ using SCET. 

Equation~\eqref{eq:Bisf} implies that the momentum fractions $\xi_{a,b}$ are determined by PDFs evaluated at the scale $\mu_B\ll Q$, which is parametrically smaller than the scale $Q$ of the partonic hard interaction. The renormalization group evolution (RGE) for the initial state now proceeds in two stages. For scales $\mu < \mu_B$, the RGE is given by the standard PDF evolution~\cite{Gribov:1972ri, Georgi:1951sr, Gross:1974cs, Altarelli:1977zs, Dokshitzer:1977sg}, which redistributes the momentum fractions in the proton to lower $\xi$ values and mixes the gluon and quark PDFs. For scales $\mu > \mu_B$, the jet-like structure of the initial state becomes relevant and its evolution is properly described by the RGE of the beam function. In contrast to the PDF, the evolution of the beam function does not involve mixing between quarks and gluons and only changes $t$. In addition to the change in evolution, the transition from PDFs to beam functions at the scale $\mu_B$ also involves explicit $\alpha_s(\mu_B)$ corrections as indicated in \eq{Bisf}. These include mixing effects, such as a gluon from the proton pair-producing a quark that goes on to initiate the hard interaction and an antiquark that is radiated into the final state.  For our observables such fluctuations are not fully accounted for by the PDF evolution. These beam effects must be taken into account, which can be done by perturbative calculations. Fortunately, the standard PDFs are still sufficient to describe the nonperturbative information required for the initial state.

One should ask whether the description of the initial state by beam functions, as well as their interplay with the soft radiation, are properly captured by current Monte Carlo programs used to simulate events at the LHC and Tevatron, such as Pythia~\cite{Sjostrand:2006za, Sjostrand:2007gs} and Herwig~\cite{Corcella:2000bw, Bahr:2008pv}. In these event generators the corresponding effects should be described at leading order by the initial-state parton shower in conjunction with models for the underlying event~\cite{Sjostrand:2004pf, Sjostrand:2004ef, Butterworth:1996zw, Bahr:2008dy}. We will see that the initial-state parton shower is in fact closer to factorization with beam functions than to the inclusive factorization formula in \eq{sigff}. In particular, the physical picture of off-shell partons that arises from the factorization with beam functions has a nice correspondence with the picture adopted for initial-state parton showers a long time ago~\cite{Sjostrand:1985xi,   Bengtsson:1986gz}. Experimentally, measurements of the isolated Drell-Yan cross section provide a simple obsevable that can rigorously test the accuracy of the initial-state shower in Monte Carlo programs, by contrasting it with our analytic results.

\section{Weak Interactions and Effective Field Theory}
\label{sec:EFT_weak}

Many physical processes involve several scales and perturbative calculations typically lead to logarithms of the ratios of these scales, e.g.~$\al_s \ln \mu_1/\mu_2$. If the scales are widely separated $\mu_1 \gg \mu_2$ the logarithms become large and perturbation theory breaks down. We will see in this section how effective field theories conveniently solve this problem by factoring a process into separate pieces, each corresponding to only one of the scales. For an introduction to effective field theories see for example Refs.~\cite{Georgi:1994qn, Manohar:1996cq, Manohar:2000dt, Rothstein:2003mp}.

\begin{figure}
\centering
\subfigure[]{\includegraphics[scale=0.75]{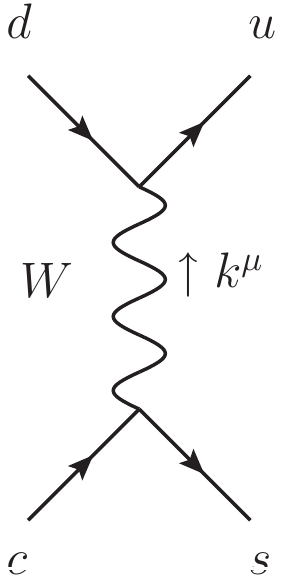}\label{fig:Wtree_a}}%
\qquad
\subfigure[]{\includegraphics[scale=0.75]{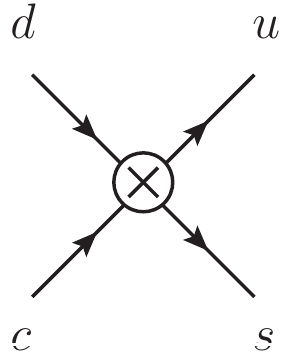}\label{fig:Wtree_b}}%
\caption[The tree-level process for $c \to s u \bar d$ in the full and effective theory.]{The tree-level process in the full theory (a) and effective theory (b).}
\label{fig:Wtree}
\end{figure}

As an example we will study the decay of a $D$ meson to a $K^-$ and a $\pi^+$ meson. A review of effective field theories for weak decays can be found in Refs.~\cite{Buchalla:1995vs,Buras:1998raa}. At the parton level we can understand this process as $c \to s u \bar d$ along with a spectator $\bar u$ quark, which is mediated by a $W$ boson of the weak force. The tree level diagram is shown in \fig{Wtree_a} and given by
\begin{equation} \label{eq:tree_decay}
  \Big(\frac{\img g_2}{\sqrt{2}}\Big)^2 V_{cs}^* V_{ud}\, 
  \frac{-\img}{k^2 - m_W^2} \Big(g^{\mu\nu} - \frac{k^\mu k^\nu}{m_W^2} \Big) 
  \bar u_s \ga_\mu P_L u_c\, \bar u_u \ga_\nu P_L u_d
  \,.
\end{equation}
Here the $g_2$ is the coupling of the weak interactions, $V_{ij}$ are elements of the CKM matrix from the weak interaction and $u_i$ is the spinor for an external quark of flavor $i$. The quarks are confined to the mesons, implying that the momentum $k^\mu \ll m_W$. We can therefore expand the propagator as
\begin{equation} \label{eq:W_exp}
  \frac{-\img}{k^2 - m_W^2} \Big(g^{\mu\nu} - \frac{k^\mu k^\nu}{m_W^2} \Big)
  = \frac{\img}{m_W^2} \Big[1+ {\cal O}\Big(\frac{k^2}{m_W^2}\Big)\Big]
  \,.
\end{equation}
This removes the $W$ boson as a dynamical degree of freedom and replaces the tree-level process in \fig{Wtree_a} by the four-fermion interaction in \fig{Wtree_b}. The $W$ propagator gets shrunk to a point because we no longer resolve the short distance physics. 

The dominant corrections to this tree-level diagram come from the strong force, which we now consider. The systematic approach to include these corrections in the Lagrangian of the effective theory is called ``matching" and is based on the operator product expansion (OPE) \cite{Wilson:OPE}. First we write down all interactions in the low energy theory that are allowed by symmetries. The matching will occur at a scale $\mu \sim m_W$, so we can treat the $c$, $s$, $d$ and $u$ quarks as massless and use the corresponding chiral symmetry to constrain possible terms. Using Fierz identities for spin and color, we are left with just two terms in our effective Lagrangian
\begin{align}
  \cL_W^\text{eff} &= - \frac{4G_F}{\sqrt{2}} V_{cs}^* V_{ud}\, [C_1(\mu)\op_1(\mu) + C_2(\mu)\op_2(\mu)] 
  \,,\nn \\
  \op_1 &= \bar s_j \ga^\mu P_L c_j \, \bar u_k \ga_\mu P_L d_k 
  \,, \quad
  \op_2 = \bar s_j \ga^\mu P_L c_k \, \bar u_k \ga_\mu P_L d_j
  \,,
\end{align}
where $G_F = \sqrt{2} g_2^2/(8 m_W^2)$ is the Fermi constant and $j$ and $k$ are color indices. Since we are working at leading order in $k^\mu/m_W$, we do not include terms with derivatives that one would get from higher order terms of the expansion in \eq{W_exp}. To determine the coefficients $C_i(\mu)$ we calculate some physical quantity in both the full theory and the effective theory and match. A common choice is renormalized matrix elements, where we are free to choose any on-shell or off-shell states since the effective theory should completely reproduce the low energy limit of the full theory. From our tree-level calculation in \eqs{tree_decay}{W_exp} we know that $C_1^\zero(\mu) =1$ and $C_2^\zero(\mu) = 0$ (the superscript $(n)$ denotes terms of order $\al_s^n$). 

\begin{figure}
\centering
\subfigure[]{\includegraphics[scale=0.75]{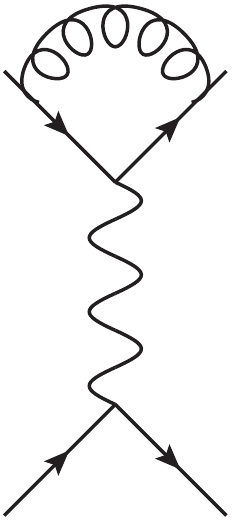}\label{fig:Wone_a}}%
\quad
\subfigure[]{\includegraphics[scale=0.75]{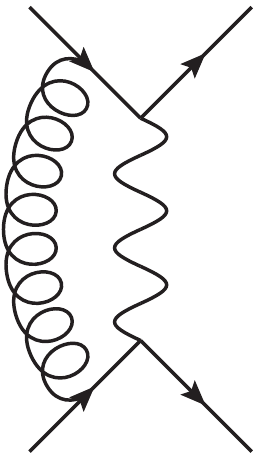}\label{fig:Wone_b}}%
\quad
\subfigure[]{\includegraphics[scale=0.75]{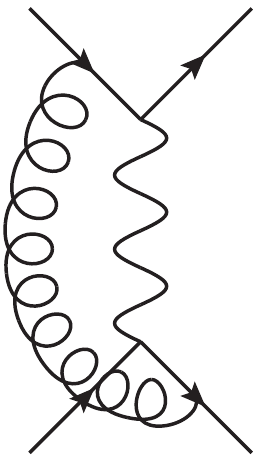}\label{fig:Wone_c}}%
\qquad
\subfigure[]{\includegraphics[scale=0.75]{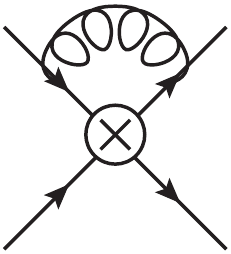}\label{fig:Wone_d}}%
\quad
\subfigure[]{\includegraphics[scale=0.75]{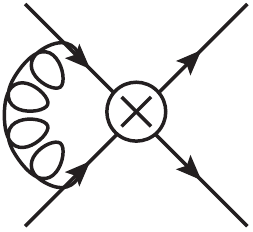}\label{fig:Wone_e}}%
\quad
\subfigure[]{\includegraphics[scale=0.75]{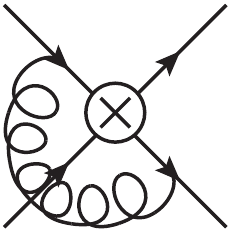}\label{fig:Wone_f}}%
\caption[The one-loop QCD corrections to $c \to s u \bar d$ in the full and effective theory.]{The one-loop QCD corrections in the full theory (a)-(c) and effective theory (d)-(f). These diagrams have symmetric counterparts.}
\label{fig:Wone}
\end{figure}

Moving on to the $\al_s$ corrections, the one-loop diagrams that contribute to the matrix element with external free quarks are shown in \fig{Wone} for the full theory and for the effective theory. The ultraviolet(UV) divergences we encounter in these calculations are regularized using dimensional regularization (DR) and renormalized in the modified minimal subtraction scheme ($\overline{\text{MS}}$). Performing the matching onto the effective theory at order $\al_s$
\begin{align} 
  \Mae{s u \bar d}{\img \cL_W^\text{full}}{c}^\one &= 
  \Mae{s u \bar d}{\img \cL_W^\text{eff}}{c}^\one 
  \nn \\
  &= -\img \frac{4G_F}{\sqrt{2}} V_{cs}^* V_{ud} 
  \sum_{i=1,2} C_i(\mu)^\zero \Mae{s u \bar d}{\op_i}{c}^\one +   
  C_i(\mu)^\one \Mae{s u \bar d}{\op_i}{c}^\zero
  \,,
\end{align}
results in
\begin{align} \label{eq:W_match}
  C_1(\mu) &= 1 + \frac{3}{N_c} \frac{\al_s(\mu)}{4\pi} \ln \frac{m_W^2}{\mu^2}
  \,, \nn \\
  C_2(\mu) &= -3\, \frac{\al_s(\mu)}{4\pi} \ln \frac{m_W^2}{\mu^2}
  \,.
\end{align}
Here $N_c = 3$ is the number of colors in QCD. It is worth pointing out that the full theory and effective theory results depend on the momenta $p^\mu$ of the external quarks (which we took to all be equal), but that this dependence cancels in the matching. This is an explicit check that effective theory reproduces the full theory in the low energy limit. 

We now take a closer look at the full theory calculation. Schematically it has the following structure
\begin{align} \label{eq:W_logs}
  \Mae{s u \bar d}{\img \cL_W^\text{full}}{c} & = 1 + \al_s L + \al_s^2 L^2 + \al_s^3 L^3 + \dots
  \nn \\ & \hspace{9ex}
  \al_s + \al_s^2 L \,+\, \al_s^3 L^2 + \dots
  \nn \\ & \hspace{16.2ex}
  \al_s^2 + \al_s^3 L \,+\, \dots\,,
\end{align}
where 1 denotes the tree level result and $L = \ln (-m_W^2/p^2)$. For our $D$ meson decay the momentum $p^\mu \ll m_W$, which implies that $L$ is large and logarithmic terms such as $\al_s L$ are more important than the finite (non-logarithmic) $\al_s$ corrections. Assuming that $L \sim 1/\al_s$, the terms on the first row would all be of $\ord{1}$ and thus need to be summed to get the correct treelevel amplitude. This is called leading logarithmic (LL) resummation. The finite $\al_s$ correction (first term on the second row) is of the same order as the logarithmic terms $\al_s^{n+1} L^n$ on the second row. Summing these terms corresponds to next-to-leading logarithmic (NLL) resummation. We will see how this resummation can be achieved in the effective field theory.

Large logarithms are a generic feature of theories with disparate scales. Our effective field theory approach allows us to separate these scales and the corresponding physics:
{\allowdisplaybreaks[0]
\begin{align}
   \Mae{s u \bar d}{\img \cL_W^\text{full}}{c} & = 1 + \al_s \ln \frac{m_W^2}{-p^2} + \dots 
   \nn \\
   &= \underbrace{\Big(1 + \al_s \ln \frac{m_W^2}{\mu^2} + \dots\Big)}_{C_i(\mu)}
   \underbrace{\Big(1 + \al_s \ln \frac{\mu^2}{-p^2} + \dots\Big)}_{\Mae{c \bar u d}{\op_i(\mu)}{b}}
   = \Mae{s u \bar d}{\img \cL_W^\text{eff}}{c}
   \,.
\end{align}}
The physics associated with $m_W$ can only enter in the Wilson coefficients. The low energy physics remains as dynamical degrees of freedom in the effective theory and drops out in the matching. The large logarithms can now be avoided by evaluating the matching coefficients in \eq{W_match} at a scale $\mu \sim m_W$ and the matrix elements at a scale $\mu \sim \sqrt{p^2}$. 

We can use the renormalization group evolution (RGE) \cite{GellMann:1954fq, Symanzik:1970rt, Callan:1970yg, tHooft:1973mm, Weinberg:1951ss} to run the Wilson coefficients $C_i(\mu)$ from a scale $\mu \sim m_W$ down to $\mu \sim \sqrt{p^2}$. The renormalization scale $\mu$ can roughly be interpreted as the scale at which the UV divergences are cut off. Lowering $\mu$ thus corresponds to reducing the degrees of freedom in the effective theory and absorbing the corresponding change in the Wilson coefficients $C_i$. With this picture in mind it makes sense that we should match at $\mu \sim m_W$ and then run down to $\mu \sim \sqrt{p^2}$. The full theory calculation contained no UV divergences in the final answer, because the mass of the W acted as a UV cutoff. In our effective theory we are expanding about the limit $m_W \to \infty$ and hence $\ln -m_W^2/p^2$ leads to UV divergences, which is what allows us to factor and resum these large logarithms.

\begin{table}
  \centering
  \begin{tabular}{l | c | c c}
  \hline \hline
  & matching & \multicolumn{2}{c}{running} \\
  & \ & anom. dim. & $\beta(\al_s)$ \\ \hline
  LO & $0$-loop & - & - \\
  NLO & $1$-loop & - & - \\
  NNLO & $2$-loop & - & - \\  
  LL & $0$-loop & $1$-loop & $1$-loop\\
  NLL & $1$-loop & $2$-loop & $2$-loop\\
  NNLL & $2$-loop & $3$-loop & $3$-loop\\  
  \hline\hline
  \end{tabular}
\caption[Order counting in perturbation theory for single logarithmic resummation.]{Order counting for fixed and resummed perturbation theory with single logarithms.}
\label{tab:counting_single}
\end{table}

We will now see explicitly how the RGE resums the large logarithms in our example. The UV divergences from the one-loop effective theory diagrams in \fig{Wone}(d)-(f) lead to the following RGE
\begin{equation}
   \mu \frac{\df}{\df\mu}\, C_i(\mu) = \ga_{ij}(\mu) C_j(\mu)
   \,, 
   \qquad \ga_{ij}(\mu) = \frac{\al_s(\mu)}{2\pi}
   \left(\begin{array}{cc}
   3/N_c & -3 \\
   -3 & 3/N_c
   \end{array}\right)
   \,.
\end{equation}
This can be solved by diagonalizing
\begin{equation}
   C_\pm = C_1 \pm C_2
   \,, \quad
   C_\pm(\mu) = C_\pm(m_W) \bigg[\frac{\al_s(m_W)}{\al_s(\mu)}\bigg]^{\ga_\pm / (2\bt_0)}
   \,, \quad
   \ga_\pm = \pm 6\, \frac{N_c \mp 1}{N_c}
   \,,
\end{equation}
where $\bt_0 = (11 C_A - 4 T_F n_f)/3$ is the lowest order coefficient in the running of $\al_s$. As a consequence, even though $C_2(m_W) = 0$ at tree level a non-zero $C_2(\mu)$ gets generated through the RGE. By expanding we explicitly see how the RGE sums the LL series
\begin{align} \label{eq:W_resum}
  \bigg[\frac{\al_s(m_W)}{\al_s(\mu)}\bigg]^{\ga_\pm/(2\bt_0)} 
  = \exp\bigg[-\ga_\pm\, \frac{\al_s(\mu)}{4\pi} \ln \frac{m_W}{\mu}\bigg]
  \sim 1 + \al_s L + \al_s^2 L^2 + \dots
   \,.
\end{align}
Resummed perturbation theory is powerful because it captures the most important pieces of higher loop diagrams, without having to calculate them. The above resummed result contains for example $\al_s^2 L^2$ which in a full theory calculation would only show up at two loops. The appropriate resummation when we match at order $\al_s$ is NLL resummation, for which the running is determined from the UV divergences of two-loop diagrams. The running is generally at one higher loop that the matching, as is shown in Table~\ref{tab:counting_single}.

After matching and running, the final step to calculate the $D \to K^+ \pi^-$ decay consists of determining the matrix elements $\Mae{K^+\pi^-}{\op_i(\mu)}{D}$ at $\mu \sim \sqrt{p^2}$. For some electroweak decays Heavy Quark Effective Theory (HQET) \cite{Eichten:1989zv, Georgi:1990um, Grinstein:1990mj} allows you to exploit some further symmetries, but at the end of the day this is a non-perturbative matrix element that needs to be determined using either lattice QCD, modeling or experiments. The good news is that in effective field theories the non-perturbative matrix elements are often universal: they can be extracted from some experiment and used in a different one. The PDFs that we encountered in the previous section are an example of this.

The weak decay we have studied captures the generic situation in effective field theories: a large separation of scales allows us to expand onto an effective theory but also leads to large logarithms. These large logarithms are factored in the effective theory. The Wilson coefficients $C_i(\mu)$ contain the effects of physics at the high scale and the operators $\op_i(\mu)$ describe the dynamics of the low energy physics. The renormalization scale $\mu$ can be thought of as the scale that separates what goes into $C_i(\mu)$ and what is described by $\op_i(\mu)$. The large logarithms are resummed by calculating $C_i(\mu)$ at the high scale (``matching") and using the RGE to run it down to the low scale (``running"). The relevant matrix elements of the operator $\op_i(\mu)$ at the low scale may be non-perturbative, in which case they should be extracted from experiment or may be determined using lattice QCD.

We conclude this section with a somewhat technical point related to our choice of renormalization scheme. For this we will look at the running of the electromagnetic coupling in $\overline{\text{MS}}$ and the off-shell momentum subtraction (ms) scheme. The one-loop contribution from the electron with mass $m$ leads to
\begin{align} \label{eq:QED_beta}
  &\mu \frac{\df}{\df\mu}\, e(\mu) = \bt(\mu) e(\mu)
  \,, \nn \\
  &\beta^{\overline{\text{MS}}}(\mu) = \frac{e^3(\mu)}{12\pi^2}
  \,, \quad
  \beta^{\text{ms}}(\mu) = \frac{e^3(\mu)}{2\pi^2} 
  \int_0^1\! \df x\, \frac{x^2(1-x)^2 \mu^2}{m^2 + x(1-x) \mu^2}
  \,.
\end{align}
For $\mu \gg m$ we have $\beta^{\text{ms}}(\mu) = \beta^{\overline{\text{MS}}}(\mu)$, but
\begin{equation}
  \beta^{\text{ms}}(\mu) = \frac{e^3(\mu)}{60\pi^2} \frac{\mu^2}{m^2} \qquad \text{for } \mu \ll m\,.
\end{equation}
So below the electron mass the effect of the electron decouples from the running, with a smooth transition through $\mu = m$. By contrast, the electron doesn't decouple in the $\overline{\text{MS}}$ scheme ($\beta^{\overline{\text{MS}}}$ doesn't depend on $m$ at all). As this example illustrates, adopting a so-called physical renormalization scheme causes high energy degrees of freedom to decouple naturally. However, the expressions become much more involved (compare $\beta^{\text{ms}}$ with $\beta^{\overline{\text{MS}}}$), which is why we prefer to use the $\overline{\text{MS}}$ scheme and manually perform the decoupling by matching and running.

\section{Soft-Collinear Effective Theory}
\label{sec:SCET_intro}

In a typical event at the LHC the final state contains photons, leptons and hadrons (see \fig{3jetLHCa}). The energetic hadrons generically come in collimated jets and there are also much less energetic hadrons everywhere. These two types are referred to as ``collinear" and ``soft" radiation and are enhanced due to the presence of collinear and soft infrared(IR) singularities in QCD. SCET is an effective theory of QCD describing these modes~\cite{Bauer:2000ew, Bauer:2000yr, Bauer:2001ct, Bauer:2001yt}. In this section we present an intuitive introduction to SCET. We start by introducing light-cone coordinates and describing the degrees of freedom in SCET. We will discuss the matching onto SCET, reparametrization invariance (RPI), the factorization of soft and collinear degrees of freedom and the resummation of large logarithms.

It is convenient to use light-cone coordinates to describe jets of energetic hadrons. Choose a light-cone vector $n^\mu$ vector pointing in the direction of one of the energetic jets and an additional light-cone vector $\bn^\mu$ such that $n \cdot \bn = 2$. For $n^\mu = (1,\vec n)$ a common choice is $\bn^\mu = (1, -\vec n)$, where of course $\vec n^2 = 1$. We can then decompose any four-vector $p^\mu = (p^+,p^-,p_\perp^\mu)$ as
\begin{equation}
  p^\mu = p^+ \frac{\bn^\mu}{2} + p^- \frac{n^\mu}{2} + p_\perp^\mu\,, \qquad
  p^+ =  n \sdt p\,, \qquad
  p^- = \bn \sdt p\,.
\end{equation}
The momenta of energetic hadrons in the jet along the $n^\mu$ direction are parametrically
\begin{equation} \label{eq:coll_mom}
 p^\mu = (p^+,p^-,p_\perp^\mu) \sim Q (\la^2,1,\la) \qquad \text{(collinear)}
 \,,
\end{equation}
where $Q$ is the scale of the hard collision and $\la \sim p_\perp/p^- \ll 1$ characterizes the size of the jet. The power counting in \eq{coll_mom} corresponds to a particle with virtuality $p^2 \sim \la^2 Q^2$ that is boosted in the $n^\mu$ direction (we consider light particles whose mass we can ignore).

\begin{figure}[t]
\centering
\includegraphics[scale=0.75]{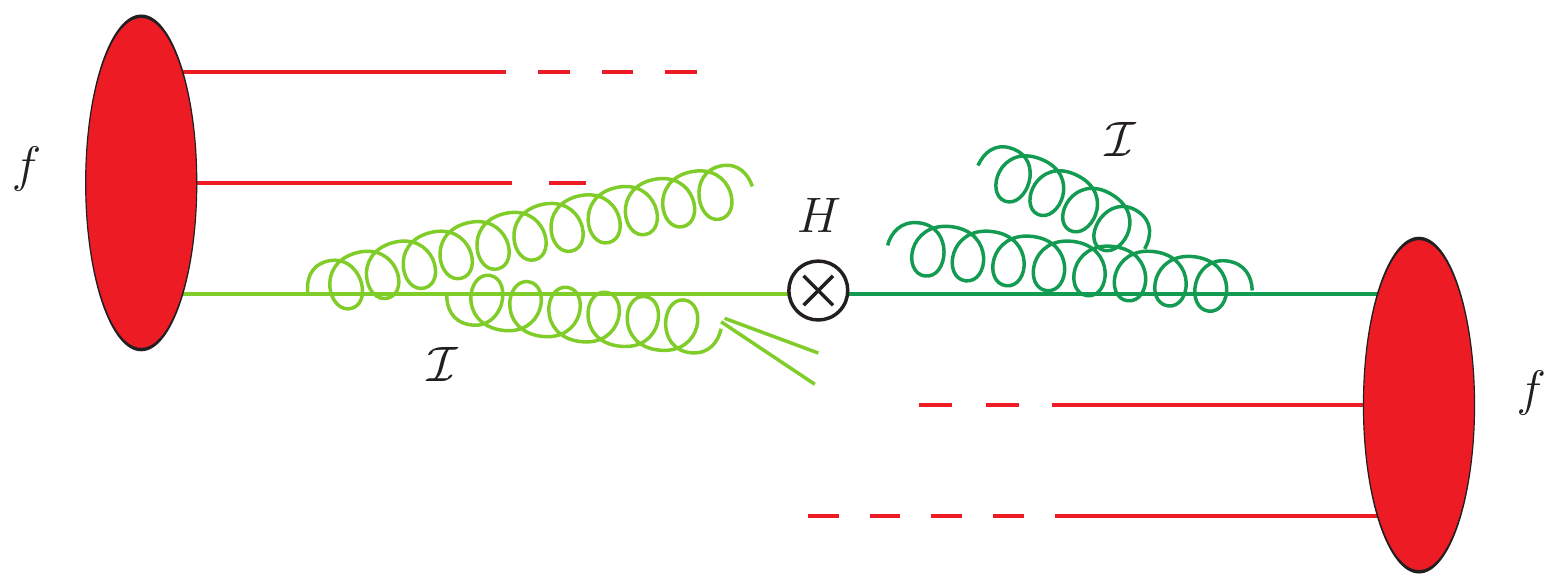}
\caption[Schematic diagram of a proton-(anti)proton collision up to the hard interaction.]{Schematic diagram of a proton-(anti)proton collision up to the hard interaction. A quark or gluon is taken out of each of the protons $f$ and propagates while emitting initial-state radiation $\cI$ until it enters the hard interaction $H$. The two shades of green correspond to the two collinear modes. For our observables the proton remnant is much more collinear and goes straight down the beam pipe.}
\label{fig:coll}
\end{figure}

For the isolated Drell-Yan process shown in \fig{3jetLHCb}, we have two jets from initial-state radiation corresponding to the directions $n_a^\mu = (1,0,0,1)$ and $n_b^\mu = (1,0,0,-1)$. We thus have two collinear modes
\begin{align} \label{eq:na_nb_coll}
 p_a^\mu = (p_a^+,p_a^-,p_{a\perp}^\mu) \sim Q (\la^2,1,\la) \qquad \text{(}n_a\text{ collinear)}
 \,, \nn \\
 p_b^\mu = (p_b^-,p_b^+,p_{b\perp}^\mu) \sim Q (1,\la^2,\la) \qquad \text{(}n_b\text{ collinear)}
 \,,
\end{align}
where we take $n_b = \bn_a$ and $n_a = \bn_b$ for convenience. 
By convention the $p_{a,b}^-$ will always correspond to the large light-cone component.
Each of the two initial-state jets originates from a single quark or gluon coming out of the proton, which propagates and radiates until it enters the hard interaction, as shown schematically in \fig{coll}. 

\begin{figure}[t]
\centering
\includegraphics[scale=0.75]{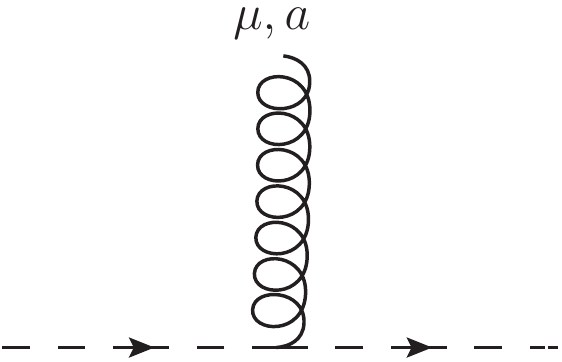}
\quad
\parbox{13ex}{
\vspace{-15ex}
$ \displaystyle = \ \img g \frac{\nslash}{2}\, n^\mu T^a$}
\caption[Feynman rule for the interaction between a collinear quark and a soft gluon.]{Feynman rule for the interaction between an n-collinear quark (dashed line) and a soft gluon, at leading order in the power counting.}
\label{fig:softcoll}
\end{figure}

Since the final state hadrons are colorless, there must be some degrees of freedom connecting the jets that carry color charge. The collinear modes cannot interact directly because that would lead to a momentum
\begin{align}
   p_a^\mu + p_b^\mu \sim Q (\la^2,1,\la) + Q (1,\la^2,\la) = Q(1,1,\la)
   \,, \qquad
   (p_a + p_b)^2 \sim Q^2
   \,.
\end{align}
In SCET we have integrated out the hard interaction and so the corresponding modes with virtuality $Q^2$ are no longer dynamical degrees of freedom. The interactions between the collinear modes is provided by additional ``soft" degrees of freedom which have the following power counting\footnote{In the literature these are sometimes referred to as ultrasoft modes, where soft is reserved for modes with power counting $(p^+,p^-,p_\perp^\mu) \sim Q (\la,\la,\la)$. As discussed below \eq{altmodes}, we do not need these modes.}
\begin{equation} \label{eq:soft_mom}
  p_s^\mu = (p_s^+,p_s^-,p_{s\perp}^\mu) \sim Q (\la^2,\la^2,\la^2) \qquad \text{(soft)}
  \,.
\end{equation}
They describe much less energetic radiation $E \sim \la^2 Q$ without a preferred direction and correspond to long range fluctuations since their virtuality $p^2 \sim Q \la^4$ is smaller. At a more technical level we expand the collinear momenta $p^\mu = \tilde p^\mu + p_r^\mu$ into a discrete large momentum $\tilde p^\mu = (0,\tilde p^-,\tilde p_\perp^\mu) \sim Q (0,1,\la)$ and a small residual momentum $p_r^\mu \sim Q \la^2$ that can be exchanged with the soft radiation. The interaction between a soft and a collinear mode at leading order in the power counting is shown in \fig{softcoll}.

SCET differs in several ways from effective theories where a heavy particle has been integrated out (such as the $W$ boson in the previous section). The expansion parameter $\la$ is not set by the mass of a heavy particle and there are different modes corresponding to the same particles, e.g.~we have both collinear gluons and soft gluons. We thus have to be careful not to double count degrees of freedom, which can be removed through so-called zero-bin subtractions~\cite{Manohar:2006nz}.

One might wonder if there are any other degrees of freedom that we should include in our effective theory. The answer to this question depends on the observable you want to study. For the weak decay in the previous section the small external momenta told us that we did not need to treat the $W$ boson as a dynamical degree of freedom in our effective theory. For the isolated Drell-Yan process the invariant mass of the leptons sets the scale of the hard interaction $Q$. Measuring the $B_a^+$ and $B_b^+$ variables introduced in \eq{Badef} and restricting $B_{a,b}^+ \ll Q$ implies that energetic radiation is restricted to be along the beam axis. This is what leads to the two collinear modes in \eq{na_nb_coll} with $\la^2 \sim B_{a,b}^+/Q$. The contribution to $B_a^+$ and $B_b^+$ from the soft radiation is of the same parametric size as that of the collinear radiation, so we need to include the soft degrees of freedom.

In principle you could imagine including modes with a scaling such as
\begin{align}\label{eq:altmodes}
(a) & \qquad p \sim Q (\la,\la,\la)
\,,\nn \\
(b) & \qquad p \sim Q (\la^4,\la^4,\la^4)
\,,\nn \\
(c) & \qquad p \sim Q (\la^4,1,\la^2)
\,,\nn \\
(d) & \qquad p \sim Q (\la^2,\la^2,\la)
\,.
\end{align}
In (a) we are considering a mode with virtuality $p^2 \sim \la^2 Q^2$ that, like the soft radiation, has no preferred direction. This mode would lead to contributions of order $\la Q$ to $B_{a,b}^+$. However, our measurement demands $B_{a,b}^+ \sim \la^2 Q$ and so (a) is ruled out in final state radiation. We now address virtual contributions from (a), by arguing that these modes cannot couple to the collinear and soft degrees of freedom. Attempting to couple (a) to collinear modes leads to a virtuality $p^2 \sim \la Q^2$, which is too large and has been integrated out. The momentum components of (a) are all larger than those of the soft modes in \eq{soft_mom}, and so these short range fluctuations are not resolved by our soft modes. There is no coupling between (a) and the soft degrees of freedom in the leading order Lagrangian.

The degrees of freedom in (b) correspond to a soft mode with virtuality $p^2 \sim \la^8 Q^2$ and is already contained in our soft degrees of freedom. Alternatively, the contribution of (b) to $B_a^+$ and $B_b^+$ is power suppressed, so there is no need to separately include them. 
The same is true for (c), which is a collinear mode with a smaller virtuality than in \eq{coll_mom} and is therefore already contained in our existing collinear degrees of freedom. 

\begin{figure}[t]
\centering
\includegraphics[scale=0.75]{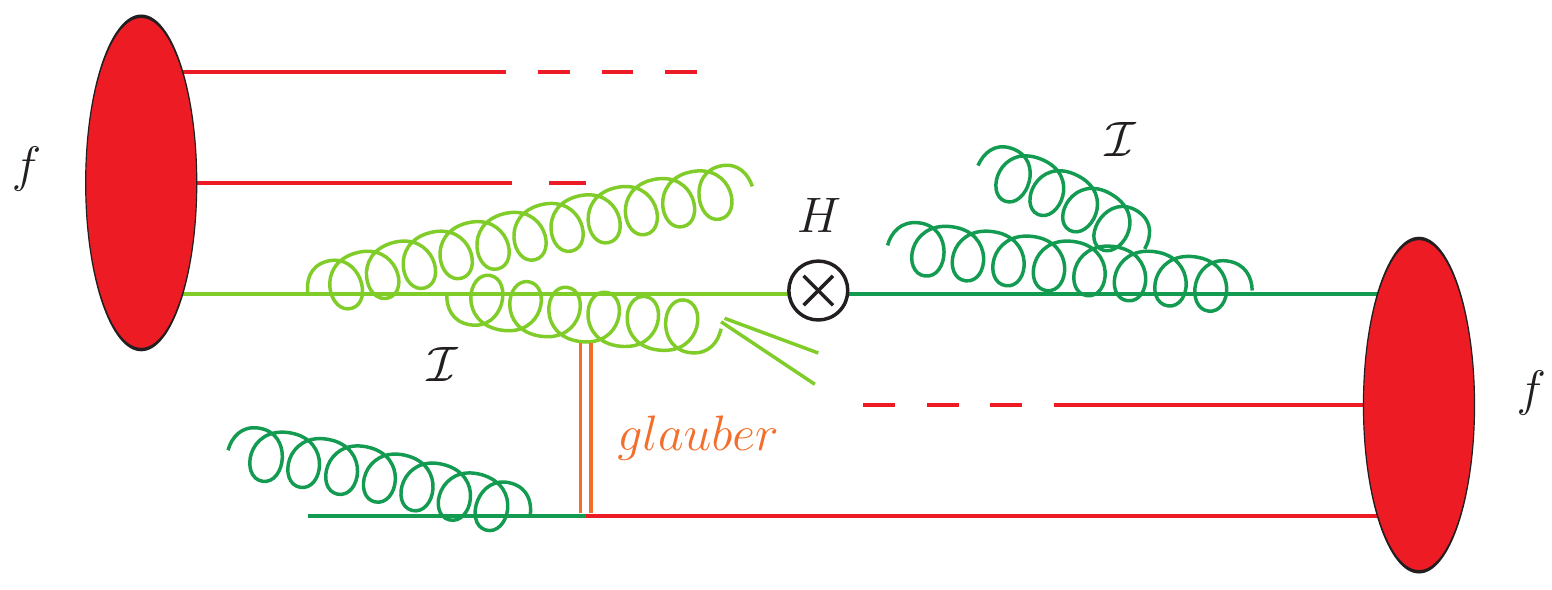}
\caption[Schematic diagram of a proton-(anti)proton collision including Glauber gluons.]{Schematic diagram of a proton-(anti)proton collision up to the hard interaction. Glauber gluons (double line) can couple an active parton with a spectator from the other proton. The spectator was very collinear and would have gone straight down the beam pipe but the Glauber gluon kicks it into active region, where it contributes to $B_{a,b}^+$.}
\label{fig:collglauber}
\end{figure}

Gluons with a scaling as in (d) are called Glauber gluons. They cannot enter in the final state since they are offshell modes, but they can mediate perpendicular momentum transfer between the two different collinear sectors in \eq{na_nb_coll}. In particular they could connect the active parton from one proton with a quark or gluon in the remnant of the other proton. As a consequence the proton remnant no longer goes straight down the beam pipe and contributes to $B_{a,b}^+$, see \fig{collglauber}. We will discuss below how to factorize the soft and the various collinear degrees of freedom. If Glauber gluons need to be taken into account, this factorization breaks down. In \subsec{glauber} we show that the contribution from Glauber gluons cancels out for our factorization theorem. 

To summarize, identifying the relevant modes is really the art of effective field theories. Including any unnecessary degrees of freedom is permitted and will not affect the outcome, as long as one avoids double counting by appropriate zero-bin subtractions. However, it will dramatically increase the computational effort required.

\begin{figure}
\centering
\subfigure[]{\includegraphics[scale=0.6]{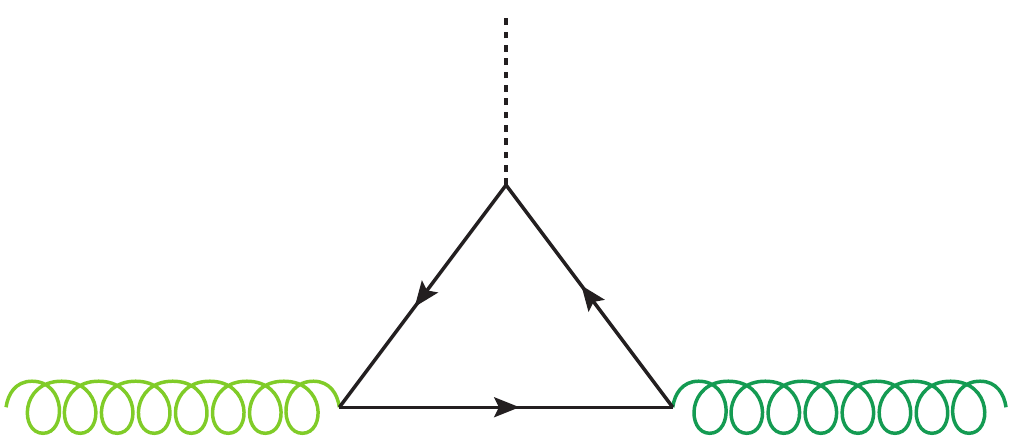}\label{fig:hardmatch_a}}%
\quad
\subfigure[]{\includegraphics[scale=0.6]{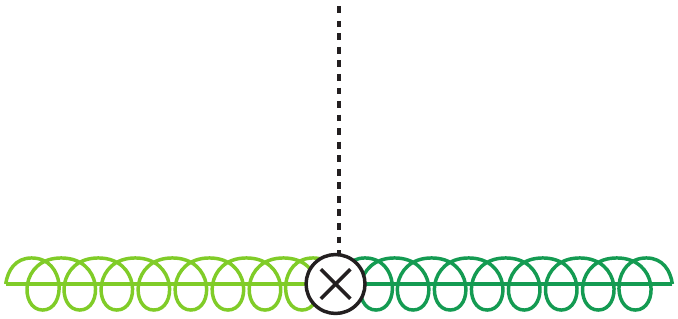}\label{fig:hardmatch_b}}%
\quad
\subfigure[]{\includegraphics[scale=0.6]{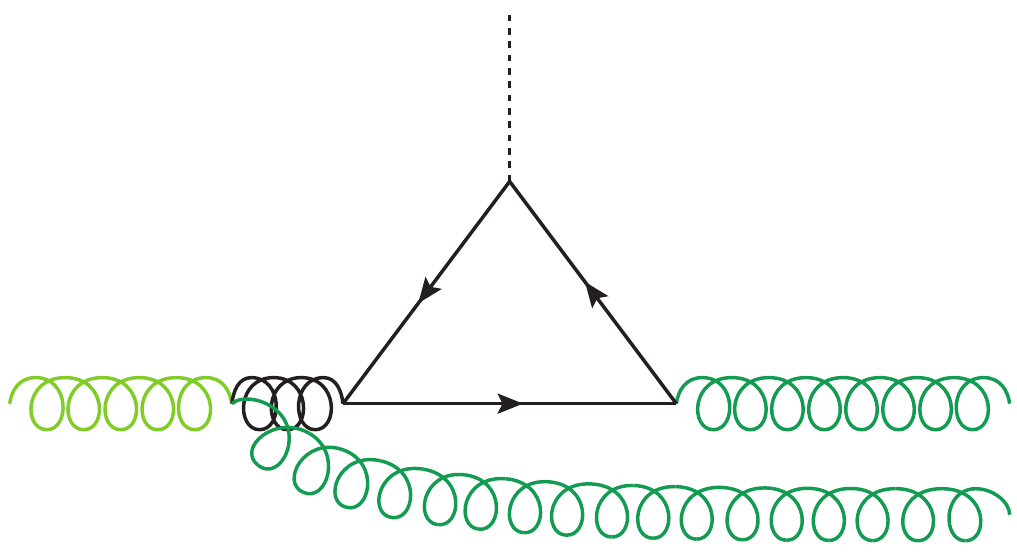}\label{fig:hardmatch_c}}%
\qquad
\subfigure[]{\includegraphics[scale=0.6]{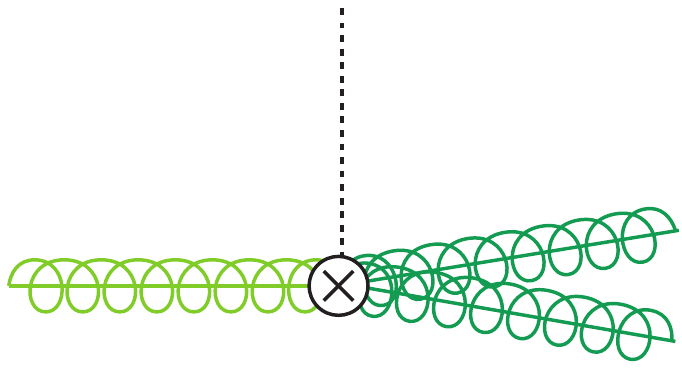}\label{fig:hardmatch_d}}%
\caption[The matching for Higgs production through gluon fusion.]{The matching for Higgs production through gluon fusion. The full theory diagram (a) gets matched onto the effective interaction (b). Springs with a line through them are collinear gluons. The different shades of green correspond to the two collinear modes and the solid black lines with the hard degrees of freedom that get integrated out. In (c) we consider the emission of an additional $\bn \sdt A$ which is at the same order in the power counting and gets matched onto (d). Including arbitrary many emissions of $\bn \sdt A$ gluons generates a collinear Wilson line $W_n$ in SCET.}
\label{fig:hardmatch}
\end{figure}

We now discuss the matching onto SCET in which we integrate out the hard interaction~\cite{Bauer:2002nz}. In the previous section our effective Lagrangian only contained two terms $\cL_W^\text{eff} = \sum_{i=1,2} C_i \op_i$, because the weak interaction is insensitive to the details of our external states. In SCET the collinear degrees of freedom have large $p^- \sim Q$ components, which enter in the hard interaction. Matching onto SCET thus yields Wilson coefficients and operators that depend on both the direction $n^\mu$ and size $\lp^-$ of these large components,
\begin{equation}
  \cL^\text{eff} = \sum_i \sum_{n_a,n_b} \int \df \lp_a^-  \int \df \lp_b^- 
  C_i(n_a,\lp_a^-,n_b,\lp_b^-) \op_i(n_a,\lp_a^-,n_b,\lp_b^-)
  \,.
\end{equation}
As an example we show the matching for Higgs production through gluon fusion in \fig{hardmatch}. The collinear directions are fixed by the directions of the protons and $\lp^-_{a,b}$ are related to the virtuality and rapidity of the Higgs by momentum conservation.

Collinear gluons have the scaling $A_n^\mu = (A_n^+,A_n^-,A_{n\perp}^\mu) \sim (\la^2,1,\la)$, which e.g.~follows from the homogenous power counting of the covariant derivative. This implies that we can add arbitrary many collinear $\bn \cdot A_n$ fields in the matching at the same order in the power counting, see \fig{hardmatch}(c) and (d). Fortunately gauge invariance restricts these operators, organizing these gluons into collinear Wilson lines $W_n$ \cite{Bauer:2001ct}. 

There was some arbitrariness in our choice of $n^\mu$ and $\bn^\mu$, since the light-cone vector $n^\mu$ only needs to point roughly in the direction of the jet and we can choose any $\bn^\mu$ that satisfies $n \cdot \bn = 2$. The precise choice of these parameters should not affect our calculations, providing us with additional symmetries known as reparametrization invariance (RPI) \cite{Manohar:2002fd}. We will often use RPI-III that transforms $n^\mu \to e^\al n^\mu$ and $\bn^\mu \to e^{-\al} \bn^\mu$. As an example, an RPI-III invariant function $f(p^+, p^-)$ can only depend on the invariant combination $p^+ p^-$.

\begin{figure}
\centering
\subfigure[]{\includegraphics[scale=0.75]{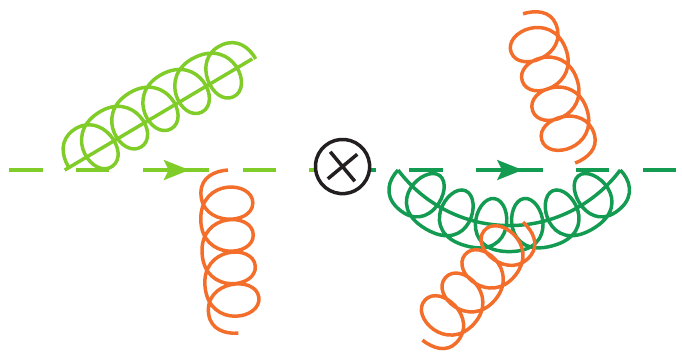}\label{fig:softfact_a}}%
\quad
\subfigure[]{\includegraphics[scale=0.75]{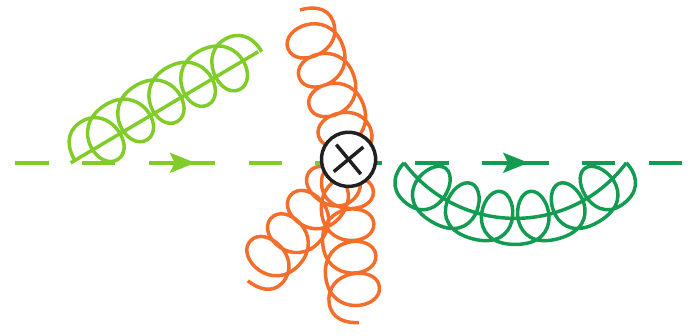}\label{fig:softfact_b}}%
\caption[The decoupling of soft radiation from collinear modes in pictures.]{The soft radiation in (a) organizes itself into emissions from soft Wilson lines shown in (b). The dashed lines correspond to collinear quarks, and the springs with (without) a solid line through them correspond to collinear (soft) gluons.}
\label{fig:softfact}
\end{figure}

We now discuss the factorization of the soft and various collinear degrees of freedom. It turns out that at leading order in the power counting the coupling of soft degrees of freedom to any collinear sector organizes itself into a Wilson line $Y$ of soft gluons \cite{Bauer:2001yt}, as shown in \fig{softfact}. This is due to the simple coupling between soft and collinear degrees of freedom in \fig{softcoll} and their relative power counting in \eqs{coll_mom}{soft_mom}. By factoring out this soft Wilson line from the collinear quark and gluon field $\xi_n$ and $A_n^\mu$
\begin{align}
\xi_{n,\lp}(x) &= Y(x)\,\xi^\zero_{n,\lp}(x)
\,,\nn\\
A^\mu_{n,\lp}(x) &= Y(x)\, A^{\mu\zero}_{n,\lp}(x)\,Y^\dagger(x)
\,,\end{align}
we get collinear fields $\xi_n^\zero \!, A_n^{\mu\zero}$ that are decoupled from the soft degrees of freedom at leading order in the power counting. This is known as the BPS field redefinition.

We conclude this section by discussing the resummation of large logarithms. Our measurement sets the expansion parameter $\la^2 \sim B_a^+/Q \ll 1$ and SCET will allow us to resum the corresponding large logarithms $L = \ln B_a^+/Q$. However, rather than single logarithms $\al_s L$ we now encounter double logarithms $\al_s L^2$ which leads to a different structure than in \eq{W_logs}
\begin{align} \label{eq:SCET_logs}
  \sigma \sim 1 + \al_s L^2 &+ \al_s^2 L^4 + \al_s^3 L^6 + \dots
  \nn \\
  \al_s L &+ \al_s^2 L^3 + \al_s^3 L^5 + \dots
  \nn \\
  \al_s &+ \al_s^2 L^2 + \al_s^3 L^4 + \dots
  \,.
\end{align}
If we assume $L \sim 1/\alpha_s$, getting the correct tree level result requires us to sum the first row, the second row, everything on the third row except for the first term etc. This looks complicated, but once these logarithms are factored in SCET the story is very similar to that in \sec{EFT_weak}. Each factor only depends on one physical scale and by evaluating it at this scale we avoid large logarithms. We then use the RGE to evolve all the factors to a common scale $\mu$, which resums the logarithms in \eq{SCET_logs}.
The evolution of a factor has the following parametric structure
\begin{align}\label{eq:SCET_resum}
  U(\mu_0,\mu) \sim \exp\Big[\sum_k (\al_s L)^k L + (\al_s L)^k + \al_s (\al_s L)^k + \dots\Big] 
\,.\end{align}
The first term, $\sum_k (\al_s L)^k L$, was not present for the weak decays in \eq{W_resum}. It is due to the cusp anomalous dimension which comes from the cusp of Wilson lines. This term is related to the different structure of the logarithms in \eq{SCET_logs}, as is clear from expanding  $U(\mu_0,\mu)$. Assuming $L \sim 1/\alpha_s$, we need the first two terms of the exponent in \eq{SCET_resum} to get the correct tree level result, which requires the two-loop cusp anomalous dimension. The non-cusp piece of the anomalous dimension does not contribute to the first term of the exponent in \eq{SCET_resum} and is only needed at one-loop. The general counting for fixed order and resummed perturbation theory is shown in Table~\ref{tab:counting_double}. We will not consider LL resummation because this does not include the $\sum_k (\al_s L)^k$ term in the exponent in \eq{SCET_resum}, which parametrically contributes at $\ord{1}$.

\begin{table}
  \centering
  \begin{tabular}{l | c | c c c}
  \hline \hline
  & matching & \ & running & \\
  & \ & non-cusp & cusp & $\beta(\al_s)$ \\ \hline
  LO & $0$-loop & - & - & - \\
  NLO & $1$-loop & - & - & - \\
  NNLO & $2$-loop & - & - & - \\
  LL & $0$-loop & $0$-loop & $1$-loop & $1$-loop\\  
  NLL & $0$-loop & $1$-loop & $2$-loop & $2$-loop\\
  NNLL & $1$-loop & $2$-loop & $3$-loop & $3$-loop\\
  \hline\hline
  \end{tabular}
\caption[Order counting in perturbation theory for double logarithmic resummation.]{Order counting for fixed and resummed perturbation theory with double logarithms.}
\label{tab:counting_double}
\end{table}

\section{Outline}

Chapter \ref{ch:general} provides an introduction to beam functions and factorization for isolated processes, in which we discuss our main results. The goal of this chapter is to give a thorough discussion of the physical picture behind our results which is accessible to non-expert readers. We define the ``beam thrust" event shape that we use to impose the central vet jeto in isolated Drell-Yan, and we present the factorization theorem. The renormalization of the beam functions and the relationship between the beam functions and the PDFs is discussed, and we make the comparison with the initial-state parton shower. We compare our factorization theorem for isolated Drell-Yan to the fixed-order calculation and discuss its RGE structure. For convenience we have included a quick overview of our notation and symbols in \app{notation}. In \app{plusdisc} we list the plus distribution definitions and identities that we use, as well as identities for discontinuties needed in the calculations in \ch{quark} and \ch{gluon}.

In chapter \ref{ch:beamf} we discuss several formal aspects of the quark and gluon beam functions, including their definition in terms of matrix elements of operators in SCET, their renormalization, their analytic structure, and the operator product expansion relating the beam functions to PDFs. In particular, we show that the beam functions have the same RGE as the jet functions to all orders in perturbation theory. (Part of the proof is relegated to \app{renorm}.)

We perform the one-loop matching of the quark beam function onto quark and gluon PDFs in \ch{quark}. Using an offshellness IR regulator we give explicit details of the calculations for the quark beam function and PDF.  We verify explicitly that the IR singularities cancel in the matching and extract results for the next-to-leading order (NLO) matching coefficients. In \app{dimreg} we repeat the matching calculation for the quark beam function in pure dimensional regularization. The calculation of the one-loop matching coefficients for the gluon beam function is performed in \ch{gluon}.
In chapter \ref{ch:beam_plots}, we show and discuss plots for the quark and gluon beam function at NLO in fixed-order perturbation theory and next-to-next-to-leading logarithmic (NNLL) order in resummed perturbation theory.

In chapter \ref{ch:fact}, we derive in detail the factorization theorem for isolated processes $pp\to XL$ using SCET, and apply it to the case of Drell-Yan. 

Plots for the beam thrust cross section for Drell-Yan at NLO and NNLL are shown and discussed in \ch{BT}. In \ch{higgs} we apply our factorization theorem to Higgs production through gluon fusion and show results for the corresponding beam thrust cross section. The necessary ingredients for evaluating these cross sections at NNLL are collected in \app{pert}.

In \ch{njet} we extend our work to final state jets. For a signal with $N$ jets we define a global event shape called ``N-jettiness" to veto any unwanted additional jets. We describe theoretical and experimental benefits of our approach and present a factorization formula for the corresponding $N$-jet cross section. We conclude in \ch{concl}.


\chapter{Introduction to Beam Functions and Isolated Drell-Yan}
\label{ch:general}

This chapter provides an extensive discussion of how factorization with beam
functions works, including the necessary kinematic definitions for the variables
that constrain the hadronic final state. In the interest of avoiding technical
details, we only discuss the physics contained in the factorization theorems.
Readers interested in the field-theoretic definitions for the beam functions are
referred to \ch{beamf}, while those interested in the derivation of the
factorization theorem in SCET and explicit definitions for all its ingredients
are referred to \ch{fact}. The work in this chapter was first presented in Ref.~\cite{Stewart:2009yx}.

In \sec{DYfact}, we review the factorization theorems for inclusive Drell-Yan
and threshold Drell-Yan, and then explain the factorization theorem for our
isolated Drell-Yan process. We use a simple setup where measurements on the
final-state hadrons use hemispheres orthogonal to the beam.  These observables
are generalized in \sec{generalobs} to uniformly account for measurements
that sample over a wide variety of boosts between the hadronic and partonic
center-of-mass frames.  We explain the relation between beam functions and
parton distribution functions in \sec{BtoF}. We compare the beam-function
renormalization group evolution to initial-state parton showers in
\sec{showercomparison}.  In \sec{fixedorder}, we show how the various
pieces in the factorization theorem arise from the point of view of a
fixed-order calculation.  In \sec{RGE}, we compare the structure of large
logarithms and their resummation for the different factorization theorems. This
yields an independent argument for the necessity of beam functions.

\section{Drell-Yan Factorization Theorems}
\label{sec:DYfact}

To describe the Drell-Yan process $pp\to X\ell^+\ell^-$ or $p\bar p\to
X\ell^+\ell^-$, we take
\begin{equation}
P_a^\mu + P_b^\mu = p_X^\mu + q^\mu
\,,\end{equation}
where $P_{a,b}^\mu$ are the incoming (anti)proton momenta,
$\ECM = \sqrt{(P_a+P_b)^2}$ is the total center-of-mass energy,
and $q^\mu$ is the total momentum of the $\ell^+\ell^-$ pair. We also define
\begin{align} \label{eq:DYvars}
\tau = \frac{q^2}{\ECM^2}
\,, \qquad
Y = \frac{1}{2} \ln\frac{P_b \cdot q}{P_a \cdot q}
\,, \qquad
x_a = \sqrt{\tau} e^Y
\,, \qquad
x_b = \sqrt{\tau} e^{-Y}
\,,\end{align}
where $Y$ is the total rapidity of the leptons with respect to the beam axis,
and $x_a$ and $x_b$ are in one-to-one correspondence with $\tau$ and $Y$.
Their kinematic limits are
\begin{align} \label{eq:Ylimit}
0 \leq \tau \leq 1
\,, \qquad
2 \abs{Y} \leq - \ln\tau
\,, \qquad
\tau \leq x_a \leq 1
\,, \qquad
\tau \leq x_b \leq 1
\,.\end{align}
The invariant mass of the hadronic final state is bounded by
\begin{equation} \label{eq:mXincl}
m_X^2 = p_X^2 \leq \ECM^2(1 - \sqrt{\tau})^2
\,.\end{equation}
In Drell-Yan
\begin{equation}
Q = \sqrt{q^2} \gg \lqcd
\end{equation}
plays the role of the hard interaction scale. In
general, for factorization to be valid at some leading level of approximation
with a perturbative computation of the hard scattering, the measured observable
must be infrared safe and insensitive to the details of the hadronic final
state.

\begin{figure*}[ht!]
\subfigure[\hspace{1ex}Inclusive Drell-Yan.]{%
\includegraphics[width=0.33\textwidth]{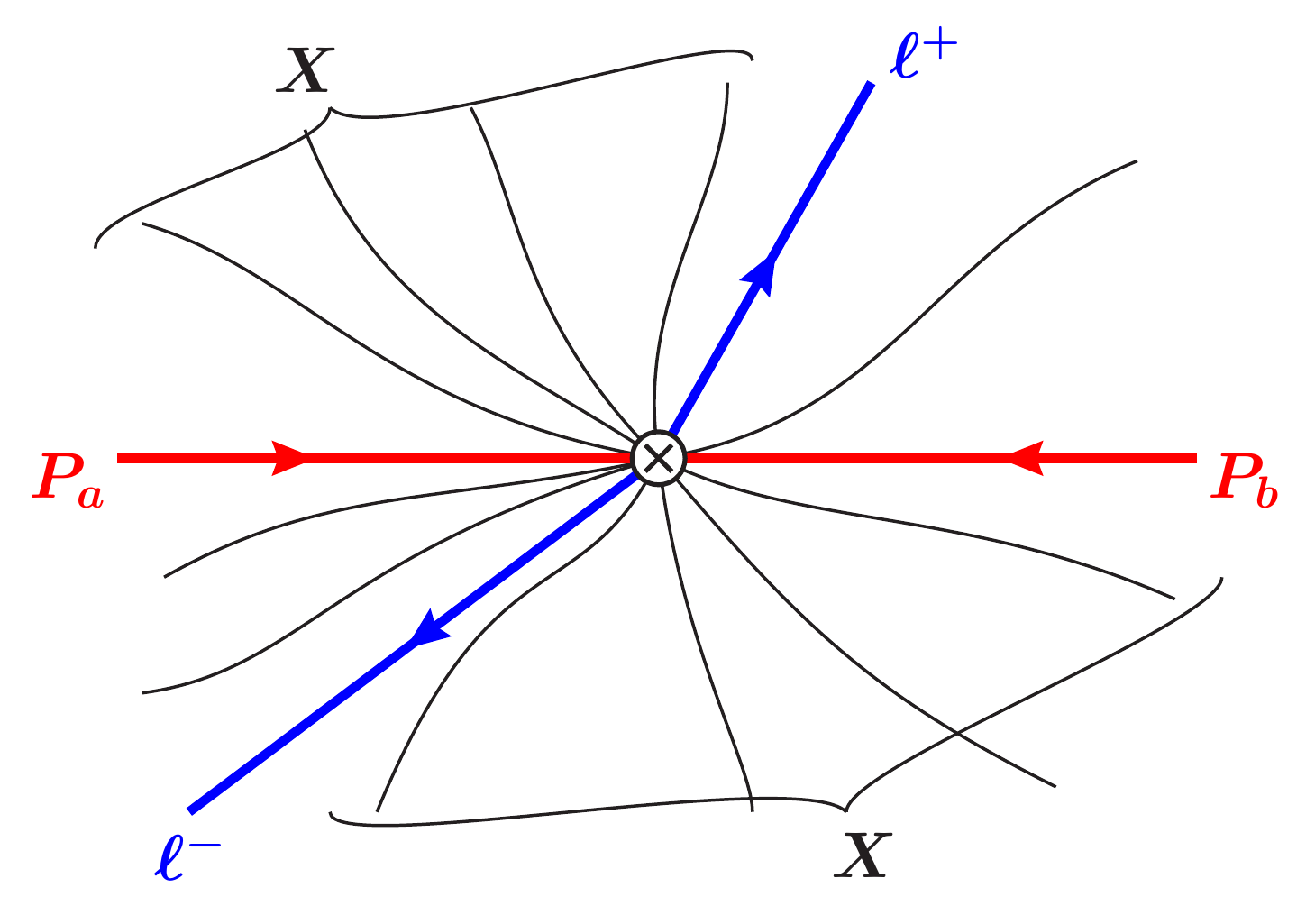}%
\label{fig:drellyan_inclusive}%
}\hfill%
\subfigure[\hspace{1ex}Drell-Yan near threshold.]{%
\includegraphics[width=0.33\textwidth]{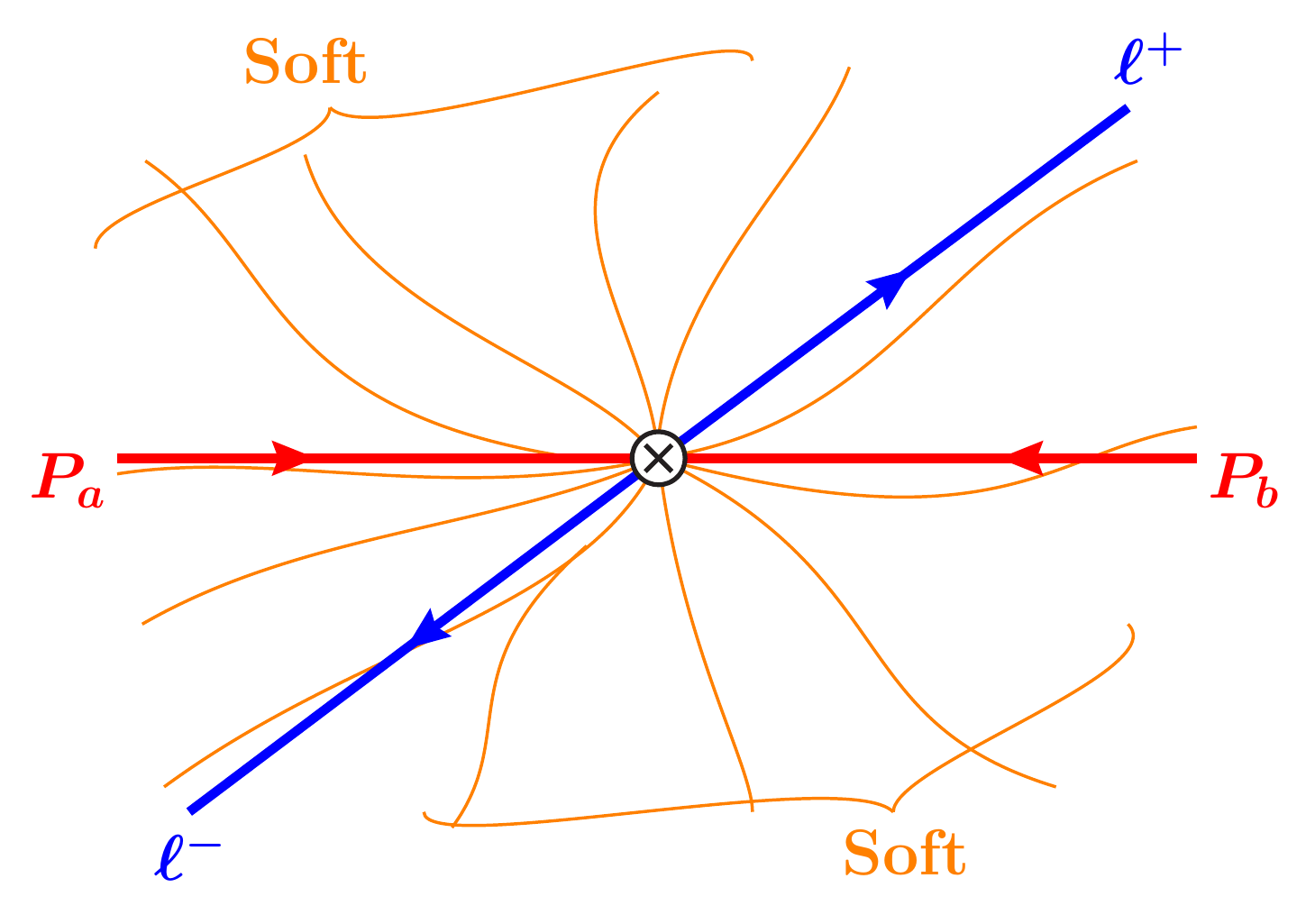}%
\label{fig:drellyan_threshold}%
}\hfill%
\subfigure[\hspace{1ex}Isolated Drell-Yan.]{%
\includegraphics[width=0.33\textwidth]{figs/drellyan_labels}%
\label{fig:drellyan_excl}%
}%
\\
\parbox{0.33\textwidth}{\ }\hfill%
\subfigure[\hspace{1ex}Dijet near threshold.]{%
\includegraphics[width=0.33\textwidth]{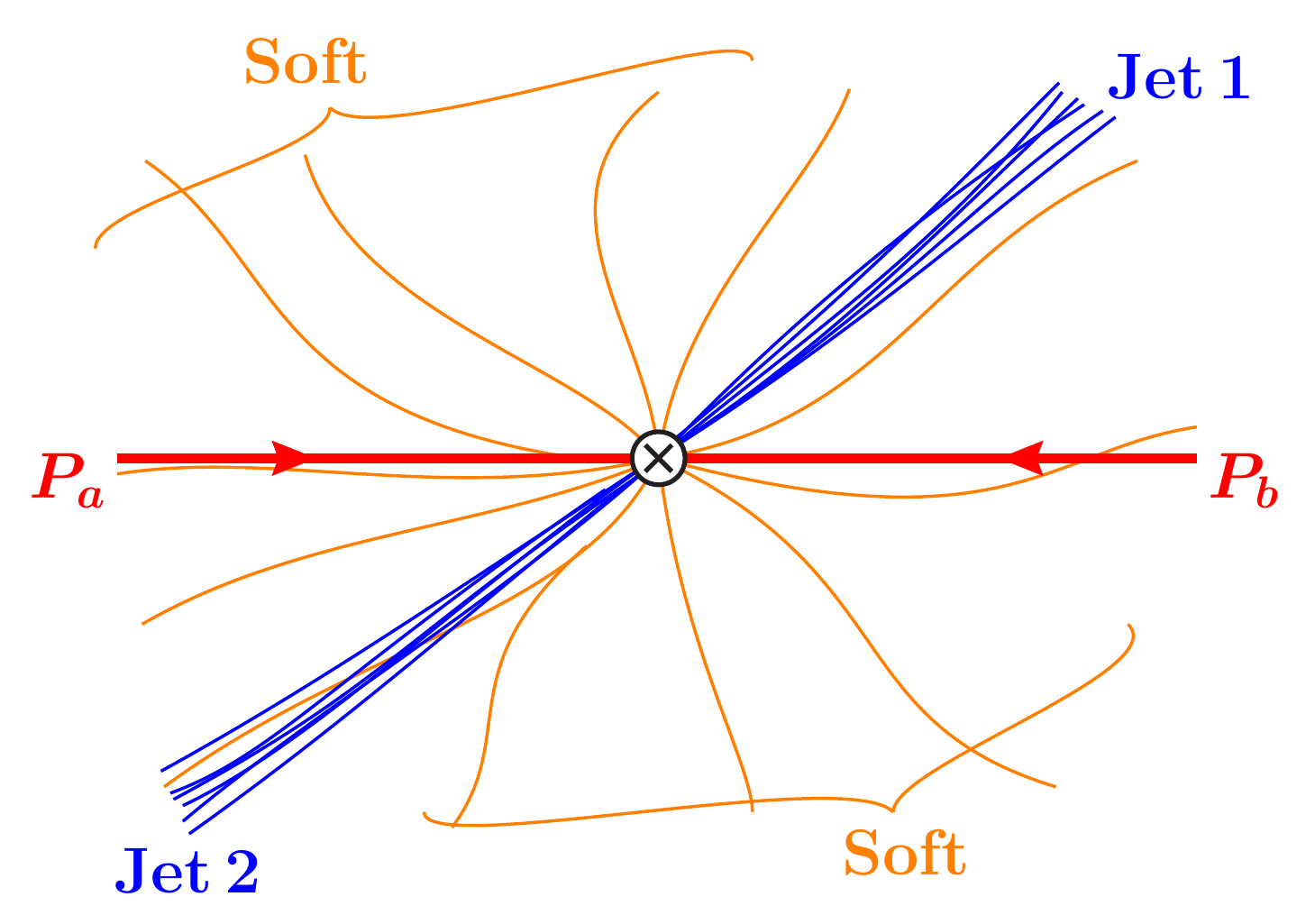}%
\label{fig:2jet_threshold}%
}%
\subfigure[\hspace{1ex}Isolated dijet production.]{%
\includegraphics[width=0.33\textwidth]{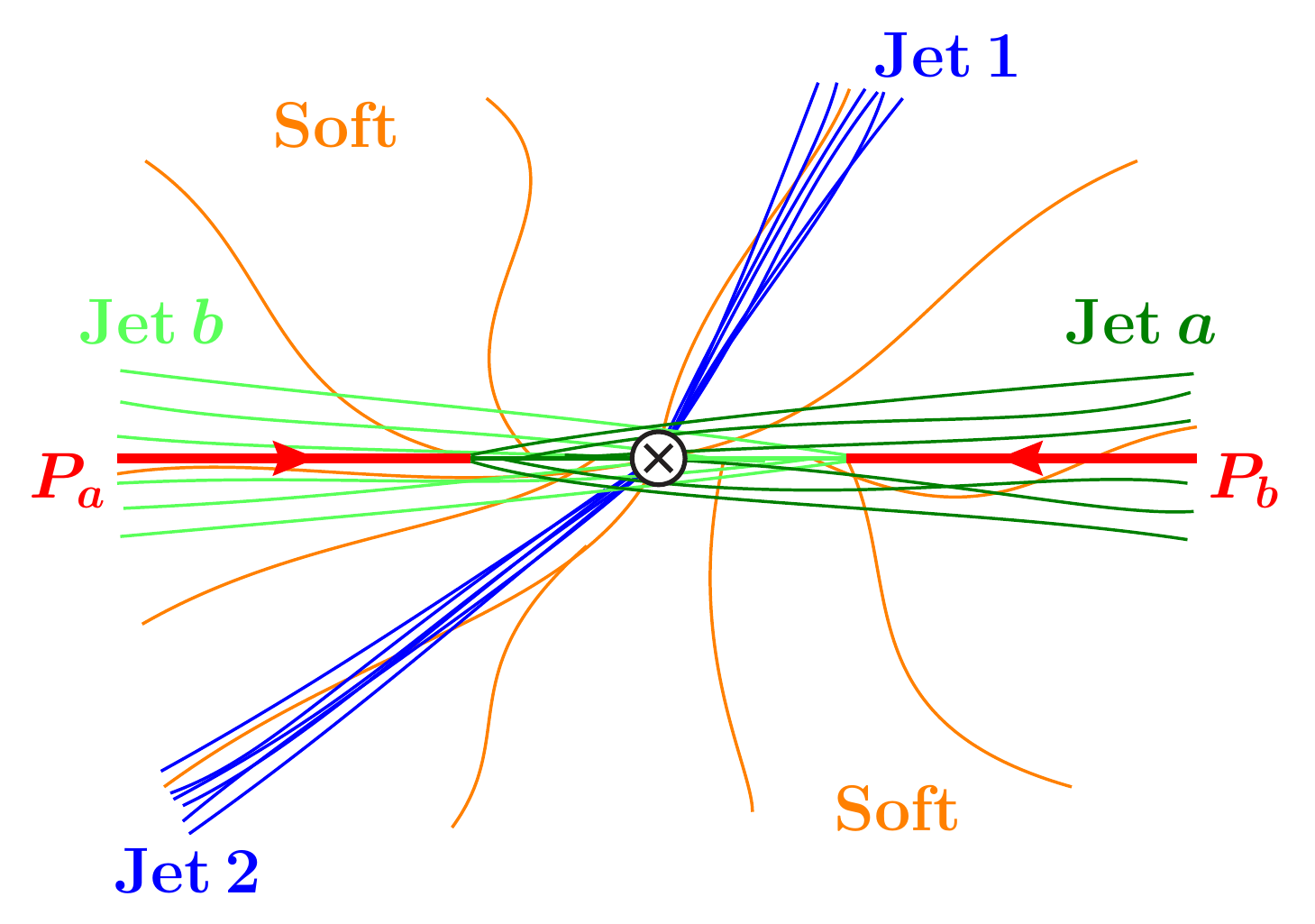}%
\label{fig:2jet_excl}%
}%
\caption[The final-state configuration for inclusive, isolated and threshold Drell-Yan and
threshold and isolated dijet production.]{\label{fig:pp} Different final-state configurations for $pp$
  collisions. The top row corresponds to Drell-Yan factorization theorems for
  the (a) inclusive, (b) threshold, and (c) isolated cases. The bottom row
  shows the corresponding pictures with the lepton pair replaced by dijets. }
\end{figure*}

For inclusive Drell-Yan, illustrated in \fig{drellyan_inclusive}, one sums over
all hadronic final states $X$ allowed by \eq{mXincl} without imposing any cuts.
Hence, the measurement is insensitive to any details of $X$ because one sums
over all possibilities. In this situation
there is a rigorous derivation of the classic factorization
theorem~\cite{Bodwin:1984hc, Collins:1985ue, Collins:1988ig}
\begin{align} \label{eq:DYincl}
\frac{1}{\sigma_0}\, \frac{\df\sigma}{\df q^2 \df Y}
 &= \sum_{i,j} \int\!\frac{\df \xi_a}{\xi_a}\, \frac{\df \xi_b}{\xi_b}\,
  H^\incl_{ij}\Bigl(\frac{x_a}{\xi_a},\frac{x_b}{\xi_b},q^2, \mu \Bigr)\,
  f_i(\xi_a, \mu)\,f_j(\xi_b, \mu)\Bigl[1 + \ORd{\frac{\lqcd}{Q}}\Bigr]
\,,
\end{align}
where $\sigma_0=4\pi\alem^2/(3N_c \ECM^2 q^2)$, and the integration limits are
$x_{a}\le \xi_{a} \le 1$ and $x_{b}\le \xi_{b} \le 1$. The sum is over partons
$i,j=\{g,u,\bar u,d,\ldots\}$, and $f_i(\xi_a)$ is the parton distribution
function for finding parton $i$ inside the proton with light-cone momentum
fraction $\xi_a$ along the proton direction. Note that $\xi_{a,b}$
are partonic variables, whereas $x_{a,b}$ are leptonic, and the two are only
equal at tree level. The inclusive hard function $H^\incl_{ij}$ can be computed in fixed-order
perturbative QCD as the partonic cross section to scatter partons $i$ and
$j$ [corresponding to $\df\sigma_{ij}^\parton$ in \eq{sigff}] and is known to
two loops~\cite{Altarelli:1979ub, Hamberg:1990np, Harlander:2002wh,
Anastasiou:2003yy, Anastasiou:2003ds}.

For threshold Drell-Yan, one imposes strong restrictions to only allow soft
hadronic final states with $m_X\ll Q$, as illustrated in
\fig{drellyan_threshold}. Using \eq{mXincl}, this can be ensured by forcing
$(1-\sqrt{\tau})^2 \ll \tau$, so that one is close to the threshold $\tau \to
1$. In this case, there are large double logarithms that are not accounted for
by the parton distributions.  Furthermore, since
\begin{align} \label{eq:threshold}
1 \geq \xi_{a,b} \geq x_{a,b} \geq \tau \to 1
\,,\end{align}
a single parton in each proton carries almost all of the energy, $\xi_{a,b}\to
1$. The partonic analog of $\tau$ is the variable
\begin{align} \label{eq:DYz}
  z = \frac{q^2}{\xi_a \xi_b \ECM^2} = \frac{\tau}{\xi_a\xi_b} \leq 1
\,,\end{align}
and $\tau\to 1$ implies the partonic threshold limit $z\to 1$.
As \eq{Ylimit} forces $Y\to 0$ for $\tau\to 1$, it is convenient to
integrate over $Y$ and consider the $\tau \to 1$ limit for $\df\sigma/\df q^2$.
The relevant factorization theorem in this limit is~\cite{Sterman:1986aj, Catani:1989ne}
\begin{align} \label{eq:DYendpt}
\frac{1}{\sigma_0}\, \frac{\df\sigma}{\df q^2}
&= \sum_{ij} H_{ij}(q^2, \mu) \int\!\frac{\df\xi_a}{\xi_a}\,\frac{\df\xi_b}{\xi_b}\,f_i(\xi_a, \mu)\, f_{j}(\xi_b, \mu)
\nn\\ &\quad \times
 Q\, S_\thr \Bigl[Q\Bigl(1-\frac{\tau}{\xi_a \xi_b}\Bigr), \mu \Bigr]
\Bigl[1 + \ORd{\frac{\lqcd}{Q}, 1-\tau} \Bigr]
\,,\end{align}
where we view \eq{DYendpt} as a hadronic factorization theorem in its own
right, rather than simply a refactorization of $H_{ij}^{\rm incl}$ in
\eq{DYincl}.  This Drell-Yan threshold limit has been studied
extensively~\cite{Magnea:1990qg, Korchemsky:1992xv, Catani:1996yz,
  Belitsky:1998tc, Moch:2005ky, Idilbi:2006dg, Becher:2007ty}.  Factorization
theorems of this type are the basis for the resummation of large logarithms in
near-threshold situations.  In contrast to \eq{DYincl}, the sum in \eq{DYendpt}
only includes the dominant $q\bar q$ terms for various flavors, $ij=\{u\bar u,
\bar u u, d\bar d, \ldots \}$.  Other combinations are power-suppressed and only
appear at $\ord{1-\tau}$ or higher.  The threshold hard function $H_{ij} \sim
\abs{C_i C^*_j}$ is given by the square of Wilson coefficients in SCET, and can
be computed from the timelike quark form factor. The threshold Drell-Yan soft
function $S_\thr$ is defined by a matrix element of Wilson lines and contains
both perturbative and nonperturbative physics.  If it is treated purely in
perturbation theory at the soft scale $Q(1-\tau)$, there are in principle
additional power corrections of $\mathcal{O}[\lqcd/Q(1-\tau)]$ in
\eq{DYendpt}~\cite{Korchemsky:1996iq}.

Our goal is to describe the isolated Drell-Yan process shown in
\fig{drellyan_excl}.  Here, the colliding partons in the hard interaction are
far from threshold as in the inclusive case, but we impose a constraint that
does not allow central jets.  Soft radiation still occurs everywhere, including
the central region.  Away from threshold, the hard interaction only carries away
a fraction of the total energy in the collision. The majority of the remaining
energy stays near the beam. The colliding partons emit collinear radiation along
the beams that can be observed in the final state, shown by the green lines
labeled ``Jet $a$'' and ``Jet $b$'' in \fig{drellyan_excl}. This radiation
cannot be neglected in the factorization theorem and necessitates the beam
functions.  In the threshold case, these jets are not allowed by the limit $\tau
\to 1$, which forces all available energy into the leptons and leaves only soft
hadronic radiation.%
\footnote{ Note that the proof of factorization for the partonic cross section
  in the partonic threshold limit $z\to 1$ is not sufficient to establish the
  factorization of the hadronic cross section, unless one takes the limit
  $\tau\to 1$.  The hadronic factorization theorem assumes that all real
  radiation is soft with only virtual hard radiation in the hard function. The
  weaker limit $z\to 1$ still allows the incoming partons to emit energetic real
  radiation that cannot be described by the threshold soft function.  Only the
  $\tau\to 1$ limit forces the radiation to be soft. This point is not related
  to whether or not the threshold terms happen to dominate numerically away from
  $\tau \to 1$ due to the shape of the PDFs or other reasons.}
In the inclusive case there are no restrictions on additional hard
emissions, in which case initial-state radiation is included in the partonic
cross section in $H_{ij}^\incl$.

Also shown in \fig{drellyan_excl} is the fact that the leptons in isolated
Drell-Yan need not be back to back, though they are still back to back in the
transverse plane [see \sec{kinematics}]. In this regard, isolated Drell-Yan
is in-between the threshold case, where the leptons are fully back to back with
$Y\approx 0$, and the inclusive case, where they are unrestricted.

In \figs{2jet_threshold}{2jet_excl} we show analogs of threshold Drell-Yan and
isolated Drell-Yan with the leptons replaced by final-state jets. We will
discuss the extension to jets in \ch{njet}.

To formulate isolated Drell-Yan we must first discuss how to veto hard emissions
in the central region. For this purpose, it is important to use an observable
that covers the full phase space. Jet algorithms are good tools to identify
jets, but not necessarily to veto them. Imagine we use a jet algorithm and
require that it does not find any jets in the central region. Although this
procedure covers the full phase space, the restrictions it imposes on the final
state depend in detail on the algorithm and its criteria to decide if something
is considered a jet or not.  It is very hard to incorporate such restrictions
into explicit theoretical calculations, and in particular into a rigorous
factorization theorem. Even if possible in principle, the resulting beam and
soft functions would be very complicated objects, and it would be difficult to
systematically resum the large logarithms arising at higher orders from the
phase-space restrictions.  Therefore, to achieve the best theoretical precision,
it is important to implement the central jet veto using an inclusive kinematic
variable.  This allows us to derive a factorization theorem with analytically
manageable ingredients, which can then be used to sum large phase-space
logarithms. For a clean theoretical description, this observable
must be chosen carefully such that it is infrared safe and sensitive
to emissions everywhere in phase space.  Observables satisfying these criteria
for hadron colliders have been classified and studied in
Refs.~\cite{Banfi:2004nk, Banfi:2005mt}, and are referred to as global event
shapes.  (Issues related to non-global observables have been discussed for
example in Refs.~\cite{Dasgupta:2001sh, Berger:2001ns, Appleby:2003sj,
  Forshaw:2006fk}.)

\begin{figure}[t!]
\centering
\includegraphics[scale=0.55]{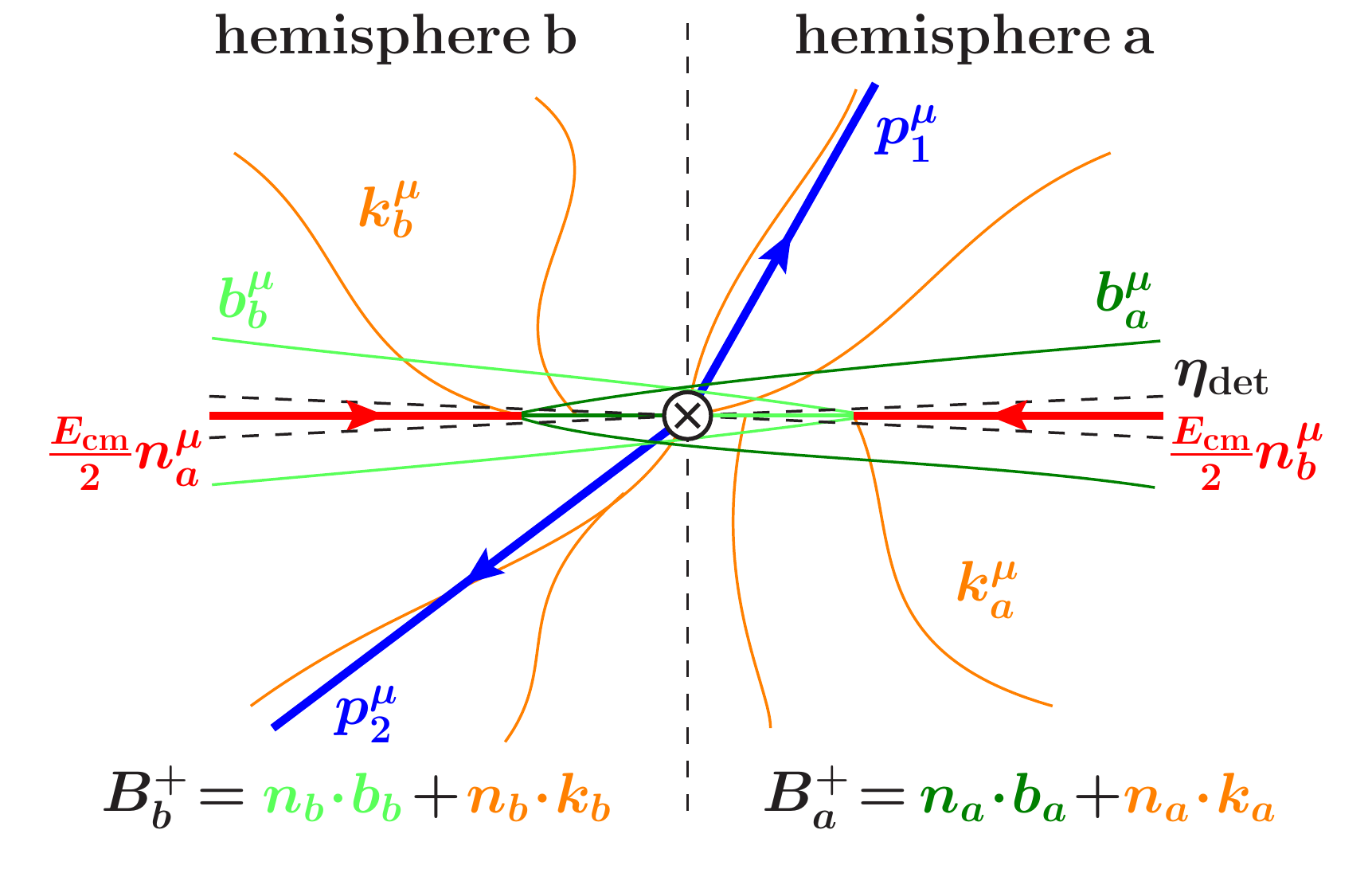}
\caption{Definition of hemispheres and kinematic variables for isolated
Drell-Yan. }
\label{fig:DYkin}
\end{figure}

We will consider a simple kinematic variable that fulfills the above criteria,
leaving the discussion of more sophisticated generalizations to the next
section.  The key variables for the isolated Drell-Yan process are shown in
\fig{DYkin}. The proton momenta $P_a^\mu$ and $P_b^\mu$ are used to define
lightlike vectors $n_a^\mu$ and $n_b^\mu$,
\begin{equation}
 P_a^\mu = \frac{\ECM}{2}\, n_a^\mu
 \,,\qquad
 P_b^\mu = \frac{\ECM}{2}\, n_b^\mu
\,,\end{equation}
where the protons are massless and $n_a^2=0$, $n_b^2=0$, and $n_a\sdt n_b =2$.
Using the beam axis, we define two hemispheres $a$ and $b$ opposite to the
incoming protons. We then divide up the total hadronic momentum as
\begin{align}
p_X^\mu = B_a^\mu + B_b^\mu
\,,\end{align}
where $B_a^\mu = p_{X_a}^\mu$ and $B_b^\mu = p_{X_b}^\mu$ are the total final-state
hadronic momenta in hemispheres $a$ and $b$. Of these, we consider the components
\begin{align}
B_a^+ = n_a \sdt B_a = B_a^0 (1 + \tanh y_a)\, e^{-2y_a}
\,,\qquad
B_b^+ = n_b \sdt B_b = B_b^0 (1 + \tanh y_b)\, e^{-2y_b}
\,,\end{align}
where $B_{a,b}^0$ are the energy components and $y_{a,b}$ are the total
rapidities of $B_{a,b}^\mu$ with respect to the forward direction $n_{a,b}$ for
each hemisphere.  Here, $\lim_{y\to \infty}(1+\tanh y) = 2$ and $1+\tanh y \geq
1.8$ for $y \geq 1$, so $B_{a,b}^+$ scale exponentially with the rapidities $y_{a,b}$.

In terms of the measured particle momenta $p_k$ in hemisphere $a$,
\begin{equation} \label{eq:Baplus_calc}
B_a^+ = \sum_{k \in a} n_a\sdt p_k = \sum_{k \in a} E_k (1 + \tanh\eta_k) e^{-2\eta_k}
\,.\end{equation}
Here, $E_k$ and $\eta_k$ are the experimentally
measured energy and pseudorapidity with respect to $\vec n_a$, and we neglect
the masses of final-state hadrons. An analogous formula applies for $B_b^+$.
Hence, $B_a^+$ and $B_b^+$ receive large
contributions from energetic particles in the central region, while contributions
from particles in the forward region are suppressed. Thus, requiring small
$B_{a,b}^+ \ll Q$ is an effective way to restrict the energetic radiation in
each hemisphere as a smooth function of rapidity, allowing forward jets and
disallowing central jets.  At the same time, soft radiation with energies $\ll
Q$ is measured, but not tightly constrained.

As an example, consider the cut
\begin{equation} \label{eq:ycut1}
B_{a,b}^+ \leq Q\, e^{-2 y_\cut}
\,.\end{equation}
This constraint vetoes any events with a combined energy deposit of more than
$Q/2$ per hemisphere in the central rapidity region $\abs{y} \leq y_\cut$.
In the smaller region $\abs{y} \leq y_\cut - 1$, the energy allowed
by \eq{ycut1} is reduced by a factor of $e^2 \simeq 7$, essentially
vetoing any jets there.  In the larger region $\abs{y} \leq y_\cut\!+\! 1$,
it is increased by the same factor, so beyond $y_\cut + 1$ the hadronic final state is
essentially unconstrained.  Thus, a typical experimental value might be $y_\cut
= 2$, which vetoes energetic jets in the central region $\abs{y} \leq 1$. The
precise value of the cut on $B_{a,b}^+$ will of course depend on the
requirements of the experimental analyses.

Note that the variable $B_a^+$ is similar to the total transverse energy in
hemisphere $a$, defined as
\begin{equation} \label{eq:ET_calc}
E_{Ta} = \sum_{k \in a} \frac{E_k}{\cosh\eta_k} = \sum_{k \in a} E_k (1+\tanh\eta_k)e^{-\eta_k}
\,.\end{equation}
$B_a^+$ has two advantages over $E_{Ta}$. First, the exponential sensitivity to
rapidity is much stronger for $B_a^+$, which means it provides a stronger
restriction on jets in the central region and at the same time is less sensitive
to jets in the forward region. Second, since $B_a^+$ is a specific four-momentum
component and linear in four-momentum, $(p_1 + p_2)^+ = p_1^+ + p_2^+$, it is
much simpler to work with and to incorporate into the factorization theorem.
It is clear that the isolated Drell-Yan factorization theorem discussed here
can be extended to observables with other exponents, $e^{-a \eta_k}$, much like the angularity
event shapes in $e^+e^-$~\cite{Berger:2003iw}.

One should ask, down to what values can $B_{a,b}^+$ be reliably measured
experimentally? In principle, particles at any rapidity contribute to
$B_{a,b}^+$, but the detectors only have coverage up to a maximum
pseudorapidity $\eta_\det$, as indicated in \fig{DYkin}. For the hadron
calorimeters at the LHC $\eta_\det \simeq 5$ and at the Tevatron
$\eta_\det\simeq 4$.  In the hadronic center-of-mass frame, the
unscattered partons inside the proton have plus components of
$\ord{\lqcd^2/\ECM}$, so any contributions from the unmeasured proton
remnants are always negligible. The question then is, what is the maximal
contribution to $B_{a,b}^+$ from initial-state radiation that is missed as it is
outside the detector? In the extreme scenario where all proton energy
is deposited right outside $\eta_\det$, we would have $B_{a,b}^+ = 14
\TeV e^{-10} = 0.6 \GeV$ at the LHC and $B_{a,b}^+ = 2\TeV e^{-8} = 0.7 \GeV$ at
the Tevatron. In more realistic scenarios, the contribution
from such radiation is suppressed by at least another factor of $10$ or more.
Therefore, the finite detector range is clearly not an issue for measuring
values $B_{a,b}^+ \gtrsim 2\,{\rm GeV}$, and the relevant limitation will be
the experimental resolution in $B_{a,b}^+$.

The factorization theorem for isolated Drell-Yan, which we prove in
\ch{fact}, reads
\begin{align} \label{eq:DYbeam}
 &\frac{1}{\sigma_0} \frac{\df\sigma}{\df q^2 \df Y \df B_a^+\df B_b^+}
 = \sum_{ij} H_{ij}(q^2, \mu) \int\!\df k_a^+\, \df k_b^+\,
 q^2 B_i[\w_a(B_a^+ -k_a^+), x_a, \mu] 
\nn\\ & \qquad \times B_j[\w_b(B_b^+ -k_b^+),x_b, \mu]
S_\hemiin(k_a^+,k_b^+, \mu)
\biggl\{1 + \mathcal{O}\biggl[\frac{\lqcd}{Q}, \frac{\w_{a,b} B_{a,b}^+}{Q^2} \biggr]\biggr\}
.\end{align}
The physical interpretation of \eq{DYbeam} is that we take partons $i$ and $j$
out of the initial-state jets $B_{i}$, $B_j$ and hard-scatter them to final
state particles with $H_{ij}$, while including $S_\hemiin$ to describe the
accompanying soft radiation.  The hard function $H_{ij}$ is identical to the one
in the threshold factorization theorem in \eq{DYendpt}, and the sum in
\eq{DYbeam} is again only over $ij=\{u\bar u, \bar u u, d\bar d, \ldots\}$. The
quark and antiquark beam functions $B_q$ and $B_{\bar q}$ describe the effects
of the incoming jets and have replaced the PDFs. The variables $\w_{a,b} =
x_{a,b}\ECM$.  The hard partons are taken from initial-state jets rather than
protons, so unlike in the threshold case the gluon PDF now contributes via the
beam functions. We will see how this works in more detail in \sec{BtoF}.
Finally, $S_\hemiin$ is the initial-state hemisphere soft function.

The kinematic variables in \eq{DYbeam} are displayed in \fig{DYkin}. The soft
function depends on the momenta $k_a^+ =n_a\sdt k_a$ and $k_b^+ =n_b\sdt k_b$ of
soft particles in hemispheres $a$ and $b$, respectively. Much like PDFs, the
beam functions $B_i(t_a, x_a, \mu)$ and $B_j(t_b, x_b, \mu)$ depend on the
momentum fractions $x_a$ and $x_b$ of the active partons $i$ and $j$
participating in the hard collision.  In addition, they depend on invariant-mass
variables
\begin{equation}
t_a = \w_a b_a^+\geq 0
\,,\qquad
t_b = \w_b\, b_b^+ \geq 0
\,,\end{equation}
where $\w_{a,b} = x_{a,b}\ECM$ are the hard momentum components and $b_a^+ =
n_a\sdt b_a$. The momentum $b_a^\mu$ is defined as the total momentum of the
energetic particles radiated into hemisphere
$a$, as shown in \fig{DYkin}, and similarly for $b_b^+$.
(The kinematics are shown in more detail in \fig{beam_kinematics}.)
Before the hard interaction, the momentum of the active quark can be written as
\begin{equation}
\w_a\,\frac{n_a^\mu}{2} - b_a^+\, \frac{n_b^\mu}{2} - b_{a\perp}^\mu
\,.\end{equation}
The first term is its hard momentum along the proton direction, and the last two
terms are from the momentum it lost to radiation, where $b_{a\perp}^2 = -
\vec{b}_{aT}^2$ contains the transverse components. The quark's spacelike
invariant mass is $-\w_a b_a^+ - \vec{b}_{aT}^2 = -t_a - \vec{b}_{aT}^2$. The
beam function $B_i$ for hemisphere $a$ depends on $t_a=\w_a b_a^+ = x_a\ECM
b_a^+$, which is the negative of the quark's transverse virtuality. (When the
distinction is unimportant we will usually refer to $t$ simply as the quark's
virtuality.) By momentum conservation $b_a^+ = B_a^+ - k_a^+$, leading to the
convolution of the beam and soft functions as shown in \eq{DYbeam}. Physically,
the reason we have to subtract the soft momentum from $B_a^+$ is that the beam
function only properly describes the collinear radiation, while the soft
radiation must be described by the soft function. An analogous
discussion applies to $B_j$ and $t_b$ for hemisphere $b$. The convolutions in
the factorization theorem thus encode the cross talk between the soft radiation
and energetic collinear radiation from the beams.

By measuring and constraining $B^+_a$ we essentially measure the virtuality of
the hard parton in the initial state. As the proton cannot contain partons with
virtualities larger than $\lqcd^2$, the initial state at that point must be
described as an incoming jet containing the hard off-shell parton. This is the
reason why beam functions describing these initial-state jets must appear in
\eq{DYbeam}. It also follows that since $t \gg \lqcd^2$ we can calculate the
beam functions perturbatively in terms of PDFs, which we discuss further in
\sec{BtoF}.

It is convenient to consider a cumulant cross section, including all events with
$B_{a,b}^+$ up to some specified value, as in \eq{ycut1}. Integrating
\eq{DYbeam} over $0 \leq B_{a,b}^+ \leq B^+_\max $ we obtain
\begin{align} \label{eq:intDYbeam}
 &\frac{1}{\sigma_0} \frac{\df\sigma}{\df q^2 \df Y}(B_\max^+)
 = \sum_{ij} H_{ij}(q^2, \mu) \int\!\df k_a^+\, \df k_b^+\,
\tB_i[\w_a(B^+_\max -k_a^+), x_a, \mu] 
\nn \\ & \qquad \times
\tB_j[\w_b(B^+_\max -k_b^+), x_b, \mu]
 S_\hemiin(k_a^+,k_b^+, \mu)
\biggl\{1 + \mathcal{O}\biggl[{\frac{\lqcd}{Q}, \frac{\omega_{a,b}B^+_\max}{Q^2}} \biggr]\biggr\}
,\end{align}
where the soft function $S_\hemiin$ is the same as in \eq{DYbeam}, and we
defined the integrated beam function
\begin{equation} \label{eq:tB_def}
\tB_i(t_\max, x, \mu) = \int\! \df t\, B_i(t, x, \mu)\,\theta(t_\max - t)
\,.\end{equation}
The cut $B_{a,b}^+ \leq B^+_\max$ implies the limit $b_{a,b}^+ \leq B^+_\max -
k_{a,b}^+$ and $t_{a,b} \leq \w_{a,b} (B^+_\max - k_{a,b}^+)$, leading to the
convolutions in \eq{intDYbeam}.

The factorization theorem \eq{DYbeam} and its integrated version \eq{intDYbeam}
are valid in the limit $t_{a,b}/Q^2 \simeq B_{a,b}^+/Q \equiv \lambda^2 \ll 1$,
and receive power corrections of $\ord{\lambda^2}$. Thus, for $B^+_\max = Q
e^{-2y_\cut}$ with $y_\cut = 1$, we expect the power corrections not to exceed
$e^{-2} \sim 10\%$. This is not a fundamental limitation, because the power 
corrections can be computed in SCET if necessary. If the soft function is treated purely
perturbatively, there are additional power corrections of
$\ord{\lqcd/B_{a,b}^+}$, which account for soft singularities as $B_{a,b}^+ \to 0$.
The variables $B_{a,b}^+$ are infrared safe with respect to collinear splittings~\cite{Sterman:1978bj}.

The hard function receives perturbative $\alpha_s$ corrections at the hard scale
$\mu_H\simeq Q$, the beam functions have $\alpha_s$ corrections at the
intermediate beam scale $\mu_B^2 \simeq t_\max \simeq Q B^+_\max$, and the soft
function at the soft scale $\mu_S \simeq B^+_\max$. For example, for $Q \simeq
1\TeV$ and $y_\cut = 2$ we have $\mu_B \simeq 140\GeV$ and $\mu_S \simeq
20\GeV$.  Even with a very small $Q \simeq 100\GeV$, perhaps for Higgs
production, $\mu_B \simeq 14\GeV$ and $\mu_S \simeq 2\GeV$ are still
perturbative (although at this point nonperturbative contributions $\sim
\lqcd/\mu_S$ to the soft function might no longer be small and may be
incorporated with the methods in Refs.~\cite{Hoang:2007vb,Ligeti:2008ac}). In
fixed-order perturbation theory, the cross section contains large single and
double logarithms, $\ln(B^+_\max/Q) \simeq -4$ and $\ln^2(B^+_\max/Q) \simeq
16$, invalidating a fixed-order perturbative expansion. The factorization
theorem allows us to systematically resum these logarithms to all orders in
perturbation theory, which is discussed in more detail in \sec{RGE}.

The factorization theorem \eq{DYbeam} also applies to other non-hadronic final
states such as $Z' \to \ell^+\ell^-$, or Higgs production with $H\to
\gamma\gamma$ or $H\to Z Z^*\to 4\ell$. In each case, $q^2$ and $Y$ are the
total non-hadronic invariant mass and rapidity, and central jets are vetoed with
a cut on $B_{a,b}^+$. The only dependence on the process is in the hard
function, which must be replaced appropriately and can be taken directly from
the corresponding threshold factorization theorem. One may also consider $W$
production with $W\to \ell\bar\nu$, with an appropriate replacement of $q^2$ and
$Y$ with the charged lepton's rapidity.  For a light Higgs with $Q\sim m_H$, the
isolated Drell-Yan factorization theorem applies to Higgs production through
gluon fusion $gg\to H$ and Higgs-strahlung $q \bar{q} \to VH$, which are the
dominant production channels at the LHC and Tevatron, respectively.%
\footnote{ In vector-boson fusion and associated production $gg\to t \bar{t} H$,
  the situation is more complicated and one has to explicitly consider the
  process $pp\to X jj H$ with two forward (top) jets.}  For a generic process
$pp\to XL$, the sum over $ij = \{gg, u\bar u, \bar u u, d\bar d,\ldots\}$
includes a gluon-gluon contribution, but still no cross terms between different
parton types, and there will be two independent soft functions $S^{q\bar
  q}_\hemiin$ and $S^{gg}_\hemiin$.  [As shown in \ch{fact}, only the
$q\bar q$ soft function contributes to isolated Drell-Yan, so the labels were
omitted in \eq{DYbeam}.] Indeed, the gluon-gluon contribution involving the
gluon beam and soft functions, $B_g$ and $S^{gg}_\hemiin$, gives the dominant
contribution in the case of Higgs production.

With the above physical picture, we can understand why the gluon beam function
appeared in $\gamma\,p\to J/\psi X$ in the analysis of
Ref.~\cite{Fleming:2006cd} in the limit where $E_{J/\psi} \to E_\gamma$. Taking
$p_X$ as the total momentum of final-state hadrons other than the $J/\psi$, one has
$n\sdt p_X \sim \ECM(1 - E_{J/\psi}/E_\gamma)$, where $n$ is the proton
direction. For $E_{J/\psi}$ close to $E_\gamma$, energetic radiation in the final state is
restricted to a jet close to the $n$ direction. Just as for our
$B_{a,b}^+$, the measurement of $E_{J/\psi}$ probes the radiation emitted by the
colliding gluon in the initial state. Thus, the proton is broken apart prior to the
hard collision, and the gluon beam function is required to describe the initial
state.

\section{Generalized Observables}
\label{sec:generalobs}

The factorization theorem in \eq{DYbeam} applies for $t_a \ll q^2$ and $t_b \ll
q^2$. This includes the situation where in the hadronic center-of-mass frame
there is a numerically significant asymmetry $\w_a = x_a\ECM > \w_b = x_b\ECM$.
This means that the boost between the hadronic and partonic center-of-mass frames,
given by the leptonic $Y = \ln\sqrt{\w_a/\w_b} = \ln\sqrt{x_a/x_b}$, is
significantly different from zero. We explore the implications of this here.

If there is no hierarchy, $\w_a \approx \w_b \approx \sqrt{\w_a \w_b} = Q$,
corresponding to $Y\approx 0$, we can define a simple variable to constrain both
hemispheres simultaneously,
\begin{equation}
\widehat{B} = \frac{B_a^+ + B_b^+}{Q}
\,.\end{equation}
From \eq{DYbeam}, this gives
\begin{align}\label{eq:fact_bhat}
\frac{1}{\sigma_0} \frac{\df\sigma}{\df q^2 \df Y \df \widehat B}
&= \sum_{ij} H_{ij}(q^2, \mu) \int\!\df t_a\, \df t_b\,
B_i(t_a, x_a, \mu)\, B_j(t_b, x_b, \mu)
\nn\\ &\quad \times
Q\, S_B\Bigl(Q\widehat{B} - \frac{t_a}{\w_a}- \frac{t_b}{\w_b}, \mu\Bigr)
\,,\end{align}
where the soft function is defined as
\begin{equation} \label{eq:SB}
S_B(k^+, \mu)
= \!\int\!\df k_a^+ \df k_b^+\, S_\hemiin(k_a^+, k_b^+, \mu )\,
\delta(k^+ - k_a^+ - k_b^+)
\,.\end{equation}
The advantage of using $\widehat B$ is that the soft function now only depends
on the single variable $k^+ = k_a^+ + k_b^+$, much like the soft function for
thrust in $e^+e^-$ collisions.

\begin{figure}[t!]
\centering
\includegraphics[scale=0.55]{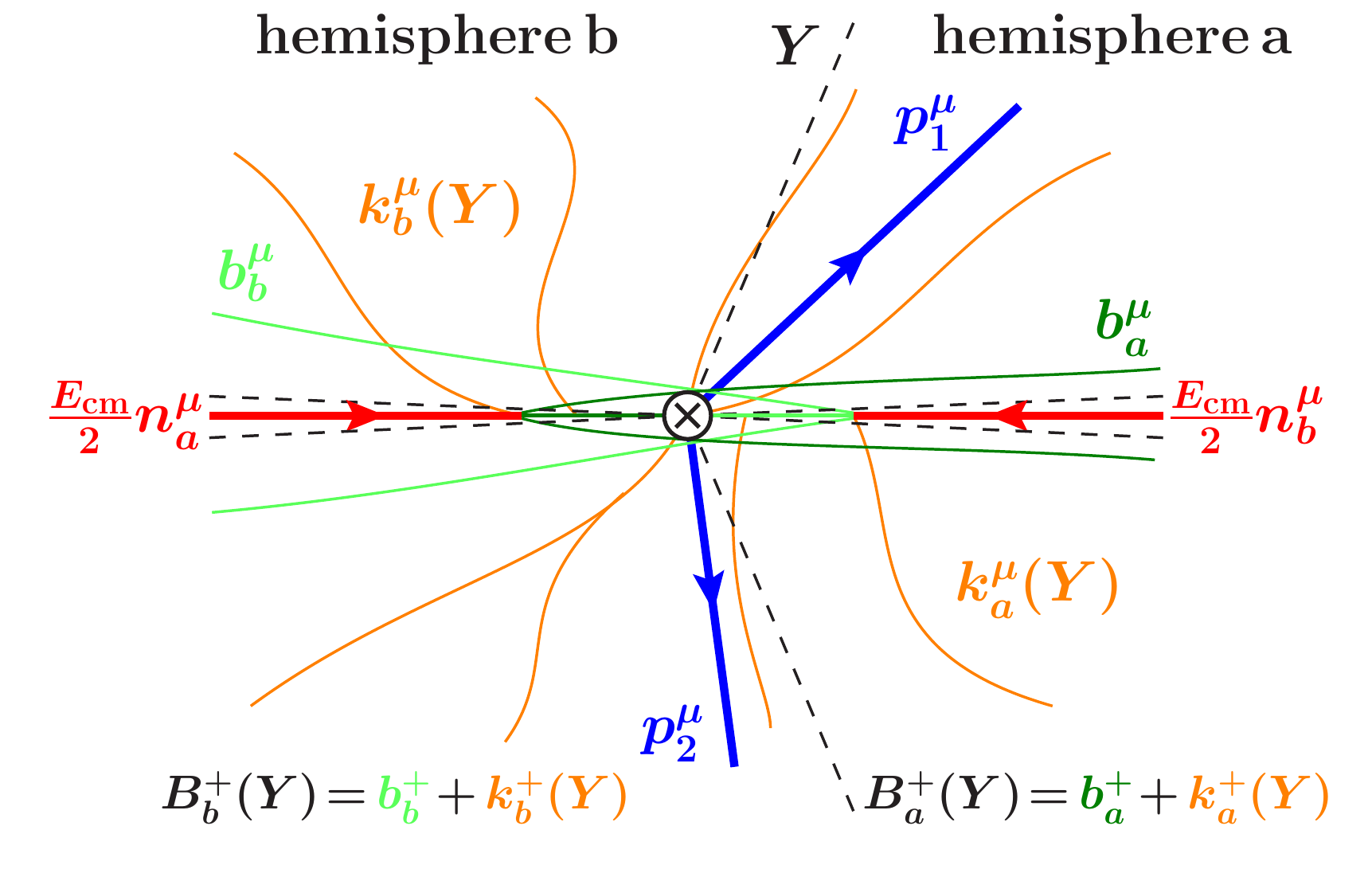}
\caption[Generalized definition of hemispheres for a boosted partonic center-of-mass.]{Generalized definition of hemispheres. The total rapidity of
the leptons is $Y$, $b_{a,b}^+ = n_{a,b}\sdt b_{a,b}$, and $k_{a,b}^+(Y) = n_{a,b}\sdt k_{a,b}(Y)$.}
\label{fig:DYkin_boosted}
\end{figure}

If we have a hierarchy $\w_b < Q < \w_a$, the final state has a substantial
boost in the $n_a$ direction, as shown in \fig{DYkin_boosted}. In this case, the
energetic radiation will generically be much closer to the beam axis in
hemisphere $a$ than in hemisphere $b$. To take this into account, it is natural
to impose different cuts on $B_a^+$ and $B_b^+$. Using the boost-invariant
combinations $\w_a B_a^+/q^2$ and $\w_b B_b^+/q^2$ to define the cut, we obtain
\begin{equation} \label{eq:ycut2}
\frac{\w_a B_a^+}{q^2} = \frac{B_a^+}{\w_b} \leq e^{-2y_\cut}
\,,\qquad
\frac{\w_b B_b^+}{q^2} = \frac{B_b^+}{\w_a} \leq e^{-2y_\cut}
\,,\end{equation}
so $B_a^+$ has a tighter constraint than $B_b^+$, as desired.  If we simply
replace $\widehat{B}$ by $B_a^+/\w_b + B_b^+/\w_a$, the soft function analogous
to $S_B$ in \eq{SB} will depend on $(\w_ak_a^+ + \w_b k_b^+)/Q^2$.

However, we should also adjust the hemispheres themselves to take into account
the significant boost of the partonic center-of-mass frame. We therefore define
a generalized hemisphere $a$ as $y > Y$ and hemisphere $b$ as $y < Y$, as shown
in \fig{DYkin_boosted}. The corresponding total hemisphere momenta are denoted as
$B_{a,b}^+(Y)$ and the soft hemisphere momenta as $k_{a,b}^+(Y)$. The original
definitions in \fig{DYkin} correspond to $B_{a,b}^+(0) \equiv B_{a,b}^+$ and
$k_{a,b}^+(0) \equiv k_{a,b}^+$. The generalization of $\widehat B$ is given by
the boost-invariant combination
\begin{equation} \label{eq:tauB}
\tau_B = \frac{\w_a B_a^+(Y) + \w_b B_b^+(Y)}{q^2}
\,.\end{equation}
With the generalized definition of the hemispheres, $B_{a,b}^+(Y)$ and $\w_{a,b}$ transform
under a boost by $y$ in the $n_a$ direction as
\begin{align}
B_a^+(Y) &\to B_a^{+\prime}(Y + y) = e^{-y} B_a^+(Y)
\,,&
B_b^+(Y) &\to B_b^{+\prime}(Y + y) = e^{y} B_b^+(Y)
\,,\nn\\
\w_a &\to \w_a' = e^{y} \w_a
\,, &
\w_b &\to \w_b' = e^{-y} \w_b
\,.\end{align}
Thus, boosting by $y = -Y$ from the hadronic to the partonic
center-of-mass frame gives
\begin{equation} \label{eq:tauB_partoniccms}
\tau_B
= \frac{\w_a' B_a^{+\prime}(0) + \w_b' B_b^{+\prime}(0)}{q^2}
= \frac{B_a^{+\prime}(0) + B_b^{+\prime}(0)}{Q}
\,.\end{equation}
In the partonic center-of-mass frame we have $\w_a' = \w_b' = Q$, so there is no hierarchy. Correspondingly, the generalized hemispheres in this frame are again perpendicular to the beam axis, so \eq{tauB_partoniccms} has the same form as $\widehat B$.

Note that for $e^+e^- \to$ jets, one can use the thrust axis to define two
hemispheres with $n_{a,b}$ analogous to our case. In the 2-jet limit, thrust is
then given by $1 - T = (Q\, n_a\sdt p_{X_a} + Q\, n_b\sdt p_{X_b})/2Q^2$.
Hence, we can think of $\tau_B$ as the analog of thrust for incoming jets. For
this reason we will call $\tau_B$ the ``beam thrust''.

In analogy to \eqs{ycut1}{ycut2}, we define the cutoff on $\tau_B$ by
\begin{equation} \label{eq:tauBycut}
\tau_B \leq e^{-2 y_B^\cut}
\,.\end{equation}
For $\tau_B \to 0$ or equivalently $y_B^\cut \to \infty$ the jets along the beam axes
become pencil-like, while for
generic $y_B^\cut$ we allow energetic particles up to rapidities $y \lesssim y_B^\cut$
(with $y$ measured in the partonic center-of-mass frame).

The beam functions are boost-invariant along the beam axis, so the different
hemisphere definitions do not affect them. The soft function is boost-invariant
up to the hemisphere definition, which defines its arguments $k_{a,b}^+$. Hence,
boosting by $-Y$ we have $S_\hemiin[e^Y k_a^+, e^{-Y} k_b^+; Y] =
S_\hemiin[k_a^+, k_b^+; 0] = S_\hemiin(k_a^+, k_b^+)$, where the third argument
denotes the definition of the hemispheres. This implies that the soft function for
$\tau_B$ is the same as in \eq{SB}. The factorization theorem for $\tau_B$
following from \eq{DYbeam} is
\begin{align} \label{eq:dsigma_tauB}
\frac{1}{\sigma_0} \frac{\df\sigma}{\df q^2 \df Y \df \tau_B}
&= \sum_{ij} H_{ij}(q^2, \mu) \int\!\df t_a\, \df t_b\,
B_i(t_a, x_a, \mu)\, B_j(t_b, x_b, \mu)
\nn\\ &\quad \times
Q\,S_B\Bigl(Q\,\tau_B - \frac{t_a + t_b}{Q}, \mu\Bigr)
\,.\end{align}
Integrating over $0 \leq \tau_B \leq \exp(-2y_B^\cut)$ we obtain
\begin{equation} \label{eq:sigma_yBcut}
\frac{\df\sigma}{\df q^2 \df Y}(y_B^\cut)
= \int_0^{\exp(-2y_B^\cut)}\!\df \tau_B\,  \frac{\df\sigma}{\df q^2 \df Y\df\tau_B}
\,.\end{equation}
We will use \eqs{dsigma_tauB}{sigma_yBcut} to show plots of our results in chapters \ref{ch:BT} and \ref{ch:higgs}.

\section{Relating Beam Functions and PDFs}
\label{sec:BtoF}

The beam functions can be related to the PDFs by performing an operator product
expansion, because $t_{a,b} \gg \lqcd^2$. This yields the factorization formula
\begin{align} \label{eq:B_fact}
B_i(t, x, \mu)
&= \sum_j\!\int_x^1 \frac{\df\xi}{\xi}\, \cI_{ij}\Bigl(t,\frac{x}{\xi},\mu \Bigr) f_j(\xi, \mu)
\biggl[1 + \ORd{\frac{\lqcd^2}{t}}\biggr]
,\end{align}
where we sum over partons $j=\{g, u,\bar u, d, \ldots\}$, $\cI_{ij}$ are
perturbatively calculable Wilson coefficients, and $f_j$ is the standard PDF
for parton $j$. The $\ord{\lqcd^2/t}$ power corrections in
\eq{B_fact} involve proton structure functions at subleading twist. Further
mathematical details on \eq{B_fact} are discussed in
\ch{beamf}, whereas here we focus on the physical ramifications.

\begin{figure*}[t!]
\centering
\includegraphics[scale=0.75]{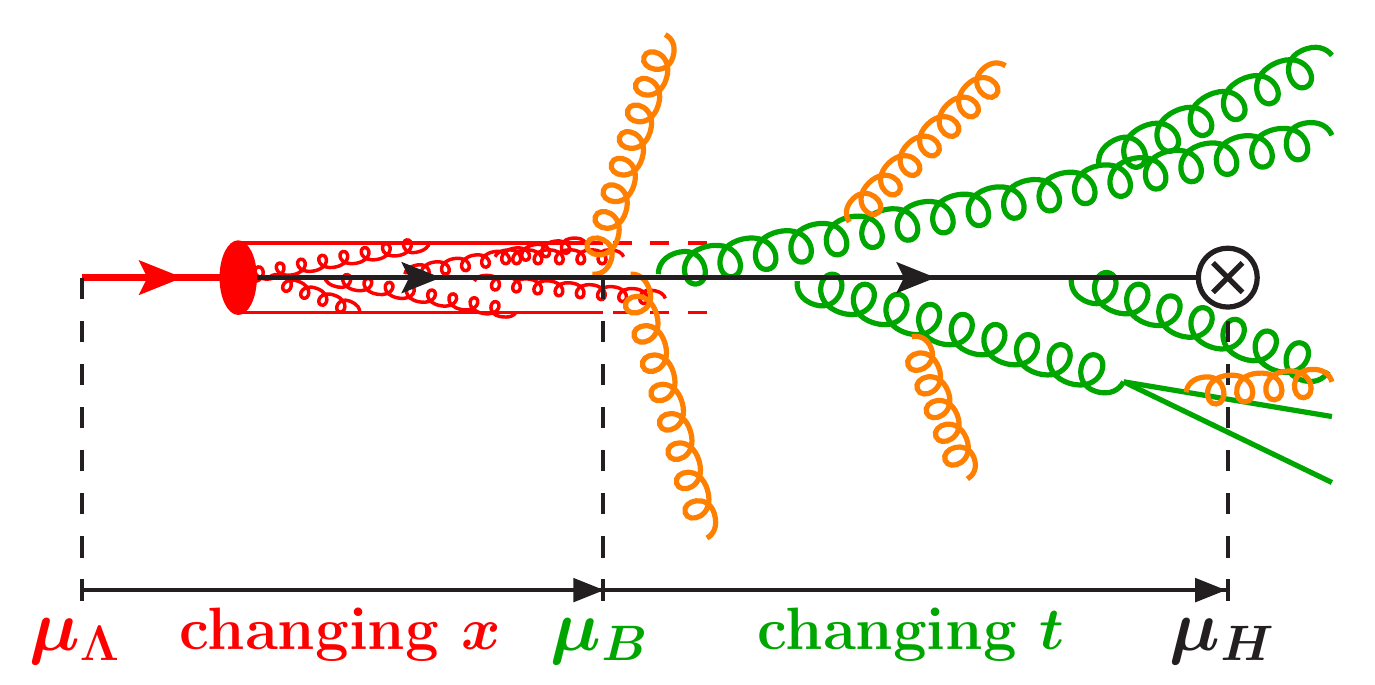}
\caption[Evolution of the initial state.]{Evolution of the initial state. Starting from the low scale
  $\mu_\Lambda$, the incoming proton is described by the $x$-dependent evolution of the PDFs, which
  redistributes the total momentum of the proton between its constituents. At
  the scale $\mu_B$, the proton is probed by measuring the radiation in the final state
  and breaks apart.
  This is the scale where the PDFs are evaluated and the $x$-dependent evolution stops.
  Above $\mu_B$, the proton has ceased to exist, and the initial state behaves
  like an incoming jet, whose evolution is governed by the virtuality $t$ of the
  off-shell spacelike parton that eventually enters the hard interaction at the
  scale $\mu_H$.}
\label{fig:beam}
\end{figure*}

The interpretation of \eq{B_fact} is illustrated in \fig{beam}. At a
hadronic scale $\mu_\Lambda\sim 1\GeV$, the initial conditions for the
PDFs $f_j$ can be specified, and one has the standard DGLAP evolution up to the
scale $\mu_B$,
\begin{align} \label{eq:fevolution}
\mu \frac{\df}{\df\mu} f_j(\xi, \mu)
= \sum_{j'} \int\!\frac{\df\xi'}{\xi'}\, P_{jj'}\Bigl(\frac{\xi}{\xi'}, \mu\Bigr) f_{j'}(\xi', \mu)
\,.\end{align}
The anomalous dimensions $P_{jj'}$ are the standard QCD splitting
functions for quarks, antiquarks, and gluons (including the color factors and coupling constant).
Equation~\eqref{eq:B_fact} applies at the scale
$\mu=\mu_B$, since this is the scale at which a measurement on the proton is
performed by observing the soft and collinear radiation
contributing to $B_{a,b}^+$.  At this scale, a parton $j$ with momentum fraction
$\xi$ is taken out of the incoming proton according to the probability
distribution $f_j(\xi, \mu)$.  As the parton continues to propagate and evolve
with $\mu>\mu_B$, it is modified by virtual radiation and by the emission of
real radiation, which forms a jet. The evolution in this region no longer depends on
$\xi$, but instead on the virtuality $t$. This evolution occurs with
fixed $x$ and fixed parton type $i$, via the beam function RGE
\begin{align} \label{eq:Bevolution}
\mu \frac{\df}{\df\mu} B_i(t, x, \mu) = \int\!\df t'\, \gamma^i_B(t - t', \mu)\, B_i(t',x, \mu)
\,.\end{align}
This result for initial-state jet evolution has the same structure as the
evolution for final-state jets.  In fact, the anomalous dimension $\gamma^q_B$ is
identical to that for the quark jet function to all orders in perturbation
theory. We further discuss the renormalization of the beam function and prove
this correspondence with the jet function in \sec{B_RGE}.

The effect of initial-state real and virtual radiation is described by the
perturbatively calculable Wilson coefficients $\cI_{ij}(t, x/\xi, \mu)$ at the
scale $\mu=\mu_B$. They encode several physical effects.  The virtual loop
corrections contribute to the $\cI_{ii}$ and modify the effective strength of
the various partons.  If the radiation is real, it has physical timelike
momentum. Hence, it pushes the active parton in the jet off shell with spacelike
virtuality $-t < 0$ and reduces its light-cone momentum fraction from
$\xi$ to $x$.

In addition, the real radiation can change the identity of the colliding parton,
giving rise to the sum over $j$ in \eq{B_fact}. For example, an incoming quark
can radiate an energetic gluon which enters the hard interaction, while the
quark itself goes into the final state.  This gives a contribution of the quark
PDF to the gluon beam function through $\cI_{gq}$.  Similarly, an incoming gluon
can pair-produce, with the quark participating in the hard interaction and the
antiquark going into the final state.  This gives a contribution of the gluon
PDF to the quark beam function through $\cI_{qg}$. There are also of course real
radiation contributions to the diagonal terms, $\cI_{qq}$ and $\cI_{gg}$, where
the parton in the PDF and the parton participating in the hard interaction have
the same identity.

At lowest order in perturbation theory, the parton taken out of the proton
directly enters the hard interaction without emitting radiation,
\begin{align}
\cI_{ij}^\tree\Bigl(t,\frac{x}{\xi},\mu \Bigr)
  = \delta_{ij}\, \delta(t)\, \delta\Bigl(1 - \frac{x}{\xi}\Bigr)
\,.\end{align}
Thus at tree level, the beam function reduces to the PDF
\begin{equation} \label{eq:Bi_tree}
B^\tree_i(t, x, \mu) = \delta(t)\, f_i(x,\mu)
\,.\end{equation}
Beyond tree level, $\cI_{ij}(t, x/\xi, \mu)$ can be determined perturbatively as
discussed in more detail in \ch{beamf}, where we give precise field-theoretic
definitions of the beam functions. We calculate the one-loop coefficients $\cI_{ij}$ for the quark
beam function in \ch{quark} and for the gluon beam function in \ch{gluon}.

Interestingly, in the threshold factorization theorem \eq{DYendpt}, cross terms
between quark and gluon PDFs are power suppressed, so the gluon PDF does not
contribute at leading order.  In the inclusive case \eq{DYincl}, such cross
terms are leading order in the power counting. For isolated Drell-Yan, there are
no cross terms between quark and gluon beam functions, but there are
leading-order cross terms between different PDFs, which appear via the
contributions of different PDFs to a given beam function in \eq{B_fact}. Thus,
the isolated case is again in-between the inclusive and threshold cases.

\section{Comparison with Initial-State Parton Shower}
\label{sec:showercomparison}

The physical situation associated with the beam evolution has an interesting
correspondence with that of initial-state parton showers. As pictured in the
region between $\mu_B$ and $\mu_H$ in \fig{beam}, the parton in the beam
function evolves forward in time while emitting a shower of radiation into the
final state governed by the anomalous dimension $\gamma_B^i(t - t',\mu)$ in \eq{Bevolution}.
This equation has no parton mixing. Each emission by the radiating parton
increases the magnitude of its spacelike virtuality
$-t<0$, pushing it further off-shell in a spacelike direction. At the time the
parton is annihilated in the hard collision, it has evolved to some $t$ with $\abs{t}\ll q^2$,
so the large momentum transfer $q^2$ guarantees that no partons in the final state
are spacelike. This description agrees quite well with
the physical picture associated with the evolution of the primary parton in an
initial-state parton shower, as summarized in Ref.~\cite{Sjostrand:2006za}.

Differences in the description arise when one considers the initial-state
parton shower in more detail (for simplicity we focus on the so-called longitudinal
evolution). The shower is based on the evolution equation for the PDFs in
\eq{fevolution}. An evolution forward in time is not practical because of the
lack of prior knowledge of the scale of the hard interaction, so the
shower uses backward evolution starting at a given partonic hard scale
$Q$~\cite{Sjostrand:1985xi}. Knowing the identity of the final parton $i$, the
shower evolves based on the probability $\df\mathcal{P}_i/\df t$ that parton
$i$ is unresolved into parton $j$ via the splitting $j\to ij'$ at an earlier
(lower) scale $t$. The evolution equation is~\cite{Sjostrand:2006za}
\begin{align} \label{eq:ishower}
\frac{\df \mathcal{P}_i(x, t_\max, t)}{\df t}
&= \biggl[\sum_{jj'} \int_x^{z_\max}\!\frac{\df z}{z}\,
  P_{j\to ij'}(z, t)\, \frac{f_j(x/z,t)}{f_i(x,t)} \biggr]
\frac{1}{t}\, \mathcal{P}_i(x,t_\max,t)
\,,\end{align}
where $\mathcal{P}_i(x,t_\max,t)$ is the shower Sudakov exponential, which is interpreted
as the probability for no emissions to occur between the initial value $t_\max$ and $t$.
The evolution variable $t$, which determines the scale of the splitting, is
usually chosen as the virtuality or transverse momentum of the parton.

The mixing of partons in the PDF evolution influences the shower. In particular,
the evolution kernel depends on the PDF $f_j(x/z, t)$, which determines the number density of
partons of type $j$ at the scale $t$, and inversely on the PDF $f_i(x, t)$.
Thus, unlike in the beam evolution in \eq{Bevolution}, the shower evolution in
\eq{ishower} still knows the identity of the initial-state hadron. Double
logarithms in the initial-state parton shower are generated in $q\to qg$ and $g\to gg$
splittings because of the soft-gluon singularity $\sim 1/(1-z)$ in the splitting
functions. This singularity is regulated~\cite{Sjostrand:2006za} by the upper
cutoff $z_\max = x/(x+x_\eps)$, where $x_\epsilon$ provides a lower cutoff on
the gluon energy in the rest frame of the hard scattering, $E_g \geq
x_\eps\gamma\ECM/2 \simeq 2\GeV$ (where $\gamma$ is the boost factor of the hard
scattering). Hence, one logarithm, $\ln x_\eps$, is generated by the $z$
integration, and one logarithm, $\ln t$, by the collinear $1/t$ singularity.  In contrast,
the beam function contains double logarithms $\ln^2 t$
similar to a final-state parton shower, where the $z$ integration yields a kernel
$\sim(\ln t)/t$ that produces a double logarithm $\ln^2 t$ via the $t$
evolution.

The above comparison is very rough. For example, the influence of soft radiation
on both the shower and on the isolated factorization theorem was not compared
and is likely to be important. Furthermore, the goal of the shower is to provide
a universal method for populating fully exclusive final states, while the beam
function applies for a more inclusive situation with a particular measurement.
Note that just the presence of mixing in the initial-state parton shower and absence of
mixing in the beam-function evolution does not imply an inconsistency. For example,
it is well known that the final-state parton shower reproduces the correct
double logarithms for $e^+e^-$ event shapes~\cite{Catani:1992ua}, even
though there is no parton mixing in the evolution of the corresponding hard,
jet, and soft functions.  In the future it would be interesting to test in
detail the correspondence between the double logarithms generated by the initial-state
parton shower and those predicted by our factorization theorem for the isolated
Drell-Yan process.

\section{Relation to Fixed-Order Calculation}
\label{sec:fixedorder}

\begin{figure*}[t!]
\vspace{-3ex}
\subfigure[]{ \label{fig:DYBtree}
\parbox{0.4\textwidth}{%
\begin{equation*}
\begin{aligned}
\parbox[c]{20ex}{\includegraphics[width=20ex]{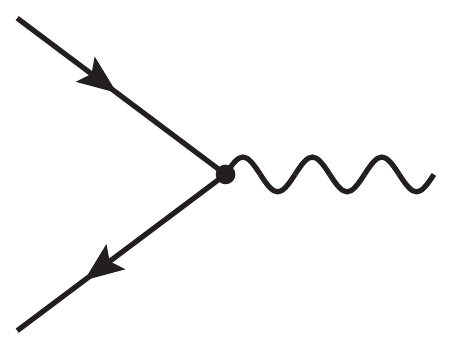}}
&=
\parbox[c]{20ex}{\includegraphics[width=20ex]{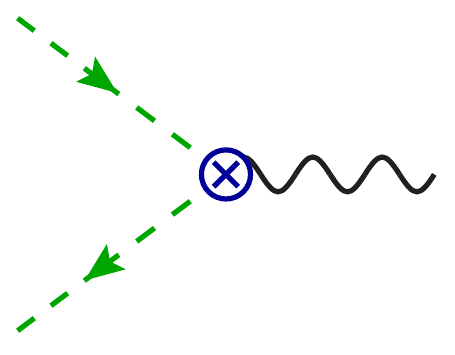}}
\\ &\qquad
\hardc{H_{q\bar q}^\zero}\,\beamc{B_q^\zero\,B_{\bar q}^\zero}\,\softc{S_{q\bar q}^\zero}
\end{aligned}
\end{equation*}}}%
\hfill\hfill\subfigure[]{ \label{fig:DYBglue}
\parbox{0.4\textwidth}{%
\begin{equation*}
\begin{aligned}
\parbox[c]{20ex}{\includegraphics[width=20ex]{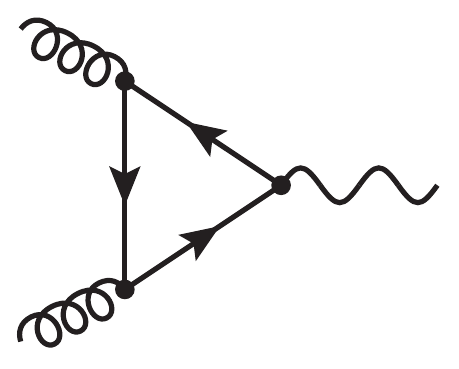}}
&=
\parbox[c]{20ex}{\includegraphics[width=20ex]{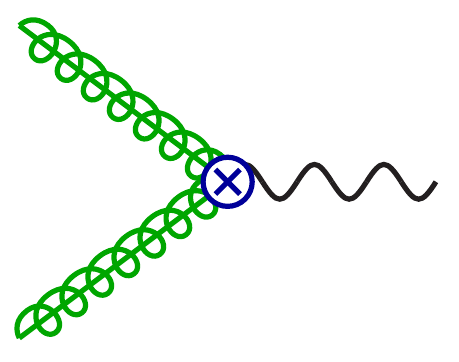}}
\\ &\qquad
\hardc{H_{gg}^\one}\,\beamc{B_g^\zero\,B_g^\zero}\,\softc{S_{gg}^\zero}
\end{aligned}
\end{equation*}}%
}\hspace*{\fill}%

\vspace{-3ex}
\subfigure[]{\label{fig:DYBvertex} \parbox{\textwidth}{%
\begin{equation*}
\begin{aligned}
\parbox[c]{20ex}{\includegraphics[width=20ex]{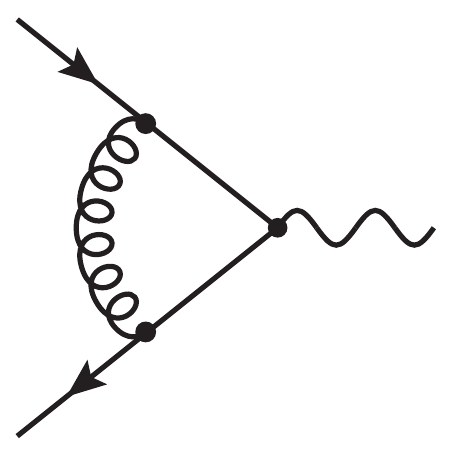}}
&=
\parbox[c]{20ex}{\includegraphics[width=20ex]{figs/feyn/qqbar_Z_0}}
\mspace{-15mu}&&+
\parbox[c]{20ex}{\includegraphics[width=20ex]{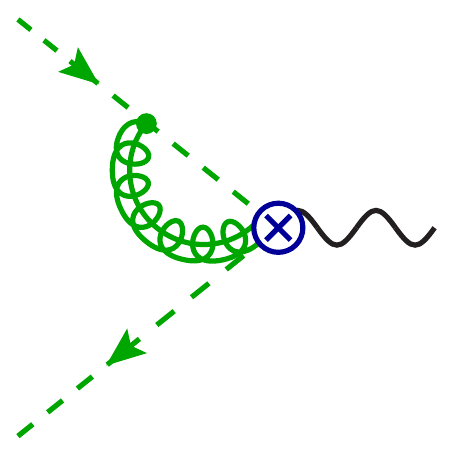}}
\mspace{-15mu}&&+
\parbox[c]{20ex}{\includegraphics[width=20ex]{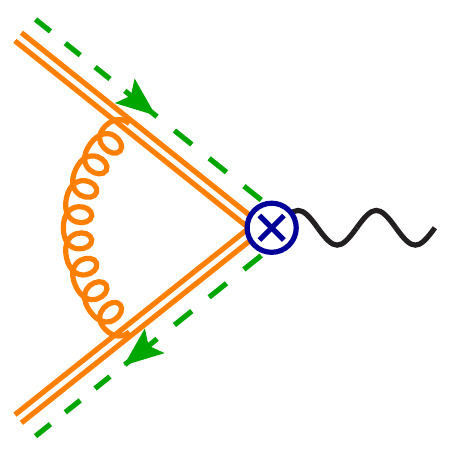}}
\\ &\qquad
\hardc{H_{q\bar q}^\one}\,\beamc{B_q^\zero\,B_{\bar q}^\zero}\,\softc{S_{q\bar q}^\zero}
&&\qquad
\hardc{H_{q\bar q}^\zero}\,\beamc{B_q^\one\,B_{\bar q}^\zero}\,\softc{S_{q\bar q}^\zero}
&&\qquad
\hardc{H_{q\bar q}^\zero}\,\beamc{B_q^\zero\,B_{\bar q}^\zero}\,\softc{S_{q\bar q}^\one}
\end{aligned}
\end{equation*}}}

\vspace{-3ex}
\subfigure[]{\label{fig:DYBbeamqg}
\parbox{0.8\textwidth}{%
\begin{equation*}
\begin{aligned}
\parbox[c]{20ex}{\includegraphics[width=20ex]{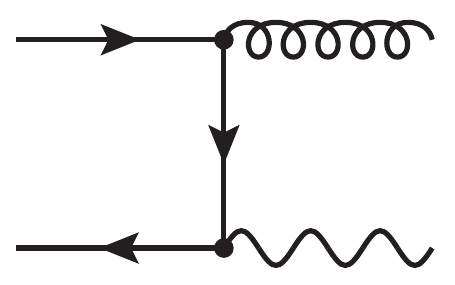}}
&=
\parbox[c]{20ex}{\includegraphics[width=20ex]{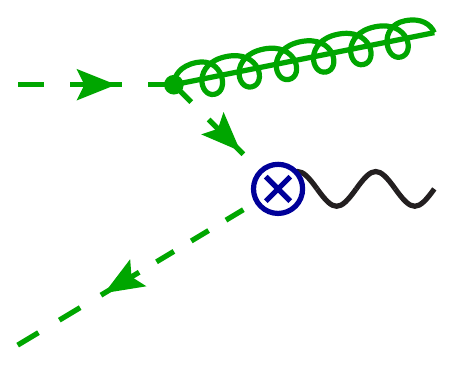}}
\mspace{-18mu}&&+
\parbox[c]{20ex}{\includegraphics[width=20ex]{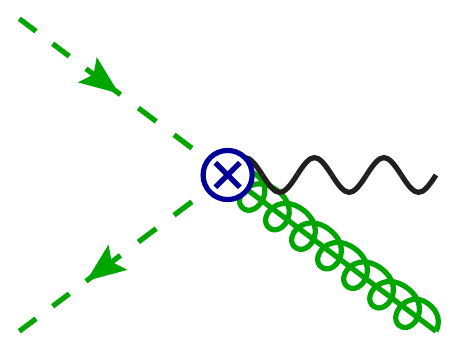}}
\mspace{-18mu}&&+
\parbox[c]{20ex}{\includegraphics[width=20ex]{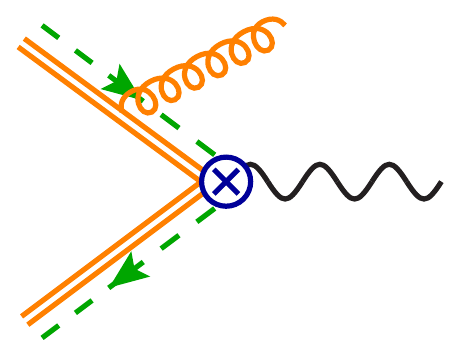}}
\\ & \qquad
\hardc{H_{q\bar q}^\zero}\,\beamc{B_q^\one\,B_{\bar q}^\zero}\,\softc{S_{q\bar q}^\zero}
&&\qquad \hardc{H_{q\bar q}^\zero}\,\beamc{B_q^\zero\,B_{\bar q}^\one}\,\softc{S_{q\bar q}^\zero}
&&\qquad \hardc{H_{q\bar q}^\zero}\,\beamc{B_q^\zero\,B_{\bar q}^\zero}\,\softc{S_{q\bar q}^\one}
\end{aligned}
\end{equation*}}}\hspace*{\fill}

\vspace{-3ex}
\subfigure[]{\label{fig:DYBbeamgq}
\parbox{0.4\textwidth}{%
\begin{equation*}
\begin{aligned}
\parbox[c]{20ex}{\includegraphics[width=20ex]{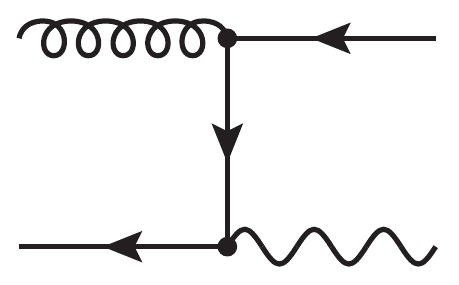}}
&=
\parbox[c]{20ex}{\includegraphics[width=20ex]{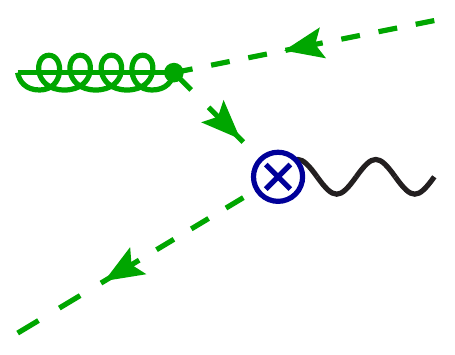}}
\\ & \qquad
\hardc{H_{q\bar q}^\zero}\,\beamc{B_q^\one\,B_{\bar q}^\zero}\,\softc{S_{q\bar q}^\zero}
\end{aligned}
\end{equation*}}}%
\hfill\hfill\subfigure[]{\label{fig:DYBschannel}
\parbox{0.4\textwidth}{%
\begin{equation*}
\parbox[c]{20ex}{\includegraphics[width=20ex]{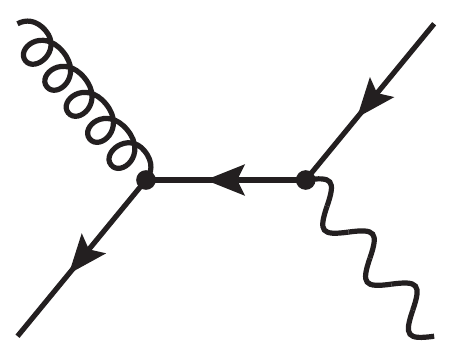}}
= \text{power correction}
\end{equation*}}}\hspace*{\fill}
\caption[Factorization for isolated Drell-Yan in pictures.]{\label{fig:DYBgraphs} Factorization for isolated Drell-Yan in pictures.
  The left-hand side of each equality are graphs in QCD, while the right-hand
  side shows the sum of the corresponding SCET diagrams. Dashed lines are collinear quarks, and springs with
  a line through them are collinear gluons. The double lines denote soft Wilson lines, and the gluons attached to them are soft. }
\end{figure*}

The factorization theorem for the cross section in \eq{DYbeam} and the
factorization for the beam function in \eq{B_fact} together allow us to describe
in more detail how various Feynman diagrams that would appear in a fixed-order
calculation contribute to the cross section in our kinematic region. Various
examples are shown in \fig{DYBgraphs}.

In \fig{DYBtree}, we have the tree-level $q\bar{q}$
annihilation producing a $\gamma$ or $Z$, which involves the
tree-level $\ord{\alpha_s^0}$ hard function, beam functions, and soft function,
denoted by a superscript $\zero$ in the figure. In \fig{DYBglue}, initial-state
gluons couple to a quark loop (e.g. a top quark), which subsequently annihilates
into a $\gamma$, $Z$, or Higgs. The quarks in this loop are far off
shell, so they can be integrated out and appear as one-loop corrections,
$H_{gg}^\one$, to the hard coefficient in the factorization theorem. Other
possibilities for this graph are power suppressed.

The situation for the vertex correction in \fig{DYBvertex} is more involved. If
the gluon in the loop is hard, all particles in the loop are far off shell and
can be integrated out, giving the one-loop hard function $H_{q\bar q}^\one$
shown as the first term on the right-hand side. In the second term, the gluon is
collinear to the incoming quark beam and gives a virtual one-loop contribution
to the quark beam function, $B_q^\one$. The third term is the analog of the
second, but now with the gluon collinear to the incoming antiquark. Finally in
the fourth term, the gluon is soft, communicating between the incoming collinear
beams. Here, the eikonal approximation holds for describing the quark
propagators. The generalization of this to all orders in $\alpha_s$ leads to the
fact that the soft function is a matrix element of Wilson lines. Although a
single loop graph contributes in several different places in the factorization
theorem, all of these contributions have a precise separation in SCET. We
will use this separation in \ch{fact} to prove the isolated Drell-Yan
factorization theorem.

An interesting contribution occurs in \fig{DYBbeamqg}, where a gluon is radiated
into the final state. Because of the kinematic restrictions in isolated
Drell-Yan, this gluon can only be collinear to the incoming quark, collinear to
the incoming antiquark, or soft, and these three possibilities are represented
by the diagrams on the right-hand side of the equality. In the first case, we
have a real-emission correction to the quark beam function, $B_q^\one$.  In the
second case, the intermediate quark is far off shell and can be integrated out,
and the gluon collinear to the antiquark arises from a collinear Wilson line
contribution in $B_{\bar q}^\one$.  The third case gives a real-emission
correction to the soft function, $S^\one_{q\bar q}$.  The full-theory graph in
\fig{DYBbeamqg} has a $t$-channel singularity.  An important fact about the
isolated Drell-Yan factorization theorem is that it fully captures the dominant
parts of this singularity, and allows a simple framework for a resummation of
higher order $\alpha_s$ corrections enhanced by large double logarithms due to
this singularity.  For threshold Drell-Yan, the kinematic restrictions are
stronger and only allow the third graph with soft initial-state radiation.  In
inclusive Drell-Yan, the gluon is treated as hard, and the graph in
\fig{DYBbeamqg} only corrects $H_{q\bar q}^\incl$, without providing a framework
for summing the large double logarithms that appear when we make a global
measurement of the radiation in each hemisphere defined by the beams.

The situation is a bit simpler for \figs{DYBbeamgq}{DYBschannel}. In
\fig{DYBbeamgq}, the incoming collinear gluon from the PDF pair-produces a quark
and antiquark both collinear to this beam direction, and the quark enters the
hard interaction.  Therefore, this is a one-loop correction to the quark beam
function, $B_q^\one$, proportional to the gluon PDF $f_g$. The beam functions
again allow us to resum the possibly large logarithms due to this $t$-channel
singularity.  Other possibilities for the final-state antiquark in
\fig{DYBbeamgq} lead to power-suppressed contributions. Similarly, the
$s$-channel graph in \fig{DYBschannel}, which has the same initial and final
states as \fig{DYBbeamgq}, has no leading-power contribution and only
contributes to \eq{DYbeam} in the power-suppressed terms.  The same is also true
for Drell-Yan in the threshold region. Only inclusive Drell-Yan receives a
leading-order hard contribution from the $s$-channel graph, which is then
treated as of the same size as the $t$-channel graphs.

\section{Renormalization Group Evolution}
\label{sec:RGE}

In this section, we discuss and compare the structure of large logarithms in
the cross sections for inclusive, threshold, and isolated Drell-Yan.  These
large logarithms may be summed using the renormalization group evolution of the
individual functions appearing in the factorization theorems. In fact, the
structure of large logarithms in the differential $B_{a,b}^+$ cross section
allows us to infer the necessity of the beam functions in the isolated
factorization theorem.  This procedure provides a method of determining whether
beam functions enter for other observables or processes than those studied here.
The consistency of the RGE was used to provide a similar consistency check in
Ref.~\cite{Fleming:2007qr} when deriving a new factorization theorem for the
invariant-mass distribution of jets initiated by a massive quark in $e^+e^-$
collisions.  In that case, the RGE consistency provided important constraints on
the structure of the factorization theorem at scales below the heavy-quark mass.

\begin{figure*}[t!]
\centering
\includegraphics[scale=0.75]{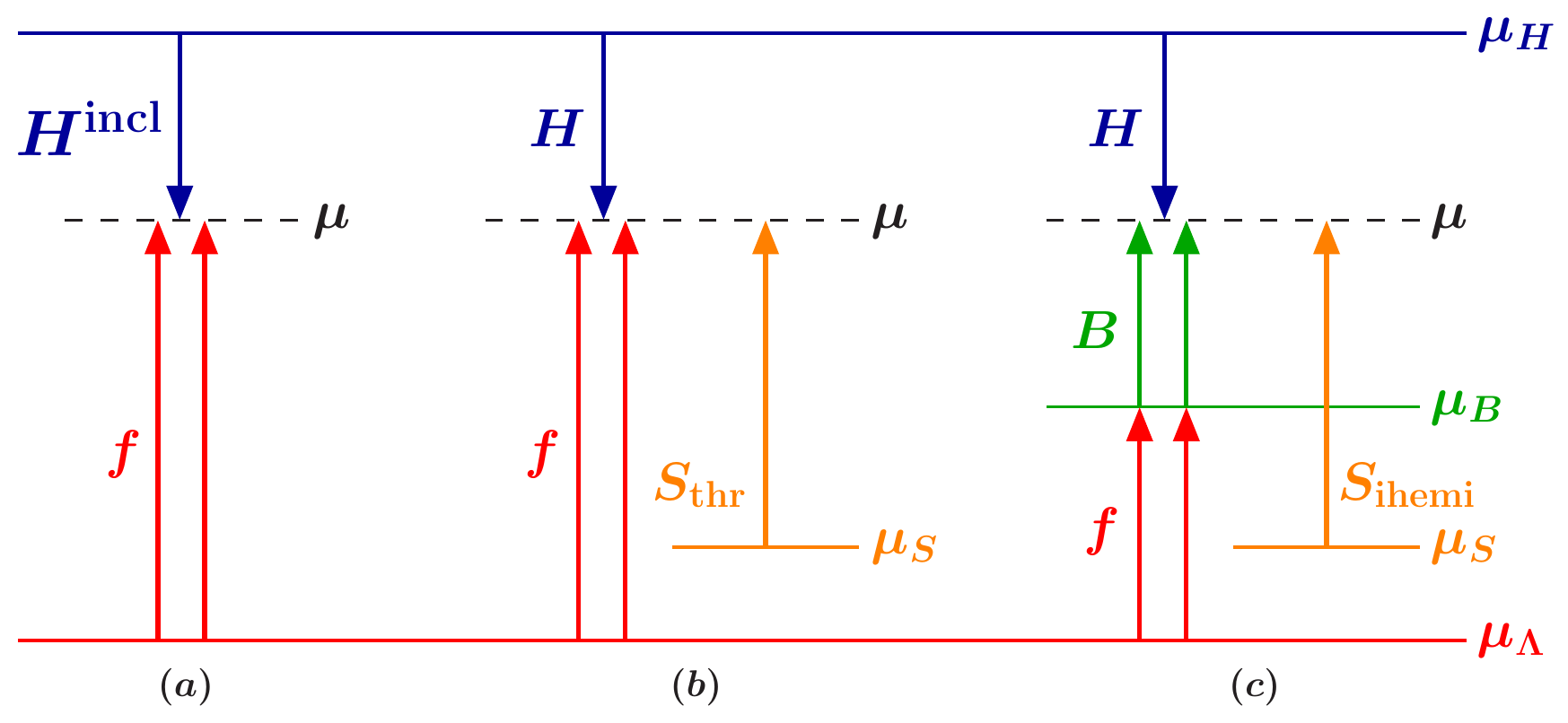}
\caption[RGE running for inclusive, threshold and isolated Drell-Yan.]{RGE running for different Drell-Yan scenarios. Case (a) corresponds to
  the inclusive case. Case (b) corresponds to the threshold case, where the kinematics
  forces all hadrons in the final state to be soft.  Case (c) corresponds to the
  isolated case. Here, the PDFs freeze out at the intermediate beam scale
  $\mu_B$, above which they are replaced by beam functions.}
\label{fig:running_drellyan}
\end{figure*}

In inclusive Drell-Yan, the hard functions $H^\incl_{ij}$ are sensitive to the
scale $\mu_H \simeq Q$ of the hard interaction, and the proton mass defines a
low scale $\mu_\Lambda \simeq 1\GeV \gtrsim \lqcd$ (which is still
large enough so perturbation theory can be applied for the PDF evolution). The measurement of $q^2$
and $Y$ in this case does not introduce additional scales, and thus does not
influence the structure of the logarithms. Thus, we have the hierarchy
$\mu_\Lambda \ll \mu_H$, and the large logarithms are $L =
\ln(\mu_\Lambda/\mu_H)$. Here, only single-logarithmic series, $(\alpha_s L)^k$,
are generated at higher orders in perturbation theory. The logarithms are
factorized as $\ln(\mu/\mu_H) + \ln(\mu_\Lambda/\mu)$ in the factorization
theorem in \eq{DYincl} and may then be resummed.  The general form of the
running is pictured in \fig{running_drellyan}(a). The logarithms
$\ln(\mu_\Lambda/\mu)$ are summed by evolving the PDFs $f_i(\xi_a, \mu)$ and
$f_j(\xi_b, \mu)$ from $\mu_\Lambda$ up to the common scale $\mu$. The inclusive
hard function, $H^\incl(x_a/\xi_a,x_b/\xi_b,q^2,\mu)$, is evolved from $\mu_H$
down to $\mu$, summing the logarithms $\ln(\mu/\mu_H)$. The choice of $\mu$ is
arbitrary.  Taking $\mu\simeq\mu_H$ corresponds to only running the PDFs up,
while for $\mu\simeq \mu_\Lambda$ only $H^\incl$ runs down. The equivalence of
these two choices implies that $H^\incl$ must be convoluted with the two PDFs
and exhibit a factorized structure for logarithms in the $a$ and $b$ variables.

For threshold Drell-Yan, the kinematic restrictions only allow soft radiation in
the final state. This induces additional large logarithms $\ln(1-\tau)$.  These
can be written in terms of a ratio of scales $\ln(\mu_S/\mu_H)$, where the soft
scale $\mu_S \simeq Q (1-\tau)$ is another important scale in the analysis. The
logarithms $L = \ln(\mu_S/\mu_H)$ appear as double-logarithmic series $(\alpha_s
L^2)^k$ in the cross section.  In the threshold factorization theorem in
\eq{DYendpt}, these double logarithms can be summed by evolving the PDFs and the
threshold soft and hard functions, $S_\thr$ and $H$, to a common scale $\mu$, as
shown in \fig{running_drellyan}(b).  Since $\xi_{a,b}\to 1$, the logarithms
$\ln(1-\xi_a)$ and $\ln(1-\xi_b)$ are also large. The RGE for the PDFs must be
expanded, and the result sums a double-logarithmic series of $\ln^2(1-\xi)$
terms.  The threshold soft function sums double logarithms $\ln^2(\mu/\mu_S)$
between $\mu_S$ and $\mu$, while the threshold hard function sums double
logarithms $\ln^2(\mu/\mu_H)$ between $\mu_H$ and $\mu$.  The RG
equations are
\begin{align}\label{eq:RGEendpt}
\mu\,\frac{\df}{\df\mu} H(q^2, \mu)
&= \gamma_H(q^2,\mu)\, H(q^2,\mu)
\,,\nn\\
\mu\,\frac{\df}{\df\mu} f_i(\xi,\mu)
&= \int\!\frac{\df\xi'}{\xi'}\, P^{\rm expanded}_{ii}\Bigl(\frac{\xi}{\xi'}, \mu\Bigr)\, f_i(\xi',\mu)
\,,\\\nn
\mu\, \frac{\df}{\df\mu} S_\thr(k,\mu)
&= \int\!\df k_s'\, \gamma_{S_\thr}(k - k',\mu)\, S_\thr(k',\mu)
\,.\end{align}
The consistency of the RGE at the scale $\mu$ shown in \fig{running_drellyan}(b)
implies that the double logarithms in $f_i$, $f_j$, and $S_\thr$ combine in such
a way that the RGE of the convolution $f_i f_j\otimes S_\thr$ is identical to
that of $H$, and hence only depends on $q^2$.

For isolated Drell-Yan, the kinematic restrictions allow both soft and collinear
initial-state radiation, and induce an invariant-mass scale for each beam
function, $\mu_B^2 \simeq x_a \ECM B_{a}^+ $ and $\mu_B^2 \simeq x_b \ECM B_b^+
$, and a soft scale $\mu_S \simeq B_{a,b}^+ $. For simplicity, we use a common
scale $\mu_B$ for both beam functions in our discussion here. (Since the
evolution of the two beam functions is independent, one can just as easily
implement two independent beam scales.) As we saw in \sec{DYfact}, at
partonic center-of-mass energies of a hundred GeV to a few TeV there is a large
hierarchy between the different scales, $\mu_\Lambda \ll \mu_S \ll \mu_B \ll
\mu_H$, and correspondingly large double and single logarithms of the ratios of
these scales.  The RGE running for this case is shown in
\fig{running_drellyan}(c).  Here, the PDFs are not restricted to their
endpoints, so their evolution is given by \eq{fevolution}, which involves the
unexpanded and nondiagonal $P_{ij}(\xi/\xi')$ and sums single logarithms,
$(\alpha_s L)^k$. For each $f_j$ this evolution joins at $\mu = \mu_B$ with the
Wilson coefficients $\cI_{ij}$ in the beam function factorization $B_i =
\cI_{ij}\otimes f_j$ of \eq{B_fact}. The $\cI_{ij}$ cancel the $\xi$-dependent
evolution of $f_j$, and turn it into the $t$-dependent evolution of $B_i$, which
sums a double-logarithmic series. The objects meeting at the common scale $\mu$
in \fig{running_drellyan}(c) are the hard function (same as threshold case) and the beam and soft functions,
\begin{align}
\mu\,\frac{\df}{\df\mu} B_i(t, x, \mu)
&= \int\!\df t'\, \gamma_B^i(t - t',\mu)\, B_i(t',x,\mu)
\,,\nn\\
\mu\,\frac{\df}{\df\mu} S_\hemiin(k_a^+, k_b^+,\mu)
&= \int\!\df k_a'\,\df k_b'\, S_\hemiin(k_a', k_b',\mu)
\gamma_{S_\hemiin}(k_a^+ - k_a', k_b^+ - k_b',\mu)\,
\,.\end{align}
The consistency of the RGE at $\mu$ now implies that the double-logarithmic running
in the different variables for $B_i$, $B_j$, and $S_\hemiin$ cancels such that the
convolution $B_i B_j \otimes S_\hemiin$ has an RGE identical to $H$, which only depends
on $q^2$.  (A detailed discussion of this consistency can be found in
Ref.~\cite{Fleming:2007xt} for the analogous case of two jet functions and the
final-state hemisphere soft function, $J J\otimes S_\hemiout$.) It is important that this
cancellation would not be possible if we tried to replace $B_i$ by $f_i$ in the
isolated factorization theorem.  Given the type of double logarithms in
the cross section, the single logarithms summed by the PDFs at generic $x$
cannot combine with the double logarithms in $S_\hemiin$ to
give a result in agreement with the double logarithms in $H$. Thus, the structure
of double logarithms necessitates the presence of beam functions in the
isolated factorization theorem.

By the same argument we can conclude that for all processes involving a
threshold-type hard function $H$ with double logarithms, and with $x_{a,b}$ away
from one, the description of the initial-state radiation will require beam
functions $B_i$.  This includes all situations where $H$ is the square of Wilson
coefficients of SCET operators, $H=\sum_k\abs{C_k}^2$ (for example when the
energetic partons in the hard collision all have distinct collinear directions).
In particular, the theoretical description of any threshold process with $x\to
1$ can be extended to a factorization theorem for the respective isolated case
with $x$ away from one. This is achieved by adding variables $B_{a,b}^+$,
replacing the PDFs by beam functions, and replacing the threshold soft function
by an appropriate soft function for the isolated case.

Thus, beam functions are quite prevalent for cross sections that one may wish to
study at the LHC. In situations where the hadronic final state is constrained
with variables that are more complicated than $B_{a,b}^+$, one
generically expects to find different beam functions and different soft
functions encoding these constraints.  This extension is analogous to how
the choice of jet algorithm modifies the definition of the jet and soft functions
for central jets produced by the hard collision~\cite{Bauer:2008jx}.  Even with
this generalization, the beam and soft functions will both
sum double-logarithmic series, and we expect that the factorization relating the beam
function to the PDFs will carry through, just with different Wilson coefficients
$\cI_{ij}$.


\chapter{Beam Functions}
\label{ch:beamf}

We discuss several formal aspects of the quark and gluon beam functions in this chapter, which were first reported in Ref.~\cite{Stewart:2010qs}. The beam functions are defined in \sec{definition} as matrix elements of operators in SCET. We include a brief review of the necessary SCET ingredients. In \sec{B_RGE} we work out the renormalization of the beam functions and show they have the same RGE as jet functions to all-orders in perturbation theory. We relate beam functions to PDFs in \sec{OPE} by performing an OPE, and calculate the tree-level matching coefficients in \sec{tree_match}. Finally, in \sec{beamT} we discuss the analytic structure of the beam functions and derive a relationship with matrix elements of time-ordered products of fields. This allows us to use the usual Feynman rules when we calculate the beam functions in the next chapters. 

\section{Definition}
\label{sec:definition}

In this section we discuss the definition of the quark and gluon beam functions in terms of matrix elements of operators in SCET, and compare them to the corresponding definition of the PDF. The operator language will be convenient to elucidate the renormalization structure and relation to jet functions in the following section. 

We first discuss some SCET ingredients that are relevant later on. We introduce light-cone vectors $n^\mu$ and $\bn^\mu$ with $n^2 = \bn^2 = 0$ and $n\cdot\bn = 2$ that are used to decompose four-vectors into light-cone coordinates $p^\mu = (p^+, p^-, p^\mu_\perp)$, where $p^+ = n\cdot p$, $p^- = \bn\cdot p$ and $p_\perp^\mu$ contains the components perpendicular to $n^\mu$ and $\bn^\mu$.

In SCET, the momentum $p^\mu$ of energetic collinear particles moving close to the $n$ direction is separated into large and small parts
\begin{equation}
p^\mu = \lp^\mu + p_r^\mu = \bn\cdot\lp\, \frac{n^\mu}{2} + \lp_{n\perp}^\mu + p_r^\mu
\,.\end{equation}
The large part $\lp^\mu = (0, \lp^-, \lp_{n\perp})$ has components $\lp^- = \bn\cdot\lp$ and $\lp_{n\perp} \sim \la \lp^-$, and the small residual piece $p_r^\mu = (p_r^+, p_r^-, p_{r\perp}^\mu) \sim \lp^-(\la^2, \la^2, \la^2)$ with $\la \ll 1$. The corresponding $n$-collinear quark and gluon fields are multipole expanded (with expansion parameter $\la$). This means particles with different large components are described by separate quantum fields, $\xi_{n,\lp}(y)$ and $A_{n,\lp}(y)$, which are distinguished by explicit momentum labels on the fields (in addition to the $n$ label specifying the collinear direction). We use $y$ to denote the position of the fields in the operators to reserve $x$ for the parton momentum fractions.

Interactions between collinear fields cannot change the direction $n$ but change the momentum labels to satisfy label momentum conservation. Since the momentum labels are changed by interactions, it is convenient to use the short-hand notations
\begin{equation} \label{eq:xi}
\xi_n(y) = \sum_{\lp \neq 0} \xi_{n,\lp}(y)
\,,\qquad
A_n^\mu(y) = \sum_{\lp \neq 0} A^\mu_{n,\lp}(y)
\,.\end{equation}
The sum over label momenta explicitly excludes the case $\lp^\mu = 0$ to avoid double-counting the soft degrees of freedom (described by separate soft quark and gluon fields). In practice when calculating matrix elements, this is implemented using zero-bin subtractions~\cite{Manohar:2006nz} or alternatively by dividing out matrix elements of Wilson lines~\cite{Lee:2006nr, Idilbi:2007ff}. The dependence on the label momentum is obtained using label momentum operators $\bnP_n$ or $\cP_{n\perp}^\mu$ which return the sum of the minus or perpendicular label components of all $n$-collinear fields on which they act.

The decomposition into label and residual momenta is not unique. Although the explicit dependence on the vectors $n^\mu$ and $\bn^\mu$ breaks Lorentz invariance, the theory must still be invariant under changes to $n^\mu$ and $\bn^\mu$ which preserve the power counting of the different momentum components and the defining relations $n^2 = \bn^2 = 0$, $n\cdot\bn = 2$. This reparametrization invariance (RPI)~\cite{Chay:2002vy, Manohar:2002fd} can be divided into three types. RPI-I and RPI-II transformations correspond to rotations of $n$ and $\bn$. We will mainly use RPI-III under which $n^\mu$ and $\bn^\mu$ transform as
\begin{equation} \label{eq:RPI}
n^\mu \to e^\alpha n^\mu
\,,\qquad
\bn^\mu \to e^{-\alpha} \bn^\mu
\,,\end{equation}
which implies that the vector components transform as $p^+ \to e^\alpha p^+$ and $p^- \to e^{-\alpha} p^-$. In this way, the vector $p^\mu$ stays invariant and Lorentz symmetry is restored within a cone about the direction of $n^\mu$. Since \eq{RPI} only acts in the $n$-collinear sector, it is not equivalent to a spacetime boost of the whole physical system.

We now define the following bare operators
\begin{align} \label{eq:tiop_def}
\tiop_q^\bare(y^-,\w)
&= e^{-\img\hp^+ y^-/2}\,
\bar{\chi}_n \Bigl(y^- \frac{n}{2}\Bigr) \frac{\bnslash}{2} \bigl[\delta(\w - \bnP_n)\chi_n(0)\bigr]
\,, \nn \\
\tiop_{\bar q}^\bare(y^-,\w)
&= e^{-\img\hp^+ y^-/2}\,
\tr \Bigl\{\frac{\bnslash}{2} \chi_n \Bigl(y^- \frac{n}{2}\Bigr) \bigl[\delta(\w - \bnP_n) \bar\chi_n(0)\bigr]\Bigr\}
\,, \nn \\
\tiop_g^\bare(y^-,\w)
& = -\w\, e^{-\img\hp^+ y^-/2}\,
 \cB_{n\perp\mu}^c \Bigl(y^- \frac{n}{2}\Bigr) \bigl[\delta(\w - \bnP_n) \cB_{n\perp}^{\mu c}(0) \bigr]
\,.\end{align}
Their renormalization will be discussed in the next section. The corresponding renormalized operators are denoted as $\tiop_i(y^-, \w, \mu)$ and are defined in \eq{op_ren_pos} below.
Here, $\hp^+$ is the momentum operator of the residual plus momentum and acts on everything to its right. The overall phase is included such that the Fourier-conjugate variable to $y^-$ corresponds to the plus momentum of the initial-state radiation, see \eq{oq_FT} below. The operator $\delta(\w - \bnP_n)$ only acts inside the square brackets and forces the total sum of the minus labels of all fields in $\chi_n(0)$ and $\cB_{n\perp}(0)$ to be equal to $\w$. The color indices of the quark fields are suppressed and summed over, $c$ is an adjoint color index that is summed over, and the trace in $\tiop_{\bar{q}}$ is over spin. The operators are RPI-III invariant, because the transformation of the $\delta(\w - \bnP_n)$ is compensated by that of the $\bnslash$ in $\tiop_{q,\bar{q}}$ and the overall $\w$ in $\tiop_g$.

The fields
\begin{equation} \label{eq:chiB}
\chi_n(y) = W_n^\dagger(y)\, \xi_n(y)
\,,\qquad
\cB_{n\perp}^\mu = \frac{1}{g} \bigl[W_n^\dagger(y)\, \img D_{n\perp}^\mu W_n(y) \bigr]
\,,\end{equation}
with $\img D_{n\perp}^\mu = \cP^\mu_{n\perp} + g A^\mu_{n\perp}$, are composite SCET fields of $n$-collinear quarks and gluons.
In \eq{tiop_def} they are at the positions $y^\mu = y^- n^\mu/2$ and $y^\mu = 0$. The Wilson lines
\begin{equation} \label{eq:Wn}
W_n(y) = \biggl[\sum_\text{perms} \exp\Bigl(-\frac{g}{\bnP_n}\,\bn\sdt A_n(y)\Bigr)\biggr]
\end{equation}
are required to make $\chi_n(y)$ and $\cB_{n\perp}^\mu(y)$ gauge invariant with respect to collinear gauge transformations~\cite{Bauer:2000yr, Bauer:2001ct}. They are Wilson lines in label momentum space consisting of $\bn\sdt A_n(y)$ collinear gluon fields. They sum up arbitrary emissions of $n$-collinear gluons from an $n$-collinear quark or gluon, which are $\ord{1}$ in the SCET power counting. Since $W_n(y)$ is localized with respect to the residual position $y$, $\chi_n(y)$ and $\cB_{n\perp}^\mu(y)$ are local operators for soft interactions. The fields in \eqs{tiop_def}{chiB} are those after the field redefinition~\cite{Bauer:2001yt} decoupling soft gluons from collinear particles. Thus at leading order in the power counting these collinear fields do not interact with soft gluons through their Lagrangian and no longer transform under soft gauge transformations. Hence, the operators in \eq{tiop_def} are gauge invariant under both soft and collinear gauge transformations. The soft interactions with collinear particles are factorized into a soft function, which is a matrix element of soft Wilson lines (see \subsec{soft_coll_fact}).

Note that our fields in \eq{tiop_def} have continuous labels and hence are not the standard SCET fields with discrete labels. They only depend on the minus coordinate, $y^-$, corresponding to the residual plus momentum, $p_r^+$, and not a full four-vector $y^\mu$. As discussed in detail in the derivation of the factorization theorem in \ch{fact}, it is convenient to absorb the residual minus and perpendicular components into the label momenta which then become continuous variables. For example, for the minus momentum (suppressing the perpendicular dependence)
\begin{align}
\sum_{\lp^-}\, e^{-\img \lp^- y^+/2} \chi_{n,\lp^-}(y^-, y^+)
&= \sum_{\lp^-} \int\!\df p_r^-\,e^{-\img (\lp^- + p_r^-) y^+/2} \chi_{n,p^-}(y^-)
\nn\\
&= \int\!\df p^-\,e^{-\img p^- y^+/2} \chi_{n, p^-}(y^-)
\,.\end{align}
In this case, $W_n(y^-n/2)$ can also be written in position space where all gluon fields sit at the same residual minus coordinate, $y^-$, and are path ordered in the plus coordinate (corresponding to the label minus momentum) from $y^+$ to infinity.

Next, we introduce the Fourier-transformed operators
\begin{equation} \label{eq:op_FT}
\op_i^\bare(\w b^+, \w)
= \frac{1}{2\pi} \int \! \frac{\df y^-}{2\abs{\w}}\, e^{\img b^+ y^-/2}\, \tiop_i^\bare(y^-,\w)
\,.\end{equation}
For example, for the quark operator
{\allowdisplaybreaks[0]
\begin{align} \label{eq:oq_FT}
\op_q^\bare(\w b^+, \w)
&= \frac{1}{2\pi} \int \! \frac{\df y^-}{2\abs{\w}}\, e^{\img (b^+ - \hp^+) y^-/2}
\Bigl(e^{\img\hp^+ y^-/2} \bar \chi_n(0) e^{-\img\hp^+ y^-/2}\Bigr) \frac{\bnslash}{2}
\bigl[\delta(\w - \bnP_n) \chi_n(0)\bigr]
\nn\\
&= \bar \chi_n(0)\, \delta(\w b^+ - \w\hp^+)\, \frac{\bnslash}{2} \bigl[\delta(\w - \bnP_n) \chi_n(0)\bigr]
\,.\end{align}}
In the first step we used residual momentum conservation to shift the position of the field. Here we see that the overall phase in \eq{tiop_def} allows us to write the $b^+$ dependence in terms of $\delta(\w b^+ - \w\hp^+)$, which means that $b^+$ measures the plus momentum of any intermediate state that is inserted between the fields. For the corresponding renormalized operators $\op_i(\w b^+, \w, \mu)$ see \eq{op_ren} below.

We divide by $\abs{\w}$ in \eq{op_FT} to make the integration measure of the Fourier transform
RPI-III invariant. (Taking the absolute value $\abs{\w}$ ensures that the definition of the Fourier transform does not depend on the sign of $\w$.) As a result, the Fourier-transformed operators are still RPI-III invariant and only depend on $b^+$ through the RPI-III invariant combination $t = \w b^+$. The beam functions are defined as the proton matrix elements of the renormalized operators $\op_i(t, \w,\mu)$,
\begin{equation} \label{eq:B_def}
B_i(t, x = \w/P^-,\mu) = \Mae{p_n(P^-)}{\theta(\w) \op_i(t,\w,\mu)}{p_n(P^-)}
\,.\end{equation}
The matrix elements are always averaged over proton spins, which we suppress in our notation. Note that part of the definition in \eq{B_def} is the choice of the direction $n$ such that the proton states have no perpendicular momentum, $P^\mu = P^- n^\mu/2$, which is why we write $\ket{p_n(P^-)}$. By RPI-III invariance, the beam functions can then only depend on the RPI-III invariant variables $t = \w b^+$ and $x = \w/P^-$. The restriction $\theta(\w)$ on the right-hand side of \eq{B_def} is included to enforce that the $\chi_n(0)$, $\bar\chi_n(0)$, or $\cB_{n\perp}(0)$ fields annihilate a quark, antiquark, or gluon out of the proton, as we discuss further at the start of \sec{beamT}.

The definition of the beam functions can be compared with that of the quark and gluon PDFs. In SCET, the PDFs are defined~\cite{Bauer:2002nz} in terms of the RPI-III invariant operators
\begin{align} \label{eq:oq_def}
\oq^\bare_q(\w')
&= \theta(\w')\, \bar{\chi}_n(0) \frac{\bnslash}{2} \bigl[\delta(\w' - \bnP_n) \chi_n(0)\bigr]
\,, \nn \\
\oq^\bare_{\bar q}(\w')
&= \theta(\w')\, \tr \Bigl\{\frac{\bnslash}{2} \chi_n(0) \bigl[\delta(\w' - \bnP_n) \bar\chi_n(0)\bigr] \Bigr\}
\,, \nn \\
\oq^\bare_g(\w')
& = -\w'\theta(\w')\, \cB_{n\perp\mu}^c(0) \bigl[\delta(\w' - \bnP_n) \cB_{n\perp}^{\mu c}(0) \bigr]
\,,\end{align}
as the proton matrix elements of the corresponding renormalized operators $\oq_i(\w', \mu)$ defined in \eq{oq_ren} below,
\begin{equation} \label{eq:f_def_SCET}
f_i(\w'/P^-,\mu) = \Mae{p_n(P^-)}{\oq_i(\w',\mu)}{p_n(P^-)}
\,.\end{equation}
By RPI-III invariance, the PDFs can only depend on the momentum fraction $\xi =
\w'/P^-$. Beyond tree level $\xi$ or $\w'$ are not the same as $x$ or $\w$,
which is why we denote them differently.  Without the additional $\theta(\w')$
in the operators in \eq{oq_def} the quark and anti-quark PDFs would combine into
one function, with the quark PDF corresponding to $\w>0$ and the antiquark PDF to
$\w<0$. We explicitly separate these pieces to keep analogous definitions for
the PDFs and beam functions.

It is important to note that the collinear fields in \eq{oq_def} do not require zero-bin subtractions, because as is well-known, the soft region does not contribute to the PDFs. If one makes the field redefinitions $\xi_n\to Y \xi_n$ and $A_n \to Y A_n Y^\dagger$ to decouple soft gluons, then the soft Wilson lines $Y$ cancel in \eq{oq_def}. Equivalently, if the fields in \eq{oq_def} include zero-bin subtractions then the subtractions will cancel in the sum of all diagrams, just like the soft gluons.
(This is easy to see by formulating the zero-bin subtraction as a field redefinition~\cite{Lee:2006nr} analogous to the soft one but with Wilson lines in a different light-cone direction.)
In contrast, the collinear fields in the beam function operator in \eq{tiop_def} must include zero-bin subtractions. We will see this explicitly at one loop in our PDF and beam function calculations in \ch{quark}.

The SCET definitions of the PDFs are equivalent to the standard definition in terms of full QCD quark fields $\psi$ in position space. For example, the quark PDF in QCD is defined as~\cite{Collins:1981uw}
\begin{equation} \label{eq:f_def_QCD}
f_q(\w'/P^-, \mu) = \theta(\w') \int\! \frac{\df y^+}{4\pi}\,
  e^{-\img \w' y^+/2}
  \MAe{p_n(P^-)}{\Bigl[\bar\psi \Bigl(y^+\frac{\bn}{2}\Bigr)
   W\Bigl(y^+\frac{\bn}{2},0\Bigr) \frac{\bnslash}{2} \psi(0) \Bigr]_\mu}{p_n(P^-)}
\,.\end{equation}
The square brackets denote the renormalized operator.
Here, the fields are separated along the $\bn$ direction and the lightlike Wilson line $W(y^+\bn/2,0)$ is required
to render the product of the fields gauge invariant. The relation to the SCET definition is that the SCET fields in \eq{oq_def} (without zero-bin subtractions) involve a Fourier transform of $\psi$ in $y^+$ to give the conjugate variable $\w'$. The corresponding Wilson lines in \eq{f_def_SCET} are precisely the $W_n$ contained in the definitions of $\chi_n$ and $\cB_{n\perp}^\mu$. Hence, the QCD and SCET definitions of the PDF are equivalent (provided of course that one uses the same renormalization scheme, which we do).

Comparing to \eq{tiop_def}, the difference between the beam functions and PDFs is that for the beam functions the fields are additionally separated along the $n$ light-cone, with a large separation $y^- \gg y^+$ corresponding to the
small momentum $b^+ \ll \w$. Thus, formulating equivalent definitions of the beam functions directly in
QCD would be more challenging, as it would require QCD fields that are simultaneously
separated in the $n$ and $\bn$ directions. For this case, it is not clear a priori how to
obtain an unambiguous gauge-invariant definition, because Wilson lines connecting the fields along different paths are not equivalent.
This ambiguity is resolved in SCET, where the multipole expansion distinguishes the different scales and divides the possible gauge transformations into global, collinear, and soft transformations, allowing one to treat the separations along the two orthogonal light-cones independently. The large $y^-$ separation corresponds to soft Wilson lines and soft gauge transformations that are independent from collinear gauge transformations corresponding to the small $y^+$ dependence. As already mentioned, the operators in \eq{tiop_def} are separately gauge invariant under both types of gauge transformations.

\section{Renormalization and RGE}
\label{sec:B_RGE}

The beam functions and PDFs are defined as the matrix elements of renormalized operators. The renormalization of the operators immediately yields that of the functions defined by their matrix elements. In this section we derive the RG equations and show that the anomalous dimensions of the beam and jet functions are the same to all orders in perturbation theory.

We start by considering the known renormalization of the PDF, but in the SCET operator language. The renormalized PDF operators are given in terms of the bare operators in \eq{oq_def} as
\begin{equation} \label{eq:oq_ren}
\oq_i^\bare(\w) = \sum_j \int\! \frac{\df \w'}{\w'}\, Z^f_{ij}\Big(\frac{\w}{\w'},\mu\Big) \oq_j(\w',\mu)
\,.\end{equation}
In general, operators with different $i$ and $\w$ can (and will) mix into each other, so the renormalization constant $Z^f_{ij}(\w/\w',\mu)$ is a matrix in $i,j$ and $\w,\w'$. RPI-III invariance then restricts the integration measure to be $\df\w'/\w'$ and $Z^f_{ij}(\w/\w',\mu)$ to only depend on the ratio $\w/\w'$. Hence, the form of \eq{oq_ren} is completely specified by the SCET symmetries. The $\mu$ independence of the bare operators $\oq_i^\bare(\w)$ yields an RGE for the renormalized operators in $\overline{\mathrm{MS}}$
\begin{align} \label{eq:oq_RGE}
\mu\frac{\df}{\df\mu} \oq_i(\w,\mu)
&= \sum_j \int\! \frac{\df \w'}{\w'}\, \gamma^f_{ij}\Bigl(\frac{\w}{\w'},\mu\Bigr)\, \oq_j(\w',\mu)
\,, \nn \\
\gamma^f_{ij}(z,\mu) &= -\sum_k \int\! \frac{\df z'}{z'}\, (Z^f)_{ik}^{-1}\Bigl(\frac{z}{z'},\mu\Bigr)\, \mu\frac{\df}{\df\mu} Z^f_{kj}(z',\mu)
\,,\end{align}
where the inverse $(Z^f)_{ik}^{-1}(z, \mu)$ is defined as
\begin{equation}
\sum_k  \int\! \frac{\df z'}{z'}\, (Z^f)_{ik}^{-1}\Bigl(\frac{z}{z'},\mu\Bigr) Z^f_{kj}(z',\mu)
= \delta_{ij}\, \delta(1-z)
\,.\end{equation}
Taking the proton matrix element of \eq{oq_RGE} yields the RGE for the PDFs
\begin{equation} \label{eq:PDF_RGE}
\mu\frac{\df}{\df\mu} f_i(\xi,\mu)
= \sum_j\int\! \frac{\df\xi'}{\xi'}\, \gamma^f_{ij}\Bigl(\frac{\xi}{\xi'},\mu\Bigr)\, f_j(\xi',\mu)
\,.\end{equation}
The solution of this RGE can be written in terms of an evolution function $U^f$ which acts on the initial PDF $f_j(\xi',\mu_0)$ and takes it to $f_i(\xi,\mu)$,
\begin{equation} \label{eq:Uf_def}
f_i(\xi,\mu) = \int\! \frac{\df\xi'}{\xi'}\, U^f_{ij}\Bigl(\frac{\xi}{\xi'},\mu,\mu_0\Bigr) f_j(\xi',\mu_0)
\,.\end{equation}

From \eq{PDF_RGE} we can identify the anomalous dimensions $\gamma_{ij}^f(z)$ in terms of the
QCD splitting functions. For example, in dimensional regularization in the $\overline{\mathrm{MS}}$ scheme, the one-loop anomalous dimensions for the quark PDF are the standard ones
\begin{equation} \label{eq:gammaf}
\gamma_{qq}^f(z, \mu) = \frac{\alpha_s(\mu)C_F}{\pi}\, \theta(z) P_{qq}(z)
\,,\qquad
\gamma_{qg}^f(z, \mu) = \frac{\alpha_s(\mu)T_F}{\pi}\, \theta(z) P_{qg}(z)
\,,\end{equation}
with the $q\to qg$ and $g\to q\bar{q}$ splitting functions
\begin{align} \label{eq:Pqq_def}
P_{qq}(z)
&= \cL_0(1-z)(1+z^2) + \frac{3}{2}\,\delta(1-z)
= \biggl[\theta(1-z)\frac{1+z^2}{1-z} \biggr]_+
\,,\nn\\
P_{qg}(z) &= \theta(1-z)\bigl[(1-z)^2+ z^2\bigr]
 \,.\end{align}
The plus distribution $\cL_0(x) = [\theta(x)/x]_+$ is defined as usual, see \eq{plusdef}. For later convenience we do not include the overall color factors in the definitions in \eq{Pqq_def}.

We now go through an analogous discussion for the beam functions. The renormalized operators $\tiop_i(y^-, \w, \mu)$ are given in terms of the bare operators in \eq{tiop_def} by
\begin{equation} \label{eq:op_ren_pos}
\tiop_i^\bare(y^-, \w) = \tZ_B^i\Bigl(\frac{y^-}{2\w}, \mu\Bigr) \tiop_i(y^-,\w,\mu)
\,,\end{equation}
where $\tZ_B^i(y^-/2\w,\mu)$ is the position-space renormalization constant. In \app{renorm}, we give an explicit proof that the beam function renormalization is multiplicative in this way to all orders in perturbation theory.
The underlying reason is that the renormalization of the theory should preserve locality, so renormalizing the nonlocal beam function operator should not affect the $y^-$ separation between the fields. For example, mixing between operators with different $y^-$ would destroy locality at distance scales within the validity range of the effective theory. RPI-III invariance then implies that $\tZ_B^i$ can only depend on the ratio $y^-/2\w$ (the factor of $1/2$ is for convenience).
In principle, one might think there could also be mixing between operators with different $i$ or $\w$ in \eq{op_ren_pos} [as was the case for the PDFs in \eq{oq_ren}]. Our derivation in \app{renorm} shows that this is not the case.

Taking the Fourier transform of \eq{op_ren_pos} according to \eq{op_FT}, we find
\begin{align} \label{eq:op_ren}
\op_i^\bare(t, \w) &= \int\! \df t'\, Z_B^i(t - t', \mu)\, \op_i(t', \w, \mu)
\,, \nn \\
Z_B^i(t, \mu) &= \frac{1}{2\pi} \int\! \frac{\df y^-}{2\abs{\w}}\, e^{\img t y^-/2\w}\, \tZ_B^i \Bigl(\frac{y^-}{2\w}, \mu\Bigr)
\,.\end{align}
Since the bare operator is $\mu$ independent, taking the derivative with respect to $\mu$, we find the RGE for the renormalized operator
\begin{align} \label{eq:op_RGE}
\mu \frac{\df}{\df \mu} \op_i(t, \w, \mu) &= \int\! \df t' \gamma_B^i(t - t', \mu)\, \op_i(t', \w, \mu)
\,,\nn\\
\gamma_B^i(t,\mu) &= - \int\! \df t'\, (Z_B^i)^{-1}(t-t', \mu)\, \mu\frac{\df}{\df\mu} Z_B^i (t', \mu)
\,,\end{align}
where the inverse of $Z_B^i(t, \mu)$ is defined as usual,
\begin{equation}
\int\! \df t'\, (Z_B^i)^{-1}(t-t',\mu)\, Z_B^i(t',\mu) = \delta(t)
\,.\end{equation}

Taking the proton matrix element of \eq{op_RGE} we obtain the corresponding RGE for the beam function,
\begin{equation} \label{eq:B_RGE}
\mu \frac{\df}{\df \mu} B_i(t, x, \mu) = \int\! \df t'\, \gamma_B^i(t-t',\mu)\, B_i(t', x, \mu)
\,.\end{equation}
As discussed in \app{renorm}, to all orders in perturbation theory the anomalous dimension has the form
\begin{equation} \label{eq:gaB_gen}
\gamma_B^i(t, \mu)
= -2 \Gamma^i_\cusp(\alpha_s)\,\frac{1}{\mu^2}\cL_0\Bigl(\frac{t}{\mu^2}\Bigr)
+ \gamma_B^i(\alpha_s)\,\delta(t)
\,,\end{equation}
where $\cL_0(x) = [\theta(x)/x]_+$ is defined in \eq{plusdef}, $\Gamma_\cusp^i(\alpha_s)$ is the cusp anomalous dimension for quarks/antiquarks ($i = q$) or gluons ($i = g$), and $\gamma_B^i(\alpha_s)$ denotes the non-cusp part. Since there is no mixing between operators $\op_i(t, \w, \mu)$ with different $i$ or $\w$, the beam function RGE only changes the virtuality $t$ but not the momentum fraction $x$ and does not mix quark and gluon beam functions. By rescaling the plus distribution,
\begin{equation}
\frac{1}{\mu^2}\cL_0\Bigl(\frac{t}{\mu^2}\Bigr)
= \frac{1}{\mu_0^2}\cL_0\Bigl(\frac{t}{\mu_0^2}\Bigr) - 2\ln\frac{\mu}{\mu_0}\, \delta(t)
\end{equation}
we can see that $\gamma_B^i(t, \mu)$ has logarithmic $\mu$-dependence, which means that the RGE sums Sudakov double logarithms.

The solution of the RGE in \eq{B_RGE} with the form of the anomalous dimension in \eq{gaB_gen} is known~\cite{Balzereit:1998yf, Neubert:2004dd, Fleming:2007xt}. It takes the form
\begin{equation} \label{eq:Brun}
B_i(t,x,\mu) =  \int\! \df t'\, B_i(t - t',x,\mu_0)\, U_B^i(t',\mu_0,\mu)
\,,\end{equation}
where the evolution kernel can be written as~\cite{Ligeti:2008ac}
\begin{equation} \label{eq:UB}
U_B^i(t, \mu_0, \mu) = \frac{e^{K_B^i -\gamma_E\, \eta_B^i}}{\Gamma(1+\eta_B^i)}\,
\biggl[\frac{\eta_B^i}{\mu_0^2} \cL^{\eta_B^i} \Bigl( \frac{t}{\mu_0^2} \Bigr) + \delta(t) \biggr]
\,.\end{equation}
The distribution $\cL^\eta(x)$ is defined in \eq{plusdef}, and the RGE functions $K_B^i \equiv K_B^i(\mu_0, \mu)$ and $\eta_B^i \equiv \eta_B^i(\mu_0, \mu)$ are given in \eq{Brun_full}.

The SCET quark, antiquark, and gluon jet functions are given by~\cite{Bauer:2001yt, Fleming:2003gt}
\begin{align} \label{eq:J_def}
&J_q(\w p^+\! + \w_\perp^2, \mu)
\\ & \qquad
= \frac{(2\pi)^2}{N_c} \!\int\! \frac{\df y^-}{2\abs{\w}}\, e^{\img p^+ y^-/2}\,
   \tr\MAe{0}{ \Bigl[ \frac{\bnslash}{2} \chi_n \Bigl(y^- \frac{n}{2}\Bigr)
   \bigl[\delta(\w + \bnP_n)\delta^2(\w_\perp + \cP_{n\perp}) \bar \chi_n(0) \bigr] \Bigr]_\mu}{0}
\,,\nn\\
&J_{\bar q}(\w p^+\! + \w_\perp^2, \mu)
\nn\\ & \qquad
= \frac{(2\pi)^2}{N_c} \!\int\! \frac{\df y^-}{2\abs{\w}}\, e^{\img p^+ y^-/2}\,
   \MAe{0}{\Bigl[ \bar \chi_n \Bigl(y^- \frac{n}{2}\Bigr)\frac{\bnslash}{2}
   \bigl[\delta(\w + \bnP_n) \delta^2(\w_\perp + \cP_{n\perp}) \chi_n(0) \bigr] \Bigr]_\mu}{0}
\,, \nn
\end{align}
\begin{align}
&J_g(\w p^+\! + \w_\perp^2, \mu)
\nn\\\nn & \qquad
= -\frac{(2\pi)^2}{N_c^2 - 1}\, \w \!\int\!\frac{\df y^-}{2\abs{\w}}\, e^{\img p^+ y^-/2}
   \MAe{0}{ \Bigl[ \cB_{n\perp\mu}^c \Bigl(y^- \frac{n}{2}\Bigr)
   \bigl[ \delta(\w + \bnP_n) \delta^2(\w_\perp + \cP_{n\perp}) \cB_{n\perp}^{\mu c}(0) \bigr] \Bigr]_\mu}{0}
\,,\end{align}
where the notation $[\ldots]_\mu$ again denotes the renormalized operators.
Here, we used the same conventions as for the beam functions where the large label momenta $\w$ and $\w_\perp$ are continuous, so the only position dependence of the fields is in the minus component. RPI invariance requires that the jet function only depends on the total invariant mass of the jet, $p^2 = \w p^+ + \w_\perp^2$. When the jet function appears in a factorization theorem, the direction of the jet is either measured (e.g. by measuring the thrust axis in $e^+e^-\to 2$ jets) or fixed by kinematics (e.g. in $B\to X_s\gamma$ the jet direction is fixed by the direction of the photon) and $n$ is chosen along the jet direction, so one typically has $\w_\perp = 0$. Taking the vacuum matrix element of $\op_q^\bare(t,\w)$, we get
\begin{align}\label{eq:qjet}
&\frac{(2\pi)^2}{N_c}\Mae{0}{\op_q^\bare(-t,-\w)}{0}
\nn\\ & \qquad
= \frac{(2\pi)^2}{N_c}
\int\! \df^2 \w_\perp \frac{1}{2\pi} \int \! \frac{\df y^-}{2\abs{\w}}\, e^{\img t y^-/(2\w)}\,
  \MAe{0}{\bar \chi_n \Bigl(y^- \frac{n}{2}\Bigr) \frac{\bnslash}{2}\, \delta(\w + \bnP_n) \delta^2(\w_\perp - \cP_{n\perp}) \chi_n(0)}{0}
\nn \\ & \qquad
= \int\! \df^2 \vec\w_\perp J^\bare_{\bar q}(t - \vec\w_\perp^2)
\equiv \hJ^\bare_{\bar q}(t) = \hJ^\bare_q(t)
\,.\end{align}
In the last step we used that the quark and antiquark jet functions are the same. The analogous relation holds for the antiquark operator, $\op_{\bar q}(t,\w,\mu)$. The $\vec\w_\perp$ integral is bounded and does not lead to new UV divergences, because the jet function only has support for nonnegative argument, $0 < \vec\w_\perp^2 < t$, and $t$ is fixed. Similarly, for the gluon operator we have
\begin{equation}\label{eq:gjet}
\frac{(2\pi)^2}{N_c^2 - 1}\Mae{0}{\op_g^\bare(-t, -\w)}{0}
= \int\! \df^2 \vec\w_\perp J_g^\bare(t - \vec\w_\perp^2, \mu) \equiv \hJ_g^\bare(t)
\,.\end{equation}
The renormalization of $J_i^\bare(t)$ does not depend on the choice of $\w_\perp$ in \eq{J_def}. Since $\hJ_i^\bare(t)$ is simply an average over different choices for $\w_\perp$ it has the same renormalization. Hence $J_i(t, \mu)$ and $\hJ_i(t, \mu)$ have the same anomalous dimension,
{\allowdisplaybreaks[0]
\begin{align}
\mu\frac{\df}{\df\mu} \hJ_i(t, \mu)
&= \int\! \df^2 \vec\w_\perp\, \df s\, \gamma_J^i(t - \vec\w_\perp^2 - s, \mu)\, J_i(s,\mu)
= \int\! \df t' \gamma_J^i(t - t', \mu) \int\! \df^2 \vec\w_\perp\, J_i(t' - \vec\w_\perp^2,\mu)
\nn \\
&= \int\!\df t'\,\gamma_J^i(t - t', \mu)\, \hJ_i(t', \mu)
\,.\end{align}}
On the other, \eq{qjet} says that $\hJ_i^\bare(t)$ is a matrix element of the same operator as the beam function, whose general renormalization was given in \eq{op_ren}. We thus conclude that the beam and jet function anomalous dimensions are identical to all orders in perturbation theory,
\begin{equation} \label{eq:gaJgaB}
\gamma_J^i(t, \mu) = \gamma_B^i(t, \mu)
\,.\end{equation}
For the cusp part this result already follows from our explicit one-loop calculation, since $\Gamma^i_\cusp$ is universal and its coefficients are the same at one loop. Our one-loop result provides a cross check for the identity of the one-loop non-cusp part of the anomalous dimension, which agree. Furthermore, $\gamma_J^q(\alpha_s)$ and hence $\gamma_B^q(\alpha_s)$ can be obtained to three loops from Refs.~\cite{Moch:2004pa, Moch:2005id}, and for completeness the result is given in \app{pert}.

\section{Operator Product Expansion}
\label{sec:OPE}

The difference between the beam function operators in \eq{tiop_def} and the PDF operators in \eq{oq_def} is the additional separation in the $y^-$ coordinate between the fields. Hence, by performing an operator product expansion about the limit $y^-\!\to 0$ we can expand the renormalized operators $\tiop_i(y^-, \w, \mu)$ in terms of a sum over $\oq_i(\w', \mu)$,
\begin{equation} \label{eq:tiop_RGE}
\tiop_i(y^-, \w, \mu)
= \widetilde{J}_i\Bigl(\frac{y^-}{2\w},\mu\Bigr) 1
 + \sum_j \int\! \frac{\df \w'}{\w'}\,
  \widetilde{\cI}_{ij}\Bigl(\frac{y^-}{2\w},\frac{\w}{\w'},\mu \Bigr) \oq_j(\w',\mu)
+ \ORd{\frac{y^-}{\w}}
\,.\end{equation}
For completeness we included the identity operator on the right-hand side. The form of the matching coefficients $\widetilde{\cI}_{ij}$ and $\widetilde{J}_i$ is again constrained by RPI-III invariance so the structure of the OPE is completely determined by the SCET symmetries.

Fourier transforming both sides of \eq{tiop_RGE} with respect to $y^-$ we get
\begin{equation} \label{eq:op_OPE}
\op_i(t, \w, \mu) = \hJ_i(-t,\mu) 1 + \sum_j \int\! \frac{\df \w'}{\w'}\,
\cI_{ij}\Bigl(t,\frac{\w}{\w'},\mu \Bigr) \oq_j(\w',\mu) + \ORd{\frac{y^-}{\w}}
\,.\end{equation}
Taking the vacuum matrix element of both sides, and using $\mae{0}{\oq_j}{0} = 0$, we just get the coefficient of the identity operator on the right-hand side, which from \eqs{qjet}{gjet} is thus given by $\hJ_i(-t,\mu)$. Taking the proton matrix element of \eq{op_OPE} with $\w > 0$ according to \eq{B_def}, this first term drops out, because the jet functions only have support for $-\w>0$ (or alternatively because the corresponding diagrams are disconnected), and we obtain the OPE for the beam function
\begin{equation} \label{eq:beam_fact}
B_i(t,x,\mu)
= \sum_j \int \! \frac{\df \xi}{\xi}\, \cI_{ij}\Bigl(t,\frac{x}{\xi},\mu \Bigr) f_j(\xi,\mu)
  \biggl[1 + \ORd{\frac{\lqcd^2}{t}}\biggr]
\,.\end{equation}
For $B_g$ this equation was already derived in Ref.~\cite{Fleming:2006cd} using a moment-space OPE for the matrix element (modulo missing the mixing contribution from the quark PDF). The higher-order power corrections in \eq{beam_fact} must scale like $1/t$ and are therefore of $\ord{\lqcd^2/t}$ where $\lqcd^2$ is the typical invariant mass of the partons in the proton. (Equivalently, one can think of the scaling as $(\lqcd^2/\w)/b^+$ where $\lqcd^2/\w$ is the typical plus momentum of the parton in the proton.) They are given in terms of higher-twist proton structure functions. Since \eq{op_OPE} is valid for $t \gg \lqcd^2$, this also means that we can calculate the matching coefficients in perturbation theory at the beam scale $\mu_B^2 \simeq t$. This matching calculation is carried out in the usual way by computing convenient matrix elements of the operators on both sides of \eq{op_OPE} and extracting the matching coefficients from the difference. This is carried out at tree level in the next section, while the full one-loop matching calculation for the quark beam function is given in \ch{quark} and for the gluon beam function in \ch{gluon}. On the other hand, for $t\sim\lqcd^2$ the beam functions are nonperturbative and the OPE would require an infinite set of higher-twist proton structure functions. In this case, the beam functions essentially become nonperturbative $b^+$-dependent PDFs.

The physical interpretation of the beam function OPE in \eq{beam_fact} leads exactly to the physical picture shown in \fig{beam} and discussed in the introduction. At the beam scale $\mu_B\simeq t$, the PDFs are evaluated and a parton $j$ with momentum fraction $\xi$ is taken out of the proton. It then undergoes further collinear interactions, which are described by the perturbative Wilson coefficients $\cI_{ij}(t, z, \mu)$. By emitting collinear radiation it looses some of its momentum, and the final momentum fraction going into the hard interaction is $x < \xi$. In addition, the sum on $j$ indicates that there is a mixing effect from terms without large logarithms, e.g. the quark beam function gets contributions from the quark, gluon, and antiquark PDFs. For example, when an incoming gluon from the proton pair-produces, with the quark participating in the hard interaction and the antiquark going into the beam remnant, then this is a mixing of the gluon PDF into the quark beam function. These are the physical effects that would usually be described by the PDF evolution. The difference is that once we are above the beam scale these effects only cause non-logarithmic perturbative corrections, which means the parton mixing and $x$-reshuffling now appears in the matching, while the RG evolution of the beam function only changes $t$, as we saw above. In \ch{beam_plots} we will see that these matching corrections are still important numerically and must be taken into account. For example, since the gluon PDF at small $\xi$ is very large compared to the quark and antiquark PDFs, it still gives an important contribution to the quark and antiquark beam functions.

The consistency of the RGE requires that the $\mu$ dependence of the Wilson coefficients $\cI_{ij}(t, z, \mu)$ turns the RG running of the PDFs into the proper RG running of the beam functions. Taking the $\mu$ derivative of \eq{beam_fact} we find the evolution equation for the Wilson coefficients
\begin{align}
\mu\frac{\df}{\df\mu}\cI_{ij}(t,z,\mu)
= \sum_k \int\!\df t'\, \frac{\df z'}{z'}\, \cI_{ik}\Bigl(t - t', \frac{z}{z'},\mu\Bigr)\Bigl[\gamma_B^i(t', \mu)\, \delta_{kj} \delta(1 - z')
- \delta(t') \gamma^f_{kj}(z',\mu) \Bigr]
\,.\end{align}
The solution to this RGE can be easily obtained in terms of the evolution factors for the PDF and beam function in \eqs{Uf_def}{UB},
\begin{align}
\cI_{ij}(t,z,\mu) &= \sum_k\int\! \df t'\, \frac{\df z'}{z'}\,
  \cI_{ik}\Bigl(t - t',\frac{z}{z'},\mu_0\Bigr)\, U_B^i(t',\mu_0,\mu)\, U^f_{kj}(z',\mu_0,\mu)
\,.\end{align}
Hence as expected, the RGE running of $\cI_{ij}(t, z, \mu)$ cancels the running of the PDFs and adds in the running of the beam function.

\section{Tree-level Matching onto PDFs}
\label{sec:tree_match}

\begin{figure}[t]
\hfill%
\subfigure[]{\includegraphics[scale=0.75]{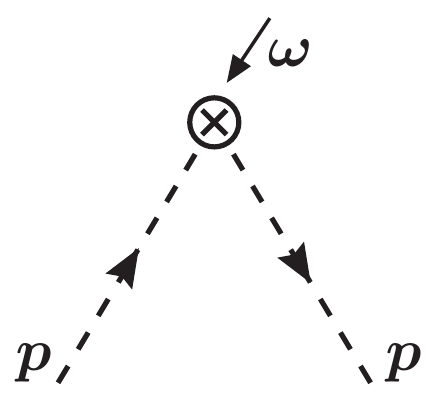}\label{fig:ftree}}%
\hfill%
\subfigure[]{\includegraphics[scale=0.75]{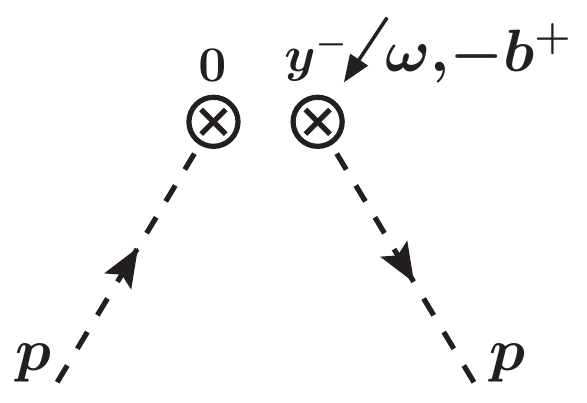}\label{fig:Btree}}%
\hspace*{\fill}%
\caption[Tree-level diagram for the quark PDF and the quark beam function.]{Tree-level diagram for the quark PDF (a) and the quark beam function (b).
For the latter, the $y^-$ coordinate separation in the operator is indicated
by drawing separated vertices for each field.}
\end{figure}

To illustrate the application of the OPE, we will calculate the Wilson
coefficients $\cI_{ij}$ at tree level, starting with $\cI_{qq}$. We can use any
external states for the computation of the Wilson coefficient as long as they
have nonzero overlap with our operator. Thus, we pick the simplest choice,
$n$-collinear quark and gluon states, $\ket{q_n(p)}$ and $\ket{g_n(p)}$, with
momentum $p^\mu = (p^+,p^-,0)$ where $p^->0$ is the large momentum. In the
following section we will use a small $p^+<0$ as an IR regulator, but otherwise
$p^+$ is set to zero. The tree-level diagrams with an external quark for the
quark PDF and beam function are shown in \figs{ftree}{Btree}. They give
\begin{align} \label{eq:op_tree}
\Mae{q_n(p)}{\oq_q(\w',\mu)}{q_n(p)}^\zero
&= \theta(\w')\,\bar u_n(p)\delta(\w'-p^-)\, \frac{\bnslash}{2} u_n(p)
= \theta(\w')\,\delta(1-\w'/p^-)
\,, \nn \\
\Mae{q_n(p)}{\op_q(t,\w,\mu)}{q_n(p)}^\zero
&=  \bar u_n(p)\,\delta(t)\,\delta(\w-p^-)\, \frac{\bnslash}{2} u_n(p)
= \delta(t) \, \delta(1 - \w/p^-)
\,.\end{align}
Here and in the following the superscript $(i)$ indicates the $\ord{\alpha_s^i}$ contribution. Note that the results in \eq{op_tree} are the same whether we use a state with fixed spin and color or whether we average over spin and color. Taking the matrix element of both sides of \eq{op_OPE} and using \eq{op_tree}, we can read off the tree-level matching coefficient
\begin{equation}
\cI_{qq}^\zero(t,z,\mu) = \cI_{\bar{q}\bar{q}}^\zero(t,z,\mu) = \delta(t)\, \delta(1-z)
\,.\end{equation}

Similarly, the tree-level results for the gluon PDF and beam function are
\begin{align} \label{eq:opg_tree}
\Mae{g_n(p)}{\oq_g(\w',\mu)}{g_n(p)}^\zero
&= \theta(\w')\,\delta(1-\w'/p^-)
\,, \nn \\
\Mae{g_n(p)}{\op_g(t,\w,\mu)}{g_n(p)}^\zero
&= -\w\, \varepsilon^*\sdt \varepsilon \, \delta(t)\, \delta(\w-p^-)
= \delta(t)\, \delta(1-\w/p^-)
\,,\end{align}
leading to
\begin{equation}
\cI_{gg}^\zero(t,z,\mu) = \delta(t)\, \delta(1-z)
\,.\end{equation}
Since at tree level the quark (gluon) matrix elements of the gluon
(quark) operators vanish,
\begin{align} \label{eq:qg_tree}
\Mae{g_n(p)}{\oq_g(\w',\mu)}{g_n(p)}^\zero  &= \Mae{q_n(p)}{\oq_q(\w',\mu)}{q_n(p)}^\zero = 0
\,,\nn\\
\Mae{g_n(p)}{\op_q(t,\w,\mu)}{g_n(p)}^\zero &= \Mae{q_n(p)}{\op_g(t,\w,\mu)}{q_n(p)}^\zero = 0
\,,\end{align}
we obtain
\begin{equation}
\cI_{qg}^\zero(t,z,\mu) = \cI_{gq}^\zero(t,z,\mu) = 0
\,.\end{equation}
To summarize, the complete tree-level results are
\begin{equation}
\cI_{ij}^\zero(t,z,\mu) = \delta_{ij} \delta(t)\, \delta(1-z)
\,,\qquad
B_i^\zero (t, x, \mu) = \delta(t) f_i(x, \mu)
\,.\end{equation}
The interpretation is simply that at tree level the parton taken out of the
proton goes straight into the hard interaction. However, even at tree level the
OPE already provides nontrivial information. From our general discussion we know
that the matching should be performed at the beam scale $\mu_B^2 \simeq t$ to avoid
large logarithms in the $\ord{\alpha_s}$ terms, and this determines the scale at
which the PDFs must be evaluated to be $\mu = \mu_B$.

\section{Analytic Structure and Time-Ordered Products}
\label{sec:beamT}

In this section we discuss the analytic structure of the beam functions. For the OPE matching calculation we want to calculate partonic matrix elements of $\op_q(t, \w, \mu)$. For this purpose it is convenient to relate the matrix elements of the products of fields in $\op_q(t, \w, \mu)$ to discontinuities of matrix elements of time-ordered products of fields, since the latter are easily evaluated using standard Feynman rules. For notational simplicity we only consider the quark operator $\op_q(t, \w)$ and suppress the spin indices and $\mu$ dependence. The discussion for the antiquark and gluon operators are analogous.

We are interested in the forward matrix element of $\op_q(t, \w)$ between some $n$-collinear state $\ket{p_n} \equiv \ket{p_n(p^+, p^-)}$ with large momentum $p^-$ and small residual momentum $p^+$. Inserting a complete set of states $\sum_X \ket{X}\bra{X}$, we get
{\allowdisplaybreaks[0]
\begin{align} \label{eq:Bkin}
\Mae{p_n}{\op_q(t, \w)}{p_n}
&= \sum_X \MAe{p_n}{\bar\chi_n(0) \frac{\bnslash}{2}\,\delta(t - \w\hp^+)}{X}
\Mae{X}{\bigl[\delta(\w - \bnP_n) \chi_n(0)\bigr]}{p_n}
\\\nn
&= \sum_X \delta(t - \w p_X^+)\, \delta(\w - p^- + p_X^-)
   \MAe{p_n}{\bar\chi_n(0) \frac{\bnslash}{2}}{X} \Mae{X}{\chi_n(0)}{p_n}
\,.\end{align}}
The $\delta(\w-\bnP_n)$ by definition only acts on the field inside the square bracket, returning its minus momentum, which by momentum conservation must be equal to the difference of the minus momenta of the external states. Since $\w = p^- - p_X^-$, requiring $\w > 0$ implies $p_X^- < p^-$. This means that the action of the field reduces the momentum of the initial state so it effectively annihilates a parton in the initial state $\ket{p_n}$. Similarly, for $\w < 0$ we would have $p_X^- > p^-$ and the field would effectively create an antiquark in $\bra{X}$. Also, since $\ket{X}$ are physical states, we have $p_X^\pm \geq 0$ so $\w \leq p^-$ and $t = \w p_X^+$ and $\w$ have the same sign.

Hence, for the beam function, where $\ket{p_n}\equiv\ket{p_n(P^-)}$ is the proton state, the restriction to $\w > 0$ in its definition, \eq{B_def}, enforces that we indeed take a quark out of the proton. (Note that $\w<0$ does not correspond to the anti-quark beam function.) Taking the states $\ket{X}$ to be a complete set of physical hadronic states, the beam function has the physical support
\begin{equation}
0 < x < 1 - \frac{p_{X\mathrm{min}}^-}{P^-} < 1
\,,\qquad
t > \w\, p_{X\mathrm{min}}^+ > 0
\,,\end{equation}
where $p_{X\min}^\pm > 0$ are the smallest possible hadronic momenta (which are strictly positive because with an incoming proton $\ket{X}$ can neither be massless nor the vacuum state).
For the jet function the external state is the vacuum $\ket{p_n} = \ket{0}$ yielding $\delta(\w + p_X^-)$, so the matrix element in \eq{Bkin} vanishes for $\w>0$.

\begin{figure}
\parbox{\columnwidth}{\centering
\includegraphics[scale=0.85]{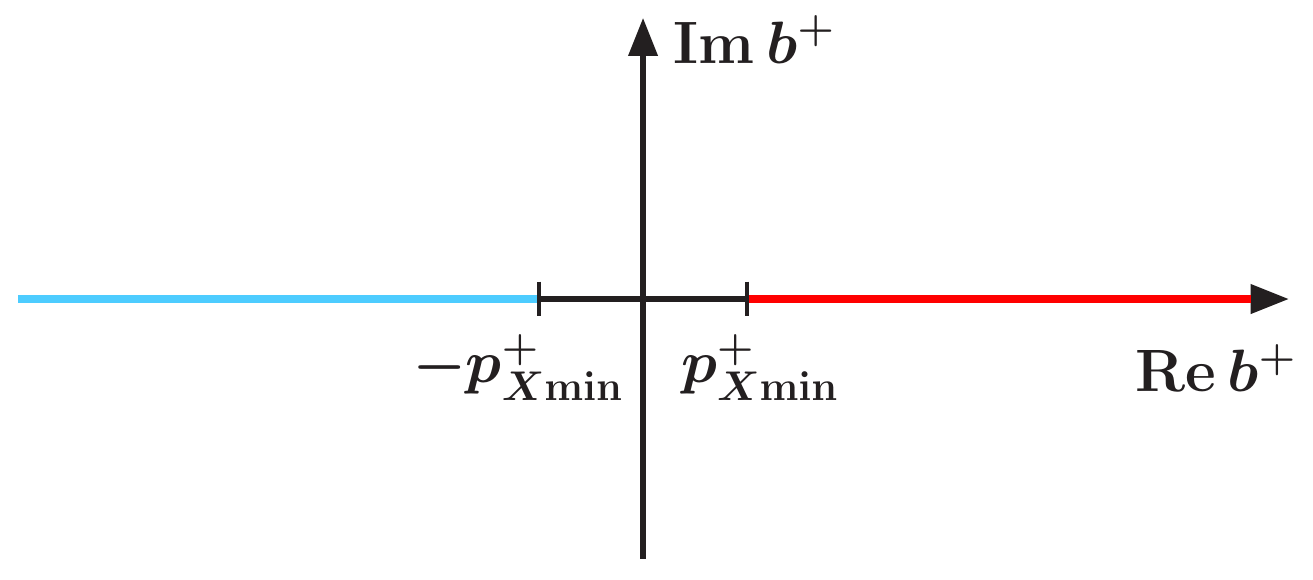}}
\caption{Cuts in the complex $b^+$ plane for the time-ordered product in \eq{Tresult}.}
\label{fig:Bcuts}
\end{figure}

Next, consider the following time-ordered analog of $\Mae{p_n}{\op_q(t, \w)}{p_n}$,
\begin{align}
\Mae{p_n}{T_q(\w b^+, \w)}{p_n}
&= \frac{1}{2\pi} \int\! \frac{\df y^-}{2\abs{\w}}\, e^{\img (b^+ - p^+) y^-/2}
   \MAe{p_n}{T\Bigl\{\bar\chi_n\Bigl(y^-\frac{n}{2}\Bigr) \frac{\bnslash}{2} \bigl[\delta(\w-\bnP_n) \chi_n(0)\bigr]\Bigr\}}{p_n}
\,.\end{align}
Writing out the time-ordering,
\begin{align}
&T\Bigl\{\bar\chi_n\Bigl(y^-\frac{n}{2}\Bigr) \frac{\bnslash}{2} \bigl[\delta(\w-\bnP_n) \chi_n(0)\bigr]\Bigr\}
\nn\\ & \quad
= \theta(y^-) \bar\chi_n\Bigl(y^-\frac{n}{2}\Bigr) \frac{\bnslash}{2} \bigl[\delta(\w-\bnP_n) \chi_n(0)\bigr]
 -\theta(-y^-) \bigl[\delta(\w-\bnP_n) \chi_n(0)\bigr] \bar\chi_n\Bigl(y^-\frac{n}{2}\Bigr) \frac{\bnslash}{2}
\,,\end{align}
using
\begin{equation}
\theta(\pm y^-) = \frac{\img}{2\pi} \int\!\df\kappa\, \frac{e^{\mp \img\kappa y^-}}{\kappa + \img 0}
\,,\end{equation}
inserting a complete set of states, and translating the fields to spacetime position zero,
\begin{align} \label{eq:Tresult}
&\Mae{p_n}{T_q(\w b^+,\w)}{p_n}
\nn\\ & \quad
=\frac{\img}{(2\pi)^2} \int\!\frac{\df y^-}{2\abs{\w}}\, \frac{\df \kappa}{\kappa+\img 0}
\sum_X \biggl[ e^{\img(b^+ - p_X^+ -\kappa)y^-/2}\, \delta(\w - p^- + p_X^-)\,
\MAe{p_n}{\bar\chi_n(0)\frac{\bnslash}{2}}{X} \Mae{X}{\chi_n(0)}{p_n}
\nn\\ &\qquad
  + e^{\img(b^+ + p_X^+ + \kappa)y^-/2}\, \delta(\w + p^- - p_X^-)\,
  \Mae{p_n}{\chi_n(0)}{X}\MAe{X}{\bar \chi_n(0)\frac{\bnslash}{2}}{p_n} \biggr]
\nn\\ & \quad
= \frac{\img}{2\pi\abs{\w}} \sum_X \biggl[
  \frac{\delta(\w - p^- + p_X^-)}{b^+ - p_X^+ + \img 0}
  \MAe{p_n}{\bar\chi_n(0)\frac{\bnslash}{2}}{X} \Mae{X}{\chi_n(0)}{p_n}
\nn\\ & \qquad\qquad\qquad\quad
- \frac{\delta(\w + p^- - p_X^-)}{b^+ + p_X^+ - \img 0}
  \Mae{p_n}{\chi_n(0)}{X}\MAe{X}{\bar \chi_n(0)\frac{\bnslash}{2}}{p_n} \biggr]
\,.\end{align}
The first term creates a cut in the complex $b^+$ plane for $b^+ \geq p_{X\mathrm{min}}^+$. This cut is shown as the dark red line in \fig{Bcuts}. The second term produces a cut at $b^+ \leq - p_{X\mathrm{min}}^+$, shown as the light blue line in \fig{Bcuts}.

The beam function matrix element in \eq{Bkin} can be identified as precisely the discontinuity of the first term in \eq{Tresult} with respect to $b^+$. Thus, for the beam function we have
\begin{equation}
B_q(\w b^+, \w) = \Disc_{b^+>0}\, \Mae{p_n(P^-)}{\theta(\w)T_q(\w b^+, \w)}{p_n(P^-)}
\,.\end{equation}
Taking the discontinuity only for $b^+>0$ ensures that we only pick out the cut due to the first term in \eq{Tresult}. Here, the discontinuity of a function $g(x)$ for $x>x_0$ is defined as
\begin{align} \label{eq:Disc_def}
\Disc_{x>x_0}\, g(x) = \lim_{\beta\to 0} \theta(x - x_0) \bigl[ g(x + \img\beta) - g(x - \img\beta) \bigr]
\,,\end{align}
and we used \eq{disc_os} to take the discontinuity of $1/(b^+ - p_X^+)$,
\begin{equation}
\Disc_{b^+>0}\,\frac{\img}{2\pi\abs{\w}}\,\frac{1}{b^+ - p_X^+} = \frac{1}{\abs{\w}}\,\delta(b^+ - p_X^+) = \delta(\w b^+ - \w p_X^+)
\,.\end{equation}
Since we explicitly specify how to take the discontinuity, we can drop the $\img 0$ prescription in the denominators. (Alternatively, we could multiply by $\img$ and take the imaginary part using the $\img 0$ prescription.) Since we are eventually only interested in the case $\w > 0$ anyway, we may as well restrict the matrix elements to $\w > 0$ from the very beginning and instead simply take the discontinuity in $t$ for $t = \w b^+ > 0$, so
\begin{equation} \label{eq:DiscTB}
\Mae{p_n}{\theta(\w)\op_q(t, \w)}{p_n} = \Disc_{t > 0}\, \Mae{p_n}{\theta(\w)T_q(t, \w)}{p_n}
\,.\end{equation}

For the matching calculation $\ket{p_n}$ is a partonic quark or gluon state. For any contributions with real radiation in the intermediate state, i.e.\ diagrams where the two $\chi_n$ or $\cB_{n\perp}$ fields in the operator $\op_i$ are joined by a series of propagators and vertices, we can use the standard Feynman rules to evaluate the time-ordered matrix element of $T_q(t, \w)$. However, with partonic external states, we can also have the vacuum state as an intermediate state, because the fields in the operator are spacetime separated. For such purely virtual contributions it is simpler to directly start from $\op_q(t, \w)$, insert the vacuum state between the fields, and then use standard Feynman rules to separately compute the two pieces $\Mae{p_n}{\bar\chi_n(0)\bnslash/2}{0}$ and $\Mae{0}{\chi_n(0)}{p_n}$. In fact, this is exactly what we already did in our tree-level calculation in \sec{tree_match}, and we will see another example in \ch{quark}. Thus, we will obtain the total partonic matrix element as
\begin{align} \label{eq:DiscTB_full}
\Mae{p_n}{\theta(\w)\op_q(t, \w)}{p_n}
&= \Mae{p_n}{\theta(\w)\op_q(t, \w)}{p_n}_\mathrm{virtual} + \Mae{p_n}{\theta(\w)\op_q(t, \w)}{p_n}_\mathrm{radiation}
\nn\\
&= \delta(t)\,\delta(\w - p^-)\, \MAe{p_n}{\bar\chi_n(0)\frac{\bnslash}{2}}{0}_\mathrm{connected} \Mae{0}{\chi_n(0)}{p_n}_\mathrm{connected}
\nn\\ & \quad
+ \Disc_{t > 0}\, \Mae{p_n}{\theta(\w)T_q(t, \w)}{p_n}_\mathrm{connected}
\,.\end{align}
The virtual contribution must be kept, since it only looks superficially disconnected because the operator itself is spacetime separated. As always, we still disregard genuinely disconnected diagrams, e.g.\ diagrams involving vacuum bubbles, when calculating the matrix elements in the second line.

\chapter{The Quark Beam Function}
\label{ch:quark}

In this chapter we compute the matching coefficients $\cI_{qq}(t, z, \mu)$ and $\cI_{qg}(t, z, \mu)$ in the OPE for the quark beam function to next-to-leading order in $\alpha_s(\mu)$, first reported in Ref.~\cite{Stewart:2010qs}. 

As explained in \sec{OPE} and \sec{tree_match}, the matching can be done by computing the partonic matrix elements of both sides of \eq{op_OPE} to NLO. We use the same $n$-collinear quark and gluon states, $\ket{q_n}\equiv \ket{q_n(p)}$ and $\ket{g_n}\equiv\ket{g_n(p)}$, as in the tree-level matching in \sec{tree_match}, with momentum $p^\mu = (p^+,p^-,0)$. Since only $\cI_{qq}(t, z, \mu)$ is nonzero at leading order, we will only need the NLO matrix elements of the quark operators, $\op_q(t, \w, \mu)$ and $\oq_q(\w, \mu)$. We will write all the matrix elements in terms of the RPI-III invariant variables
\begin{equation} \label{eq:defA}
t = b^+ \w
\,,\qquad
t' = -p^+\w = - zp^+p^-
\,,\qquad
z=\frac{\w}{p^-}
\,.\end{equation}
Here, $z$ is the partonic momentum fraction of the quark annihilated by the operator relative to the momentum of the incoming quark or gluon, and will coincide with the argument of the Wilson coefficients $\cI_{ij}(t, z, \mu)$.

To regulate the UV we use dimensional regularization with $d=4-2\eps$ dimensions and renormalize using the $\overline{\text{MS}}$ scheme. Since the matching coefficients in the OPE must be IR finite, the matrix elements of $\op_q$ and $\oq_q$ must have the same IR divergences, i.e., the beam function must contain the same IR divergences as the PDF. To explicitly check that this is the case, we separate the UV and IR divergences by regulating the IR with a small $p^+ < 0$. This forces the external states to have a small offshellness $p^+ p^- < 0$, and since $p^+p^- = -t'/z$ the IR divergences will appear as $\ln t'$. This also allows us to directly obtain the one-loop renormalization constants and anomalous dimensions for $\op_q$ and $\oq_q$ from their one-loop matrix elements.

We first compute the renormalized one-loop matrix elements of the quark PDF operator $\oq_q$ in \sec{NLO_PDF}. This calculation of the PDF for general $x$ using the SCET operator definition and with an offshellness IR regulator is quite instructive, both by itself and in comparison to the beam function calculation, which is why we give it in some detail. In \sec{NLO_B}, we compute the renormalized one-loop matrix elements of the quark beam function operator $\op_q$. We use these results to extract expressions for $\cI_{qq}(t, z, \mu)$ and $\cI_{qg}(t, z, \mu)$ valid to NLO in \sec{NLO_matching}. 

Assuming that the IR divergences in the beam function and PDF will cancel, the matching calculation can be performed more easily using dimensional regularization for both UV and IR. We do this as an illustrative exercise in \app{dimreg}, which, as it should, yields the same result for the matching coefficients. 

\section{Quark PDF with Offshellness IR Regulator}
\label{sec:NLO_PDF}

\begin{figure}
\subfigure[]{\includegraphics[scale=0.75]{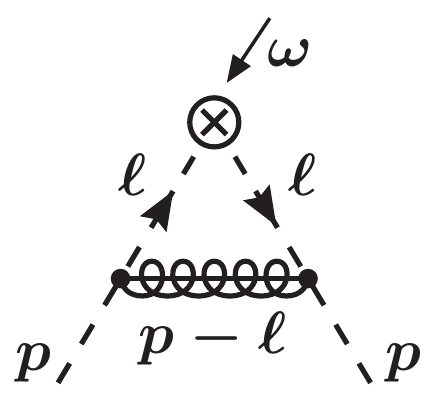}\label{fig:fone_a}}%
\hfill%
\subfigure[]{\includegraphics[scale=0.75]{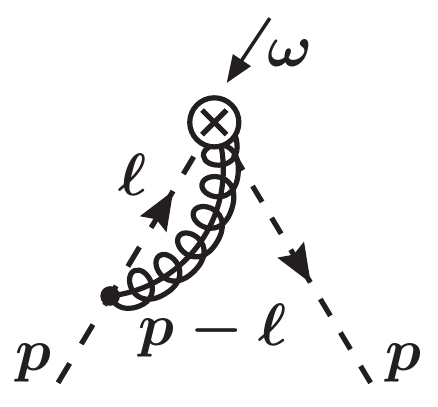}\label{fig:fone_b}}%
\hfill%
\subfigure[]{\includegraphics[scale=0.75]{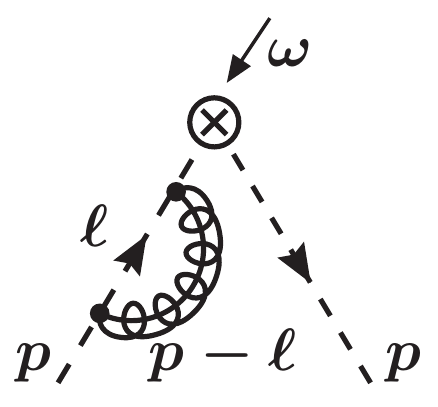}\label{fig:fone_c}}%
\hfill%
\subfigure[]{\includegraphics[scale=0.75]{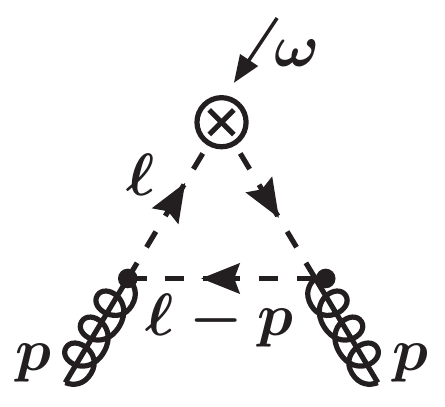}\label{fig:fone_d}}%
\caption[One-loop diagrams for the quark PDF.]{Nonzero one-loop diagrams for the quark PDF. The minus momentum $\w$ enters the vertex through its outgoing fermion line and leaves through its incoming fermion line. Diagram (c) represents the inclusion of the wave-function renormalization constant for the renormalized fields together with the corresponding
residue  factor in the LSZ formula for the $S$-matrix. Diagrams (b) and (c) have symmetric counterparts which are included in their computation.}
\label{fig:fone}
\end{figure}

We start by calculating the bare $S$-matrix elements
\begin{equation}
\Mae{q_n(p)}{\oq_q^\bare(\w)}{q_n(p)}
\,,\qquad
\Mae{g_n(p)}{\oq_q^\bare(\w)}{g_n(p)}
\,,\end{equation}
using Feynman gauge to compute the gauge-invariant sum of all diagrams. The relevant one-loop diagrams are shown in \fig{fone}. Since $\oq_q$ is a local SCET operator, we can use the usual time-ordered Feynman rules in SCET (without any of the complications discussed in \sec{beamT} for $\op_q$). The collinear $q_n q_n g_n$ vertex factor is
\begin{equation}
\img g\, T^a V_n^\mu(p, \ell)\,\frac{\bnslash}{2}
\qquad\text{with}\qquad
V_n^\mu(p, \ell)
= n^\mu + \frac{\pslash_\perp\gamma_\perp^\mu}{p^-} + \frac{\gamma_\perp^\mu \ellslash_\perp}{\ell^-}
- \frac{\pslash_\perp\ellslash_\perp}{p^-\ell^-} \bn^\mu
\,,\end{equation}
where $p^\mu$ and $\ell^\mu$ are the label momenta of the outgoing and incoming quark lines.
(Because we have a single collinear direction the computation can also be done with QCD Feynman rules, still accounting for zero-bin subtractions, with the only difference being the Dirac algebra in the numerator of the loop integral. We checked that the final results for each diagram are indeed the same either way.)

The diagram in \fig{fone_a} is
\begin{align}
 & \Mae{q_n}{\oq_q^\bare(\w)}{q_n}^{(a)}
 \nn \\ & \quad
 = -\img \Bigl(\frac{e^{\gamma_E} \mu^2}{4\pi}\Bigr)^{\eps} g^2 C_F \int \frac{\df^d\ell}{(2\pi)^d}
 \frac{\bar u_n(p) V_n^\mu(p, \ell) V_{n\mu}(\ell, p) \frac{\bnslash}{2} u_n(p) (\ell^-)^2
}{(\ell^2+\img 0)^2 [(\ell-p)^2+\img 0] } \,
 \delta(\ell^- -\w)
\,,\end{align}
where $g \equiv g(\mu)$ is the renormalized $\overline{\textrm{MS}}$ coupling.
The Dirac algebra for the numerator gives
\begin{equation}
\bar u_n(p) V_n^\mu(p, \ell) V_{n\mu}(\ell, p) \frac{\bnslash}{2} u_n(p) (\ell^-)^2
= \bar u_n(p) \gamma_\perp^\mu \ellslash_\perp \ellslash_\perp \gamma_{\perp\mu} \frac{\bnslash}{2} u_n(p)
= p^- (d-2) \ell_\perp^2
\,.\end{equation}
To compute the loop integral we write $\df^d\ell = \df\ell^+ \df\ell^- \df^{d-2} \vec\ell_\perp/2$, where $\vec\ell_\perp$ is Euclidean, so $\ell_\perp^2 = -\vec\ell_\perp^2$. The $\ell^+$ integral is done by contour integration as follows. For $\ell^-<0$ all poles are above the axis and for $\ell^- > p^-$ all poles are below the axis, so both cases give zero. Hence, the $\ell^-$ integration range is restricted to $0 < \ell^- < p^-$, where there is a double pole below the axis from the $1/(\ell^2 + \img 0)^2$ and a single pole above the axis from the $1/[(\ell - p)^2 + \img 0]$. Taking the single pole above amounts to replacing the second denominator by $2\pi\img/(\ell^-\! - p^-)$ and setting
$\ell^+ = p^+ - \vec \ell_\perp^2/(p^-\! -\ell^-)$ everywhere else. After performing the contour integral the $\img 0$ have served their purpose and can be set to zero everywhere. The $\ell^-$ integral is trivial using the $\delta(\ell^-\! - \w)$ and turns the $\ell^-$ limits into an overall $\theta(\w)\theta(p^-\! - \w)$. The remaining $\vec\ell_\perp$ integration is done in $d - 2 = 2(1 - \eps)$ Euclidean dimensions as usual. Putting everything together, we obtain
\begin{align} \label{eq:PDFa1}
&\Mae{q_n}{\oq_q^\bare(\w)}{q_n}^{(a)}
\nn\\ & \qquad
= \Bigl(\frac{e^{\gamma_E} \mu^2}{4\pi}\Bigr)^{\eps} g^2 C_F\, \theta(\w) \theta(p^-\! - \w)\, \frac{(d - 2) (p^-\! - \w)}{4\pi\, p^-} \!\int\! \frac{\df^{d-2}\vec\ell_\perp}{(2\pi)^{d-2}}\,
\frac{\vec\ell_\perp^2\, }{[\vec\ell_\perp^{2} + (1-z)t']^2 }
\nn\\ &\qquad
= \frac{\alpha_s(\mu)C_F}{2\pi}\,
  \theta(z) \theta(1-z)\, \Gamma(\eps) \Bigl(\frac{e^{\gamma_E}\mu^2}{t'}\Bigr)^{\eps} (1-z)^{1-\eps} (1-\eps)^2
\nn \\ &\qquad
= \frac{\alpha_s(\mu) C_F}{2\pi} \,\theta(z) \theta(1-z)\,(1-z) \biggl\{ \frac{1}{\eps} - \ln\frac{t'}{\mu^2} - \ln(1-z) -2 \biggr\}
\,,\end{align}
where in the last line we expanded in $\eps$.

In the diagram in \fig{fone_b}, the gluon is annihilated by the Wilson line inside one of the $\chi_n$ fields. The contraction with the one in $\bar{\chi}_n$ is $\propto\delta(\ell^-\! - \w)$ and the contraction with the one in $\chi_n$ is $\propto\delta(p^-\! - \w)$. The $1/\bnP_n$ in the Wilson lines [see \eq{Wn}] contributes a factor $1/(\ell^-\!-p^-)$ with a relative minus sign between the two contractions. (There is also a diagram where the gluon connects both Wilson lines which vanishes because the Wilson lines only contain $\bn \sdt A$ gluons and we use Feynman gauge.) Adding \fig{fone_b} and its mirror graph, which gives an identical contribution, we get
\begin{align} \label{eq:PDFb1}
& \Mae{q_n}{\oq_q^\bare(\w)}{q_n}^{(b)}\,
  \nn \\ & \
  = 2\img \Bigl(\frac{e^{\ga_E} \mu^2}{4\pi}\Bigr)^{\eps} g^2 C_F \int\! \frac{\df^d\ell}{(2\pi)^d}\,
  \frac{
\bn_\mu \bar u_n(p) V_n^\mu \frac{\bnslash}{2} u_n(p) \ell^-
}{(\ell^-\! - p^-)(\ell^2 + \img 0) [(\ell-p)^2 + \img 0] } \,
  \bigl[\delta(\ell^-\! -\w) - \delta(p^- \! -\w) \bigr]
\nn\\ & \
 = \frac{\alpha_s(\mu) C_F }{\pi}\,\Gamma(\eps)\biggl(\frac{e^{\gamma_E} \mu^2}{-p^+p^-} \biggr)^\eps
\! \int\! \df\ell^-\,\theta(\ell^-)\,\theta(p^-\!-\ell^-) \frac{(\ell^-/p^-)^{1-\eps}}{(1 -\ell^-/p^- )^{1+\eps} }
  \bigl[\delta(\ell^-\! -\w) - \delta(p^- \! -\w) \bigr]
\nn\\ & \
 =\frac{\alpha_s(\mu) C_F}{\pi}\,
 \Gamma(\eps) \Bigl(\frac{e^{\gamma_E}\mu^2}{t'}\Bigr)^\eps
\biggl\{ \frac{\theta(z)\theta(1-z) z}{(1-z)^{1+\eps} } - \delta(1 -z ) \, \frac{\Gamma(2-\eps)\Gamma(-\eps)}{\Gamma(2-2\eps)}
  \biggr\}
\,.\end{align}
In the first step we used $\bn_\mu V_n^\mu = 2$ and $\bar u_n(p) \bnslash u_n(p) = 2p^-$, performed the $\ell^+$ integral by contours and did the $\vec\ell_\perp$ integral as usual. The $\ell^+$ integral has the same pole structure as in \fig{fone_a} (except that the double pole at $\ell^+ = 0$ is now a single pole), which restricts the $\ell^-$ integral to the finite range $0 < \ell^- < p^-$. Expanding \eq{PDFb1} in $\eps$, using the distribution identity in \eq{distr_id}, we get
\begin{align} \label{eq:PDFb2}
\Mae{q_n}{\oq_q^\bare(\w)}{q_n}^{(b)}
&=\frac{\alpha_s(\mu) C_F}{\pi}\, \Gamma(\eps)\Bigl(\frac{e^{\gamma_E}\mu^2}{t'}\Bigr)^\eps
\biggl\{ \theta(z)
 \biggl[-\frac{1}{\eps}\,\delta(1-z) + \cL_0(1-z)z -\eps \cL_1(1-z)z\biggr]
\nn\\ &\quad\hspace{25ex}
 +\delta(1 -z ) \biggl[\frac{1}{\eps} + 1 + \eps\Bigl(2-\frac{\pi^2}{6}\Bigr) \biggr]
  \biggr\} \,
\nn\\
&=\frac{\alpha_s(\mu) C_F}{\pi}\, \theta(z) \biggl\{
  \Bigl(\frac{1}{\eps} - \ln\frac{t'}{\mu^2} \Bigr) \bigl[\cL_0(1-z)z + \delta(1-z)  \bigr] - \cL_1(1-z)z
\nn\\* & \quad\hspace{15ex}
+ \delta(1 -z) \Bigl(2-\frac{\pi^2}{6} \Bigr)
  \biggr\}
\,,\end{align}
where $\cL_n(x) = [\theta(x)(\ln^n x)/x]_+$ are the usual plus distributions defined in \eq{plusdef}.

In the last step in \eq{PDFb1}, the $\ell^-$ integral produces an additional $1/\eps$ pole in each of the two terms corresponding to real and virtual radiation from the two different Wilson line contractions. It comes from the singularity at $\ell^- = p^-$, where the gluon in the loop becomes soft. (This soft IR divergence appears as a pole in $\eps$ because the offshellness only regulates the collinear IR divergence here.) The soft IR divergences cancel in the sum of the virtual and real contributions, as can be seen explicitly in the first line of \eq{PDFb2} where the $1/\eps$ poles in curly brackets cancel between the two terms. One can already see this in the $\ell^-$ integral in \eq{PDFb1}, because for $\ell^- = p^-$ the two $\delta$ functions cancel so there is no soft divergence in the total integral. Thus, in agreement with our discussion in \sec{definition}, we explicitly see that contributions from the soft region drop out in the PDF. As a consequence, the PDF only contains a single $1/\eps$ pole and correspondingly its RGE will sum single logarithms associated with this purely collinear IR divergence.

Since the gluon in the loop is supposed to be collinear, the soft gluon region must be explicitly removed from the collinear loop integral, which is the condition $\lp \neq 0$ in \eq{xi}. For continuous loop momenta this is achieved by a zero-bin subtraction. However, since the soft region does not contribute to the PDF, it also does not require zero-bin subtractions in SCET. (If we were to include separate zero-bin subtractions for the virtual and real contributions, they would simply cancel each other.) We will see shortly that the situation for the beam function is quite different.

The last diagram with external quarks, \fig{fone_c}, is
\begin{align}  \label{eq:PDFc1}
\Mae{q_n}{\oq_q^\bare(\w)}{q_n}^{(c)} &= \delta(1-z) (Z_\xi - 1)
  = -\frac{\alpha_s(\mu) C_F}{4\pi}\, \delta(1-z) \biggl\{\frac{1}{\eps} - \ln\frac{t'}{\mu^2} + 1 \biggr\}\,
\,.\end{align}
Here we used the result for the one-loop on-shell wave-function renormalization with an offshellness IR regulator, which is the same in SCET and QCD.

Adding up the results in Eqs.~\eqref{eq:PDFa1}, \eqref{eq:PDFb2}, and \eqref{eq:PDFc1} we obtain for the bare one-loop quark matrix element
\begin{align} \label{eq:fqq_bare}
\Mae{q_n}{\oq_q^\bare(\w)}{q_n}^\one
&= \frac{\alpha_s(\mu) C_F}{2\pi}\,  \theta(z) \biggl\{
\Bigl(\frac{1}{\eps} - \ln\frac{t'}{\mu^2} \Bigr) P_{qq}(z) - \cL_1(1-z)(1 + z^2)
\nn\\ &\quad\hspace{15ex}
+ \delta(1- z ) \Bigl(\frac72 -\frac{\pi^2}{3} \Bigr) - \theta(1-z) 2 (1-z) \biggr\}
\,,\end{align}
where
\begin{equation}
P_{qq}(z) = \cL_0(1-z)(1 + z^2) + \frac{3}{2}\, \delta(1-z) = \biggl[\theta(1-z)\frac{1+z^2}{1-z} \biggr]_+
\end{equation}
is the $q\to qg$ splitting function, see \eq{Pqq_def}.

Next, we consider the matrix element of $\oq_q$ between gluon states $\ket{g_n} \equiv \ket{g_n(p)}$. The only relevant diagram is shown in \fig{fone_d},
\begin{align}
& \Mae{g_n}{\oq_q^\bare(\w)}{g_n}^{(d)}
\nn \\ & \quad
= \img\Bigl(\frac{ e^{\gamma_E} \mu^{2\eps}}{4\pi}\Bigr)^\eps g^2 T_F
 \int \frac{\df^d \ell}{(2\pi)^d} \,
 \frac{(-\varepsilon_\mu^* \varepsilon_\nu)
\tr\bigl[V_n^\mu V_n^{\nu} \frac{\bnslash\nslash}{4}\bigr]
 (\ell^-)^2(\ell^- - p^-)}{(\ell^2+\img 0)^2 [(\ell-p)^2+\img 0] } \, \delta(\ell^- -\w)
\,.\end{align}
Here $\varepsilon \equiv \varepsilon(p)$, $V_n^\mu \equiv V_n^\mu(\ell-p, \ell)$  and $V_n^\nu\equiv V_n^\nu(\ell, \ell-p)$. Since the physical polarization vector is perpendicular, $n\cdot\varepsilon(p) = \bn\cdot\varepsilon(p) = 0$, we only need the perpendicular parts of the collinear vertices. The numerator then becomes
\begin{align}
& \tr\Bigl[V_n^\mu V_n^{\nu} \frac{\bnslash\nslash}{4} \Bigr]
 (\ell^-)^2(\ell^-\! - p^-) 
\nn \\ & \quad =
\frac{1}{2}\tr\biggl[
\biggl( \frac{\ellslash_\perp \gamma_\perp^\mu}{\ell^- - p^-} + \frac{\gamma_\perp^\mu \ellslash_\perp}{\ell^-} \biggr)
\biggl( \frac{\ellslash_\perp \gamma_\perp^\nu}{\ell^-} + \frac{\gamma_\perp^\nu \ellslash_\perp}{\ell^- - p^-} \biggr)
\biggr] (\ell^-)^2(\ell^-\! - p^-)
\nn\\ & \quad = 
2\frac{(p^-)^2}{\ell^- - p^-}\,\ell_\perp^2 g_\perp^{\mu\nu} + 8 \ell^- \ell_\perp^\mu \ell_\perp^\nu
= 2 g_\perp^{\mu\nu} p^-\Bigl(\frac{1}{1-z} - \frac{4z}{d - 2}\Bigr) \vec\ell_\perp^2
\,.\end{align}
In the last step we used that under the integral we can replace $\ell^- = \w = z p^-$ and $\ell_\perp^\mu \ell_\perp^\nu = \ell_\perp^2 g_\perp^{\mu\nu}/(d-2)$. The remaining loop integral is exactly the same as in \fig{fone_a}, so the bare one-loop gluon matrix element becomes
\begin{align} \label{eq:fqg_bare}
\Mae{g_n}{\oq_q^\bare(\w)}{g_n}^\one
&= \frac{\alpha_s(\mu) T_F}{2\pi}\, \theta(z)\,\theta(1 - z)
\Gamma(\eps) \Bigl(\frac{e^{\gamma_E}\mu^2}{t'}\Bigr)^\eps (1-z)^{-\eps}(1- 2z + 2z^2 - \eps)
\nn\\
&= \frac{\alpha_s(\mu) T_F}{2\pi}\,\theta(z)
\biggl\{ \Bigl[\frac{1}{\eps} - \ln\frac{t'}{\mu^2} - \ln(1-z) \Bigr] P_{qg}(z) - \theta(1 - z) \biggr\}
\,.\end{align}
Here
\begin{equation}
P_{qg}(z) = \theta(1-z)\,(1 - 2z + 2z^2)
\end{equation}
is the $g\to q\bar{q}$ splitting function from \eq{Pqq_def}.

Note that the diagram analogous to \fig{fone_d} with the two gluons crossed can be obtained from \fig{fone_d} by taking $p^\mu \to -p^\mu$, which takes $z\to -z$. The limits resulting from the $\ell^+$ integral are then $-1\leq z\le 0$ or $-p^- < \w < 0$, and since we require $\w > 0$ for $\oq_q$, this diagram vanishes. The diagram involving the SCET vertex with two collinear gluons vanishes because the $\ell^+$ integral does not have poles on both sides of the axis.

From the bare matrix elements in \eqs{fqq_bare}{fqg_bare} we can obtain the renormalization of $\oq_q$. Taking parton matrix elements of \eq{oq_ren} and expanding to NLO,
\begin{align} \label{eq:fbare2}
& \Mae{q_n}{\oq_q^\bare(\w)}{q_n}^\one
\nn \\ & \quad
= \sum_j \int\! \frac{\df \w'}{\w'}\,
\biggl[Z^{f\one}_{qj}\Bigl(\frac{\w}{\w'}, \mu \Bigr) \Mae{q_n}{\oq_j(\w',\mu)}{q_n}^\zero +
Z^{f\zero}_{qj}\Bigl(\frac{\w}{\w'}, \mu\Bigr) \Mae{q_n}{\oq_j(\w',\mu)}{q_n}^\one \biggr]
\nn \\ & \quad
= Z_{qq}^{f\one}(z, \mu) + \Mae{q_n}{\oq_q(\w,\mu)}{q_n}^\one
\,, \nn \\
&\Mae{g_n}{\oq_q^\bare(\w)}{g_n}^\one
\nn \\ & \quad
= \sum_j \int\! \frac{\df \w'}{\w'}\,
\biggl[Z^{f\one}_{qj}\Bigl(\frac{\w}{\w'}, \mu\Bigr) \Mae{g_n}{\oq_j(\w',\mu)}{g_n}^\zero +
Z^{f\zero}_{qj}\Bigl(\frac{\w}{\w'}, \mu\Bigr) \Mae{g_n}{\oq_j(\w',\mu)}{g_n}^\one \biggr]
\nn \\ & \quad
= Z_{qg}^{f\one}(z, \mu) + \Mae{g_n}{\oq_q(\w,\mu)}{g_n}^\one
\,,\end{align}
where we used the tree-level matrix elements in \eqs{op_tree}{qg_tree} and $Z^{f\zero}_{ij}(z, \mu) = \delta_{ij}\,\delta(1 - z)$. The $\overline{\mathrm{MS}}$ counter terms required to cancel the $1/\eps$ poles in the bare PDF matrix elements are then
\begin{equation} \label{eq:Zpdf}
  Z_{qq}^f(z) = \delta(1-z) + \frac{1}{\eps}\, \frac{\alpha_s(\mu) C_F}{2\pi}\,
  \theta(z) P_{qq}(z)
  \,,\qquad
  Z_{qg}^f(z) = \frac{1}{\eps}\, \frac{\alpha_s(\mu) T_F}{2\pi}\,
  \theta(z) P_{qg}(z)
 \,.\end{equation}
Expanding \eq{oq_RGE} to NLO, the one-loop anomalous dimensions are obtained by
\begin{equation}
\gamma_{ij}^f(z, \mu) = -\mu \frac{\df}{\df\mu} Z^{f\one}_{ij}(z, \mu)
\,,\qquad
\mu \frac{\df}{\df\mu} \alpha_s(\mu) = -2\eps\, \alpha_s(\mu) + \beta[\alpha_s(\mu)]
\,,\end{equation}
which with \eq{Zpdf} yields the anomalous dimension for the quark PDF in \eq{gammaf},
\begin{equation}
  \gamma_{qq}^f(z, \mu) = \frac{\alpha_s(\mu) C_F}{\pi}\, \theta(z) P_{qq}(z)
\,,\qquad
 \gamma_{qg}^f(z, \mu) = \frac{\alpha_s(\mu)T_F}{\pi}\,\theta(z) P_{qg}(z)
\,.\end{equation}
Finally, the renormalized NLO PDF matrix elements, which we will need for the matching computation in \sec{NLO_matching} below, are
\begin{align} \label{eq:fren}
\Mae{q_n}{\oq_q(\w, \mu)}{q_n}^\one
&= -\frac{\alpha_s(\mu) C_F}{2\pi}\,  \theta(z) \biggl\{
P_{qq}(z) \ln\frac{t'}{\mu^2} + \cL_1(1-z)(1 + z^2)
\nn\\ &\quad\hspace{17ex}
- \delta(1- z ) \Bigl(\frac72 -\frac{\pi^2}{3} \Bigr) + \theta(1-z) 2 (1-z) \biggr\}
\,, \nn \\
\Mae{g_n}{\oq_q(\w, \mu)}{g_n}^\one
&= -\frac{\alpha_s(\mu) T_F}{2\pi}\,\theta(z)
\biggl\{P_{qg}(z) \Bigl[\ln\frac{t'}{\mu^2} + \ln(1-z)\Bigr] + \theta(1 - z) \biggr\}
\,.\end{align}

\section{Quark Beam Function with Offshellness IR Regulator}
\label{sec:NLO_B}

\begin{figure}
\subfigure[]{\includegraphics[scale=0.75]{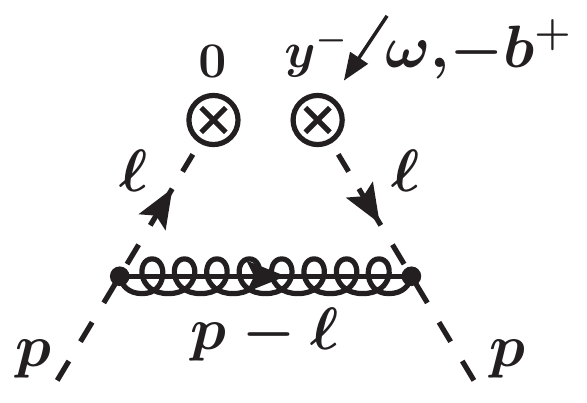}\label{fig:Bone_a}}%
\hfill%
\subfigure[]{\includegraphics[scale=0.75]{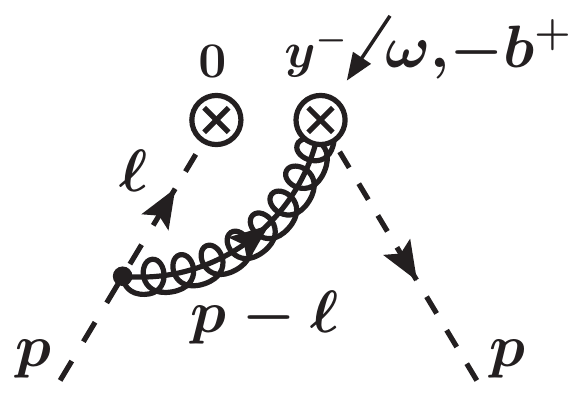}\label{fig:Bone_b}}%
\hfill%
\subfigure[]{\includegraphics[scale=0.75]{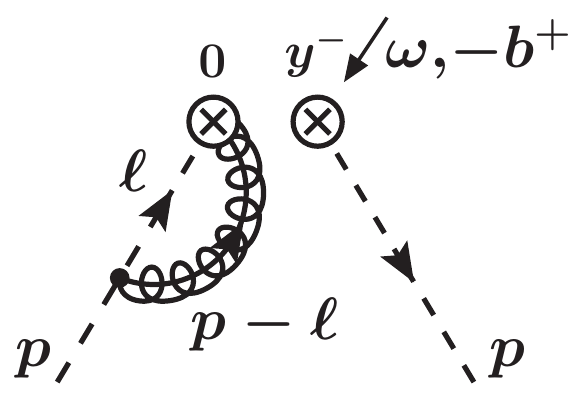}\label{fig:Bone_c}}%
\\
\subfigure[]{\includegraphics[scale=0.75]{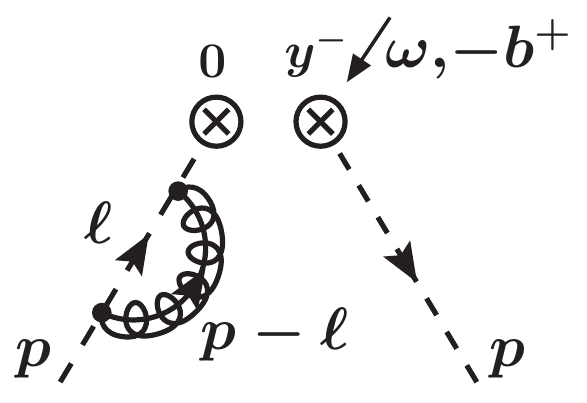}\label{fig:Bone_d}}%
\hfill%
\subfigure[]{\includegraphics[scale=0.75]{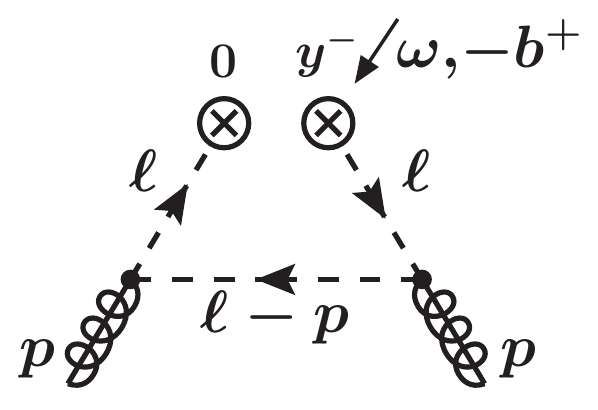}\label{fig:Bone_e}}%
\hfill%
\subfigure[]{\includegraphics[scale=0.75]{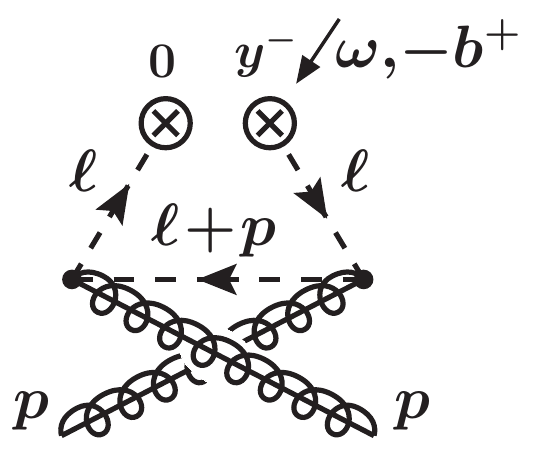}\label{fig:Bone_f}}%
\caption[One-loop diagrams for the quark beam function.]{One-loop diagrams for the quark beam function. The minus momentum $\w$ is incoming at the vertex and the $b^+$ momentum is outgoing. Diagram (d) denotes the wave-function contribution. Diagrams (b), (c), and (d) have symmetric counterparts which are equal to the ones shown and included in the computation. Diagram (f) and the diagram with the gluon connecting both vertices vanish.}
\label{fig:Bone}
\end{figure}

Next, we calculate the bare beam function $S$-matrix elements,
\begin{equation}
\Mae{q_n(p)}{\theta(\w) \op_q^\bare(t, \w)}{q_n(p)}
\,,\qquad
\Mae{g_n(p)}{\theta(\w) \op_q^\bare(t, \w)}{g_n(p)}
\,,\end{equation}
to NLO. The corresponding one-loop diagrams are shown in \fig{Bone}. The matrix elements are calculated as explained in \sec{beamT} in \eq{DiscTB_full}: For the virtual diagrams with vacuum intermediate state we explicitly insert the vacuum state, while for the real-emission diagrams we use \eq{DiscTB}. In the latter case, we first take the $\Disc$, then expand in $\eps$ to extract the UV divergences, and at last take the $t' \to 0$ limit to isolate the IR divergences into $\ln t'$ terms. Some helpful formulas for calculating the discontinuity and taking the limit $t'\to 0$ are given in \app{plusdisc}.

For the beam function calculation the $p^+ < 0$ actually plays a dual role: For the UV divergent piece we can treat the calculation as in \SCETa, and so $p^+ \sim b^+ \sim \la^2 p^-$, which allows us to explicitly check the structure of the convolution in \eq{op_ren}. The renormalized result contributes to the matching onto PDFs, matching from \SCETa onto \SCETb. In the matching, $-p^+ \ll b^+$ plays the role of the IR regulator, since we are required to use the same states as in the PDF calculation. We will see that the IR divergences $\ln t'$ match up with those present in the PDF calculation, and hence drop out in the coefficients $\cI_{ij}$.

The diagrams in \fig{Bone} have the same Dirac and propagator structure and overall factors as the corresponding PDF diagrams in \fig{fone}, so we can reuse those parts from the previous section. The difference compared to the PDF calculation is that for the real-emission diagrams, instead of doing the $\ell^+$ integral by contours, $\ell^+$ is fixed by the additional $\delta$ function in $b^+$, and since we use time-ordered perturbation theory we must now take the discontinuity. This also alters the structure of the remaining $\vec\ell_\perp$ integral, for which we now use Feynman parameters to combine the denominators. After carrying out the $\vec\ell_\perp$ integration, we will need the following two Feynman parameter integrals
\begin{align} \label{eq:I12}
I_1(A, B, \eps)
&= \int_0^1\! \df \alpha\, [(1 - \alpha)A - \alpha B]^{-1-\eps}
= \frac{(-B)^{-\eps} - A^{-\eps} }{\eps(A+B)}
\,, \\
I_2(A, B, \eps)
&= \int_0^1\! \df \alpha\, (1 - \alpha) [(1 - \alpha)A - \alpha B]^{-1-\eps}
= -\frac{(-B)^{1-\eps} - A^{1-\eps}}{\eps(1-\eps)(A+B)^2} - \frac{A^{-\eps} }{\eps(A+B)}
\,. \nn\end{align}

The first diagram, \fig{Bone_a}, has real radiation in the final state, so we use \eq{DiscTB}
\begin{align} \label{eq:Ba}
 & \Mae{q_n}{\theta(\w)\op^\bare_q(t,\w)}{q_n}^{(a)}
 \nn \\ & \
\!= -\img \Bigl(\frac{e^{\gamma_E} \mu^2}{4\pi}\Bigr)^{\eps}g^2 C_F \frac{\theta(\w)}{\w}\, \Disc_{t>0}\! \int\! \frac{\df^d\ell}{(2\pi)^d}
 \frac{p^- (d-2)\ell_\perp^2
}{(\ell^2+\img 0)^2 [(\ell-p)^2+\img 0] } \,
 \delta(\ell^-\! -\w) \delta(\ell^+\! + b^+\! - p^+)
\nn \\ & \
= -\img \Bigl(\frac{e^{\gamma_E} \mu^2}{4\pi}\Bigr)^{\eps}g^2 C_F\,\frac{\theta(z)(d-2)}{(2\pi)^2 z}\, \Disc_{t>0}
   \int_0^1 \!\df\alpha \int\! \frac{\df^{d-2}\vec\ell_\perp}{(2\pi)^{d-2}}\,
    \frac{(1-\alpha)\, \vec\ell_\perp^2 }{[\vec \ell_\perp^2  +  (1 - \alpha)A - \alpha B]^3 }
\nn \\ & \
= \frac{\alpha_s(\mu)C_F}{2\pi}\,\frac{\theta(z)}{z}\, \Gamma(1+\eps) (e^{\gamma_E} \mu^2)^\eps (1-\eps)^2
  \Bigl[-\frac{\img}{2\pi}\Disc_{t>0}\, I_2(A, B, \eps) \Bigr]
\,,\end{align}
where we abbreviated
\begin{align}
A= t + t'
\,,\qquad
B= \frac{1 - z}{z}\,t
\,,\qquad
A+B = \frac{t}{z} + t'
\,.\end{align}
Since $t' > 0$ and $z > 0$, the only discontinuity in $I_2(A, B, \eps)$ for $t > 0$ arises from $(-B)$. Using \eq{disc_os} to take the $\Disc$, we obtain
\begin{align}
-\frac{\img}{2\pi}\Disc_{t>0}\,I_2(A, B, \eps)
&= \frac{\img}{2\pi}\Disc_{t>0} \frac{(-B)^{1-\eps}}{\eps(1-\eps)(A+B)^2}
= \theta(t)\frac{\sin\pi\eps}{\pi\eps(1-\eps)}\, \frac{\theta(B) B^{1-\eps}}{(A+B)^2}
\,,\nn\\
&=  \theta\Bigl(\frac{1-z}{z}\Bigr)\theta(t)\frac{\sin\pi\eps}{\pi\eps(1-\eps)}\, \frac{[(1-z)t]^{1-\eps}z^{1+\eps}}{(t + zt')^2}
\,.\end{align}
Note that there is only a discontinuity for $B > 0$, so taking the discontinuity for $t > 0$ requires $(1-z)/z > 0$, and since $z > 0$ we obtain the expected limit $z < 1$. Since there are no UV divergences, we can let $\eps \to 0$, and \eq{Ba} becomes
\begin{equation} \label{eq:DiscBa}
\Mae{q_n}{\theta(\w)\op^\bare_q(t,\w)}{q_n}^{(a)}
= \frac{\alpha_s(\mu)C_F}{2\pi}\,\theta(z)\theta(1-z)(1-z)\frac{\theta(t)\,t}{(t + zt')^2}
\,.\end{equation}
The above result has a collinear IR singularity for $t\to 0$ which is regulated by the nonzero $t'$. We can isolate the IR singularity using \eqs{plusdef}{limits} by letting $\beta\equiv z t'/\mu^2 \to 0$ while holding $\tilde t = t + zt'$ fixed%
\footnote{We keep the dependence on $\tilde t$ in our calculation as it will be useful for checking the
structure of the renormalization in the following section.},
\begin{equation} \label{eq:Ba_limit}
\lim_{t'\to 0} \frac{\theta(t)\,t}{(t + zt')^2}
= \lim_{z t'/\mu^2 \to 0}\biggl[
  \frac{\theta(\tilde t - z t')}{\tilde t} - \frac{\theta(\tilde t - z t') zt'}{\tilde t^2} \biggr]
= \frac{1}{\mu^2}\cL_0\Bigl(\frac{\tilde t}{\mu^2}\Bigr)
- \delta(\tilde t) \Bigl(\ln \frac{z t'}{\mu^2} + 1\Bigr)
\,.\end{equation}
The final result for \fig{Bone_a} is thus
\begin{equation} \label{eq:Ba_final}
\Mae{q_n}{\theta(\w)\op^\bare_q(t,\w)}{q_n}^{(a)}
= \frac{\alpha_s(\mu)C_F}{2\pi}\,\theta(z)\theta(1-z)(1-z)
\biggl\{\frac{1}{\mu^2}\cL_0\Bigl(\frac{\tilde t}{\mu^2}\Bigr) - \delta(\tilde t) \Bigl(\ln \frac{z t'}{\mu^2} + 1\Bigr)\biggr\}
\,.\end{equation}

Next, we consider the real-emission diagram in \fig{Bone_b}. It corresponds to the $\delta(\ell^- \!- \w)$ term in \eq{PDFb1}. Together with its mirror graph, giving an identical contribution, we obtain
\begin{align} \label{eq:Bb1}
 & \Mae{q_n}{\theta(\w)\op_q^\bare(t,\w)}{q_n}^{(b)}
 \nn \\ &\qquad
= 2\img \Bigl(\frac{e^{\gamma_E} \mu^2}{4\pi}\Bigr)^{\eps}g^2 C_F \frac{\theta(\w)}{\w}\, \Disc_{t>0}\!
   \int\!\frac{\df^d\ell}{(2\pi)^d}\,\frac{2p^-\ell^-\,  \delta(\ell^-\! -\w)\, \delta(\ell^+\! + b^+\! - p^+)}
   {(\ell^-\! - p^-)(\ell^2+\img 0) [(\ell-p)^2+\img 0] } \,
\nn \\ &\qquad
= \frac{\alpha_s(\mu)C_F}{\pi}\, \frac{\theta(z)}{1 -z}\, \Gamma(1+\eps)(e^{\gamma_E} \mu^2)^\eps
  \Bigl[-\frac{\img}{2\pi}\Disc_{t>0}\, I_1(A, B, \eps) \Bigr]
\nn \\ &\qquad
= \frac{\alpha_s(\mu)C_F}{\pi}\,\theta(z) \Gamma(1+\eps)\Bigl(\frac{e^{\gamma_E} \mu^2}{t}\Bigr)^\eps
\frac{\sin\pi\eps}{\pi\eps} \frac{\theta(t) }{t + zt'}\, \frac{\theta(1-z)z^{1+\eps}}{(1-z)^{1+\eps}}
\,,\end{align}
where in the second step we performed the loop integral as before, and in the last step we used \eq{disc_os} to take the discontinuity. As for \fig{Bone_a}, the loop integral produces no UV divergence. However, as in the PDF calculation for \fig{fone_b}, there is a soft gluon IR divergence at $z\to 1$ or $\ell^-\to p^-$ producing a $\delta(1-z)/\eps$ IR pole when expanding the last factor using \eq{distr_id}. In contrast to the PDF calculation, the soft gluon region must now be explicitly excluded from the collinear loop integral. In dimensional regularization with an offshellness IR regulator the relevant zero-bin integral is scaleless and vanishes. Thus, including the zero-bin subtraction removes the $1/\eps$ IR divergence and replaces it by an equal $1/\eps$ UV divergence such that all $1/\eps$ poles in the final result are UV divergences.
Expanding in $\eps$, we have
\begin{equation} \label{eq:Bb2}
\Mae{q_n}{\theta(\w)\op_q^\bare(t,\w)}{q_n}^{(b)}\,
 =  \frac{\alpha_s(\mu) C_F }{\pi} \, \theta(z)\,\frac{\theta(t)}{t + zt'}
  \Bigl\{ \delta(1-z) \Bigl(-\frac{1}{\eps}  + \ln\frac{t}{\mu^2}\Bigr) + \cL_0(1-z)z \Bigr\}
\,,\end{equation}
and taking the same limit as in \eq{Ba_limit} to isolate the IR divergences,
\begin{align}
\lim_{t'\to 0} \frac{\theta(t)}{t + zt'}
&= \lim_{z t'/\mu^2\to 0} \frac{\theta(\tilde t - z t')}{\tilde t}
= \frac{1}{\mu^2}\cL_0\Bigl(\frac{\tilde t}{\mu^2}\Bigr) - \delta(\tilde t) \ln \frac{z t'}{\mu^2}
\,,\\\nn
\lim_{t'\to 0} \frac{\theta(t)}{t + zt'}\ln\frac{t}{\mu^2}
&= \lim_{z t'/\mu^2\to 0} \frac{\theta(\tilde t - z t')}{\tilde t}\ln\frac{\tilde t - zt'}{\mu^2}
= \frac{1}{\mu^2}\cL_1\Bigl(\frac{\tilde t}{\mu^2}\Bigr) - \delta(\tilde t) \Bigl(\frac{1}{2}\ln^2\frac{z t'}{\mu^2} + \frac{\pi^2}{6}\Bigr)
\,,\end{align}
the final result for \fig{Bone_b} is
\begin{align} \label{eq:Bb_final}
&\Mae{q_n}{\theta(\w)\op_q^\bare(t,\w)}{q_n}^{(b)}
\nn\\ & \qquad
= \frac{\alpha_s(\mu) C_F}{\pi}\, \theta(z) \biggl\{
 \biggl[
 \frac{1}{\mu^2} \cL_0 \Bigl(\frac{\tilde t}{\mu^2}\Bigr) -\delta(\tilde t) \ln \frac{zt'}{\mu^2} \biggr]
\Bigl[-\frac{1}{\eps}\, \delta(1-z) + \cL_0(1 - z)z \Bigr]
\nn\\ & \qquad\hspace{18ex}
 + \biggl[\frac{1}{\mu^2} \cL_1\Bigl(\frac{\tilde t}{\mu^2}\Bigr) -\delta(\tilde t) \Bigl(\frac{1}{2}\ln^2\frac{t'}{\mu^2} + \frac{\pi^2}{6}\Bigr)
   \biggr] \delta(1 - z)
 \biggr\}
\,.\end{align}

For the diagram in \fig{Bone_c} (and its mirror diagram) we insert the vacuum intermediate state between the fields in $\op_q$ as in \eq{DiscTB_full}, resulting in a one-loop virtual diagram involving a single field. The calculation is exactly the same as for the $\delta(p^-\! -\w)$ term in \eq{PDFb1} times an overall $\delta(t)$,
\begin{align} \label{eq:Bc_final}
  &\Mae{q_n}{\theta(\w) \op_q^\bare(t,\w)}{q_n}^{(c)}\,
  \nn \\ & \qquad
  = - 2\img \Bigl(\frac{e^{\gamma_E} \mu^2}{4\pi}\Bigr)^\eps g^2 C_F\,
  \delta(t) \delta(p^-\! -\w) \int\! \frac{\df^d\ell}{(2\pi)^d}\,
  \frac{2p^-\ell^-}{(\ell^-\! - p^-)(\ell^2+\img 0) [(\ell-p)^2+\img 0] }
\nn \\ & \qquad
 = -\frac{\alpha_s(\mu) C_F}{\pi}\,
 \Gamma(\eps) \Bigl(\frac{e^{\gamma_E}\mu^2}{t'}\Bigr)^\eps
  \delta(t) \delta(1 -z )\, \frac{\Gamma(2-\eps)\Gamma(-\eps)}{\Gamma(2-2\eps)}
\nn \\ & \qquad
 = \frac{\alpha_s(\mu) C_F}{\pi}\, \delta(\tilde t)\delta(1 -z ) \biggl\{
 \frac{1}{\eps^2} + \frac{1}{\eps} \Bigl(1 - \ln\frac{t'}{\mu^2}\Bigr)
  + \frac12 \ln^2\frac{t'}{\mu^2} - \ln\frac{t'}{\mu^2} + 2 -\frac{\pi^2}{12} \biggr\}
\,.\end{align}
In the last step we expanded in $\eps$ and took the IR limit. To be consistent we have to use the same IR limit in the virtual diagrams as in the real-emission diagrams above, which simply turns the overall $\delta(t)$ into a $\delta(\tilde t)$,
\begin{equation} \label{eq:Bc_limit}
\lim_{t'\to 0} \delta(t) = \lim_{zt'/\mu^2\to 0} \delta(\tilde t - zt') = \delta(\tilde t)
\,.\end{equation}
As in the PDF calculation, the UV divergence in the loop produces a $\Gamma(\eps)$ and the soft IR divergence a $\Gamma(-\eps)$. The latter is converted by the zero-bin subtraction into a UV divergence, producing the $1/\eps^2$ pole. The $1/\eps^2$ poles do not cancel anymore between Figs. \ref{fig:Bone_b} and \ref{fig:Bone_c} as they did for the PDF in \fig{fone_b}, because the phase space of the real emission in \fig{Bone_b} is now restricted by the measurement of $b^+$ via the $\delta(\ell^+ + b^+ - p^+)$. For the same reason \fig{Bone_a} has no UV divergence anymore, while \fig{fone_a} did. The $(1/\eps)\ln t'$ terms in \eqs{Bb_final}{Bc_final}, which are a product of UV and collinear IR divergences, still cancel between the real and virtual diagrams, ensuring that the UV renormalization is independent of the IR, as should be the case.

The final one-loop contribution to the quark matrix element, \fig{Bone_d} and its mirror diagram, comes from wave-function renormalization,
\begin{align} \label{eq:Bd_final}
\Mae{q_n}{\theta(\w)\op_q^\bare(t,\w)}{q_n}^{(d)}
&= \delta(t) \delta(1-z) (Z_\xi -1) \nn \\
&= -\frac{\alpha_s(\mu) C_F}{4\pi}\, \delta(\tilde t)\delta(1-z)
  \biggl\{\frac{1}{\eps} - \ln\frac{t'}{\mu^2} + 1 \biggr\}
\,.\end{align}
Adding up the results in Eqs.~\eqref{eq:Ba_final}, \eqref{eq:Bb_final}, \eqref{eq:Bc_final}, and \eqref{eq:Bd_final}, we obtain the bare beam function quark matrix element at one loop,
\begin{align} \label{eq:Bqbare}
   \Mae{q_n}{\theta(\w)\op_q^\bare(t,\w)}{q_n}^\one
  &= \frac{\alpha_s(\mu) C_F}{2\pi}\,\theta(z) \biggl\{
   \biggl[ \delta(\tilde t) \Bigl(\frac{2}{\eps^2} +
    \frac{3}{2\eps} \Bigr) -
    \frac{2}{\eps}\, \frac{1}{\mu^2} \cL_0\Bigl(\frac{\tilde t}{\mu^2}\Bigr)
  \biggr]\delta(1-z)
  \nn \\ & \quad 
  +\frac{2}{\mu^2} \cL_1\Bigl(\frac{\tilde t}{\mu^2}\Bigr)\delta(1-z) +
  \frac{1}{\mu^2} \cL_0\Bigl(\frac{\tilde t}{\mu^2}\Bigr)\cL_0(1-z)(1 + z^2)
  \\\nn & \quad 
  -\delta(\tilde t) \biggl[ P_{qq}(z)\ln\frac{zt'}{\mu^2} - \delta(1-z) \Bigl(\frac{7}{2} - \frac{\pi^2}{2}\Bigr) + \theta(1-z)(1-z) \biggr]
  \biggr\}
\,.\end{align}

We now consider the beam function matrix element with external gluons. The corresponding diagrams are shown in \figs{Bone_e}{Bone_f}. For \fig{Bone_e}, which is analogous to \fig{fone_d}, we find
\begin{align}  \label{eq:Be}
 & \Mae{g_n}{\theta(\w)\op_q^\bare(t,\w)}{g_n}^{(e)}
\nn \\ & \quad
 = \img\Bigl(\frac{ e^{\gamma_E} \mu^{2\eps}}{4\pi}\Bigr)^\eps g^2 T_F \frac{\theta(\w)}{\w}\,
2p^- \Bigl(\frac{1}{1-z} - \frac{4z}{d - 2}\Bigr)
\Disc_{t>0}\! \int\! \frac{\df^d\ell}{(2\pi)^d}\,
 \frac{\vec\ell_\perp^2\, \delta(\ell^-\! -\w) \delta(\ell^+\! + b^+\! - p^+)}{(\ell^2+\img 0)^2 [(\ell-p)^2+\img 0]}
\nn\\ & \quad
= \frac{\alpha_s(\mu)T_F}{2\pi}\,\frac{\theta(z)}{z}\,
 \Gamma(1+\eps) (e^{\gamma_E} \mu^2)^\eps \Bigl(\frac{1-\eps}{1-z} - 2z\Bigr)
  \Bigl[-\frac{\img}{2\pi}\Disc_{t>0}\, I_2(A, B, \eps) \Bigr]
\nn\\ & \quad
= \frac{\alpha_s(\mu)T_F}{2\pi}\,\theta(z) P_{qg}(z)\,\frac{\theta(t)t}{(t + zt')^2}
\,.\end{align}
The loop integral and discontinuity are exactly the same as for \fig{Bone_a}. The diagram in \fig{Bone_f} does not contribute to the quark beam function. It can be obtained from \eq{Be} by replacing $p^\mu \to -p^\mu$, which takes $t'\to -t'$ and $z\to -z$. Doing so, the only contribution to the discontinuity is still from $B = -(1 + z)t/z$ for $B > 0$, which for $t>0$ requires $-1 < z < 0$. Hence, \fig{Bone_f} does not contribute. Using \eq{Ba_limit} to take $t'\to 0$ in \eq{Be}, we get the final result for the bare one-loop gluon matrix element
\begin{equation} \label{eq:Bgbare}
\Mae{g_n}{\theta(\w) \op_q^\bare(t,\w)}{g_n}^\one
 = \frac{\alpha_s(\mu)  T_F}{ 2\pi }\, \theta(z) P_{qg}(z)
\biggl\{
  \frac{1}{\mu^2}\cL_0\Bigl(\frac{\tilde t}{\mu^2}\Bigr) - \delta(\tilde t) \Bigl(\ln \frac{z t'}{\mu^2} + 1 \Bigr)
\biggr\}
\,.\end{equation}
As for \fig{Bone_a}, it has no UV divergences because of the measurement of $b^+$, which means that the renormalization does not mix $\op_q$ and $\op_g$.

\section{Renormalization and Matching}
\label{sec:NLO_matching}

Using the bare matrix elements calculated in the previous section, we can extract the renormalization of $\op_q$. We first take $\tilde t = t + zt'\to t$ in the bare matrix elements. Then, expanding the quark matrix element of \eq{op_ren} to one-loop order,
\begin{align} \label{eq:Bbare_ren}
&\Mae{q_n}{\op_q^\bare(t,\w)}{q_n}^\one
\nn \\ &\qquad
= \int\! \df t'\,
\biggl[Z^{q\one}_B(t-t', \mu) \Mae{q_n}{\op_q(t',\w,\mu)}{q_n}^\zero + Z_B^{q\zero}(t-t', \mu) \Mae{q_n}{\op_q(t',\w,\mu)}{q_n}^\one \biggr]
\nn \\ &\qquad
= Z_B^{q\one}(t, \mu)\, \delta(1-z) + \Mae{q_n}{\op_q(t,\w,\mu)}{q_n}^\one
\,,\end{align}
we can then read off the $\overline{\mathrm{MS}}$ renormalization constant from \eq{Bqbare}
\begin{equation} \label{eq:ZB}
 Z_B^q(t,\mu) = \delta(t) +
  \frac{\alpha_s(\mu) C_F}{2\pi}
   \biggl[ \delta(t) \Bigl(\frac{2}{\eps^2} +
    \frac{3}{2\eps} \Bigr) -
    \frac{2}{\eps} \frac{1}{\mu^2} \cL_0\Bigl(\frac{t}{\mu^2}\Bigr)
  \biggr]
\,.\end{equation}
The fact that the gluon matrix element is UV finite and the UV divergences in the quark matrix element are proportional to $\delta(1-z)$ confirms at one loop our general result that the renormalization of the beam function does not mix quarks and gluons or change the momentum fraction.

In \eqs{Bbare_ren}{ZB} we used that we already know the structure of the renormalization from our general arguments in \sec{B_RGE}, i.e.\ that $Z_B^q$ only depends on the difference $t - t'$. Alternatively, we can also use the dependence on $z$ and the finite dependence on $t'$ via $\tilde t$ to explicitly check the structure of the renormalization. In this case, we must use the same IR limit also for the tree-level result in \eq{op_tree}, which using \eq{Bc_limit} becomes
\begin{equation} \label{eq:Btree}
\Mae{q_n}{\theta(\w)\op_q(t,\w, \mu)}{q_n}^\zero
=\lim_{t'\to 0}\delta(t)\, \delta(1-z) = \delta(\tilde t)\,\delta(1-z)
\,.\end{equation}
Taking $Z_B^q(t, t', \w/\w', \mu)$ to be a general function of $t$, $t'$ and $\w/\w'$, we now get for \eq{Bbare_ren}
\begin{align}
&\Mae{q_n}{\op_q^\bare(t,\w)}{q_n}^\one
\nn\\ & \qquad
= \int\!\df t''\frac{\df\w'}{\w'}\, Z_B^{q\one}\Bigl(t, t'', \frac{\w}{\w'}, \mu\Bigr)\, \delta(t'' + z' t')\, \delta\Bigl(1- \frac{\w'}{p^-}\Bigr) + \Mae{q_n}{\op_q(t,\w,\mu)}{q_n}^\one
\nn\\ & \qquad
= Z_B^{q\one}(t, -t', z) + \Mae{q_n}{\op_q(t,\w,\mu)}{q_n}^\one
\,.\end{align}
In the first step we used \eq{Btree} and $Z_B^{q\zero}(t, t', z) = \delta(t - t')\delta(1 - z)$. From \eq{Bqbare} we now find
\begin{equation}
 Z_B^q(t, t', z, \mu)
 = \biggl\{\delta(t - t') + \frac{\alpha_s(\mu) C_F}{2\pi}
   \biggl[ \delta(t - t') \Bigl(\frac{2}{\eps^2} +
    \frac{3}{2\eps} \Bigr) - \frac{2}{\eps} \frac{1}{\mu^2} \cL_0\Bigl(\frac{t - t'}{\mu^2}\Bigr)
  \biggr] \biggr\}\, \delta(1-z)
\,,\end{equation}
thus explicitly confirming at one loop that $Z_B^q(t, t', z, \mu) \equiv Z_B^q(t - t', \mu)\, \delta(1-z)$.

The one-loop anomalous dimension for the quark beam function follows from \eq{ZB},
\begin{align} \label{eq:gaB1}
\gamma_B^q(t,\mu) &=
-\mu\, \frac{\df}{\df\mu} Z_B^{q\one}(t,\mu)
= \frac{\alpha_s(\mu) C_F}{\pi}
  \biggl[ - \frac{2}{\mu^2} \cL_0\Bigl(\frac{t}{\mu^2}\Bigr) + \frac{3}{2}\, \delta(t)
  \biggr]
\,.\end{align}
It is identical to the one-loop anomalous dimension of the quark jet function. The coefficient of $\cL_0(t/\mu^2)/\mu^2$ can be identified as the one-loop expression for $-2\Gamma_\cusp^q$. Thus, \eq{gaB1} explicitly confirms the general results in \eqs{gaB_gen}{gaJgaB} at one loop.

Taking the bare matrix elements in \eqs{Bqbare}{Bgbare} and subtracting the UV divergences using \eqs{Bbare_ren}{ZB} gives the renormalized one-loop beam function matrix elements,
\begin{align} \label{eq:Bren}
\Mae{q_n}{\theta(\w) \op_q(t,\w, \mu)}{q_n}^\one
  &= \frac{\alpha_s(\mu) C_F}{2\pi}\,
 \theta(z) \biggl\{
  \frac{2}{\mu^2} \cL_1\Bigl(\frac{t}{\mu^2}\Bigr) \delta(1\!-\!z) \!+\!
  \frac{1}{\mu^2} \cL_0\Bigl(\frac{t}{\mu^2}\Bigr) \cL_0(1\!-\!z)(1 \!+\! z^2)
  \nn\\ & \quad
  -\delta(t) \biggl[ P_{qq}(z)\ln\frac{zt'}{\mu^2}
  -\delta(1-z) \Bigl(\frac{7}{2} - \frac{\pi^2}{2}\Bigr) + \theta(1-z)(1-z) \biggr]
   \biggr\}
  \,, \nn \\
\Mae{g_n}{\theta(\w) \op_q(t,\w, \mu)}{g_n}^\one
 &= \frac{\alpha_s(\mu) T_F}{ 2\pi }\, \theta(z) P_{qg}(z)
\biggl\{
  \frac{1}{\mu^2}\cL_0\Bigl(\frac{t}{\mu^2}\Bigr) - \delta(t) \Bigl(\ln \frac{z t'}{\mu^2} + 1 \Bigr)
\biggr\}
\,.\end{align}
For the matching onto the PDFs, we must take $t' \to 0$ and have therefore set $\tilde t = t$ everywhere, only keeping $t'$ in the IR divergent $\ln t'$ terms.

Expanding the OPE for the quark beam function, \eq{beam_fact}, to one loop, we have
\begin{align}
&\Mae{q_n}{\theta(\w) \op_q(t,\w,\mu)}{q_n}^\one
\nn \\&\qquad
  = \sum_j \int\! \frac{\df \w'}{\w'}\,
  \bigg[\cI_{qj}^\one\Big( t,\frac{\w}{\w'},\mu \Big) \Mae{q_n}{\oq_j(\w',\mu)}{q_n}^\zero +
  \cI_{qj}^\zero\Big( t,\frac{\w}{\w'},\mu \Big) \Mae{q_n}{\oq_j(\w',\mu)}{q_n}^\one \bigg]
\nn \\ & \qquad
  = \cI_{qq}^\one (t,z,\mu) + \delta(t) \Mae{q_n}{\oq_q(\w,\mu)}{q_n}^\one
  \,, \nn \\
&\Mae{g_n}{\theta(\w) \op_q(t,\w,\mu)}{g_n}^\one
\nn\\ & \qquad
= \sum_j \int\! \frac{\df \w'}{\w'}\,
  \bigg[\cI_{qj}^\one\Big( t,\frac{\w}{\w'},\mu \Big) \Mae{g_n}{\oq_j(\w',\mu)}{g_n}^\zero +
  \cI_{qj}^\zero\Big( t,\frac{\w}{\w'},\mu \Big) \Mae{g_n}{\oq_j(\w',\mu)}{g_n}^\one \bigg]
\nn \\ & \qquad
= \cI_{qg}^\one (t,z,\mu) + \de(t) \Mae{g_n}{\oq_q(\w,\mu)}{g_n}^\one
\,.\end{align}
Thus, the one-loop matching coefficients, $\cI^\one_{qi}(t,z,\mu)$, are obtained by subtracting the renormalized PDF matrix elements in \eq{fren} from those in \eq{Bren}. Doing so, we see that the $\ln t'$ IR divergences in \eqs{fren}{Bren} precisely cancel, as they must, such that the matching coefficients are independent of the IR regulator and only involve large logarithms that are minimized at the scale $\mu^2\simeq t$. 
\newpage
The final result for the NLO matching coefficients is given by
\begin{align} \label{eq:Iresult}
\cI_{qq}(t,z,\mu)
&= \delta(t)\, \delta(1 - z)
\\ & \quad
  + \frac{\alpha_s(\mu)C_F}{2\pi}\, \theta(z) \biggl\{
  \frac{2}{\mu^2} \cL_1\Bigl(\frac{t}{\mu^2}\Bigr) \delta(1 - z) +
  \frac{1}{\mu^2} \cL_0\Bigl(\frac{t}{\mu^2}\Bigr) \Bigl[P_{qq}(z) - \frac{3}{2}\,\delta(1-z)\Bigr]
  \nn \\ & \quad
  + \delta(t) \biggl[
  \cL_1(1 - z)(1 + z^2)
  -\frac{\pi^2}{6}\, \delta(1 - z)
  + \theta(1 - z)\Bigl(1 - z - \frac{1 + z^2}{1 - z}\ln z \Bigr) \biggr]
  \biggr\}
  \,, \nn\\\nn
\cI_{qg}(t,z,\mu)
 &= \frac{ \alpha_s(\mu) T_F }{2\pi}\, \theta(z) \biggl\{
\frac{1}{\mu^2} \cL_0\Bigl(\frac{t}{\mu^2}\Bigr) P_{qg}(z)
 + \delta(t) \biggl[P_{qg}(z)\Bigl(\ln\frac{1-z}{z} - 1\Bigr) +  \theta(1-z) \biggr]
\biggr\}
\,.\end{align}


\chapter{The Gluon Beam Function}
\label{ch:gluon}

In this chapter we compute the matching coefficients for the gluon beam function at one loop. We determine the renormalization group equation for the gluon beam function and verify that its anomalous dimension equals that of the gluon jet function. These calculations will appear in Ref.~\cite{Berger:2010BgNLO}.

The matrix element which defines the gluon beam function first appeared in Ref.~\cite{Fleming:2006cd} and was named the gluon beam function in Ref.~\cite{Stewart:2009yx}. In Ref.~\cite{Mantry:2009qz} the impact-parameter gluon beam function was introduced, which differs from the gluon beam function in \eqs{tiop_def}{B_def} because the Lorentz indices on the two $\cB_{n\perp}^\mu$ fields are not contracted and the fields have an additional separation in the residual perpendicular coordinate $y_\perp$. In Refs.~\cite{Fleming:2006cd} and \cite{Mantry:2009qz} only the matching onto gluon PDFs was considered and there is a discrepancy in one of the terms. Our results agree with those presented in \cite{Mantry:2009qz} and we also determine the matching onto quark PDFs.

In our calculation we will regularize both the UV and the IR using dimensional regularization and employ the \MSbar renormalization scheme. To perform the renormalization of the beam function and calculate the matching coefficients we replace the external proton states by free quarks or gluons, as was explained in \sec{B_RGE} for the tree-level matching. In our partonic calculations we will abbreviate the matrix element notation used in the calculation of the quark beam function in \ch{quark} by writing $f_{i/j}(z,\mu)$ and $B_{i/j}(t,z,\mu)$ to indicate that the proton is replaced by a parton of type $j$ in the PDF or beam functions, i.e.
\begin{align}
B_{g/j}(\w b^+,\w/p^-)
&= -\theta(\w) \!\int\! \frac{\df y^-}{4\pi}\,e^{\img b^+ y^-/2}
   \MAe{j_n(p^-)}{\cB_{n\perp\mu}^c\Bigl(y^- \frac{n}{2}\Bigr)
   \delta(\w-\bnP_n)\, \cB_{n\perp}^{ c \,\mu}(0)}{j_n(p^-)}
   \,, \nn \\
f_{g/j}(\w'/p^-)
&= -\theta(\w') \w' \MAe{ j_n(p^-)}{\cB_{n\perp\mu}^c(0)\,
  \delta(\w' - \bnP_n)\, \cB_{n\perp}^{c \,\mu }(0) }{j_n(p^-)}
  \,.
\end{align}

We start in \sec{gluon_PDF} with recalling the known results for the partonic PDF for our choice of regulator and scheme. In \sec{feynmanrules} we list the necessary Feynman rules, for which we introduce some shorthand notation. We calculate the one-loop gluon beam function in \sec{gluoncalc} and extract the results in \sec{gluonresults}. 

\section{The Gluon PDF at One Loop}
\label{sec:gluon_PDF}

At tree level the partonic PDF is normalized to
\begin{align} \label{eq:f_tree}
  f_{i/j}^\zero(z,\mu) = \de_{ij}\, \de(1-z)
  \,.
\end{align}
The renormalization of the PDF in the \MSbar scheme is given by
\begin{align} \label{eq:f_ren}
  & f_i^\bare(\xi)
 = \sum_j \int\! \frac{\df \xi'}{\xi'}\, Z^f_{ij}\Big(\frac{\xi}{\xi'},\mu\Big) f_j(\xi',\mu)
 = \sum_j \int\! \frac{\df z}{z}\, Z^f_{ij}(z,\mu) f_j\Big(\frac{\xi}{z},\mu\Big)
 \,,
\end{align}
where we sum over parton species $j = q, \bar q, g,$  and the entries in the matrix $Z^f$ are a series of $1/\eps$ poles with coefficients in terms of the renormalized coupling $\al_s(\mu)$. At one-loop, the results for the standard gluon PDF are~\cite{Altarelli:1977zs},
\begin{align} \label{eq:f_Z_1loop}
  Z^f_{gg}(z,\mu) &= \delta(1-z) + \frac{1}{\eps} \frac{\alpha_s(\mu)}{2\pi}\,
  \theta(z)\, \Bigl[C_A P_{gg}(z) + \frac{1}{2} \bt_0\, \de(1-z)\Bigr]
  \,, \nn \\
  Z^f_{gq}(z,\mu) &= \frac{1}{\eps} \frac{\alpha_s(\mu) C_F}{2\pi}\,
  \theta(z)\, P_{gq}(z)
  \,,
\end{align}
where $\bt_0 = (11 C_A - 4 n_f T_f)/3$, is the lowest order coefficient of the QCD $\beta$ function. The splitting functions are given by
\begin{align} \label{eq:split}
P_{gg}(z)
&= 2 \cL_0(1-z)z + 2\theta(1-z)\Bigl[\frac{1-z}{z} +  z(1-z)\Bigr]
\nn\\
&= 2\theta(1-z)\Bigl[\frac{z}{(1-z)_+} + \frac{1-z}{z} +  z(1-z)\Bigr]
\,,\nn\\
P_{gq}(z) &= \theta(1-z)\, \frac{1+(1-z)^2}{z}
\,.\end{align}

We will extract the one-loop partonic gluon PDF from the above results. With our choice of dimensional regularization as IR regulator, the only Lorentz invariant quantity they can depend on is $z$. This is dimensionless, so all diagrams contributing to the PDF calculation vanish. Hence, the one-loop PDF has IR divergences with signs opposite to the UV divergences. The UV divergences follow from \eq{f_Z_1loop} by looking at the renormalization of the partonic gluon PDF at one-loop
\begin{align} \label{eq:f_ren_1loop}
  f_{g/g}^{\bare\, \one}(z)\,
  &= \sum_j \int\! \frac{\df z'}{z'}\,
  \Big[ Z^{f\one}_{gj}\Big(\frac{z}{z'},\mu\Big) f_{j/g}^\zero(z',\mu) +
  Z^{f\zero}_{gj}\Big(\frac{z}{z'},\mu\Big) f_{j/g}^\one(z',\mu) \Big]
  \nn \\
  & = Z^{f\one}_{gg}(z,\mu) + f_{g/g}^\one(z,\mu)
  \,, \nn \\
  f_{g/q}^{\bare\, \one}(z)\,
  &= \sum_j \int\! \frac{\df z'}{z'}\,
  \Big[ Z^{f\one}_{gj}\Big(\frac{z}{z'},\mu\Big) f_{j/q}^\zero(z',\mu) +
  Z^{f\zero}_{gj}\Big(\frac{z}{z'},\mu\Big) f_{j/q}^\one(z',\mu) \Big]
  \nn \\
  &= Z^{f\one}_{gq}(z,\mu) + f_{g/q}^\one(z,\mu)
  \,.
\end{align}
This implies that the UV divergent part of the partonic $f_{i/j}^\one(z,\mu)$ is just $Z^{f\one}_{ij}(z,\mu)$. We thus obtain the renormalized PDF
\begin{align}\label{eq:f_g}
f^\one_{g/g}(z,\mu) &= - \frac{1}{\eps} \frac{\al_s(\mu)}{2\pi}\, \theta(z)
\Bigl[C_A P_{gg}(z) + \frac{1}{2} \bt_0\, \de(1-z)\Bigr]
\,, \nn \\
f^\one_{g/q}(z,\mu) &= - \frac{1}{\eps} \frac{\al_s(\mu) C_F}{2\pi}\, \theta(z) P_{qg}(z)
\,.
\end{align}

\section{SCET Feynman Rules}
\label{sec:feynmanrules}

\begin{figure}[t!]
\hfill%
\subfigure[]{\includegraphics[scale=0.75]{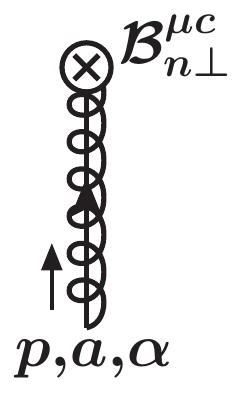}\label{fig:Brules0}}%
\hfill%
\subfigure[]{\includegraphics[scale=0.75]{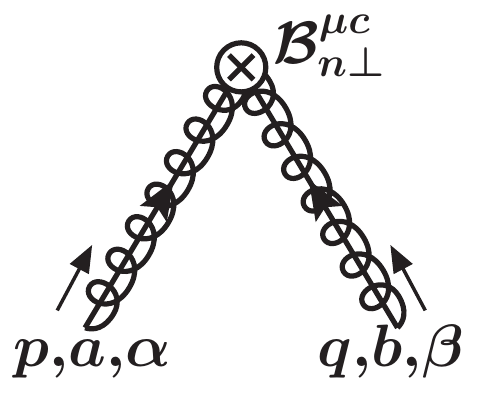}\label{fig:Brules1}}%
\hspace*{\fill}%
\caption[The SCET Feynman diagrams for the gluon field strength.]{The SCET Feynman diagrams for the gluon field strength at $\ord{g^0}$ (a) and $\ord{g}$ (b) given by the Feynman rules in 
\eq{gluon0} and \eq{gluon1}, respectively.}
\label{fig:Brules}
\end{figure}
We now list the SCET Feynman rules necessary for the calculations.
The Feynman rules for the gluon field strength, $\cB_{n\perp}^\mu$,
\begin{equation}
\cB_{n\perp}^\mu = \frac{1}{g} \Big[ W_n^\dagger\, \img D_{n\perp}^\mu W_n \Big]
\end{equation}
are illustrated in \fig{Brules}. The Feynman rule at order $\ord{g^0}$ is
\begin{equation} \label{eq:gluon0}
\text{\fig{Brules0}} = \delta^{ca} \Bigl(g_\perp^{\mu \al} -
\frac{p_\perp^\mu \bn^\al}{\bn \sdt p}  \Bigr) \equiv \delta^{ca} \cB^{(0)\,\mu \al}_{n\perp}(p)
\,,\end{equation}
and at $\ord{g}$ it is given by
\begin{align} \label{eq:gluon1}
\text{\fig{Brules1}}  =
 \img g f^{cab} \biggl[
   \frac{g_\perp^{\mu \bt} \bn^\al}{\bn \sdt p}
   -\frac{g_\perp^{\mu \al} \bn^\bt}{\bn \sdt q}
   +\Bigl(\frac{p_\perp^\mu}{\bn \sdt q} -\frac{q_\perp^\mu}{\bn \sdt p}\Bigr)
   \frac{\bn^\al \bn^\bt}{\bn \sdt (p+q)}
      \biggr]
 \equiv ig f^{cab} \cB^{(1)\,\mu \al \bt}_{n\perp}(p, q)
\,. \end{align}

We abbreviate the collinear quark-gluon vertex as
\begin{equation} \label{eq:quarkglue1}
\img g\, T^a \Bigl(
n^\mu + \frac{\pslash_\perp \gamma_\perp^\mu}{\bn \sdt p}
+ \frac{\gamma_\perp^\mu \qslash_\perp }{\bn \sdt q}
- \frac{\pslash_\perp \qslash_\perp}{\bn \sdt p \, \bn \sdt q}\, \bn^\mu
\Bigr) \frac{\bnslash}{2}
\equiv \img g\, T^a V^\mu_n (p,q)\, \frac{\bnslash}{2}
\,,\end{equation}
where $p$ and $q$ are the momenta of the outgoing and incoming quark lines, so the gluon carries incoming momentum $p - q$.
Finally, we abbreviate the triple gluon vertex by
\begin{align}
g f^{abc} V_3^{\mu \nu \rho} (p_1, p_2, p_3) & \equiv g f^{abc} [g^{\mu \nu} (p_1 - p_2)^\rho + g^{\nu \rho} (p_2 - p_3)^\mu + g^{\rho \mu} (p_3 - p_1)^\nu]
\,,
\end{align}
where all momenta are incoming and momentum conservation holds, $p_1 + p_2 + p_3 = 0$.

\section{The Gluon Beam Function at One Loop}
\label{sec:gluoncalc}

\begin{figure}
\subfigure[]{\includegraphics[scale=0.75]{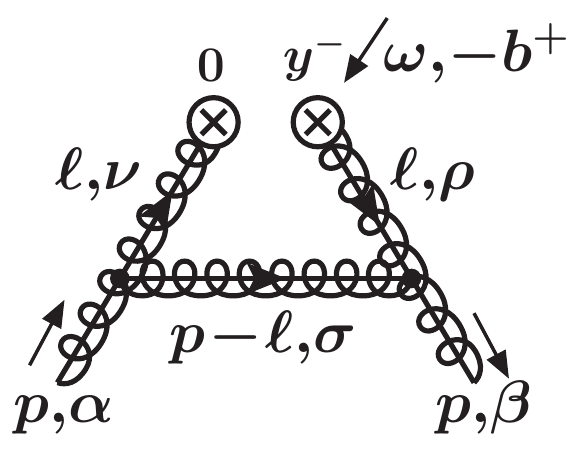}\label{fig:Bgone_a}}%
\hfill%
\subfigure[]{\includegraphics[scale=0.75]{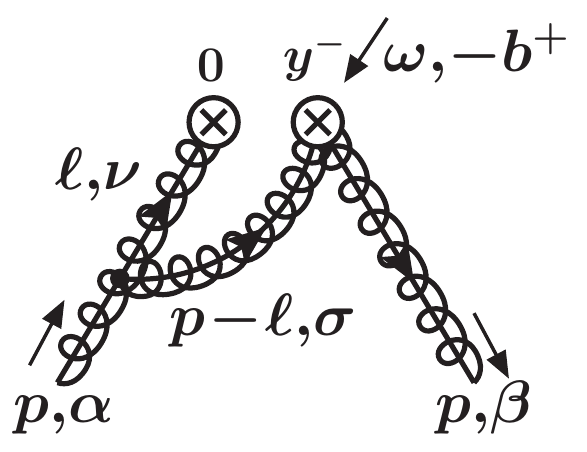}\label{fig:Bgone_b}}%
\hfill%
\subfigure[]{\includegraphics[scale=0.75]{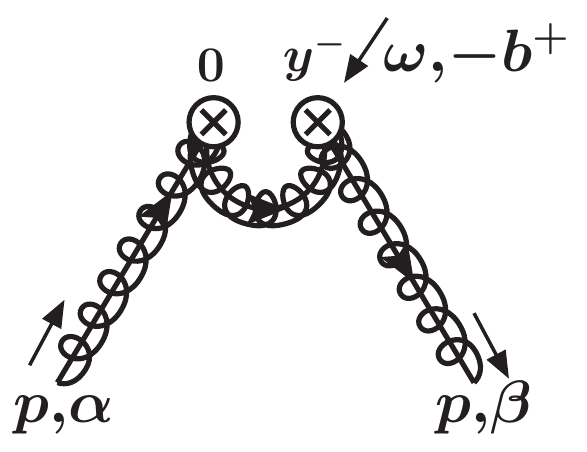}\label{fig:Bgone_c}}%
\\
\subfigure[]{\includegraphics[scale=0.75]{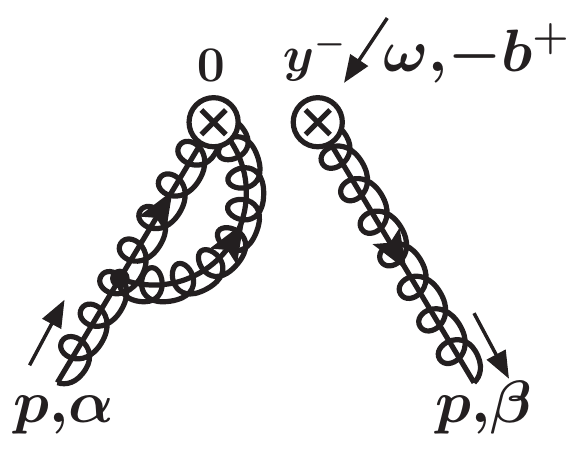}\label{fig:Bgone_d}}%
\hfill%
\subfigure[]{\includegraphics[scale=0.75]{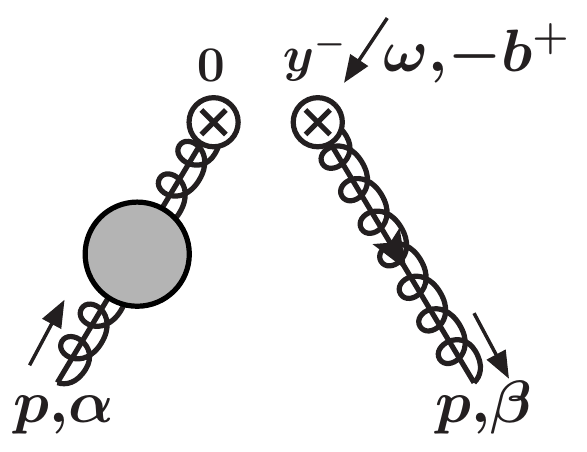}\label{fig:Bgone_e}}%
\hfill%
\subfigure[]{\includegraphics[scale=0.75]{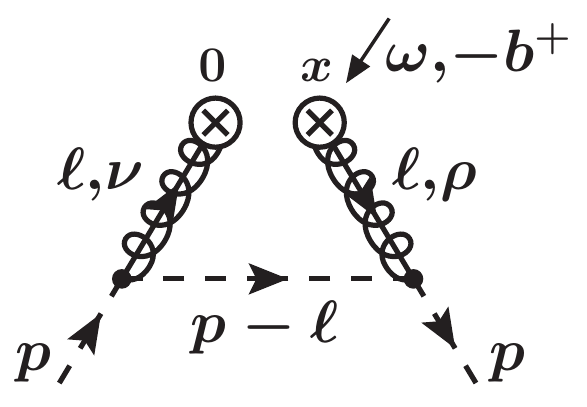}\label{fig:Bgone_f}}%
\caption[One-loop diagrams for the gluon beam function.]{One-loop diagrams for the gluon beam function. The minus momentum $\w$ is incoming at the vertex and the $b^+$ momentum is outgoing. Graphs (b), (d), and (e) have symmetric counterparts which are equal to the ones shown and included in the computation. The diagrams with the external lines crossed vanish and are not shown.}
\label{fig:Bgone}
\end{figure}

We now turn to the calculation of the partonic gluon beam function. The beam function was defined in \eq{B_def} such that at tree level it is normalized to
\begin{equation}
B^\zero_{i/j} (t, z, \mu) = \de_{ij}\, \delta(t) \delta(1 - z)
\,,
\end{equation}
The one-loop Feynman diagrams for the partonic calculation are shown in \fig{Bgone}. Since we regulate both the UV and IR with dimensional regularization, the virtual diagrams in \figs{Bgone_d}{Bgone_e} vanish because there is only the $p^-$ momentum of the external gluon flowing into the loop, which is insufficient to give a nonzero Lorentz-invariant result for the loop integral. This means that the UV divergences cancel the IR divergences. Performing the wave function renormalization in the on shell scheme, both the wave function renormalization $Z_\psi$ as well as the residue $R_\psi$ that enters in LSZ are equal to one. (A different scheme would lead to contributions to $Z_\psi$ and $R_\psi$ that cancel each other in the final result.)

To compute the one-loop real radiation diagrams for the gluon beam function, we use the relationship with the matrix element of the time-ordered product in \eq{DiscTB}. We always average over the polarizations of the external gluons, and since their momentum is $p^\mu=p^- n^\mu/2$ this gives us the rule
\begin{equation}
\frac{1}{d-2} \sum_\text{pol}\, \ve^\al\, {\ve^*}^\bt \to -\frac{g_\perp^{\al\bt}}{d-2}
\,,
\end{equation}
where $d = 4 - 2\eps$. We also average over the color of the external gluons (which is trivial). This choice for the polarizations and colors of the external states is just a matter of calculational convenience, in order to determine the renormalization and matching all we need is states with non-zero overlap with the operator.

Let us start with the diagram in \fig{Bgone_a}. To make our expressions more palatable, we first perform the color algebra, which simply yields
\begin{equation}
\frac{\delta^{ab}}{N_c^2-1} f^{aec} f^{bde}\, \delta^{fc}\, \delta^{fd} = -C_A
\,.\end{equation}
The diagram is then given by
\begin{align}
  &\text{\fig{Bgone_a}} 
  \\ & \quad =
  -\img \Bigl(\frac{e^{\gamma_E} \mu^2}{4\pi}\Bigr)^{\eps} g^2 C_A\, \theta(\w)\,
  \Disc_{t>0} \int\! \frac{\df^d \ell}{(2\pi)^d}\,
  \frac{\de(\ell^-\! - \w) \de(\ell^+\! + b^+)}{(\ell^2 + \img 0)^2[(\ell-p)^2 + \img 0]}
  \nn \\ & \qquad\times
  \frac{g_{\perp\al}^\bt}{d-2}\,
  V_3^{\al \si \nu}(p,\ell-p,-\ell) V_{3\,\bt \rho \si}(-p,\ell,p-\ell)\,
   \cB^\zero_{n\perp \mu \nu}(\ell)\, \cB^{\zero\,\mu\rho}_{n\perp}(-\ell)
  \nn \\ & \quad
  = -\img \Bigl(\frac{e^{\gamma_E} \mu^2}{4\pi}\Bigr)^{\eps}
  \frac{g^2 C_A}{(2\pi)^2}\, \theta(z)\,
  \Disc_{t>0} \int_0^1\! \df \al\, (1-\al) \int\! \frac{\df^{d-2} \vec \ell_\perp}{(2\pi)^{d-2}}\,
  \frac{(5 + 4/z^2 ) \, \vec\ell_\perp^2 + (1 + 1/z ) t}{(\vec \ell_\perp^2 + \De)^3}
  \nn \\ & \quad
  = \frac{\al_s(\mu) C_A}{2\pi}\, \theta(z)\theta(1-z)
   2\Bigl[\frac{1-z}{z} + z(1-z) + \frac{z}{2}\Bigr]
  \Ga(1+\eps) (e^{\ga_E} \mu^2)^\eps
  \frac{\sin \pi\eps}{\pi \eps} \frac{\theta(t)}{t^{1+\eps}} \Bigl(\frac{z}{1-z}\Bigr)^\eps
,\nn\end{align}
with $\De = t(1-\al/z)$. To derive the second equality we integrated $l^+$ and $l^-$ using the delta functions and applied a Feynman parametrization. The last line follows from performing the $\vec{\ell}_\perp$ integral and using the discontinuities listed in \app{plusdisc}.
By expanding in $\eps$ we find
\begin{align}
 & \text{\fig{Bgone_a}} 
 \\ & \quad
  = \frac{\al_s(\mu)C_A}{2\pi}\, \theta(z) \theta(1-z)
   2\Bigl[\frac{1-z}{z} + z(1-z) + \frac{z}{2}\Bigr]
  \biggl[\delta(t)\Bigl(-\frac{1}{\eps} + \ln\frac{1-z}{z} \Bigr)
    + \frac{1}{\mu^2} \cL_0\Bigl(\frac{t}{\mu^2}\Bigr) \biggr]
\,, \nn \end{align}
where the plus distribution $\cL_0(x)$ is defined in \eq{plusdef}.
The color structure for the diagram in \fig{Bgone_b} yields
\begin{equation}
\frac{\delta^{ab}}{N_c^2-1} f^{aec}\, \de^{fc} f^{feb} = -C_A
\,.\end{equation}
Including a factor of two for its mirror graph, we get
\begin{align}
  \text{\fig{Bgone_b}} & = -2\img \Bigl(\frac{e^{\gamma_E} \mu^2}{4\pi}\Bigr)^{\eps}
  g^2 C_A\, \theta(\w)\, \Disc_{t>0} \int\! \frac{\df^d \ell}{(2\pi)^d}\,
  \frac{\de(\ell^- - \om) \de(\ell^+ + b^+)}{(\ell^2 + \img 0) \,[(\ell-p)^2 + \img 0]}
  \nn \\ & \quad \times
  \frac{g_{\perp \al}^\bt}{d-2}\,
   V_3^{\al \si \nu}(p,\ell-p,-\ell) \,
   \cB^{\zero}_{n\perp\, \mu \nu}(\ell)\,
   {\cB^{\one \mu}_{n\perp}}_{ \si \bt}(p-\ell,-p)
  \nn \\
  & = -\img \Bigl(\frac{e^{\gamma_E} \mu^2}{4\pi}\Bigr)^{\eps} \frac{g^2 C_A}{(2\pi)^2}\,
   \theta(z)\, \frac{1+z}{1-z}\, \Disc_{t>0} \int_0^1\! \df \al
  \int\! \frac{\df^{d-2} \vec\ell_\perp}{(2\pi)^{d-2}}\, \frac{1}{(\vec \ell_\perp^2 + \De)^2}
 \nn \\
  & = \frac{\al_s(\mu)C_A}{2\pi}\, \theta(z) (1+z)z^{1+\eps} \frac{\theta(1-z)}{(1-z)^{1+\eps}}\,
  \Ga(1+\eps)(\eps^{\ga_E} \mu^2)^\eps \frac{\sin \pi \eps}{\pi \eps}\,
  \frac{\theta(t)}{t^{1+\eps}}
\,,\end{align}
using the relations listed in \app{plusdisc}. Expanding in $\eps$ yields
\newpage
\begin{align}
  &\text{\fig{Bgone_b}}
  \\ & \quad =
  \frac{\al_s(\mu) C_A}{2\pi}\, \theta(z)
  \biggl\{ \biggl[-\frac{1}{\eps} \de(t) + \frac{1}{\mu^2} \cL_0\Bigl(\frac{t}{\mu^2}\Bigr)\biggr]
  \Bigl[-\frac{2}{\eps} \de(1 - z) + \cL_0(1 - z) z(1+z)\Bigr]
  \nn \\ & \qquad +
  \frac{2}{\mu^2} \cL_1\Bigl(\frac{t}{\mu^2}\Bigr) \de(1 - z) +
  \de(t) z(1+z) \Bigl[\cL_1(1 - z)  - \cL_0(1 - z)\ln z - \frac{\pi^2}{12} \de(1 - z) \Bigr] \biggr\}
\,. \nn
\end{align}
The diagram in \fig{Bgone_c} is zero. Because the external gluons have perpendicular polarization, contracting the two $\cB_{n\perp}^\one$ in \eq{gluon1} leads to $\bn \cdot \bn = 0$ in the numerator.

Crossing the external lines of these diagrams corresponds to $p\to-p$ (the polarization and color are symmetric and hence unaffected). By changing $\ell \to -\ell$ this yields the same as the original diagram but with $\de(\ell^- - \om) \de(\ell^+ + b^+) \to \de(\ell^- + \om) \de(\ell^+ - b^+)$ and thus
\begin{align} \label{eq:cross}
  &\De = \Big(1-\frac{\al}{z}\Big) t \to \Big(1+\frac{\al}{z}\Big) t
  \,.
\end{align}
Since $\al, z>0$ the discontinuity for $t>0$ vanishes for these graphs.
Other one-loop real radiation diagrams, like the one involving a four gluon vertex, vanish because their discontinuity vanishes. All cuts of these diagrams involve cutting one of the external lines, which sets $\Delta = t$ in the loop integral and leads to a vanishing discontinuity for $t>0$.

Adding up the two non-vanishing diagrams in \figs{Bgone_a}{Bgone_b}, we find
\begin{align}\label{eq:B_gg}
  B_{g/g}^{\bare\one}(t,z) &
  = \frac{\alpha_s(\mu) C_A}{2\pi}\, \theta(z)
  \bigg( \bigg[ \frac{2}{\eps^2}\, \de(t)
  - \frac{2}{\eps}\, \frac{1}{\mu^2} \cL_0\Big(\frac{t}{\mu^2}\Big) \bigg]   \de(1-z)
  - \frac{1}{\eps}\, \de(t) P_{gg}(z)
  \\ & \quad
  + \frac{2}{\mu^2} \cL_1\Big(\frac{t}{\mu^2}\Big) \de(1-z)
  + \frac{1}{\mu^2} \cL_0\Big(\frac{t}{\mu^2}\Big) P_{gg}(z)
  \nn \\ & \quad
  + \de(t)\Big\{ \Big[\cL_1(1-z) - \frac{\theta(1-z) \ln z}{1-z}\Big] \frac{2[1-z(1-z)]^2}{z} - \frac{\pi^2}{6} \de(1-z) \Big\} \bigg)
\,. \nn
\end{align}
This expression agrees with the expression found in Ref.~\cite{Mantry:2009qz}. Transforming to moment space, we find agreement with the expression quoted in Ref.~\cite{Fleming:2006cd} up to an extra term of $-\al_s C_A \delta(1-z) \pi^2/8$.

We now determine the mixing contribution of the quark PDF to the gluon beam function, which was not calculated in Refs.~\cite{Fleming:2006cd, Mantry:2009qz}. The relevant matrix element has external quarks instead of gluons, and the corresponding diagram is shown \fig{Bgone_f}. Using fixed spin and color or averaging over them yields again identical results. Here, we average over spins, using $\frac{1}{2} \sum_\text{spins}\, u_n(p)\bar{u_n}(p) = \frac{1}{2} \pslash$. The color average gives $(1/N_c) \tr[T^a T^b] \de^{ab} = C_F$. The diagram is thus given by
\begin{align} \label{eq:Bgone_f}
  & \text{\fig{Bgone_f}} 
  \\ & \quad 
  = -\img \Bigl(\frac{e^{\gamma_E} \mu^2}{4\pi}\Bigr)^{\eps}
  g^2 C_F\, \theta(\w)\,\Disc_{t>0} \int\! \frac{\df^d \ell}{(2\pi)^d} \,
  \frac{(p^-\! - \ell^-)\, \de(\ell^-\! - \w) \de(\ell^+\! + b^+)}{(\ell^2 + \img 0)^2 [(p-\ell)^2 + \img 0]}\,
  \nn \\ & \qquad \times
  \bar u_n(p) V^\rho_n (p, p-\ell) V_{n\, \nu}(p-\ell, p) \frac{\bnslash}{2}\, u_n(p)\,
 \cB^{\zero \mu \nu}_{n\perp}(\ell)\,  \cB^{\zero}_{n\perp \mu \rho}(-\ell)
  \nn \\ & \quad
  = \img \Bigl(\frac{e^{\gamma_E} \mu^2}{4\pi}\Bigr)^{\eps} \frac{g^2 C_F}{(2\pi)^2}\,
  \theta(z) \Bigl(\frac{d-2}{1-z} + \frac{4}{z^2} \Bigr)
  \Disc_{t>0} \int_0^1\! \df \al\, (1-\al)
  \int\! \frac{\df^{d-2} \vec\ell_\perp}{(2\pi)^{d-2}}\, \frac{\vec \ell_\perp^2}{(\vec \ell_\perp^2 + \De)^3}
\nn \\ & \quad
= \frac{\al_s(\mu) C_F}{2\pi}\, \theta(z)\theta(1-z) \Ga(1+\eps) (e^{\ga_E} \mu^2)^\eps
 \Bigl[\frac{1 + (1-z)^2}{z} - \eps z\Bigr] \frac{\sin \pi\eps}{\pi\eps}\, \frac{\theta(t)}{t^{1+\eps}}\,
 \Bigl(\frac{z}{1-z}\Bigr)^\eps
 \,. \nn
\end{align}
This is the same loop integral and discontinuity as in \fig{Bgone_a}. For the crossed graph, where the external lines are interchanged, we change $\ell \to -\ell$ to find the same expression as in \eq{Bgone_f} with $\de(\ell^- - \om) \de(\ell^+ + b^+) \to \de(\ell^- + \om) \de(\ell^+ - b^+)$. This leads to \eq{cross} and a vanishing discontinuity. Expanding in $\eps$, we obtain for the bare one-loop quark matrix element
\begin{align}\label{eq:B_gq}
  & B^{\bare\one}_{g/q}(t,z)
  \\ & \quad
  =  \frac{\al_s(\mu) C_F}{2\pi}\,
   \theta(z) \biggl\{ \frac{1}{\mu^2} \cL_0\Bigl(\frac{t}{\mu^2}\Bigr) P_{gq}(z)
   + \delta(t)\biggl[P_{gq}(z)\Bigl(-\frac{1}{\eps} + \ln \frac{1-z}{z} \Bigr) + \theta(1-z) z \biggr] \biggr\}
\,.\nn
\end{align}
The matrix element with external antiquarks gives the same result, so the mixing contributions from quarks and antiquarks are identical.

\section{Renormalization, Running and Matching}
\label{sec:gluonresults}

We know that the anomalous dimension of the gluon beam function equals that of the gluon jet function, and that the IR divergences of the gluon beam function cancel those of the gluon PDF in the matching. Since we regulate both the UV and IR using dimensional regularization, we cannot separate the UV and IR divergences and we can only verify one of these statements in our calculation. We will assume the cancellation of IR divergences, which then allows us to verify that the anomalous dimension equals that of the gluon jet function. Equivalently, we could have taken the anomalous dimension of the gluon jet function for granted and have checked that the IR divergences cancel in the matching.

We denote the counterterm that renormalizes the beam function by $Z_B^g(t-t',\mu)$
\begin{align} \label{eq:B_ren}
   B_g^\bare(t, x, \mu) = \int\! \df t'\, Z_B^g(t-t',\mu) B_g(t', x, \mu)
   \,.
\end{align}
In order to extract the renormalization for the beam function from our partonic calculation,
we expand this to one-loop order for gluon states
\begin{align} \label{eq:B_ren_1loop}
   B_{g/g}^{\bare\one}(t, z)
   &= \int\! \df t'\, \Big[Z^{g\one}_B(t-t',\mu) B_{g/g}^\zero(t', z, \mu) +
   Z^{g\zero}_B(t-t',\mu) B_{g/g}^\one(t', z, \mu)\Big]
   \nn \\
   &= Z^{g\one}_B(t,\mu) \de(1-z) + B_{g/g}^\one(t,z,\mu)
   \,.
\end{align}
Similarly expanding \eq{beam_fact} for the matching onto PDFs
\begin{align} \label{eq:beam_fact_1loop}
  B_{g/g}^\one(t,z,\mu)\,
  &= \sum_j \int \! \frac{\df z'}{z'}\,
  \Big[\cI_{gj}^\one(t,z',\mu) f_{j/g}^\zero\Big(\frac{z}{z'},\mu\Big)
  + \cI_{gj}^\zero(t,z',\mu) f_{j/g}^\one\Big(\frac{z}{z'},\mu\Big)\Big]
  \nn \\
  &= \cI_{gg}^\one(t,z,\mu) + \de(t) f_{g/g}^\one(z,\mu)
  \, \nn \\
  B_{g/q}^\one(t,z,\mu)\,
  &= \sum_j \int \! \frac{\df z'}{z'}\,
  \Big[\cI_{gj}^\one(t,z',\mu) f_{j/q}^\zero\Big(\frac{z}{z'},\mu\Big)
  + \cI_{gj}^\zero(t,z',\mu) f_{j/q}^\one\Big(\frac{z}{z'},\mu\Big)\Big]
  \nn \\
  &= \cI_{gq}^\one(t,z,\mu) + \de(t) f_{g/q}^\one(z,\mu)
  \,.
\end{align}

We know that the IR divergences of the PDF cancel those of the beam function in the matching \eq{beam_fact_1loop}.
Hence, after subtracting the IR divergences of the gluon PDF in \eq{f_g} from the gluon
beam function in \eqs{B_gg}{B_gq}, all remaining divergences are ultraviolet and enter
into the renormalization. There is no UV divergence in the mixing graph and
hence the RG evolution does not mix the quark and gluon beam functions. Using \eq{B_ren_1loop} we conclude that
\begin{align} \label{eq:ga_1loop}
  & Z_B^g(t,\mu) = \de(t) + \frac{\al_s(\mu)}{2\pi} \bigg\{2C_A \bigg[\frac{1}{\eps^2}
    - \frac{1}{\eps} \frac{1}{\mu^2}\cL_0\Big(\frac{t}{\mu^2}\Big)\bigg]
  + \frac{1}{2\eps} \bt_0\, \de(t) \bigg\}
  \,, \nn \\
  & \ga_B^{g\one}(t,\mu) =  - \mu\frac{\df}{\df\mu} Z_B^{g\one} (t,\mu)
  = -\frac{\al_s(\mu) C_A }{\pi} \frac{2}{\mu^2}\cL_0\Big(\frac{t}{\mu^2}\Big)
  + \frac{\al_s(\mu) \bt_0}{2\pi}\de(t)
\,.\end{align}
We see that our one-loop calculation agrees with the anomalous dimension of the gluon jet function, as it should according to \eq{gaB_gen}.

Since the PDFs in \eq{f_g} contain no finite pieces, we infer from \eq{beam_fact_1loop} that the remaining terms are simply the matching coefficients
\begin{align} \label{eq:I_results}
 \mathcal{ I}_{gg}(t,z,\mu) & = \de(t)\de(1-z) + \frac{\al_s(\mu) C_A}{2\pi}\, \theta(z)
  \bigg( \frac{2}{\mu^2} \cL_1\Big(\frac{t}{\mu^2}\Big) \de(1-z)
  + \frac{1}{\mu^2} \cL_0\Big(\frac{t}{\mu^2}\Big) P_{gg}(z)
  \nn \\ & \quad
  + \de(t)\Big\{\Big[\cL_1(1-z) - \frac{\theta(1-z) \ln z}{1-z} \Big] \frac{2[1-z(1-z)]^2}{z}
  - \frac{\pi^2}{6} \de(1-z) \Big\} \bigg)
  \,, \nn \\
  \mathcal{I}_{gq}(t,z,\mu) & = \frac{\al_s(\mu) C_F}{2\pi}\, \theta(z)
  \bigg\{P_{gq}(z)
  \bigg[\frac{1}{\mu^2} \cL_0\Big(\frac{t}{\mu^2}\Big) + \de(t) \ln \frac{1-z}{z}\bigg] + \theta(1-z) z\, \de(t) \bigg\}
\,.
\end{align}


\chapter{Beam Function Plots}
\label{ch:beam_plots}

\begin{table}
  \centering
  \begin{tabular}{l | c c c c}
  \hline \hline
  & matching & $\gamma_x$ & $\Gamma_\cusp$ & $\beta$  \\ \hline
  LO & $0$-loop & - & - & - \\
  NLO & $1$-loop & - & - & - \\
  NLL & $0$-loop & $1$-loop & $2$-loop & $2$-loop\\
  NNLL & $1$-loop & $2$-loop & $3$-loop & $3$-loop\\
  \hline\hline
  \end{tabular}
\caption{Order counting in fixed-order and resummed perturbation theory.}
\label{tab:counting}
\end{table}

We will now compare results for the beam functions at LO and NLO in fixed-order perturbation theory as well as at NLL and NNLL in resummed perturbation theory, which were first reported in Refs. \cite{Stewart:2010qs,Berger:2010BgNLO}. Our conventions for the $\alpha_s$ loop counting are given in Table~\ref{tab:counting}. To evaluate the required convolutions of plus distributions at NNLL we use the identities from App.~B of Ref.~\cite{Ligeti:2008ac}. We always use the MSTW2008~\cite{Martin:2009bu} parton distributions at NLO for $\alpha_s(m_Z) = 0.117$ and with two-loop, five-flavor running for $\alpha_s(\mu)$. The uncertainty bands in the plots show the perturbative uncertainties, which are estimated by varying the appropriate scales as explained in each case. They do not include the additional uncertainties from the PDFs and $\alpha_s(m_Z)$.

The order of the running of $\alpha_s(\mu)$ deserves some comment. Working consistently to NLO in the matching corrections requires us to use NLO PDFs, for which the two-loop running of $\alpha_s$ was used in Ref.~\cite{Martin:2009bu}. On the other hand, the double-logarithmic running of the beam functions at NNLL requires the three-loop running of $\alpha_s$, which poses a slight dilemma. Ideally, we would need NLO PDFs using three-loop running for $\alpha_s(\mu)$, which as far as we know are not available. The numerical difference between $\alpha_s$ run at two and three loops is very small, at most $2\%$. Hence, we use the following compromise. To be consistent with our PDF set, we use the above $\alpha_s(m_Z)$ and two-loop, five-flavor running to obtain the numerical value of $\alpha_s$ at some required scale, and to be consistent with the RGE, we use the two- and three-loop expression for the QCD $\beta$ function in the RGE solutions at NLL and NNLL. (For simplicity we use the same NLO PDFs and $\alpha_s$ also at NLL.)

To illustrate the importance of the various contributions to the quark and gluon beam functions, we also consider the beam functions in the threshold limit and without the mixing contribution. In the threshold limit we only keep the terms in \eq{Iresult} and \eq{I_results} which are singular as $z \to 1$,
\begin{align} \label{eq:Ithres}
  \cI_{qq}^\text{thresh}(t,z,\mu) &= \delta(t) \delta(1-z)
  + \frac{\alpha_s(\mu)C_F}{2\pi}\, \theta(z) \biggl\{
  \frac{2}{\mu^2} \cL_1\Bigl(\frac{t}{\mu^2}\Bigr) \delta(1-z) +
  \frac{2}{\mu^2} \cL_0\Bigl(\frac{t}{\mu^2}\Bigr) \cL_0(1-z)
  \nn \\ & \hspace{32ex}
  + \delta(t) \Bigl[  2 \cL_1(1-z) -\frac{\pi^2}{6} \delta(1-z) \Bigr]
  \biggr\}
\,,\nn\\
 \mathcal{I}^\text{thresh}_{gg}(t,z,\mu) & = \de(t)\de(1-z) + \frac{\al_s(\mu) C_A}{2\pi}\, \theta(z)
  \bigg\{ \frac{2}{\mu^2} \cL_1\Big(\frac{t}{\mu^2}\Big) \de(1-z)
  + \frac{2}{\mu^2} \cL_0\Big(\frac{t}{\mu^2}\Big) \cL_0(1-z)
  \nn \\ & \hspace{32ex}
  + \de(t)\Big[ 2 \cL_1(1-z) - \frac{\pi^2}{6} \de(1-z) \Big] \bigg\}
  \,, \nn \\
  \cI_{qg}^\text{thresh}(t,z,\mu) &= \mathcal{I}^\text{thresh}_{gq}(t,z,\mu) = 0
\,.\end{align}
The mixing terms $\cI_{qg}$ and $\cI_{gq}$ contain no threshold terms (which reflects the fact that in threshold Drell-Yan the gluon PDF does not contribute). For the results without mixing contributions we keep the full $\cI_{ii}$ but set $\cI_{ij}$ to zero for $i \neq j$, which corresponds to adding the remaining non-threshold terms in $\cI_{ii}$ to the threshold result. In the plots below, the results in the threshold limit are shown by a dotted line and are labeled ``$x\to 1$'', and the results without the mixing contribution are shown by a dashed line and are labeled ``no $g$'' for the quark beam functions and ``no $q$" for the gluon beam function. The full results are shown by a solid line. Hence, the size of the non-threshold terms in $\cI_{ii}$, and therefore the applicability of the threshold limit, is seen by the shift from the dotted to the dashed line, and the effect of the mixing is given by the shift from the dashed to the solid line.

To be able to plot the beam functions as a function of the momentum fraction $x$ including the virtual terms proportional to $\delta(t)$, we integrate over $t$ up to some maximum $t_\max$,
\begin{equation}
 \tB_i(t_\max,x,\mu) = \int\! \df t\, B_i(t,x,\mu) \theta(t_\max - t)
\,,\end{equation}
where $B_i(t, x, \mu)$ is given by \eqs{Brun}{beam_fact}. In the plots, we always choose $t_\max = (x e^{-2} 7 \TeV)^2$, which one should think of as $t_\max = (e^{-y^\cut} x \Ecm)^2$. Hence, this choice of $t_\max$ corresponds to a rapidity cut $y^\cut = 2$ for $\Ecm = 7\TeV$ or equivalently $y^\cut = 2.4$ for $\Ecm = 10\TeV$. This is motivated by the upper bound $y^\cut = y_B^\cut \pm Y$, which follows from the factorization theorem \eq{dsigma_tauB} when we integrate $\tau_B \leq \exp(-2y_B^\cut)$.

\begin{figure}[t]
\includegraphics[width=0.49\textwidth]{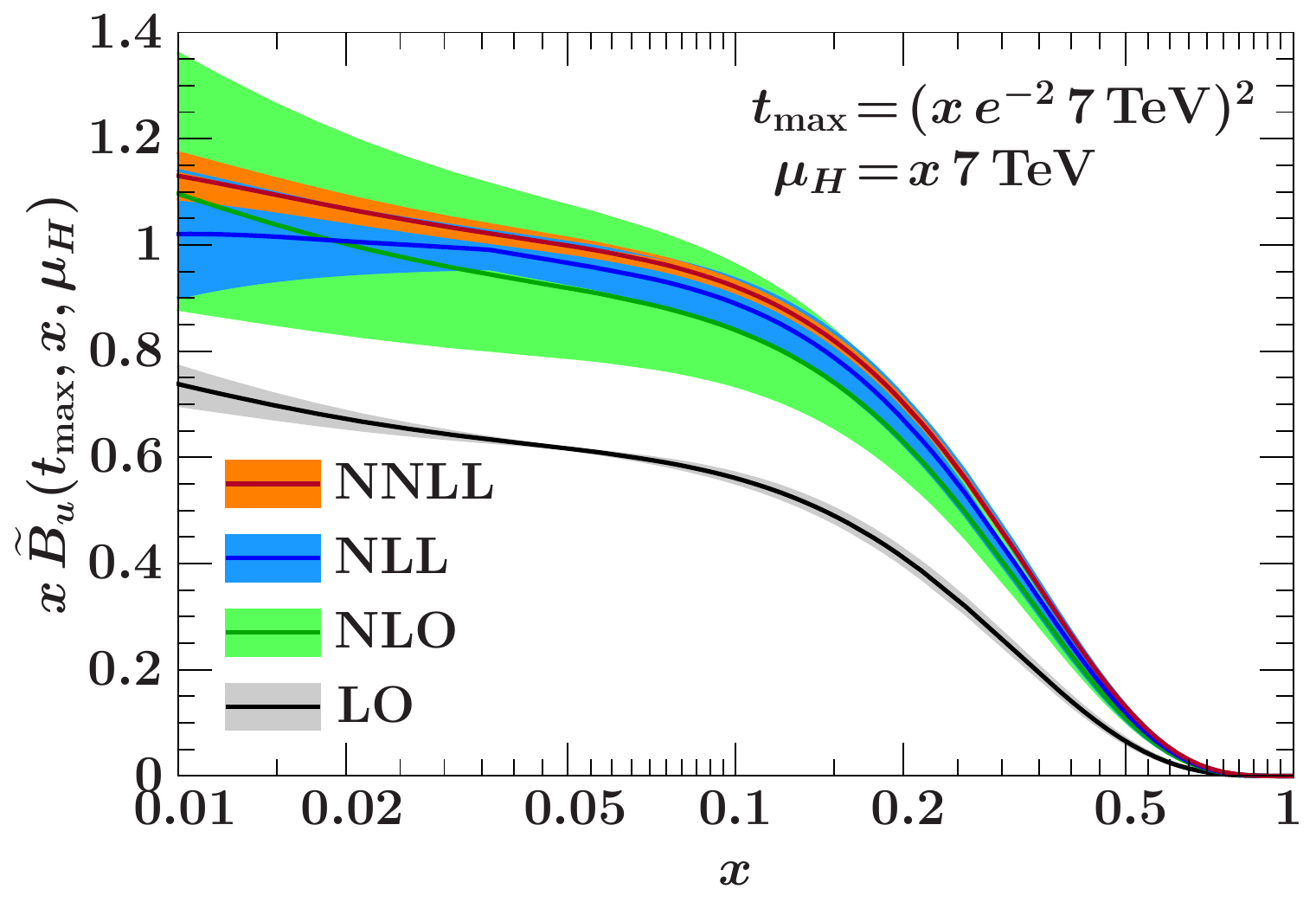}%
\hfill%
\includegraphics[width=0.49\textwidth]{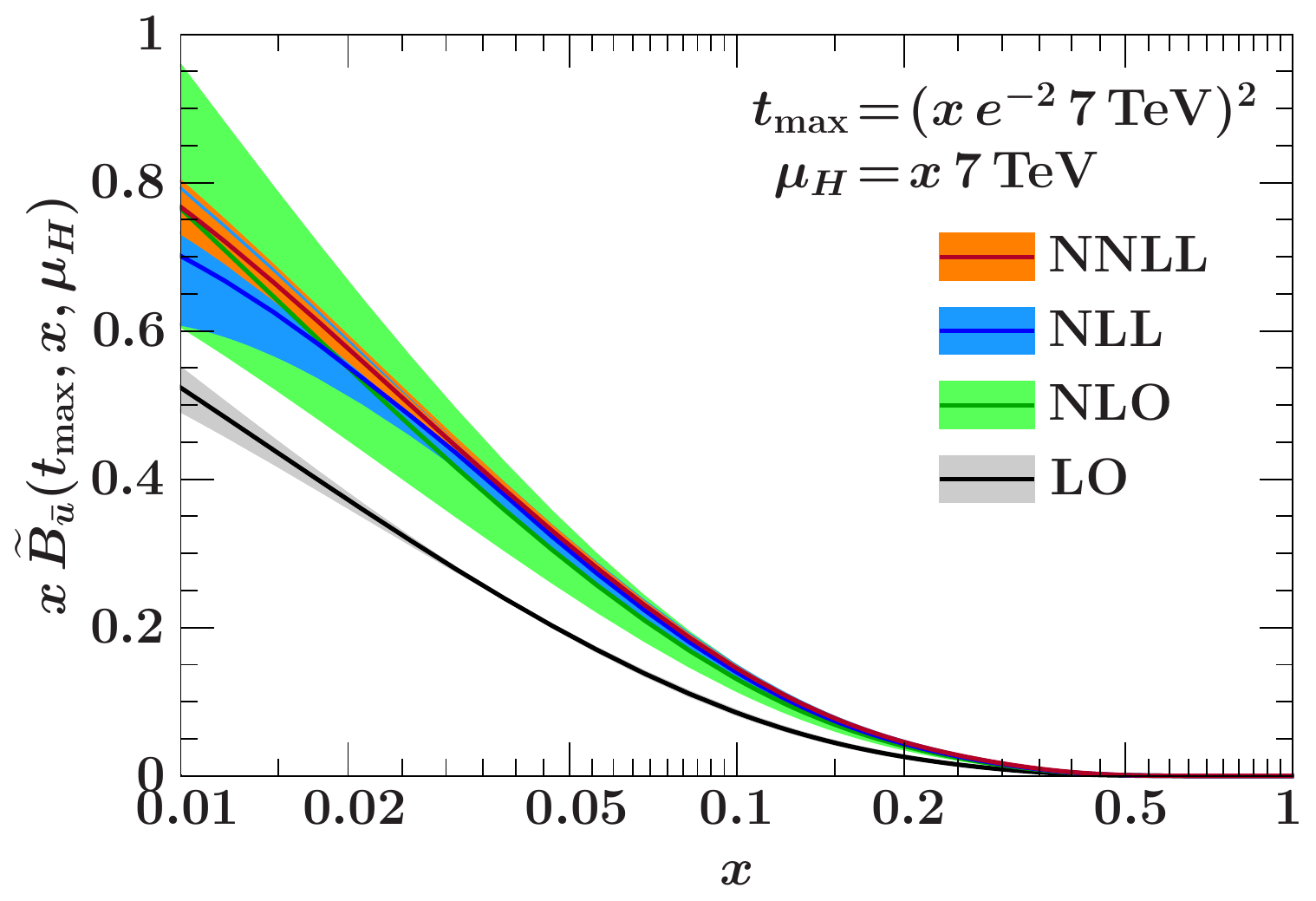}%
\\
\includegraphics[width=0.48\columnwidth]{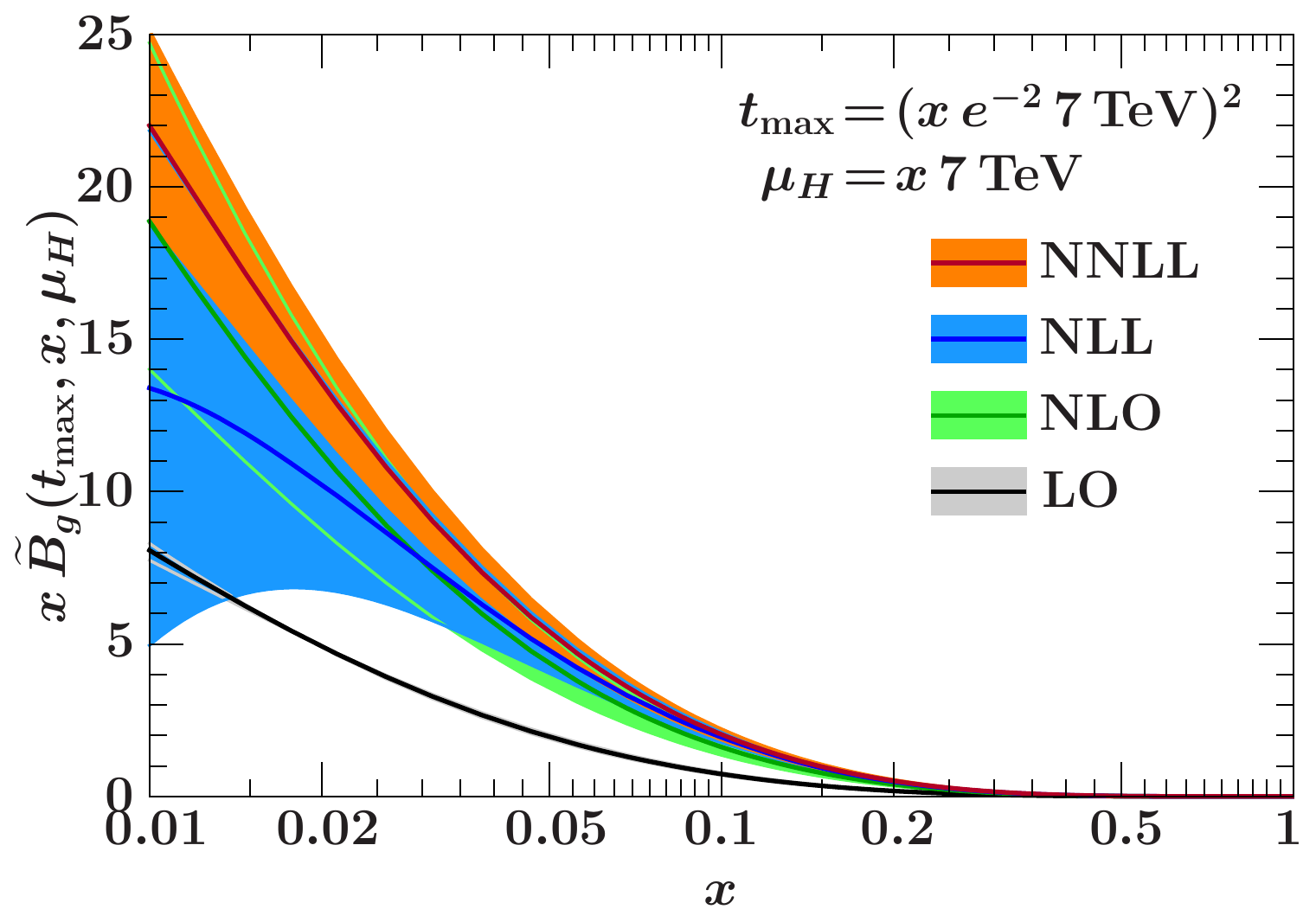}%
\caption[The $u$, $\bar u$ and $g$ beam functions at the hard scale.]{The $u$ (upper left), $\bar u$ (upper right) and $g$ (lower left) beam functions at the hard scale $\mu_H = x\, 7 \TeV$ at LO, NLO, NLL, and NNLL, integrated up to $t_\textrm{max} = (x\, e^{-2}\, 7 \TeV)^2$. The bands show the perturbative uncertainties estimated by varying  $\mu_H$ for the fixed-order results and the matching scale $\mu_B^2 \simeq t_\mathrm{max}$ for the resummed results, explained in the text.}
\label{fig:Bu_muH}
\end{figure}

Figure~\ref{fig:Bu_muH} shows the integrated $u$, $\bar u$ and gluon beam function $x \tB_i(t_\max,x,\mu_H)$ evaluated at the hard scale $\mu_H = Q = x\, 7 \TeV$. For the fixed-order results at LO (lowest gray band) and NLO (light green band) the perturbative uncertainties are obtained by varying $\mu_H$ by factors of two, since this is the scale at which the perturbation series for the matching coefficients in \eq{beam_fact} is evaluated. At LO, the resulting variation is entirely due to the scale dependence of the PDF. This is clearly an underestimation, since the perturbative corrections in the NLO beam function contain large single and double logarithms of $t_\max/Q^2$, causing a large shift in the NLO central value with large uncertainties. For the gluon beam function the shift and uncertainties are particularly large, due to differences between $\cI_{gg}^\one$ and $\cI_{qq}^\one$ such as an overall coefficient of $C_A$ instead of $C_F$.

For the resummed results at NLL (dark blue band) and NNLL (medium orange band) the beam function OPE in \eq{beam_fact} is evaluated at the beam scale $\mu_B^2 \simeq t_\max$, and the beam function is then evolved to $\mu_H$ using its RGE, \eq{Brun}. In this way, the large logarithms of $\mu_B^2/\mu_H^2 \simeq t_\max/\mu_H^2 = e^{-4}$ are resummed. The perturbative uncertainties are now evaluated by varying the matching scale $\mu_B$, where the perturbation series is evaluated, while keeping $\mu_H$ fixed. The uncertainty bands show the minimum and maximum variation in the interval $\sqrt{t_\max}/2 \leq \mu_B \leq 2\sqrt{t_\max}$ (which due to the double-logarithmic series do not occur at the edges of the interval) with the central value given by the center of the bands. The NLL result is close to the NLO result, showing that the large logarithms make up the biggest part of the NLO corrections. Consequently, the corrections from NLL to NNLL are within the NLL uncertainties. Resummed perturbation theory is thus well-behaved and should be used for the beam function at the hard scale, rather than fixed-order perturbation theory. The large uncertainty bands for the resummed gluon beam function at small $x$ are due to the strong double logarithmic running. The effect is much larger than for the quark beam functions because $C_F$ gets replaced by $C_A$, which for the running shows up in the exponent. The effect of the running is enhanced at small $x$ because $\al_s(\mu_B)$ becomes large. For $x=0.01$ we are varying the matching scale $\mu_B$ between $5\GeV$ and $20\GeV$ where $\al_s$ is much larger than for the corresponding variation for the fixed order results, where $\mu_H$ varies between $35\GeV$ and $140\GeV$.

\begin{figure}[t]
\includegraphics[width=0.49\textwidth]{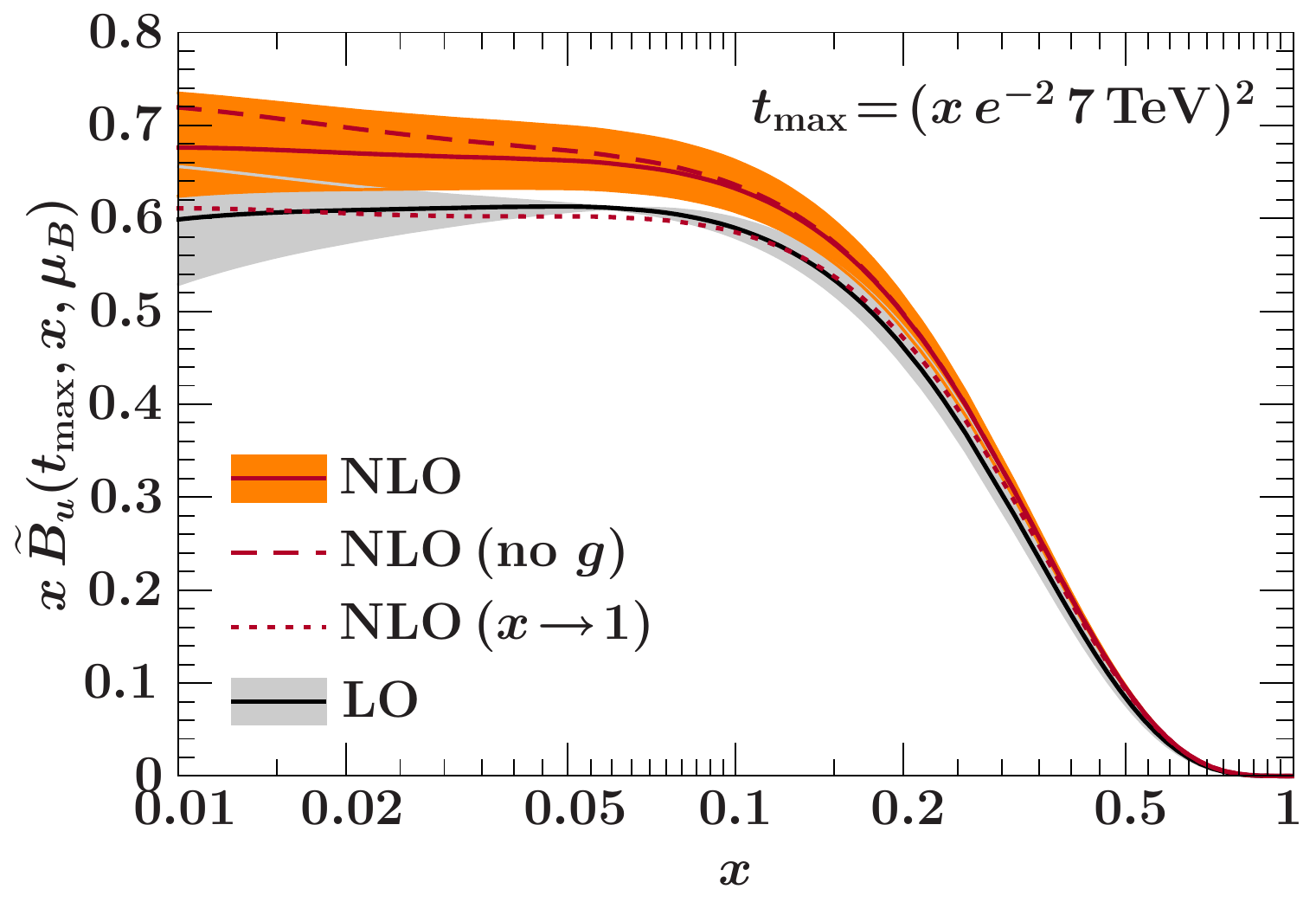}%
\hfill%
\includegraphics[width=0.49\textwidth]{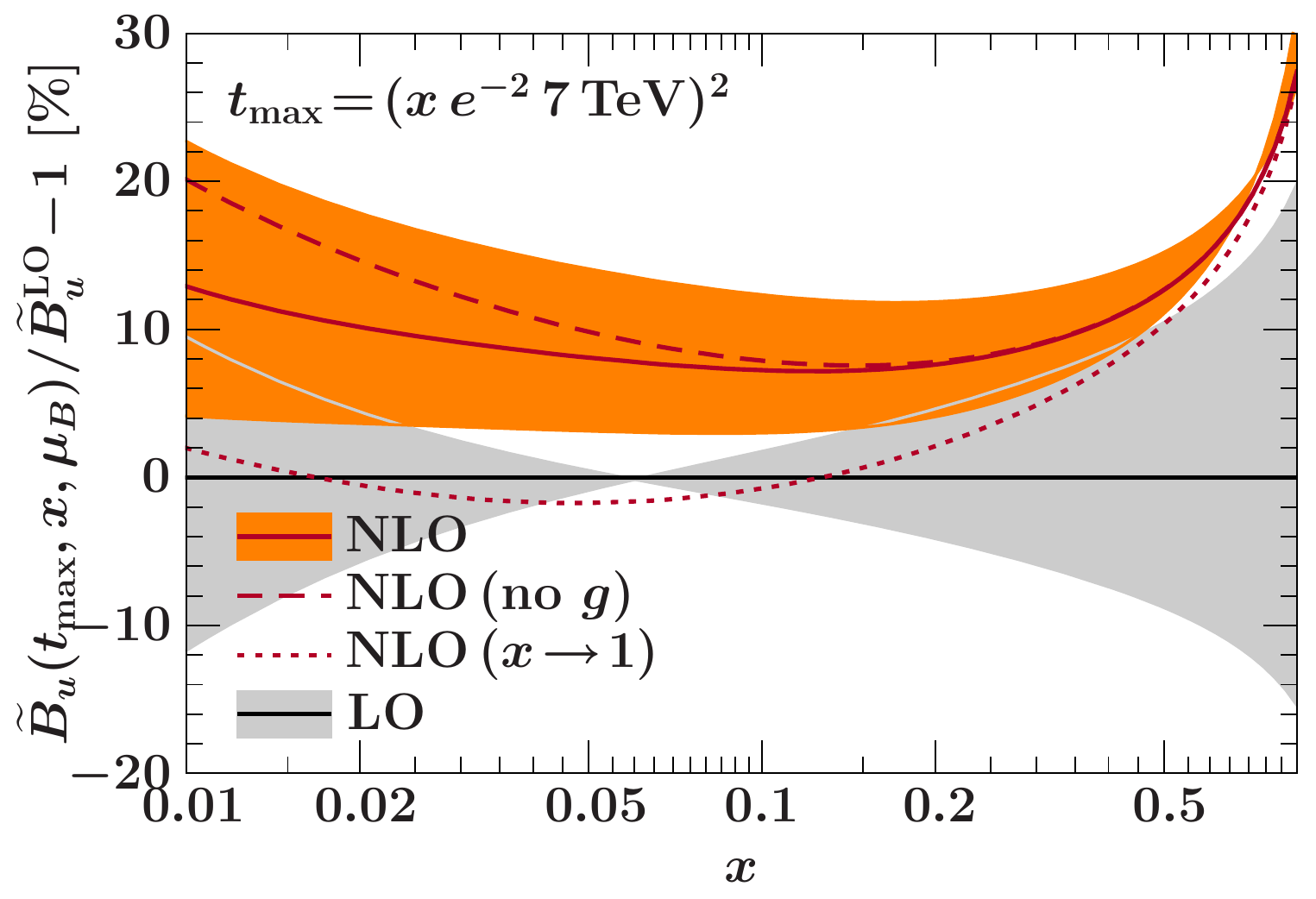}%
\\
\includegraphics[width=0.49\textwidth]{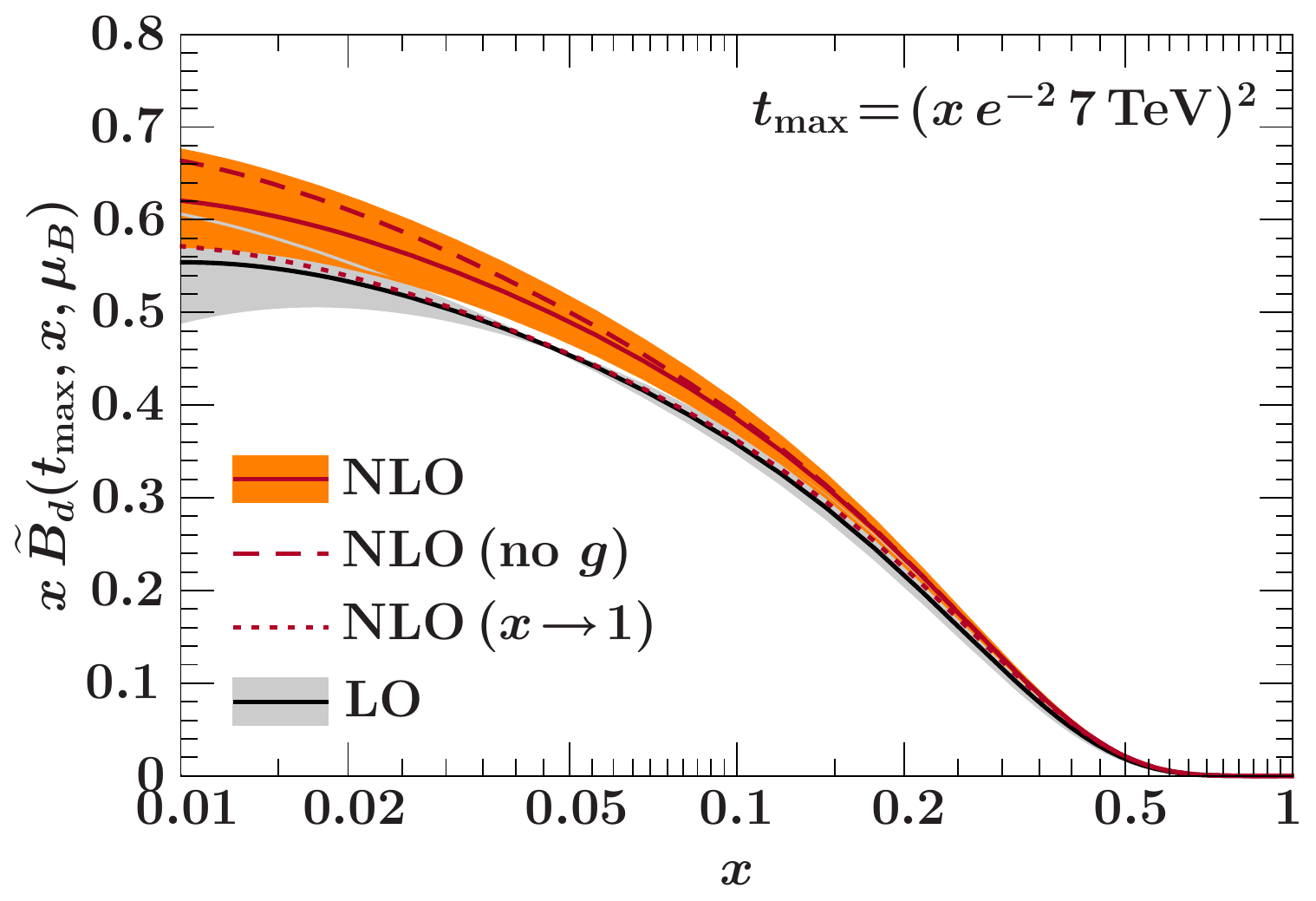}%
\hfill%
\includegraphics[width=0.49\textwidth]{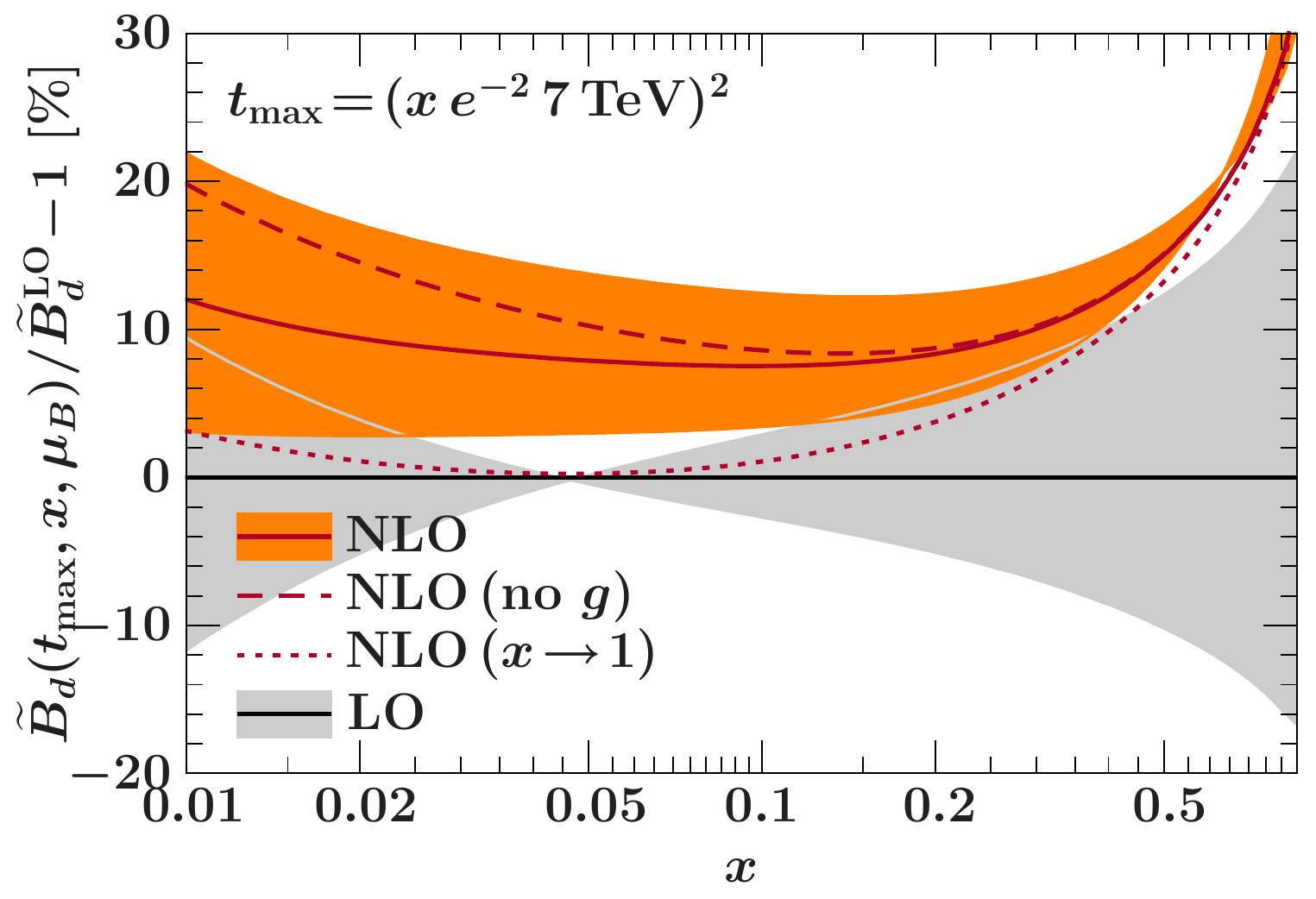}%
\caption[The $u$ and $d$ beam functions at the beam scale.]{The $u$ (top row) and $d$ (bottom row) beam functions at the beam scale $\mu_B^2 \simeq t_\mathrm{max}$ at LO and NLO, integrated up to $t_\mathrm{max} = (x e^{-2} 7 \TeV)^2$. The left column shows the functions times $x$. The right column shows the relative differences compared to the LO result. Also shown are the NLO beam functions in the threshold limit (dotted) and without the gluon contribution (dashed). The bands show the perturbative scale uncertainties as explained in the text.}
\label{fig:Bq_muB}
\end{figure}

\begin{figure}[t]
\includegraphics[width=0.49\textwidth]{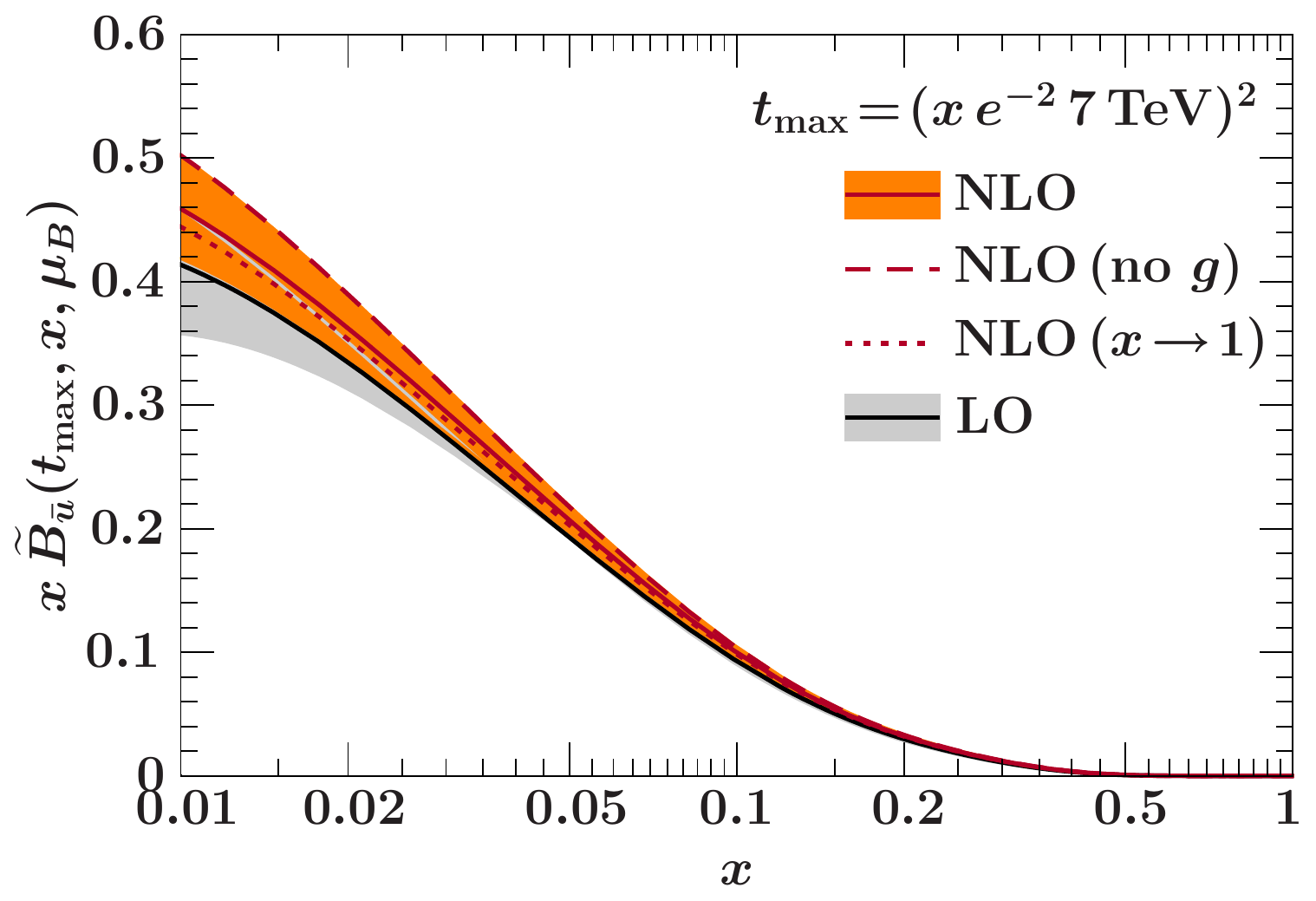}%
\hfill%
\includegraphics[width=0.49\textwidth]{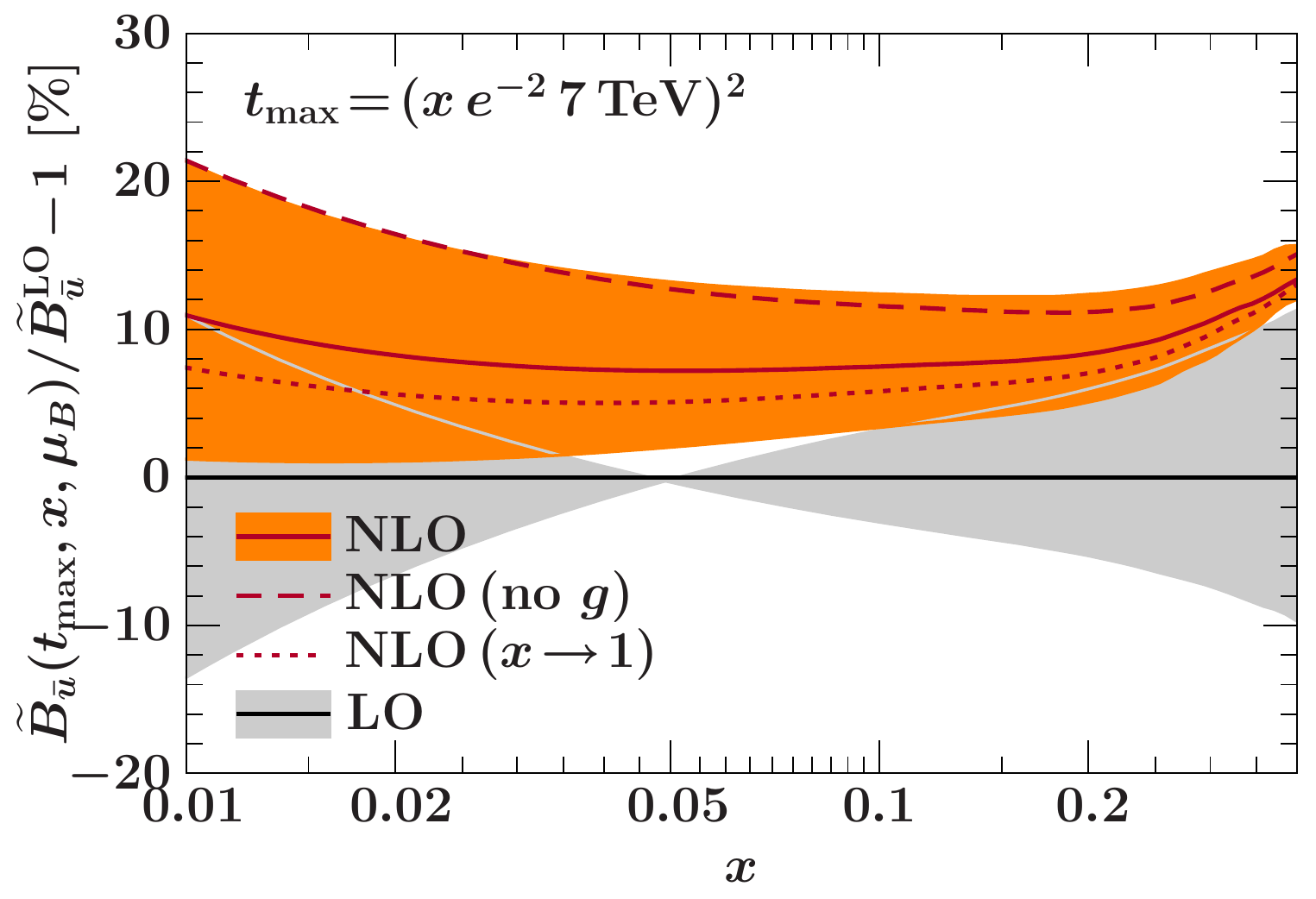}%
\\
\includegraphics[width=0.49\textwidth]{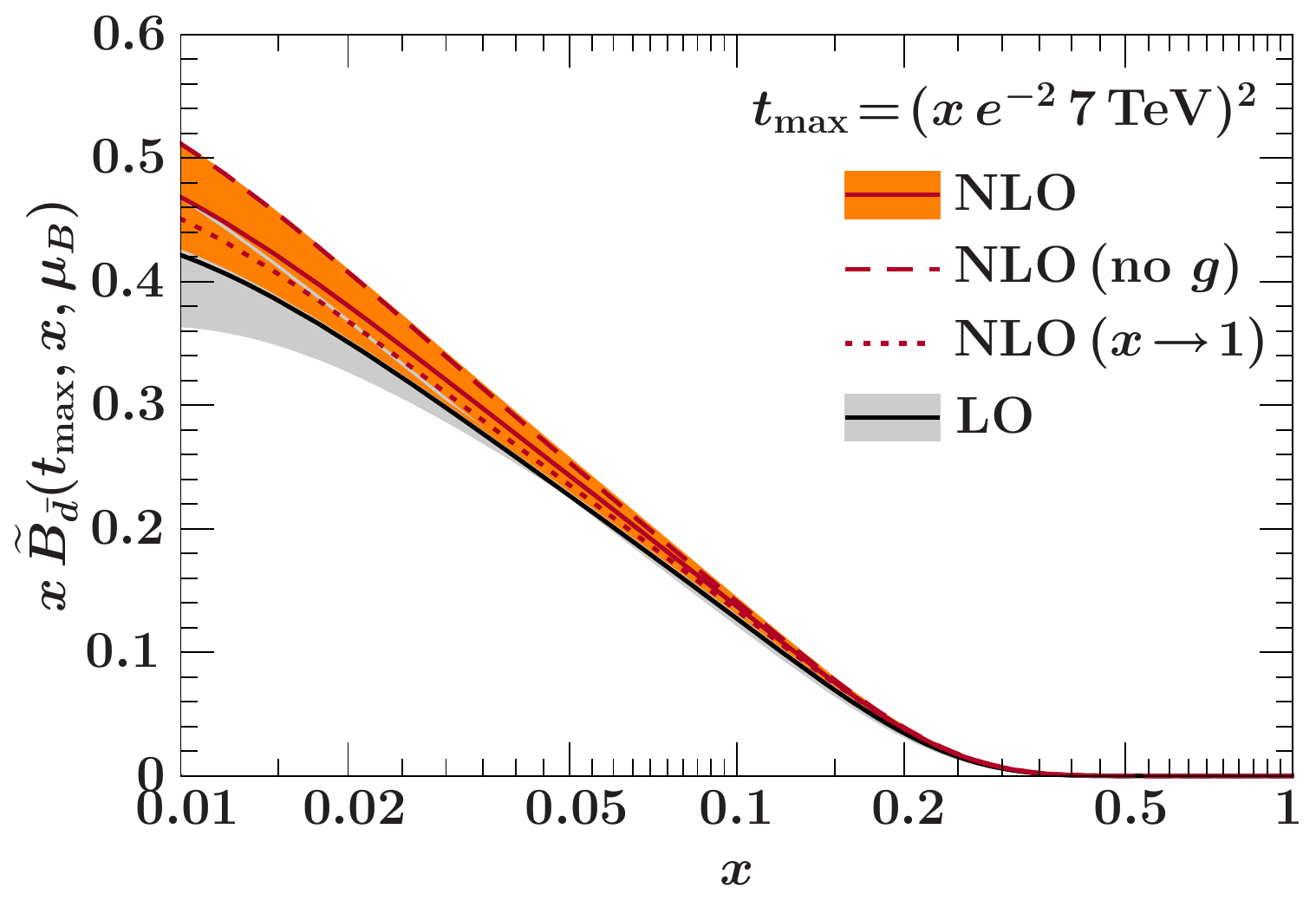}%
\hfill%
\includegraphics[width=0.49\textwidth]{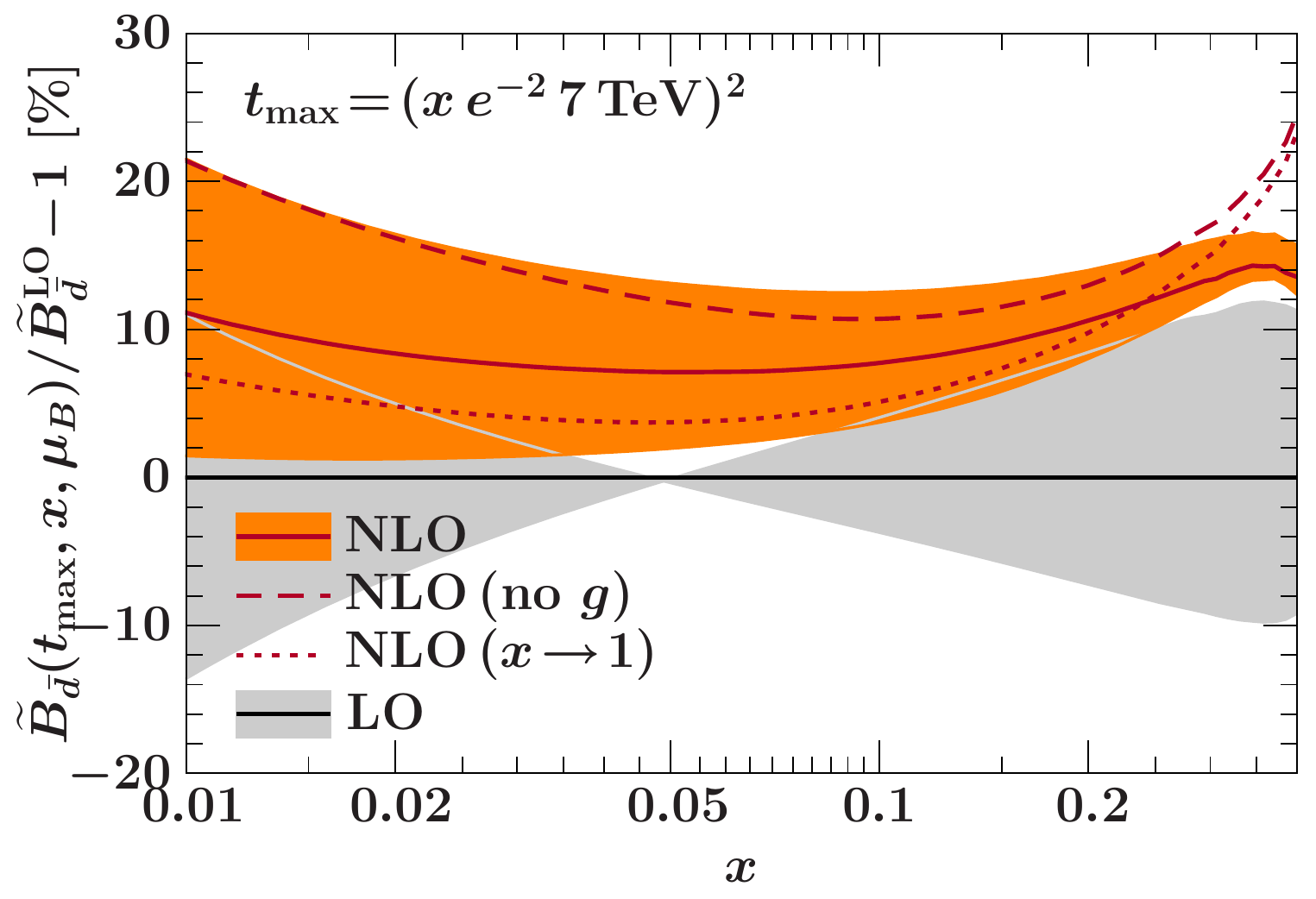}%
\caption[The $\bar u$ and $\bar d$ beam functions at the beam scale.]{The $\bar u$ (top row) and $\bar d$ (bottom row) beam functions at the beam scale. The meaning of the curves is analogous to \fig{Bq_muB}.}
\label{fig:Bqbar_muB}
\end{figure}

\begin{figure*}[t!]
\includegraphics[width=0.48\columnwidth]{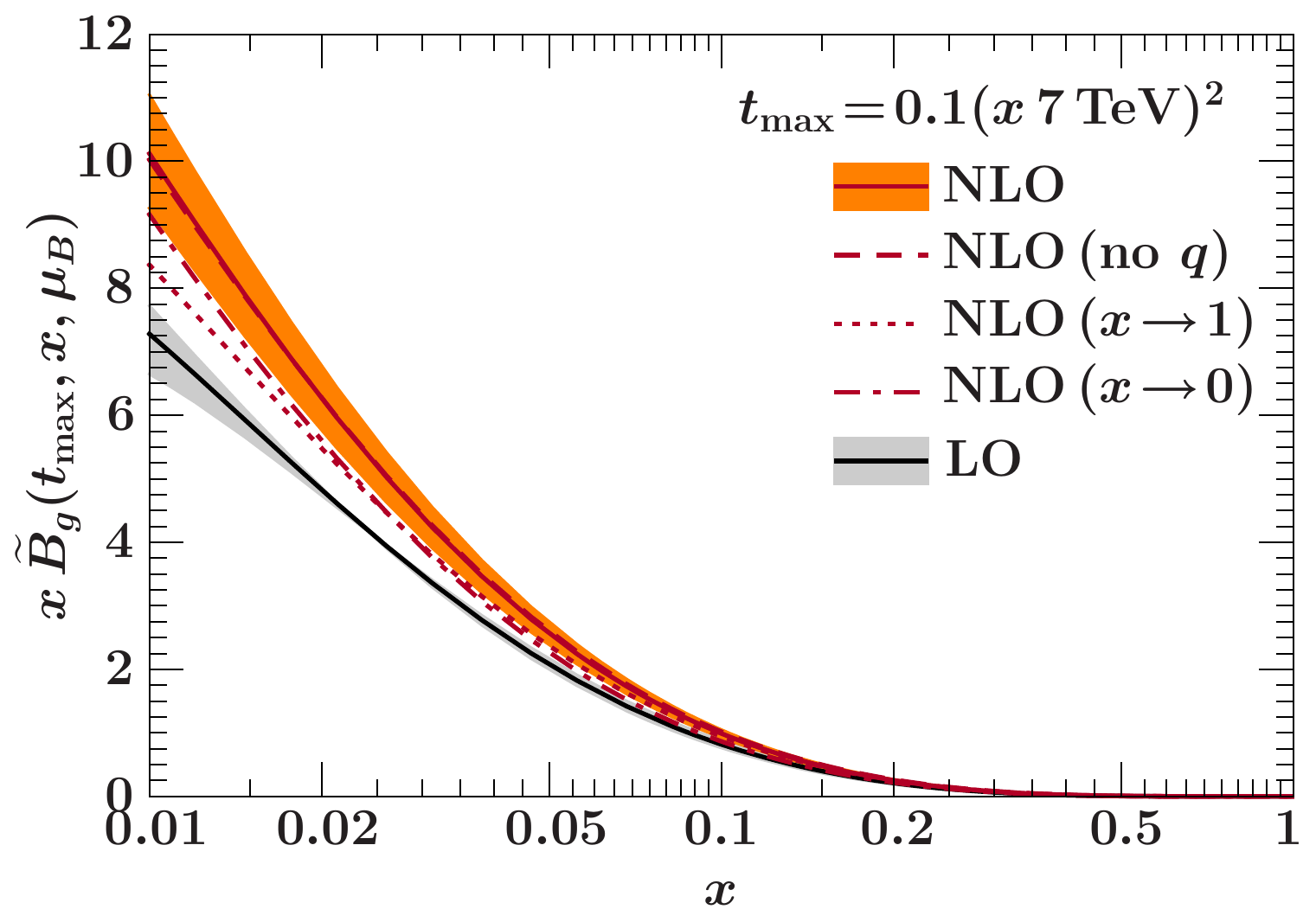}%
\hspace{0.04\columnwidth}%
\includegraphics[width=0.48\columnwidth]{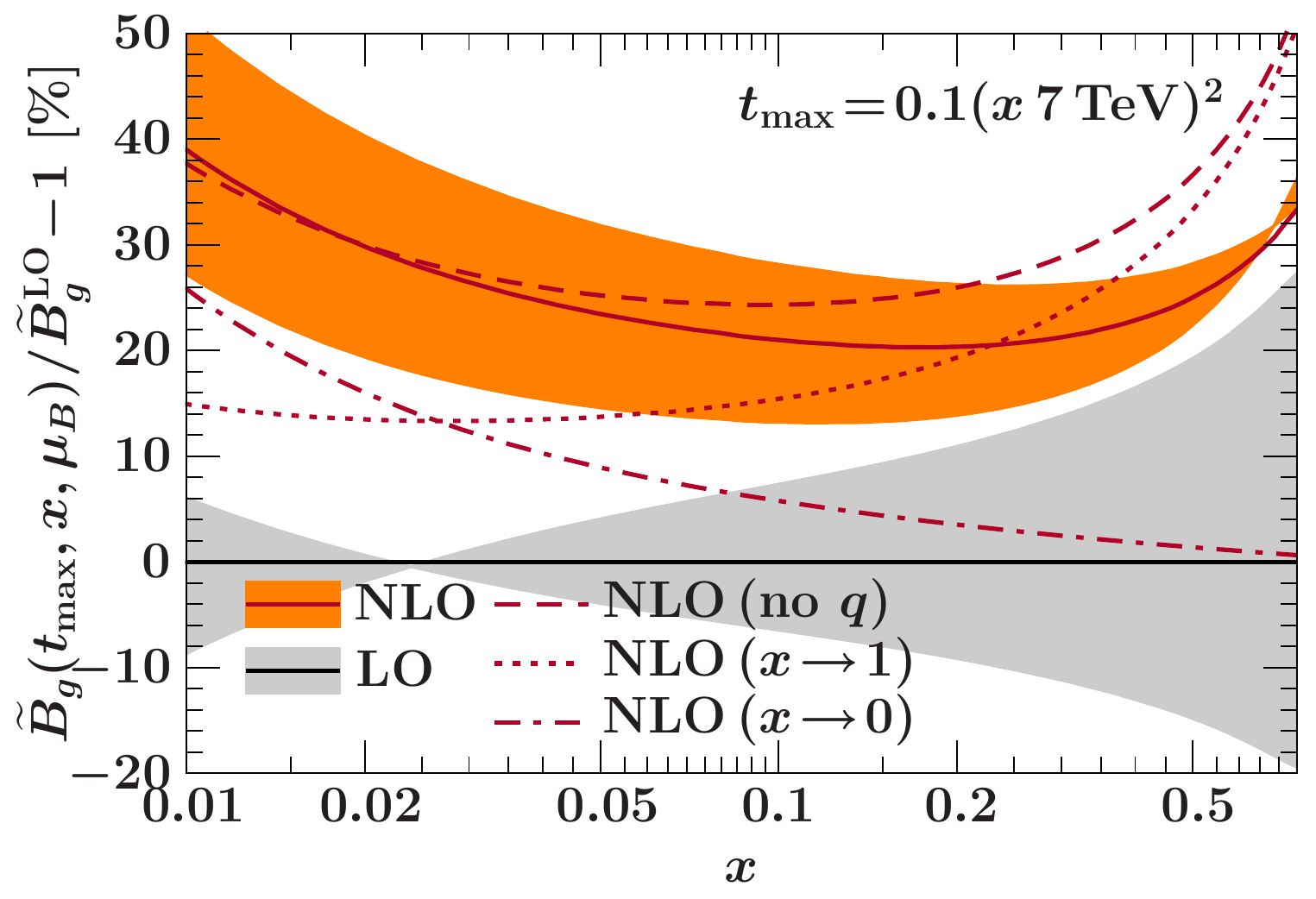}
\caption[The gluon beam function at the beam scale.]{The gluon beam function at the beam scale. The meaning of the curves is analogous to \fig{Bq_muB}.}
\label{fig:Bg_muB}
\end{figure*}

To study the perturbative corrections to the beam functions in more detail, we consider them at the scale $\mu_B^2 \simeq t_\max$, where there are no large logarithms and we can use fixed-order perturbation theory. The $u$ and $d$ beam functions at LO and NLO are shown in \fig{Bq_muB}, the $\bar u$ and $\bar d$ beam functions in \fig{Bqbar_muB} and the gluon beam function in \fig{Bg_muB}. The left panels show $x \tB_i(t_\max, x, \mu_B)$. The right panels show the same results but as relative corrections with respect to the LO results. At LO, the only scale variation comes from the PDFs and the minimum and maximum variations are obtained for $\mu_B = \{\sqrt{t_\max}/2, 2\sqrt{t_\max}\}$ with the central value at $\mu_B = \sqrt{t_\max}$. For the NLO results, the maximum variation in the range $\sqrt{t_\max}/2 \leq \mu_B \leq 2\sqrt{t_\max}$ is approximately attained for $\mu_B =\{0.7 \sqrt{t_\max},2.0\sqrt{t_\max} \}$ and the corresponding central value for $\mu_B = 1.4\sqrt{t_\max}$. To be consistent we use the same central value $\mu_B = 1.4\sqrt{t_\max}$ for the NLO results in the threshold limit and without the mixing contribution. For the quark and anti-quark beam functions the NLO perturbative corrections are of $\ord{10 \%}$ and exhibit reasonable uncertainties. The NLO corrections for the gluon beam function are $\ord{30 \%}$, which is much bigger but still reasonable.

The integration limits $x \leq \xi \leq 1$ in the beam function OPE, \eq{beam_fact}, force $z = x/\xi \to 1$ in the limit $x\to 1$. Hence, the threshold terms in \eq{Ithres} are expected to dominate over the non-threshold terms at large values of $x$. This can be seen in \figs{Bq_muB}{Bqbar_muB}, where  the threshold results shown by the dotted lines approach the full results towards large $x$ values where the beam functions vanish. For the quark beam functions in \fig{Bq_muB}, away from the endpoint, $x\lesssim 0.5$, the threshold corrections give a poor approximation to the full NLO corrections. For the antiquark beam functions in \fig{Bqbar_muB}, the threshold result turns out to be relatively close to the full result even for small $x$. However, the reason for this is a relatively strong cancelation between the non-threshold terms in the quark and gluon contributions $\cI_{qq}$ and $\cI_{qg}$ at one loop. As shown by the result without the gluon contribution (dashed lines) the non-threshold terms in the quark and gluon contributions each by themselves are of the same size or larger than the threshold contributions. This cancellation appears to be accidental, since it depends on both the relative size of the antiquark and gluon PDFs as well as the relative size of the non-threshold terms in $\cI_{qq}$ and $\cI_{qg}$. Note also that for the $\bar d$ beam function the threshold result approaches the no-gluon result rather than the full result at large $x$. This is even more visible for the gluon beam function, where the threshold terms might seem a good approximation around $x \approx 0.2$. However, for large $x$ the threshold terms approach the no-quark result which is very different from the full result. 

It has been argued~\cite{Appell:1988ie, Catani:1998tm} that the steep fall-off of the PDFs causes a systematic enhancement of the partonic threshold region $z\to 1$ even away from the hadronic threshold limit $x\to 1$. This likely explains why the threshold terms in \figs{Bq_muB}{Bqbar_muB} start to dominate already close to the $x$ values where the PDFs are close to zero, rather than strictly near $x = 1$~\cite{Becher:2007ty}. However, our results show that the same arguments do not apply in the relevant region of $x$ where the PDFs and beam functions are substantially nonzero. 


\chapter{Isolated Factorization Theorem}
\label{ch:fact}

In this chapter we derive the isolated factorization theorem in \eq{DYbeam}, which was
first presented in Ref.~\cite{Stewart:2009yx}.
Our analysis is based on factorization in SCET, which rigorously and
systematically separates hard, soft and collinear
contributions~\cite{Bauer:2001ct,Bauer:2001yt,Bauer:2002nz}.  We make use of a
setup with \SCETa and \SCETb~\cite{Bauer:2002aj}, carrying out the factorization
in two stages at the scales $Q^2$ and $\omega_{a,b}B_{a,b}^+$ respectively.  We
have an \SCETa analysis to factorize initial-state jets from soft radiation.
The initial-state jets described by beam functions in \SCETa are then matched
onto initial-state PDFs with lower offshellness for the collinear particles in
\SCETb. In this chapter, we carry out the \SCETa computation, while the matching
onto \SCETb was discussed in \ch{beamf}.  Our analysis below uses
similar tools as used in the derivation of the factorization theorem for
hemisphere invariant masses for $e^+e^-\to $ dijets in
Ref.~\cite{Fleming:2007qr}, but differs significantly due to the kinematics, and
the fact that we have initial-state rather than final-state jets and a further
matching onto \SCETb.  The soft dynamics of $e^+e^-\to $ dijets was studied
earlier in SCET in Refs.~\cite{Bauer:2002ie,Bauer:2003di}.  We start with a
brief overview of the necessary SCET ingredients in \sec{SCET} and describe
the relevant kinematics in \sec{kinematics}.  We derive the factorization
theorem for isolated $pp\to XL$ in \sec{facttheorem}, including arguments to
rule out contributions from so-called Glauber degrees of freedom. Finally in
\sec{DY_final}, we apply the factorization theorem to $pp\to X\ell^+\ell^-$
and quote final results for the beam thrust cross section with one-loop
corrections and logarithmic resummation.

\section{SCET}
\label{sec:SCET}

Soft-collinear effective theory is an effective field theory of QCD that
describes the interactions of collinear and soft particles~\cite{Bauer:2000ew,
  Bauer:2000yr, Bauer:2001ct, Bauer:2001yt}.  Collinear particles are
characterized by having large energy and small invariant mass. To separate the
large and small momentum components, it is convenient to use light-cone
coordinates. We define two light-cone vectors
\begin{equation}
n^\mu = (1, \vec{n})
\,,\qquad
\bn^\mu = (1, -\vec{n})
\,,\end{equation}
with $n^2 = \bn^2 = 0$, $n\cdot\bn = 2$, and $\vec{n}$ is a unit three-vector.
Any four-momentum $p$ can then be decomposed as
\begin{equation}
p^\mu = \bn\sdt p\,\frac{n^\mu}{2} + n\sdt p\,\frac{\bn^\mu}{2} + p^\mu_{n\perp}
\,.\end{equation}
Choosing $\vec{n}$ close to the direction of a collinear particle, its momentum
$p$ scales as $(n\cdot p, \bn\cdot p, p_{n\perp}) \sim \bn\cdot p\,(\la^2,1,\la)$,
with $\la \ll 1$ a small parameter. For example, for a jet of collinear
particles in the $\vec{n}$ direction with total momentum $p_X$, $\bn \cdot p_X
\simeq 2E_X$ corresponds to the large energy of the jet, while $n \cdot p_X
\simeq p_X^2/E_X \ll E_X$, so $\la^2 \simeq p_X^2/E_X^2 \ll 1$.

To construct the fields of the effective theory, the momentum is written as
\begin{equation}
p^\mu = \lp^\mu + k^\mu = \bn\sdt\lp\, \frac{n^\mu}{2} + \lp_{n\perp}^\mu + k^\mu
\,\end{equation}
where $\bn\cdot\lp \sim Q$ and $\lp_{n\perp} \sim \la Q$ are the large momentum
components, where $Q$ is the scale of the hard interaction, while $k\sim \la^2
Q$ is a small residual momentum. The effective theory expansion is in powers of
the small parameter $\la$.

The SCET fields for $n$-collinear quarks and gluons, $\xi_{n,\lp}(x)$ and
$A_{n,\lp}(x)$, are labeled by the collinear direction $n$ and their large
momentum $\lp$. They are written in position space with respect to the residual
momentum and in momentum space with respect to the large momentum components.
Frequently, we will only keep the label $n$ denoting the collinear direction,
while the momentum labels are summed over and suppressed. Derivatives acting on
the fields pick out the residual momentum dependence, $\img\partial^\mu \sim k
\sim \la^2 Q$. The large label momentum is obtained from the momentum operator
$\cP_n^\mu$, e.g. $\cP_n^\mu\, \xi_{n,\lp} = \lp^\mu\, \xi_{n,\lp}$. If there
are several fields, $\cP_n$ returns the sum of the label momenta of all
$n$-collinear fields. For convenience, we define $\bnP_n = \bn\cdot\cP_n$, which
picks out the large minus component.

Collinear operators are constructed out of products of fields and Wilson lines
that are invariant under collinear gauge
transformations~\cite{Bauer:2000yr,Bauer:2001ct}.  The smallest building blocks
are collinearly gauge-invariant quark and gluon fields, defined as
\begin{align} \label{eq:chiB2}
\chi_{n,\w}(x) &= \Bigl[\delta(\w - \bnP_n)\, W_n^\dagger(x)\, \xi_n(x) \Bigr]
\,,\nn\\
\cB_{n,\w\perp}^\mu(x)
&= \frac{1}{g}\Bigl[\delta(\w + \bnP_n)\, W_n^\dagger(x)\,\img D_{n\perp}^\mu W_n(x)\Bigr]
\,,\end{align}
where
\begin{equation}
\img D_{n\perp}^\mu = \cP^\mu_{n\perp} + g A^\mu_{n\perp}
\end{equation}
is the collinear covariant derivative and
\begin{equation} \label{eq:Wn2}
W_n(x) = \biggl[\sum_\text{perms} \exp\Bigl(\frac{-g}{\bnP_n}\,\bn\sdt A_n(x)\Bigr)\biggr]
\,.\end{equation}
The label operators in \eqs{chiB2}{Wn2} only act inside the square brackets. Here,
$W_n(x)$ is a Wilson line of $n$-collinear gluons in label momentum space. It
sums up arbitrary emissions of $n$-collinear gluons from an $n$-collinear quark
or gluon, which are $\ord{1}$ in the power counting. Since $W_n(x)$ is
localized with respect to the residual position $x$, we can treat
$\chi_{n,\w}(x)$ and $\cB_{n,\w}^\mu(x)$ as local quark and gluon fields. The
label momentum $\w$ is treated as a continuous variable, which is why we
use a $\delta$-function operator in \eq{chiB2}. It is set equal to the sum of the
minus label momenta of all fields that the $\delta$ function acts on, including
those in the Wilson lines, while the label momenta of the individual fields are
summed over.

In general, the effective theory can contain several collinear sectors, each
containing collinear fields along a different collinear direction. To have a
well-defined power expansion in this case, the different collinear directions
$n_i$ have to be well separated~\cite{Bauer:2002nz},
\begin{equation} \label{eq:nijsep}
  n_i\sdt n_j \gg \la^2 \qquad\text{for}\qquad i\neq j
\,,\end{equation}
which is simply the requirement that different collinear sectors are distinct
and do not overlap. For $pp\to X \ell^+\ell^-$, we need two collinear sectors,
$n_a$ and $n_b$, along the directions of the two beams. We use a bar to denote
the conjugate lightlike vector, so $n_i\cdot\bar n_i=2$. As the beams are
back to back, we have $n_a \sim \bn_b$, so $n_a\cdot n_b \sim 2$ and \eq{nijsep}
is easily satisfied.

Particles that exchange large momentum of $\ord{Q}$ between collinear particles
moving in different directions have to be off shell by an amount of
$\ord{n_i\cdot n_j Q^2}$. These modes can be integrated out of the theory at the
hard scale $Q$ by matching full QCD onto SCET, which yields the hard function.
The effective theory below the scale $Q$ then splits into several distinct
collinear sectors, where particles in the same collinear sector can still
interact with each other, while at leading order in the power counting particles
from different collinear sectors can only interact by the exchange of soft
particles. This means that before and after the hard interaction takes place,
the jets described by the different collinear sectors evolve independently from
each other with only soft but no hard interactions between them.

The soft degrees of freedom, responsible for the  radiation between collinear
jets, are described in the effective theory by soft%
\footnote{In some situations it is necessary to distinguish two types of soft
  sectors, referred to as soft and ultrasoft in the SCET literature. In this
  paper we only need what are usually called ultrasoft particles, so we will
  simply refer to these as soft.}  quark and gluon fields, $q_\soft(x)$ and
$A_\soft(x)$, which only have residual soft momentum dependence
$\img\partial^\mu \sim \la^2Q$.
They couple to the collinear sectors via the soft covariant derivative
\begin{equation}
\img D_\soft^\mu = \img \partial^\mu + g A_\soft^\mu
\end{equation}
acting on the collinear fields. At leading order in $\la$, $n$-collinear particles only couple to the $n\cdot A_\soft$ component of soft gluons, so the leading-order $n$-collinear Lagrangian only depends on $n\cdot D_\soft$. For $n$-collinear quarks~\cite{Bauer:2000yr, Bauer:2001ct}
\begin{equation}
\cL_n = \bar{\xi}_n \Bigl(\img n\sdt D_\soft + g\,n\sdt A_n + \img\Dslash_{n\perp} W_n \frac{1}{\bnP_n}\, W_n^\dagger\,\img\Dslash_{n\perp} \Bigr)\frac{\bnslash}{2} \xi_n
\,.\end{equation}
The leading-order $n$-collinear Lagrangian for gluons is given in Ref.~\cite{Bauer:2001yt}.

The coupling of soft gluons to collinear particles can be removed at leading order by defining new collinear fields~\cite{Bauer:2001yt}
\begin{align} \label{eq:BPS}
\chi^\zero_{n,\w}(x) &= Y_n^\dagger(x)\,\chi_{n,\w}(x)
\,,\\\nn
\cB^{\mu\zero}_{n,\w\perp}(x) &= Y_n^\dagger(x)\,\cB^\mu_{n,\w\perp}(x)\,Y_n(x)
= \cB^{\mu d}_{n,\w\perp }(x) \cY_n^{dc}(x)\,T^c
,\end{align}
where $Y_n(x)$ and $\cY_n(x)$ are soft Wilson lines in the fundamental and adjoint representations,
\begin{align} \label{eq:Yin}
Y_n(x) = P\exp\biggl[\img g\intlim{-\infty}{0}{s} n\sdt A_\soft(x + s\,n) \biggr]
\,, \qquad
T^c \cY^{cd}_n(x) = Y_n(x)\, T^d\, Y_n^\dagger(x)
\,.\end{align}
The symbol $P$ in \eq{Yin} denotes the path ordering of the color generators
along the integration path. The integral limits in \eq{Yin} with the reference
point at $-\infty$ are the natural choice for incoming
particles~\cite{Chay:2004zn}. The final results are always independent of the
choice of reference point, and with the above choice the interpolating fields
for the incoming proton states do not introduce additional Wilson
lines~\cite{Arnesen:2005nk}.

After the field redefinition in \eq{BPS}, the leading-order SCET Lagrangian
separates into the sum of independent $n_i$-collinear and soft Lagrangians,
\begin{equation} \label{eq:LSCET}
\cL_\mathrm{SCET} = \sum_{n_i} \cL^\zero_{n_i} + \cL_\soft + \dotsb
\,,\end{equation}
with no interactions between any of the collinear and soft sectors. The
ellipses denote terms that are subleading in the power counting. This decoupling
is what will allow us to factorize the cross section into separate beam and soft
functions. The field redefinition in \eq{BPS} introduces
soft Wilson lines in the operators, which because of \eq{LSCET} can be factored out
of the matrix element and will make up the soft function.

\section{Kinematics}
\label{sec:kinematics}

\begin{figure*}[t!]
\centering
\includegraphics[scale=0.75]{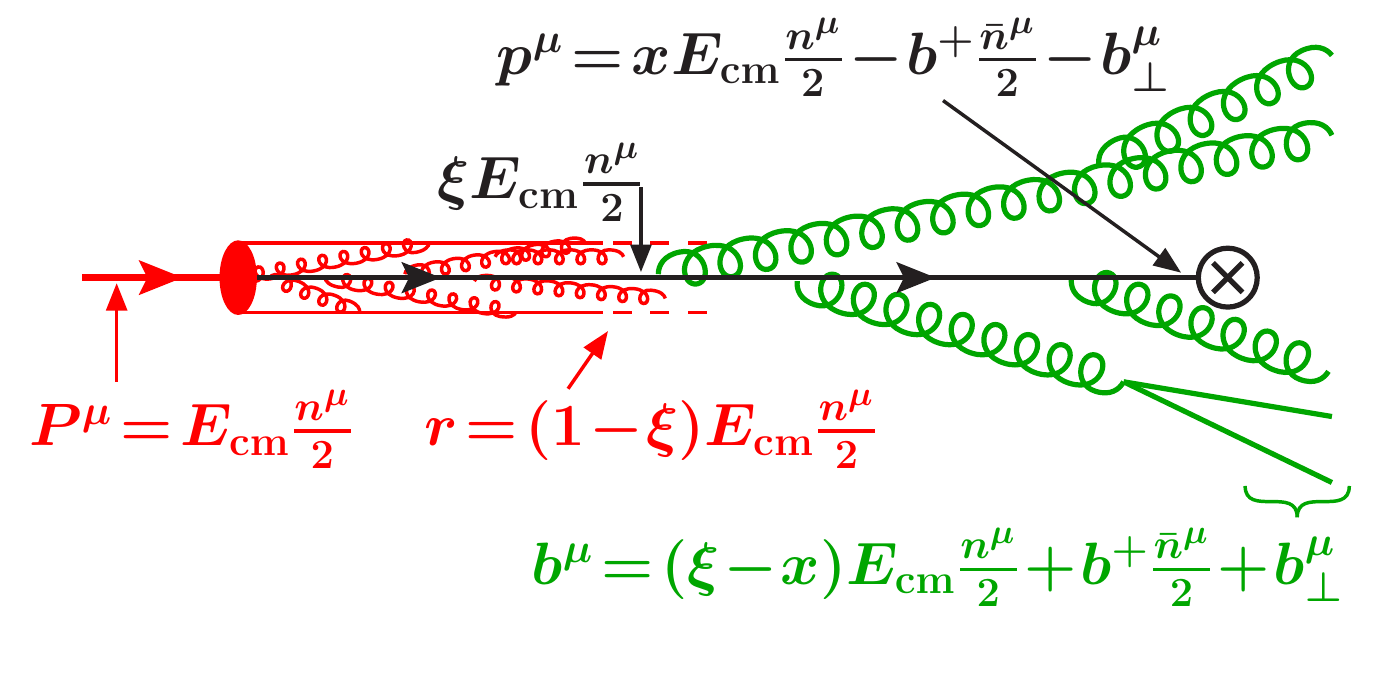}
\caption[Definition of the different collinear momenta related to the incoming beams.]{Definition of the different collinear momenta related to the incoming beams. The soft radiation
is not shown.}
\label{fig:beam_kinematics}
\end{figure*}

Before deriving the factorization theorem, we discuss the relevant kinematics,
as illustrated in \fig{beam_kinematics}. As already mentioned, we introduce a
separate set of collinear fields for each of the beams, with the light-cone
vectors $n_a$ and $n_b$ aligned with the beam directions. To derive the
factorization theorem we work in the center-of-mass frame of the hadronic
collision, so the momenta of the incoming protons are (neglecting the proton
mass)
\begin{equation} \label{eq:nab_choice}
P_a^\mu = \ECM\, \frac{n_a^\mu}{2}
\,,\qquad
P_b^\mu = \ECM\, \frac{n_b^\mu}{2}
\,,\end{equation}
with $\vec{n}_a = - \vec{n}_b$. In particular, $n_b = \bn_a$ and $n_a \cdot n_b = 2$. We will mostly keep the dependence on the two beam directions explicit, but one should keep in mind that $n_a$ and $n_b$ are related.

The collinear fields in the $n_a$ and $n_b$ directions describe the interactions within each of the beams before and after the collision, and are also responsible for initiating the hard interaction. We define the momenta of the spacelike off-shell partons that go into the hard interaction as
\begin{align} \label{eq:bain}
p_a^\mu = x_a \ECM\, \frac{n_a^\mu}{2} - b_a^+\,\frac{\bn_a^\mu}{2} - b_{a\perp}^\mu
\,, \qquad
p_b^\mu = x_b \ECM\, \frac{n_b^\mu}{2} - b_b^+\,\frac{\bn_b^\mu}{2} - b_{b\perp}^\mu
\,,\end{align}
where $x_a$ and $x_b$ are the light-cone momentum fractions at which the beam functions will be evaluated.
The power-counting parameters for the collinear sectors are
\begin{equation} \label{eq:laJab}
\la_a^2
\sim \frac{b_a^+}{x_a\ECM}
\,,\qquad
\la_b^2
\sim \frac{b_b^+}{x_b\ECM}
\,,\end{equation}
where the relevant momenta are those of the off-shell partons in \eq{bain}, because these are the momenta carried by the $n_a$- and $n_b$-collinear fields.

We write the momentum of the incoming partons that are taken out of the proton as
\begin{equation}
\xi_a \ECM\, \frac{n_a^\mu}{2} + \ord{\lqcd}
\,,\qquad
\xi_b \ECM\, \frac{n_b^\mu}{2} + \ord{\lqcd}
\,,\end{equation}
which defines the light-cone momentum fractions $\xi_{a,b}$ at which the PDFs
are evaluated.  The typical $\perp$-momenta of partons in the proton are
$\ord{\lqcd}$, while the small plus components are $\ord{\lqcd^2/\ECM}$. These
momenta are much smaller than any soft or residual momenta in \SCETa and are
expanded, which precisely corresponds to the OPE for the beam functions in
\eq{B_fact} when matching them onto \SCETb.

The momentum of the final-state remnant of the proton is thus given by
\begin{equation} \label{eq:remnant}
r_a^\mu = (1-\xi_a)\ECM\, \frac{n_a^\mu}{2}
\,,\end{equation}
while the remnant of the initial-state jet radiated into the final state by the beam function has momentum
\begin{equation} \label{eq:initialjet}
b_a^\mu = (\xi_a - x_a) \ECM\, \frac{n_a^\mu}{2} + b_a^+\,\frac{\bn_a^\mu}{2} + b_{a\perp}^\mu
\,,\end{equation}
and similarly for the $n_b$ direction. The total $n_a$-collinear momentum in the final state is the sum of \eqs{remnant}{initialjet}, or equivalently, the difference between the proton momentum and \eq{bain},
\begin{equation}
b_a^\mu + r_a^\mu = P_a^\mu - p_a^\mu = (1 - x_a) \ECM\, \frac{n_a^\mu}{2} + b_a^+\,\frac{\bn_a^\mu}{2} + b_{a\perp}^\mu
\,.\end{equation}

In addition to the collinear momenta, we define $k_s^\mu$ as the total four-momentum of the soft radiation in the final state. Hence, the total hadronic momentum in the final state is given by
\begin{equation}
p_X^\mu = (P_a^\mu - p_a^\mu) + (P_b^\mu - p_b^\mu) + k_s^\mu
\,,\end{equation}
and we can write total momentum conservation $P_a^\mu + P_b^\mu = p_X^\mu + q^\mu$ as
\begin{equation} \label{eq:momcons}
p_a^\mu + p_b^\mu = q^\mu + k_s^\mu
\,,\end{equation}
where $q^\mu$ is the total leptonic momentum.

The collinear and soft momenta, $b_a^\mu$, $b_b^\mu$, $k_s^\mu$ are not
experimentally measurable quantities. Instead, the experiments can only measure
hadronic quantities, such as the hemisphere momenta $B_a^+ =n_a\sdt B_a$ and
$B_b^+ = n_b\sdt B_b$ introduced in \sec{DYfact}. Splitting the total soft
momentum into its contributions from each hemisphere, $k_s^\mu = k_a^\mu +
k_b^\mu$ as shown in \fig{DYkin}, we then have
\begin{equation}
B_a^\mu = b_a^\mu + r_a^\mu + k_a^\mu
\,,\qquad
B_b^\mu = b_b^\mu + r_b^\mu + k_b^\mu
\,,\end{equation}
and defining $k_a^+ = n_a\cdot k_a$, $k_b^+ = n_b\cdot k_b$, we get
\begin{equation}
B_a^+ = n_a\sdt B_a = b_a^+ + k_a^+
\,,\qquad
B_b^+ = n_b\sdt B_b = b_b^+ + k_b^+
\,.\end{equation}
In particular, the remnant momenta $r_{a,b}^\mu$ do not contribute to
$B_{a,b}^+$. A physical argument for this was discussed in \sec{DYfact}.

Next, we decompose the total leptonic momentum as
\begin{equation} \label{eq:qpmperp}
q^\mu = q^-\,\frac{n_a^\mu}{2} + q^+\,\frac{n_b^\mu}{2} + q_\perp^\mu
\,,\end{equation}
where $q_\perp^\mu$ contains the two components of $q^\mu$ transverse to the
beam direction. Taking the $z$-axis along the $\vec{n}_a$ beam direction, we
have
\begin{equation}
q^\pm = q^0 \mp q_z
\,,\qquad
q_\perp^\mu = (0, \vec{q}_T, 0)
\,,\end{equation}
where $\vec{q}_T = (q_x, q_y)$ is a two-vector in the transverse $x$-$y$-plane. The total leptonic invariant mass and rapidity are
\begin{equation} \label{eq:q2Y_def}
q^2 = q^+ q^- + q_\perp^2 = q^+ q^- - \vec{q}_T^2
\,,\qquad
Y = \frac{1}{2} \ln\frac{q^-}{q^+}
\,,\end{equation}
with
\begin{align}
q^\mp = e^{\pm Y}\,\sqrt{q^2 + \vec{q}_T^2}
\,, \qquad
\df^4 q = \frac{1}{2}\, \df q^+\df q^-\, \df^2 \vec{q}_T = \frac{1}{2}\, \df q^2\,\df Y\,\df^2 \vec{q}_T
\,.\end{align}

As we will see in the next section, the derivation of the factorization
theorem requires us to be insensitive to the transverse components $\vec{q}_T$
such that we can freely integrate over them. Therefore, we have to expand the
kinematics in the limit $\vec{q}_T = 0$. This expansion is justified because
from \eq{momcons} we have
\begin{equation} \label{eq:qperpsmall}
q_\perp^\mu = - p_{X\perp}^\mu = -b_{a\perp}^\mu -b_{a\perp}^\mu - k_{\soft\perp}^\mu \sim \lambda Q
\,.\end{equation}
A parametrically large $q_\perp^\mu \sim Q$ would require a separate jet at large $p_T \sim Q$ to balance the transverse momentum, which is not allowed in our setup. The kinematics of the hard matrix element in the factorization theorem is then given by the tree-level partonic kinematics, with the partonic momentum conservation
\begin{equation} \label{eq:partonmomcons}
x_a \ECM\,\frac{n_a^\mu}{2} + x_b \ECM\,\frac{n_b^\mu}{2} = q = q^- \frac{n_a^\mu}{2} + q^+ \frac{n_b^\mu}{2}
\,,\end{equation}
which implies
\begin{align} \label{eq:xab}
x_a \ECM &= q^- = \sqrt{q^2}\, e^Y
\,, &
q^2 &= q^+ q^- = x_a x_b \ECM^2
\,,\nn\\
x_b \ECM &= q^+ = \sqrt{q^2}\, e^{-Y}
\,, &
Y &= \frac{1}{2} \ln\frac{q^-}{q^+} = \frac{1}{2} \ln\frac{x_a}{x_b}
\,.\end{align}
Equations~\eqref{eq:qperpsmall} and \eqref{eq:partonmomcons} imply that parametrically the leptons are
back to back in the transverse plane. Since $q^+$ and $q^-$ can differ
substantially, the leptons do not need to be back to back in three dimensions.

\section{Derivation of Isolated Factorization Theorem}
\label{sec:facttheorem}

We now proceed to derive the isolated factorization theorem for generic processes $pp\to
XL$, where the hadronic final state $X$ has a restriction on the hemisphere
momenta $B_{a,b}^+$. The derivation is carried out using SCET
without Glauber degrees of freedom. The proof that Glauber effects
are not required is given at the  end of this section.

\subsection{Cross Section in QCD}

We will generically refer to properties of $L$ as ``leptonic'', even though
$L$ can contain any non-strongly interacting particles.
We only consider processes where the hard
interaction couples the strong and electroweak sectors through one two-particle
QCD current.  (This includes for example Drell-Yan or Higgs production through
gluon fusion with the Higgs decaying non-hadronically, but does not include
electroweak Higgs production via vector-boson fusion.) Then, at leading order in
the electroweak interactions, we can factorize the full-theory matrix element
into its leptonic and hadronic parts
\begin{align} \label{eq:M}
\mathcal{M}(pp\to X L) = \sum_{J} L_J\,\Mae{X}{J}{pp}
\,.\end{align}
The sum runs over all relevant color-singlet two-particle QCD currents $J$, and
$L_J$ contains the corresponding electroweak matrix element, including the
electroweak propagator coupling to $J$. For example, for Drell-Yan with $L =
\ell^+\ell^-$, the relevant currents are
\begin{equation} \label{eq:JDY_QCD}
J_{Vf}^\mu = \bar{q}_f \gamma^\mu q_f
\,,\qquad
J_{Af}^\mu = \bar{q}_f \gamma^\mu \gamma_5 q_f
\,,\end{equation}
so in this case the sum over $J$ in \eq{M} includes the sums over the two Dirac
structures, the vector index $\mu$, and the quark flavor $f = \{u, d, \ldots\}$.
The corresponding $L_{Vf}^\mu$ and $L_{Af}^\mu$ are given below in \eq{LmuDY}.

The cross section for some hadronic observable $\Obs$ in the center-of-mass frame of the collision, averaged over proton spins, is
\begin{align} \label{eq:dsigma_dO}
\frac{\df\sigma}{\df q^2\df Y\df\Obs}
&= \frac{1}{2\ECM^2} \int\!\frac{\df^2 \vec{q}_T}{2(2\pi)^4}
\int\! \df \Phi_L\, (2\pi)^4 \delta^4(q - p_L)\,
\\ & \quad\times
\frac{1}{4}\sum_\mathrm{spins}\sum_{X}\Abs{\mathcal{M}(pp\to XL)}^2 \,
\delta[\Obs - f_\Obs(X)]\,
(2\pi)^4 \delta^4(P_a + P_b - q - p_X)
\,.\nn\end{align}
Here, $P_{a,b}$ are the incoming proton momenta, $p_X$ and $p_L$ are the total
hadronic and leptonic momenta, $\df \Phi_L$ denotes the leptonic phase space,
and the phase-space integrations for the hadronic final states are included in
the sum over $X$. The last $\delta$ function is overall momentum conservation.
The function $f_\Obs(X)$ inside the second $\delta$ function returns the value
of the hadronic observable $\Obs$ for a given hadronic state $X$, so the
$\delta$ function picks out all final states that contribute to a certain value
of $\Obs$. The $\delta^4(q - p_L)$ under the leptonic phase-space integral
defines the measured $q$ as the total leptonic momentum. Expanding this $\delta$
function for $\vec{q}_T = 0$, the leptonic part does not depend on $\vec{q}_T$
at leading order, and using \eq{M}, we can rewrite \eq{dsigma_dO} as
\begin{align} \label{eq:dsigmadO_LW}
\!\!\frac{\df\sigma}{\df q^2\df Y \df\Obs}
&= \frac{1}{2\ECM^2} \sum_{J,J'} L_{JJ'}(q^2, Y)\, W_{JJ'}(q^2, Y, \Obs)
\,.\end{align}
The leptonic tensor is defined as
\begin{equation} \label{eq:Lmunu_def}
L_{JJ'}(q^2, Y)
= \!\int\! \df \Phi_L\, L_J^\dagger L_{J'}\, (2\pi)^4 \delta^4\Bigl(q^-\frac{n_a}{2} + q^+\frac{n_b}{2} - p_L \Bigr)
,\end{equation}
where $q^\pm = \sqrt{q^2} e^{\mp Y}$. The hadronic tensor contains the square of the hadronic matrix element
\begin{align} \label{eq:Wmunu_def}
W_{JJ'}(q^2, Y, \Obs)
& = \int\!\frac{\df^2\vec{q}_T}{2(2\pi)^4}
\sum_X \Mae{pp}{J^\dagger(0)}{X}\Mae{X}{J'(0)}{pp}
\nn \\ & \quad\times
(2\pi)^4 \delta^4(P_a + P_b - q - p_X)\,\delta[O - f_O(X)]
\,,\end{align}
where as in \ch{beamf} we keep the average over proton spins implicit in
the matrix element. Since $W_{JJ'}$ is integrated over $\vec{q}_T$, it can only
depend on $q^2$ and $Y$, as well as the hadronic observable $\Obs$.

We are interested in the hadronic observables $B_{a}^+ = n_{a}\sdt B_{a}$
and $B_{b}^+ = n_{b}\sdt B_{b}$.  The hemisphere hadronic momenta
$B_{a,b}^\mu(X)$ can be obtained from the states $\ket{X}$ using the hemisphere
momentum operators $\hp_{a,b}^\mu$
\begin{equation} \label{eq:hatpab}
\hp_a^\mu \ket{X} = B_a^\mu(X) \ket{X}
\,,\qquad
\hp_b^\mu \ket{X} = B_b^\mu(X) \ket{X}
\,.\end{equation}
A field-theoretic definition of $\hp_{a,b}^\mu$ in terms of the energy-momentum tensor of the field theory was given in Ref.~\cite{Bauer:2008jx}. The hadronic tensor for $O \equiv \{B_a^+, B_b^+\}$ is
\begin{align} \label{eq:Wmunu_QCD}
&W_{JJ'}(q^2, Y, B_a^+, B_b^+)
\\ & \
= \!\int\!\frac{\df^2\vec{q}_T}{2(2\pi)^4} \int\!\df^4x\,e^{-\img q\cdot x}
\sum_X \Mae{pp}{J^\dagger(x)}{X}\Mae{X}{J'(0)}{pp}
\,\delta[B_a^+ \!\!-\! n_a\sdt B_a(X)]\,\delta[B_b^+ \!\!-\! n_b\sdt B_b(X)]
\nn\\ & \
= \!\int\!\frac{\df x^+\df x^-}{(4\pi)^2}\,e^{-\img (q^+ x^- + q^- x^+)/2}\,
\MAe{pp}{J^\dagger\Bigl(x^-\!\frac{n_a}{2} \!+\! x^+\!\frac{n_b}{2}\Bigr)
\delta(B_a^+ \!\!-\! n_a\sdt\hp_a)\,\delta(B_b^+ \!\!-\! n_b\sdt\hp_b) J'(0)}{pp}
\,.\nn\end{align}
In the first line we used momentum conservation to shift the position of $J^\dagger$, and in the second line we performed the integral over $\vec{q}_T$, which sets $\vec{x}_T$ to zero. We also used \eq{hatpab} to eliminate the explicit dependence on $X$, allowing us to carry out the sum over all states $X$. The restriction on the states $X$ is now implicit through the operator $\delta$ functions inside the matrix element.

\subsection{Matching QCD onto SCET}
\label{subsec:matchQCD}

In the next step, we match the QCD currents $J$ onto SCET currents by integrating out fluctuations at the hard scale $Q$. At leading order in the power counting, the matching takes the form
\begin{align} \label{eq:J_matching}
J(x) 
= \sum_{n_1, n_2} \int\!\df \w_1\,\df\w_2\, e^{-\img (\lb_1 + \lb_2)\cdot x}
\biggl[&\sum_q C_{J q\bar{q}}^{\alpha\beta}(\lb_1, \lb_2)\, O_{q\bar{q}}^{\alpha\beta}(\lb_1, \lb_2;x)
\nn \\
&+ C_{J gg}^{\mu\nu}(\lb_1, \lb_2)\, O_{gg\,\mu\nu}(\lb_1, \lb_2;x) \Bigr]
\,,\end{align}
where $\alpha$, $\beta$ are spinor indices, $\mu$, $\nu$ are vector indices, and the sum over $q$ runs over all quark flavors $\{u, d, \ldots\}$. The Wilson coefficients and operators depend on the large label momenta
\begin{equation} \label{eq:labels}
\lb_1^\mu = \w_1\,\frac{n_1^\mu}{2}
\,,\qquad
\lb_2^\mu = \w_2\,\frac{n_2^\mu}{2}
\,.\end{equation}
They will eventually be set to either $q^- n_a^\mu/2$ or $q^+ n_b^\mu/2$ by
momentum conservation, but at this point are unspecified, and the sums and
integrals over $n_1$, $n_2$ and $\w_1$, $\w_2$ in \eq{J_matching} run over all
sets of distinct collinear directions and large label momenta.  On the
right-hand side of \eq{J_matching}, the full $x$ dependence of the current is
separated into the $x$ dependence appearing in the overall phase factor with
large label momenta and the residual $x$ dependence of the SCET operators.

The SCET operators $O_{q\bar{q}}^{\alpha\beta}(x)$ and $O_{gg}^{\mu\nu}(x)$ are constructed out of the collinear fields in \eq{chiB2}. At leading order in the power counting they contain one field for each collinear direction. Since the QCD currents are color singlets, the leading operators that can contribute are
\begin{equation} \label{eq:Oi_SCET}
O_{q\bar{q}}^{\alpha\beta}(\lb_1,\lb_2; x)
= \bar\chi_{n_1,-\w_1}^{\alpha j}(x)\,\chi_{n_2,\w_2}^{\beta j}(x)
\,,\qquad
O_{gg}^{\mu\nu}(\lb_1,\lb_2; x)
= \sqrt{\w_1\,\w_2}\,\cB_{n_1,-\w_1\perp }^{\mu c}(x)\, \cB_{n_2,-\w_2\perp}^{\nu c}(x)
\,,\end{equation}
where $j$ and $c$ are color indices in the fundamental and adjoint representations. We included appropriate minus signs on the labels, such that we always have $\w_{1,2} > 0$ for incoming particles. Here, $\chi \equiv \chi_q$ is a quark field of flavor $q$, which for simplicity we keep implicit in our notation. Note that the entire spin and flavor structure of the current $J$ is hidden in the label $J$ on the matching coefficients in \eq{J_matching}. The gluon operator is symmetric under interchanging both $\mu\lra\nu$ and $\lb_1\lra \lb_2$, so its matching coefficient must have the same symmetry,
\begin{equation} \label{eq:Cgg_symmetry}
C_{J gg}^{\nu\mu}(\lb_2, \lb_1) = C_{J gg}^{\mu\nu}(\lb_1, \lb_2)
\,.\end{equation}
We define the conjugate quark operator and matching coefficient with the usual factors of $\gamma^0$, i.e.,
\begin{equation}
O^{\dagger \beta\alpha}_{q\bar q}(\lb_1, \lb_2)
= \bar\chi_{n_2,\w_2}^{\beta j}(x)\,\chi_{n_1,-\w_1}^{\alpha j}(x)
\,,\qquad
\bC^{\beta\alpha}_{J q\bar q}(\lb_1, \lb_2)
= [\gamma^0 C^\dagger_{Jq\bar q}(\lb_1, \lb_2) \gamma^0]^{\beta\alpha}
\,.\end{equation}

The matching coefficients are obtained by computing the renormalized matrix
elements $\mae{0}{...}{q\bar q}$ and $\mae{0}{...}{gg}$ on both sides of
\eq{J_matching} and comparing the results. In pure dimensional regularization
for UV and IR divergences all loop graphs in SCET are scaleless and vanish,
which means the UV and IR divergences in the bare matrix elements
precisely cancel each other. The renormalized matrix elements of the right-hand
side of \eq{J_matching} are then given by their tree-level expressions plus pure
$1/\eps$ IR divergences, which cancel against those of the full-theory
matrix elements $\mae{0}{J}{q\bar q}$ and $\mae{0}{J}{gg}$ of the left-hand
side. Hence, the matching coefficients in $\overline{\mathrm{MS}}$ are given in
terms of the IR-finite parts of the renormalized full-theory matrix
elements computed in pure dimensional regularization.

\subsection{Soft-Collinear Factorization}
\label{subsec:soft_coll_fact}

The field redefinitions in \eq{BPS} introduce soft Wilson lines into the operators in \eq{Oi_SCET},
\begin{align} \label{eq:operators_BPSed}
O_{q\bar{q}}^{\alpha\beta}(x)
&= \bar\chi_{n_1,-\w_1}^{\zero \alpha j}(x)\,
T\bigl[Y^\dagger_{n_1}(x)\,Y_{n_2}(x) \bigr]^{jk} \chi_{n_2,\w_2}^{\zero \beta k}(x)
\,,\nn\\
O_{gg}^{\mu\nu}(x)
&= \sqrt{\w_1\,\w_2}\,\cB_{n_1,-\w_1\perp}^{\zero \mu c}(x) T\bigl[\cY_{n_1}^\dagger(x) \cY_{n_2}(x) \bigr]^{cd} \cB_{n_2, -\w_2\perp}^{\zero \nu d}(x)
\,.\end{align}
The time ordering is required to ensure the proper ordering of the soft gluon fields inside the Wilson lines. It only affects the ordering of the field operators, while the ordering of the color generators is still determined by the (anti)path ordering of the Wilson lines. In the remainder, we use these redefined fields and drop the $(0)$ superscript for convenience.

Since the momentum operator is linear in the Lagrangian, \eq{LSCET} allows us to write the hemisphere momentum operators as the sum of independent operators acting in the separate collinear and soft sectors,
\begin{equation}
\hp_a = \hp_{a,n_a} + \hp_{a,n_b} + \hp_{a,\soft}
\,,\qquad
\hp_b = \hp_{b,n_a} + \hp_{b,n_b} + \hp_{b,\soft}
\,.\end{equation}
The $n_a$ ($n_b$) collinear sector cannot contribute momentum in the $n_b$ ($n_a$) hemisphere. Thus, $\hp_{a, n_b} = \hp_{b, n_a} = 0$, while $\hp_{a, n_a} = \hp_{n_a}$ and $\hp_{b, n_b} = \hp_{n_b}$ reduce to the total momentum operators for each of the collinear sectors. For the soft sector, the distinction between the two hemisphere operators is important. We can now write
\begin{align} \label{eq:hpab_fact}
\delta(B_a^+ - n_a\sdt\hp_a)
&= \int\!\df b_a^+\,\df k_a^+ \,\delta(B_a^+ - b_a^+ - k_a^+)\,\delta(b_a^+ - n_a\sdt\hp_{n_a})\,\delta(k_a^+ - n_a\sdt\hp_{a,\soft})
\,,\nn\\
\delta(B_b^+ - n_b\sdt\hp_b)
&= \int\!\df b_b^+\,\df k_b^+ \,\delta(B_b^+ - b_b^+ - k_b^+)\,\delta(b_b^+ - n_b\sdt\hp_{n_b})\,\delta(k_b^+ - n_b\sdt\hp_{b,\soft})
\,.\end{align}
Using \eq{J_matching} in the hadronic tensor in \eq{Wmunu_QCD}, the forward matrix element of the product of currents turns into the forward matrix element of the product of the operators in \eq{operators_BPSed}. Since the Lagrangian in \eq{LSCET} contains no interactions between the collinear and soft sectors after the field redefinition, we can use \eq{hpab_fact} to factorize the resulting matrix element into a product of independent $n_a$-collinear, $n_b$-collinear, and soft matrix elements.

We first look at the contribution from $O_{q\bar{q}}$. The $x$ integral of the forward matrix element of $O_{q\bar{q}}$ becomes
\begin{align} \label{eq:me_fact}
&\int\!\frac{\df x^+\df x^-}{(4\pi)^2}\,e^{-\img (q^+ x^- + q^- x^+)/2}\,e^{\img (\lb_1 + \lb_2)\cdot x}
\nn\\ & \quad
\MAe{p_{n_a} p_{n_b}}{O_{q\bar{q}}^{\dagger \beta\alpha}(x)\,
\delta(B_a^+ - n_a\sdt \hp_a)\,\delta(B_b^+ - n_b\sdt \hp_b)\, O_{q\bar{q}}^{\alpha'\beta'}(0)}{p_{n_a} p_{n_b}}
\nn\\ & \quad
= \int\!\frac{\df x^+\df x^-}{(4\pi)^2}\,e^{-\img (q^+ x^- + q^- x^+)/2}
\int\!\df b_a^+\,\df b_b^+\,\df k_a^+\,\df k_b^+\, \delta(B_a^+ - b_a^+ - k_a^+)\,\delta(B_b^+ - b_b^+ - k_b^+)
\nn\\ & \qquad\times
\int\!\df \w_a\, \df\w_b\, e^{\img (\w_a x^+ + \w_b x^-)/2}
\nn\\ & \qquad\times
\biggl\{
\delta_{n_2 n_a}\, \delta(\w_2 - \w_a)\,\delta_{n_2' n_a}\, \delta(\w_2' - \w_a)\,
\delta_{n_1 n_b}\, \delta(\w_1 - \w_b)\,\delta_{n_1' n_b}\, \delta(\w_1' - \w_b)\,
\nn\\ & \qquad\quad\times
\MAe{0}{\bT\bigl[Y^\dagger_{n_a}(x)\, Y_{n_b}(x) \bigr]^{kj} \delta(k_a^+ - n_a\sdt\hp_{a,s})\, \delta(k_b^+ - n_b\sdt\hp_{b,s})\,T\bigl[Y^\dagger_{n_b}(0)\,Y_{n_a}(0) \bigr]^{j'k'} }{0}
\nn\\ & \qquad\quad\times
\theta(\w_a)\MAe{p_{n_a}}{\bar\chi_{n_a}^{\beta k}(x)\,\delta(b_a^+ - n_a\sdt \hp_{n_a})\,
\delta(\w_a - \bnP_{n_a})\, \chi_{n_a}^{\beta' k'}(0)}{p_{n_a}}
\nn\\ & \qquad\quad\times
\theta(\w_b) \MAe{p_{n_b}}{\chi_{n_b}^{\alpha j}(x)\,\delta(b_b^+ - n_b\sdt \hp_{n_b})\,
\delta(\w_b - \bnP_{n_b})\,\bar\chi_{n_b}^{\alpha' j'}(0)}{p_{n_b}}\,
 + (a\lra b) \biggr\}
\,.\end{align}
Here, $\ket{p_{n_a}}$ and $\ket{p_{n_b}}$ are the proton states with momenta
$P_{a,b}^\mu = \ECM n_{a,b}^\mu/2$ as in \eq{nab_choice}. The two terms in
brackets in \eq{me_fact} arise from the different ways of matching up the fields
with the external proton states. The restriction to have positive labels $\w$
requires the fields in $O_{q\bar{q}}$ to be matched with the incoming proton
states and the fields in $O_{q\bar{q}}^\dagger$ with the outgoing proton states.
In principle, there are two more ways to match the fields and external states,
yielding matrix elements with the structure $\mae{p}{\chi\chi}{p}$ and
$\mae{p}{\bar{\chi}\bar{\chi}}{p}$, which vanish due to quark flavor number
conservation in QCD. For the same reason, in the full product $(\sum_q
O^\dagger_{q\bar q})(\sum_{q'} O_{q'\bar q'})$ only the flavor-diagonal term
with $q = q'$ survives.

We abbreviate the collinear and soft matrix elements in the last three lines of \eq{me_fact} as $M_{\w_a}(x^-)$, $M_{\w_b}(x^+)$, $M_\soft(x^+, x^-)$. The collinear matrix elements only depend on one light-cone coordinate because the label momenta $\w_{a,b}$ are defined to be continuous. We could have also started with discrete label momenta, $\lw_{a,b}$, and then convert to continuous labels by absorbing the residual $k_{n_a}^-$ dependence as follows:
\begin{align}
\sum_{\lw_a}\, e^{\img \lw_a x^+/2}\,M_{\lw_a}(x^+, x^-)
&= \sum_{\lw_a} \int\!\df k_a^-\,e^{\img (\lw_a + k_{n_a}^-) x^+/2}\,M_{\lw_a + k_{n_a}^-}(x^-)
\nn \\
&= \int\!\df\w_a\,e^{\img \w_a x^+/2}\,M_{\w_a}(x^-)
\,,\end{align}
and analogously for $M_{\w_b}(x^+)$. In the second step we used that by
reparameterization invariance the Fourier-transformed matrix element can only
depend on $\lw_a + k_{n_a}^- = \w_a$.

As an aside, note that in the well-studied case where the collinear matrix
elements are between vacuum states, giving rise to jet functions, the
distinction between discrete and continuous labels is not as relevant. In that
case, the SCET Feynman rules imply that the collinear matrix elements do not
depend on the residual $k^-$ (and $k_\perp$) components, and therefore the label
momenta can be treated in either way. In our case, momentum conservation with
the external state forces the collinear matrix elements to depend on $k^-$.
Therefore, the only way to eliminate the residual $k^-$ dependence is to absorb
it into continuous $\w$ labels. One can easily see this already at tree level.
Replacing the proton states by quark states with momentum $p = \lp + p_r$, we
get
\begin{align}
&\int\!\frac{\df x^+\,\df x^-}{(4\pi)^2}\,e^{-\img(k^-x^+ + k^+ x^-)/2}\,\Mae{q(p)}{\bar\chi_n(x^+, x^-)\, \delta_{\lw,\bnP_n}\,\chi_n(0)}{q(p)}
\nn \\ & \qquad
= \bar u u\, \delta_{\lw, \lp^-}\,\delta(k^- - p_r^-)\,\delta(k^+ - p_r^+)
\,,\end{align}
and the label and residual minus momenta are combined using $\delta_{\lw,
  \lp^-}\,\delta(k^- - p_r^-) = \delta(\w - p^-)$. Our continuous $\w$ is
physical and corresponds to the momentum fraction of the quark in the proton.

Returning to our discussion, to perform the $x$ integral in \eq{me_fact}, we
take the residual Fourier transforms of the matrix elements,
\begin{align} \label{eq:tildeMi}
M_{\w_a}(x^-)
&= \int\!\frac{\df k^+}{2\pi}\,e^{\img k^+ x^-/2}\,\tM_{\w_a}(k^+)
\,,\qquad
M_{\w_b}(x^+)
= \int\!\frac{\df k^-}{2\pi}\,e^{\img k^- x^+/2}\,\tM_{\w_b}(k^-)
\,,\nn\\
M_\soft(x^-, x^+)
&= \int\!\frac{\df k_s^+\,\df k_s^-}{(2\pi)^2}\,e^{\img(k_s^+ x^+ + k_s^- x^+)/2}\,\tM_\soft(k_s^+, k_s^-)
\,.\end{align}
Just as $x^\pm$, the residual momenta $k^\pm$ and $k_s^\pm$ here are all defined
with respect to the common $n = n_a$. The $x$ integral in \eq{me_fact} now
becomes
\begin{align} \label{eq:xintegral}
&\int\!\frac{\df x^+\df x^-}{(4\pi)^2}\,e^{\img (\w_a - q^-) x^+/2}\, e^{\img (\w_b - q^+)x^-/2}
M_{\w_a}(x^-)\,M_{\w_b}(x^+)\,M_\soft(x^+, x^-)
\nn\\ &\qquad
= \int\!\frac{\df k^+}{2\pi}\,\frac{\df k^-}{2\pi}\, \frac{\df k_s^+\,\df k_s^-}{(2\pi)^2}\,
\tM_{\w_a}(k^+)\,\tM_{\w_b}(k^-)\, \tM_\soft(k_\soft^+, k_\soft^-)
\nn\\ &\qquad \quad \times
\delta(\w_a - q^- + k^- + k_\soft^-)\,\delta(\w_b - q^+ + k^+ + k_\soft^+)
\nn\\ &\qquad
= \delta(\w_a - q^-)\,\delta(\w_b - q^+)\,M_{\w_a}(0)\,M_{\w_b}(0)\,M_\soft(0)
\,.\end{align}
In the last step we expanded $q^\pm - k^\pm - k_\soft^\pm = q^\pm[1 +
\ord{\la^2}]$. The remaining residual integrations are then simply the Fourier
transforms of the matrix elements at $x = 0$.

Note that without the integration over $\vec{q}_T$ in the hadronic tensor
\eq{Wmunu_QCD}, the currents would depend on $x_\perp$, which would require us
to include perpendicular components $b_{a,b\perp}$ in the label momenta, and the
soft matrix element would depend on $x_\perp$, too. (The residual $k_{\perp}$
dependence in the collinear matrix elements can again be absorbed into
continuous $b_{a,b\perp}$.) The corresponding $x_\perp$ integration in
\eq{xintegral} would yield an additional $\delta$ function
$\delta^2(\vec{b}_{a\perp} + \vec{b}_{b\perp} + \vec{q}_T -
\vec{k}_{\soft\perp})$.  Integrating over $\vec{q}_T$ effectively eliminates
this $\delta$ function, which would otherwise force us to introduce an explicit
dependence on $b_{a,b\perp}$ in the beam functions. If one considers the $q_T$
spectrum of the dileptons for $q_T^2\ll q^2$, our analysis here provides a starting
point but requires further study. One cannot just use $p_T$ in place of $B_{a,b}^+$
with our arguments to impose an analogous
restriction on the final state, because at $\ord{\alpha_s^2}$ one can have two jets at high $\vec{p}_T$ that
still have small total $\vec{p}_T$.

The $n_a$-collinear matrix element now reduces to the quark beam functions
defined in \eq{B_def},
\begin{align}
M_{\w_a}(0) &=
\theta(\w_a) \MAe{p_{n_a}}{\bar\chi_{n_a}^{\beta k}(0)\,
  \delta(b_a^+ - n_a\sdt \hp_{n_a})\,\delta(\w_a - \bnP_{n_a})\, \chi_{n_a}^{\beta' k'}(0)}{p_{n_a}}
\nn\\ & \quad
= \frac{\nslash_a^{\beta'\beta}}{4}\, \frac{\delta^{k'k}}{N_c}
  \theta(\w_a) \MAe{p_{n_a}}{\bar\chi_{n_a}(0)\,\delta(b_a^+ - n_a\sdt \hp_{n_a})\,
\delta(\w_a - \bnP_{n_a})\, \frac{\bnslash_a}{2}  \chi_{n_a}(0)}{p_{n_a}}
\nn\\ & \quad
= \frac{\nslash_a^{\beta'\beta}}{4}\, \frac{\delta^{k'k}}{N_c}\, \theta(\w_a)
 \int\!\frac{\df y^-}{4\pi}\,e^{\img b_a^+ y^-/2}
\nn\\ & \qquad \times
 \MAe{p_{n_a}}{e^{-\img \hp_{n_a}^+ y^-/2}\,e^{\img \hp_{n_a}^+ y^-/2}\,\bar\chi_{n_a}(0)\,
 e^{-\img \hp_{n_a}^+ y^-/2}\,
 \delta(\w_a - \bnP_{n_a})\,\frac{\bnslash_a}{2}\chi_{n_a}(0)}{p_{n_a}}
\nn\\ & \quad
= \frac{\nslash_a^{\beta'\beta}}{4}\, \frac{\delta^{k'k}}{N_c}\, \w_a B_q(\w_a b_a^+, \w_a/P_a^-)
\,.\end{align}
We abbreviated $\hp_{n_a}^+ = n_a\cdot \hp_{n_a}$, and
in the last step we used $e^{\img \hp_n^+ y^-/2}\,\bar\chi_n(0)\,e^{-\img\hp_n^+ y^-/2} = \bar\chi_n(y^-n/2)$ and $\hp_n^+\ket{p_n} = 0$. Similarly, for the antiquark beam function we have
\begin{align}
M_{\w_b}(0) &=
\theta(\w_b)\MAe{p_{n_b}}{\chi_{n_b}^{\alpha j}(x)\, \delta(b_b^+ - n_b\sdt \hp_b)\,
\delta(\w_b - \bnP_{n_b})\,\bar\chi_{n_b}^{\alpha' j'}(0)}{p_{n_b}}
\nn \\
&= \frac{\nslash_b^{\alpha\alpha'}}{4}\,\frac{\delta^{jj'}}{N_c}\, \w_b B_{\bar q}(\w_b b_b^+ , \w_b/P_b^-)
\,.\end{align}

Since the collinear matrix elements are color diagonal, the soft matrix element reduces to an overall color-singlet trace, which defines the $q\bar q$ incoming hemisphere soft function,
\begin{equation} \label{eq:Sqq_def}
S^{q\bar{q}}_\hemiin(k_a^+, k_b^+)
= \frac{1}{N_c}\, \tr\,\Mae{0}{ \bT\bigl[Y^\dagger_{n_a}(0)\, Y_{n_b}(0) \bigr]
 \delta(k_a^+ - n_a\sdt\hp_{a,s})\, \delta(k_b^+ - n_b\sdt\hp_{b,s})\,
  T\bigl[Y^\dagger_{n_b}(0)\,Y_{n_a}(0) \bigr]}{0}
\,.\end{equation}
The trace is over color and the factor of $1/N_c$ is included by convention,
such that at tree level we have $S^{q\bar q,\tree}_\hemiin(k_a^+, k_b^+) =
\delta(k_a^+)\,\delta(k_b^+)$.  The soft matrix element in the second term of
\eq{me_fact} with $a \lra b$ interchanged is equal to the above one due to
charge conjugation invariance of QCD. Under charge conjugation, the Wilson lines
transform as $\mathrm{C}^{-1} Y_n^{ij} \mathrm{C} = T[Y_n^{\dagger ji}]$. The
explicit time ordering is required because the fields in $Y_n$ are time-ordered
by default, and charge conjugation only changes the ordering of the color
generators but not of the field operators. For us this is not relevant, because
the ordering of the fields is determined by the overall (anti-)time ordering in
the matrix element. Thus, for the soft matrix element with $a\lra b$
interchanged, we find
\begin{align}
&\tr \Mae{0}{\bT\bigl[Y^\dagger_{n_b} Y_{n_a} \bigr]
\delta(k_a^+ - n_a\sdt\hp_{a,s})\,\delta(k_b^+ - n_b\sdt\hp_{b,s})\,
  T\bigl[Y^\dagger_{n_a} Y_{n_b} \bigr]}{0}
\\ &\qquad
\stackrel{\mathrm{C}}{=}
\tr \Mae{0}{\bT\bigl[Y^T_{n_b} Y_{n_a}^{\dagger T} \bigr]
\delta(k_a^+ - n_a\sdt\hp_{a,s})\, \delta(k_b^+ - n_b\sdt\hp_{b,s})\,
  T\bigl[Y^T_{n_a} Y_{n_b}^{\dagger T} \bigr] }{0}
= S^{q\bar{q}}_\hemiin(k_a^+, k_b^+)
\,,\nn\end{align}
where the transpose refers to the color indices. In the last step we used
$\tr[A^T B^T C^T D^T] = \tr[BADC]$ and the fact that the fields in
$Y^\dagger_{n_b}$ and $Y_{n_a}$ are spacelike separated and thus commute.  Under
parity, we have $\mathrm{P}^{-1} Y_{n_a} \mathrm{P} = Y_{n_b}$ and
$\mathrm{P}^{-1} n_a\cdot\hp_{a,s} \mathrm{P} = n_b\cdot\hp_{b,s}$. Therefore, CP
invariance implies that $S_\hemiin^{q\bar q}$ is symmetric in its arguments,
\begin{align}
S^{q\bar{q}}_\hemiin(k_a^+, k_b^+)
&\stackrel{\mathrm{CP}}{=} \frac{1}{N_c}\, \tr\,\Mae{0}{ \bT\bigl[Y^\dagger_{n_a} Y_{n_b}\bigr]
 \delta(k_a^+ - n_b\sdt\hp_{b,s})\, \delta(k_b^+ - n_a\sdt\hp_{a,s})\,
  T\bigl[Y^\dagger_{n_b} Y_{n_a}\bigr]}{0}
\nn \\
&= S^{q\bar{q}}_\hemiin(k_b^+, k_a^+)
\,.\end{align}

Having worked out the different terms in \eq{me_fact}, we are ready to include
the remaining pieces from \eqs{Wmunu_QCD}{J_matching}. The $q\bar q$
contribution to $W$ becomes
\begin{align} \label{eq:Wqqfactorized}
&W_{JJ' q\bar q}(q^2, Y, B_a^+, B_b^+)
\nn\\ &\qquad
= \int\!\df\w_a\,\df\w_b\,\delta(\w_a - q^-)\,\delta(\w_b - q^+)\,
\!\!\!\sum_{n_1, n_2, n_1', n_2'} \int\!\df\w_1\,\df\w_2\,\df\w_1'\,\df\w_2'
\,\bC_{Jq\bar q}^{\beta\alpha}(\lb_1,\lb_2)\, C_{J' q\bar q}^{\alpha'\beta'} (\lb_1',\lb_2')
\nn\\ & \qquad\quad\times
\biggl\{
\delta_{n_2 n_a}\, \delta(\w_2 - \w_a)\,\delta_{n_2' n_a}\, \delta(\w_2' - \w_a)\,
\delta_{n_1 n_b}\, \delta(\w_1 - \w_b)\,\delta_{n_1' n_b}\, \delta(\w_1' - \w_b)\,
\nn\\ &\qquad\qquad\times
\frac{\nslash_a^{\beta'\beta}}{4}\, \frac{\nslash_b^{\alpha\alpha'}}{4}\,\frac{1}{N_c}\,
\int\! \df k_a^+\,\df k_b^+\,
q^2 B_q[x_a \ECM (B_a^+ - k_a^+), x_a]\, B_{\bar q}[x_b\ECM(B_b^+ - k_b^+), x_b]
\nn\\ &\qquad\qquad\times
S^{q \bar q}_\hemiin(k_a^+,k_b^+) + (a \leftrightarrow b) \biggr\}
\nn \\ &\qquad
= H_{JJ'q \bar q}(\lb_a, \lb_b) \int\!\df k_a^+ \df k_b^+\,
q^2 B_q[x_a \ECM(B_a^+ - k_a^+), x_a]\, B_{\bar q}[x_b \ECM(B_b^+ - k_b^+), x_b]
\nn \\ &\qquad\qquad\times
 S^{q \bar q}_\hemiin(k_a^+,k_b^+) + (q \lra \bar q)
\,.\end{align}
All label sums and integrations from \eq{J_matching} eliminate the label $\delta$'s from \eq{me_fact}. In the second step we defined
\begin{align} \label{eq:lbxab}
\lb_a^\mu &= x_a\ECM\,\frac{n_a^\mu}{2}
\,, &
\lb_b^\mu &= x_b\ECM\,\frac{n_b^\mu}{2}
\nn \\
x_a & \equiv \frac{\w_a}{\ECM} = \frac{q^-}{\ECM} = \frac{\sqrt{q^2}\, e^Y}{\ECM}
\,, &
x_b &\equiv \frac{\w_b}{\ECM} = \frac{q^+}{\ECM} = \frac{\sqrt{q^2}\, e^{-Y}}{\ECM}
\,,\end{align}
as in \eq{xab}, and introduced the hard functions
\begin{align} \label{eq:Hqq_def}
H_{JJ'q \bar q}(\lb_a, \lb_b)
&= \frac{1}{N_c}\, \frac{1}{4}\,\tr_\mathrm{spins}\Bigl[
\frac{\nslash_a}{2}\, \bC_{J q\bar q}(\lb_b, \lb_a)\,
\frac{\nslash_b}{2}\, C_{J' q\bar q}(\lb_b, \lb_a) \Bigr]
\,,\nn \\
H_{JJ'\bar q q}(\lb_a, \lb_b) &= H_{JJ' q\bar q}(\lb_b, \lb_a)
\,.\end{align}
Equation~\eqref{eq:Wqqfactorized} is the final factorized result for the $O_{q\bar{q}}$ contribution
to the hadronic tensor.

Repeating the same steps for $O_{gg}$, we obtain for the forward matrix element
\begin{align} \label{eq:me_fact_gg}
&\int\!\frac{\df x^+\df x^-}{(4\pi)^2}\,e^{-\img (q^+ x^- + q^- x^+)/2}\,e^{\img (\lb_1 + \lb_2)\cdot x}\,
\nn \\ &\quad\times
\MAe{p_{n_a} p_{n_b}}{O_{gg}^{\dagger \nu\mu}(x)\,
\delta(B_a^+ - n_a\sdt \hp_{a,s})\,\delta(B_b^+ - n_b\sdt \hp_{b,s})\, O_{gg}^{\mu'\nu'}(0)}{p_{n_a} p_{n_b}}
\nn\\ & \qquad
= \int\!\df \w_a\, \df\w_b\,\delta(\w_a - q^-)\,\delta(\w_b - q^+)
\nn\\ & \qquad\quad\times
\int\!\df b_a^+\,\df b_b^+\,\df k_a^+\,\df k_b^+\, \delta(B_a^+ - b_a^+ - k_a^+)\,\delta(B_b^+ - b_b^+ - k_b^+)
\nn\\ & \qquad\quad\times
\bigl[\delta_{n_1 n_a}\, \delta(\w_1 - \w_a)\, \delta_{n_2 n_b}\, \delta(\w_2 - \w_b) + (a \lra b)\bigr]
\nn\\ & \qquad\quad\times
\bigl[\delta_{n_1'n_a}\, \delta(\w_1' - \w_a)\, \delta_{n_2' n_b}\, \delta(\w_2' - \w_b) + (a \lra b)\bigr]
\nn\\ & \qquad\quad\times
\MAe{0}{\overline{T}\bigl[\cY^\dagger_{n_a}(0)\, \cY_{n_b}(0)\bigr]^{cd}\,
\delta(k_a^+ - n_a\sdt\hp_{a,s})\, \delta(k_b^+ - n_b\sdt\hp_{b,s})\,
T\bigl[\cY^\dagger_{n_b}(0)\,\cY_{n_a}(0) \bigr]^{d'c'} }{0}
\nn\\ & \qquad\quad\times
\w_a\,\theta(\w_a)\,\MAe{p_{n_a}}{\cB_{n_a\perp}^{\mu c}(0)\,\delta(b_a^+ - n_a\sdt \hp_{n_a})\,\delta(\w_a - \bnP_{n_a})\, \cB_{n_a\perp}^{\mu' c'}(0)}{p_{n_a}}\,
\nn\\ & \qquad\quad\times
\w_b\,\theta(\w_b)\,\MAe{p_{n_b}}{\cB_{n_b\perp}^{\nu d}(0)\,\delta(b_b^+ - n_b\sdt \hp_{n_b})\,\delta(\w_b - \bnP_{n_b})\,  \cB_{n_b\perp}^{\nu' d'}(0)}{p_{n_b}}
\,,\end{align}
where we already performed the integral over $x$. The four terms in the third line correspond
to the four different ways to match up the gluon fields with the incoming proton states.
The collinear matrix elements reduce to the gluon beam function defined in \eq{B_def},
\begin{align}
&\w_a\,\theta(\w_a)\,\MAe{p_{n_a}}{\cB_{n_a\perp}^{\mu c}(0)\,\delta(b_a^+ - n_a\sdt \hp_{n_a})\,
\delta(\w_a - \bnP_{n_a})\, \cB_{n_a\perp}^{\mu'c'}(0)}{p_{n_a}}
\nn\\ &\qquad
= \frac{g_\perp^{\mu \mu'} }{2}\, \frac{\delta^{cc'}}{N_c^2-1}\, \w_a B_g(\w_a b_a^+, \w_a/P_a^-)
\,.\end{align}
Including the color traces from the beam functions, the soft matrix element
defines the gluonic incoming hemisphere soft function,
\begin{align}
&S^{gg}_\hemiin(k_a^+, k_b^+)
\\ &\
= \frac{1}{N_c^2 - 1}
\MAe{0}{\tr_\mathrm{color}\bigl\{\overline{T}\bigl[\cY^\dagger_{n_a}(0) \cY_{n_b}(0) \bigr]\,
\delta(k_a^+ \!- n_a\sdt\hp_{a,s})\, \delta(k_b^+ \!- n_b\sdt\hp_{b,s})\, T\bigl[\cY^\dagger_{n_b}(0) \cY_{n_a}(0) \bigr] \bigr\}}{0}
\,,\nn\end{align}
where the normalization is again convention. Putting everything together, the
gluon contribution to the hadronic tensor becomes
\begin{align}
W_{JJ'gg}(q^2, Y, B_a^+, B_b^+)
&= H_{JJ'gg}(\lb_a, \lb_b) \!\int\!\df k_a^+ \df k_b^+\,
q^2 B_g[x_a \ECM(B_a^+ - k_a^+), x_a] 
\nn \\ & \quad \times
B_g[x_b \ECM (B_b^+ - k_b^+), x_b]\, S^{gg}_\hemiin(k_a^+,k_b^+)
\,,\end{align}
with the hard function
\begin{equation}
H_{JJ'gg}(\lb_a, \lb_b)
= \frac{1}{N_c^2-1}\,
\frac{1}{2}\,(g_{\perp\,\mu\mu'}\, g_{\perp\,\nu\nu'} + g_{\perp\,\mu\nu'}\, g_{\perp\,\nu\mu'})\,
C_{J gg}^{\dagger\,\nu\mu}(\lb_a, \lb_b)\, C_{J' gg}^{\mu'\nu'}(\lb_a, \lb_b)
\,.\end{equation}
Here we have used the symmetry of the Wilson coefficients in \eq{Cgg_symmetry}
to simplify the four terms that arise from interchanging $a\lra b$ in
\eq{me_fact_gg}.

To obtain the full result for the hadronic tensor all we have to do now is to
add up the contributions from the different quark flavors and the gluon,
\begin{equation}
W_{JJ'}(q^2, Y, B_a^+, B_b^+)
= \sum_q W_{JJ' q\bar q}(q^2, Y, B_a^+, B_b^+) + W_{JJ'gg}(q^2, Y, B_a^+, B_b^+)
\,.\end{equation}

\newpage
Inserting this back into \eq{dsigmadO_LW}, the final result for the factorized cross section becomes
\begin{align} \label{eq:dsigma_final}
\frac{\df\sigma}{\df q^2\df Y \df B_a^+\df B_b^+}
&= \sum_{ij} H_{ij}(q^2, Y) \int\!\df k_a^+\, \df k_b^+\,
q^2 B_i[x_a \ECM(B_a^+ - k_a^+), x_a] 
\nn \\ & \quad \times
B_j[x_b \ECM(B_b^+ - k_b^+), x_b]\, S^{ij}_\hemiin(k_a^+,k_b^+)
\,,\end{align}
with $x_{a,b}\ECM = \sqrt{q^2} e^{\pm Y}$ as in \eqs{xab}{lbxab} and the hard function
\begin{align} \label{eq:Hij}
H_{ij}(q^2, Y)
&= \frac{1}{2\ECM^2} \sum_{J,J'} L_{JJ'}(q^2, Y)
\nn\\ & \quad\times
H_{JJ' ij}\Bigl(x_a\ECM \frac{n_a}{2}, x_b\ECM \frac{n_b}{2}\Bigr)
\,.\end{align}
The sum in \eq{dsigma_final} runs over parton species $ij= \{gg, u\bar u, \bar u
u, d \bar d, \ldots\}$, where $B_i$ is the beam function for parton $i$ in beam
$a$ and $B_j$ for parton $j$ in beam $b$. Equation~\eqref{eq:dsigma_final} is
the final factorization theorem for the isolated $pp\to XL$ and $p\bar{p} \to
XL$ processes. In \sec{DY_final} below we will apply it to the case of
Drell-Yan, which will yield \eq{DYbeam}.

The beam functions in \eq{dsigma_final} are universal and take into account
collinear radiation for isolated processes with $x$ away from one.
Since the soft function only depends on the color
representation, but not on the specific quark flavor, there are only two
independent soft functions $S^{q\bar q}_\hemiin$ and $S^{gg}_\hemiin$. In the
sum over $ij$ in \eq{dsigma_final}, there are no mixed terms with $ij$ corresponding to
beam functions of two different quark flavors.
Likewise, there are no mixed terms with quark and gluon beam functions.
For example, a graph like \fig{DYBbeamgq} is part of the $ij= q\bar{q}$ term in the
sum. Thus, cross terms between quark and gluon PDFs only appear via the
contributions of different PDFs to a given beam function, as shown in
\eq{B_fact}.

The only process dependence in \eq{dsigma_final} arises through the hard
functions $H_{ij}(q^2, Y)$, and one can study any desired leptonic observables
by inserting the appropriate projections in the leptonic phase-space
integrations inside $L_{JJ'}(q^2, Y)$. Since the hard function $H_{JJ'ij}$
corresponds to the partonic matrix element $\mae{ij}{J^\dagger}{0}\mae{0}{J'}{ij}$ and
$L_{JJ'}$ is given by the square of the relevant electroweak matrix elements
$L_J^\dagger L_{J'}$, $H_{ij}(q^2, Y)$ can be determined from calculations of the
partonic cross section $ij\to L$. Furthermore, $H_{ij}(q^2,Y)$ is identical to
the hard function in threshold factorization theorems and hence in many cases is
known from existing computations.

\subsection{Cancellation of Glauber Gluons}
\label{subsec:glauber}

In the above derivation we have implicitly assumed that contributions from
Glauber gluons cancel in the final cross section, so that we do not need Glauber
interactions in the effective theory. To complete the proof of factorization, we
now argue that this is indeed the case.

In principle, Glauber interactions add an additional term $\cL_G$ to the SCET
Lagrangian
\begin{align} \label{eq:LwithG}
\cL_\mathrm{SCET}
&= \cL_{n_a}(\chi_{n_a}, A_\soft) + \cL_{n_b}(\chi_{n_b}, A_\soft)
 + \cL_\soft(A_\soft)
 + \cL_G(A_G, \chi_{n_a}, \chi_{n_b}, A_\soft)
\,.\end{align}
Glauber interactions in SCET have been considered in Refs.~\cite{Idilbi:2008vm,
  Donoghue:2009cq}, but we will not require an explicit construction of $\cL_G$
here.  Our arguments will be based on the one hand, on the consistency with
processes where it has been proven that Glauber interactions cancel, and on the
other hand on systematic scale separation in the language of effective field
theory. The scale separation is valid independently of whether it leads to a
factorization into simple matrix elements, or whether it leads to a
non-factorizable matrix element with complicated dynamics.

The possible danger of the Glauber modes comes from the fact that they couple
the two collinear sectors $n_a$ and $n_b$ with momentum scaling $Q(\lambda^2, 1,
\lambda)$ and $Q(1, \lambda^2, \lambda)$. With $\cL_G$, there will still be
interactions between soft and collinear modes present in the Lagrangian even
after the field redefinition, so we cannot a priori factorize the full matrix
element into independent soft and collinear matrix elements. Therefore, we have
to revisit each step in our derivation with $\cL_G$ in mind.

\begin{figure}[t!]
\centering
\includegraphics[scale=0.75]{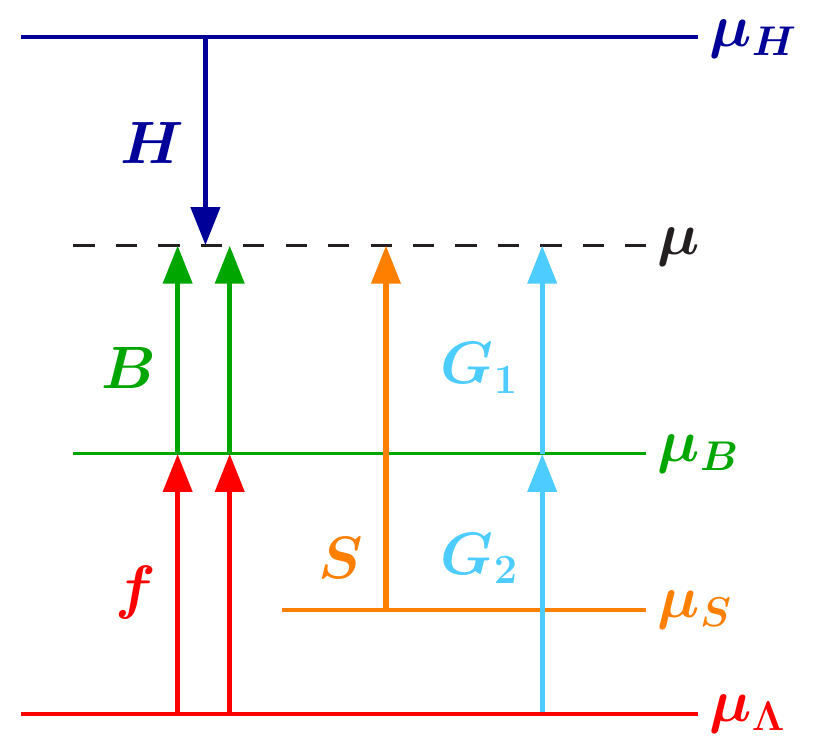}
\caption{RGE running including potential Glauber modes.}
\label{fig:running_drellyan_glauber}
\end{figure}

Our argument will be divided into three steps: (i) above the scale $\mu_B$, (ii)
at the scale $\mu_B$, and (iii) below the scale $\mu_B$. For (i) and (ii) we
have to consider Glauber modes with momentum scaling $Q(\lambda^2, \lambda^2,
\lambda)$, which we call $G_1$ modes. Since they have virtuality $\sim\mu_B^2$
they are integrated out at this scale. Any residual effects of Glauber
interactions below $\mu_B$ could occur from modes with momentum scaling
$Q(\lambda^4, \lambda^4, \lambda^2)$, which we call $G_2$. These modes are
illustrated in \fig{running_drellyan_glauber}.

Above the hard scale $\mu_H \simeq Q$, we have full QCD and no distinction
between different modes is required, so in step (i) we are concerned with
contributions of $G_1$ in the region $\mu_H > \mu > \mu_B$. At the scale
$\mu_H$, we integrate out hard modes with virtualities $Q^2$ or higher by
matching the QCD currents onto SCET currents. For our process, the leading
operators are given in \eq{Oi_SCET}, which contain only one field per collinear
direction. For the theory in \eq{LwithG}, all other possible operators
are power suppressed. The matching onto these currents is valid at an operator
level and can be performed with quark and gluon states. It is independent of the
hadronic matrix element we are going to take later on. The key point is that the
exact same matching calculation and resulting Wilson coefficients $C$ occur for
threshold Drell-Yan and $e^+e^-\to 2$ jets.  For these cases it is
known~\cite{Collins:1981ta, Collins:1989gx} that $G_1$ modes do not
affect the matching of the hard function $H\sim\abs{C}^2$ at $\mu_H$ or the
running of $H$ in the region $\mu_H> \mu > \mu_B$ shown in
Figs.~\ref{fig:running_drellyan}(b) or~\ref{fig:running_drellyan}(c). The hard
function $H$ gives a complete description of the physics down to the scale
$\mu_B$ whether or not the modes in the SCET matrix elements factorize further.
In \fig{running_drellyan_glauber}, this corresponds to taking the scale $\mu = \mu_B$.
Therefore, the $G_1$ modes can give neither large $\ln(\mu_B/\mu_H)$ terms nor
finite contributions above $\mu_B$.

In step (ii), we integrate out modes with virtualities $Q^2\lambda^2$ at the
scale $\mu_B$, which may involve matrix elements with $G_1$ modes exchanged.
This matching affects the $n_a$-collinear, $n_b$-collinear, and $G_1$ modes,
whose momentum scaling below $\mu_B$ changes to $Q(\lambda^4, 1, \lambda^2)$,
$Q(1, \lambda^4, \lambda^2)$, and $Q(\lambda^4, \lambda^4, \lambda^2)$,
respectively. Here we consider the theory right above $\mu_B$ including $G_1$
modes, leaving the discussion of the theory just below $\mu_B$ and $G_2$ modes
to step (iii). Thus, we have to consider the matrix element of the composite operator
\begin{align}
&\big[\bar\chi_{n_a} \chi_{n_b}\big](x^+, x^-)\,
\delta(B_a^+ - n_a\sdt\hp_a)\,\delta(B_b^+ - n_b\sdt\hp_b)
\delta(\w_a - \bnP_{n_a})\, \delta(\w_b - \bnP_{n_b})\,
\big[\bar\chi_{n_b} \chi_{n_a} \big](0)
\,,\end{align}
where we suppressed all spin and color indices for simplicity, and these
collinear fields still couple to soft fields in their Lagrangians. Since $\mu_B$
is a perturbative scale, we can carry out the matching onto the theory below
$\mu_B$ at the operator level and do not yet have to consider proton states.
Since the Glauber gluons are spacelike, they cannot cross the final-state cut
indicated by the $\delta$ functions and only appear in virtual subdiagrams.
We can therefore make a correspondence with the calculation in step (i) as follows.
For any given final state with collinear and soft particles, the SCET
computation for (ii) is identical to the SCET computation carried out for the
matching in step (i) but using this particular choice of external
states.\footnote{In practice one would never make such a complicated choice, but
  if one does, it must give the same result as picking a minimal state for the
  matching.} Since that SCET computation cannot induce any dependence on $G_1$
in step (i), there can also be no contributions from $G_1$ for the forward
matrix-element computation here. The result of the step (ii) matching is thus
given by a Wilson coefficient times an operator of the form
\begin{align} \label{eq:G1matching}
&\int\!\df k_a^+\, \df k_b^+\,
C(x^+, x^-, B_a^+ - k_a^+, B_b^+ - k_b^+)
\bar\chi_{n_a}'(0)\, \bT\bigl[Y^\dagger_{n_a} Y_{n_b} \bigr] \chi_{n_b}'(0)\,
\delta(k_a^+ - n_a\sdt\hp_a)\,
\nn\\ & \qquad \times
\delta(k_b^+ - n_b\sdt\hp_b)\, \delta(\w_a - \bnP_{n_a})\,
  \delta(\w_b - \bnP_{n_b})
\bar\chi_{n_b}'(0)\, T\bigl[Y^\dagger_{n_b} Y_{n_a} \bigr]\chi_{n_a}'(0)
\,,\end{align}
where the primed collinear fields have scaling $Q(\lambda^4, 1, \lambda^2)$ and
$Q(1, \lambda^4, \lambda^2)$, and the soft fields in the $Y$ Wilson lines have
scaling $Q(\lambda^2, \lambda^2, \lambda^2)$.

For step (iii) below $\mu_B$, we have to consider the $\mae{pp}{\dotsb}{pp}$
matrix element of \eq{G1matching} and possible contributions from $G_2$ Glauber
gluons, which can now also connect to spectator lines in the proton (which are
primed collinear modes). The $G_2$ gluons may spoil the factorization of the two
collinear sectors. To argue that this is not the case, we rely heavily on the
original proof of the cancellation of Glauber gluons for inclusive Drell-Yan in
Ref.~\cite{Collins:1988ig}. By construction, for our observables the $k_{a,b}^+$
variables in \eq{G1matching} are of $\ord{Q\lambda^2}$ and thus only get contributions from the
soft gluons. Hence, we are fully inclusive in the Hilbert space of the primed
collinear fields. Therefore, the $G_2$ modes as well as possible ``ultrasoft''
$Q(\lambda^4, \lambda^4, \lambda^4)$ gluons cancel in the sum over states, just
as in the inclusive case. This discussion for the cancellation of $G_2$ modes is
identical to Ref.~\cite{Aybat:2008ct}, where arguments were presented for the
cancellation of $G_2$ gluons up to the scale induced by the measurement on the
final state, which in our case is $\mu_B$.

Physically, one could imagine that Glauber modes kick the spectators in the
proton remnant such that they can contribute to $B_{a,b}^+$. The above
arguments show that this is not the case, so that our treatment of
the proton and its remnant in the derivation of the factorization
is correct.

Note that the above arguments do not suffice to show that Glauber interactions cancel
when there are additional hard central jets in the final state.

\section{Final Results for Drell-Yan}
\label{sec:DY_final}

In this section, we present the final results for the isolated Drell-Yan
cross section. Our discussion is split into four parts: the leptonic tensor,
the hard function, the soft function, and the final cross section for beam thrust.

\subsection{The Leptonic Tensor}

To give an explicit example, we now apply the final factorization result in
\eq{dsigma_final} to the Drell-Yan process with $L = \ell^+\ell^-$. The relevant
QCD currents are the vector and axial-vector currents $J_{hf}^\mu$ with $h =
\{V,A\}$, already given in \eq{JDY_QCD}. The corresponding leptonic
contributions are
\begin{align} \label{eq:LmuDY}
L^\mu_{Vf}(p_1, p_2) &= \frac{4\pi\alem}{q^2} \Bigl[
- Q_f\, \bar{u}(p_2) \gamma^\mu v(p_1)
+ \frac{v_f}{1 - m_Z^2/q^2}\, \bar{u}(p_2)
  \gamma^\mu (v_\ell - a_\ell\gamma_5) v(p_1) \Bigr]
\,,\nn\\
L^\mu_{Af}(p_1, p_2) &= \frac{4\pi\alem}{q^2}\,
\frac{-a_f}{1 - m_Z^2/q^2}
\bar{u}(p_2) \gamma^\mu (v_\ell - a_\ell\gamma_5) v(p_1)
\,,\end{align}
where in this section $p_1 = p_{\ell^+}$ and $p_2 = p_{\ell^-}$ are the
lepton momenta, $Q_f$ is the quark charge (in units of $\abs{e}$), and
$v_{\ell,f}$, $a_{\ell,f}$ are the standard vector and axial couplings of the
leptons and quarks of flavor $f$ to the $Z$ boson. We will include the
width of the $Z$ later in \eq{H_oneloop}.

The leptonic phase space integral is
\begin{align} \label{eq:leptonPS}
&\int\!\frac{\df^4 p_1\,\df^4 p_2}{(2\pi)^2}\,\delta(p_1^2) \,\delta(p_2^2)\,
\delta^4\Bigl(q^-\frac{n_a}{2} + q^+\frac{n_b}{2} - p_1 - p_2 \Bigr)
= \frac{1}{32\,\pi^2}\int\!\frac{\df\Delta y\,\df\varphi}{1+\cosh\Delta y}
\,,\end{align}
where $\varphi$ is the azimuthal angle of the leptons in the transverse plane and $\Delta y$ is the rapidity difference of the two leptons:
\begin{equation}
y_i = \frac{1}{2}\ln\frac{n_b\sdt p_i}{n_a\sdt p_i}
\,,\quad
\Delta y = y_1 - y_2
\,.\end{equation}
Since we expanded $\vec{q}_T = 0$, the leptons are back to back in the
transverse plane, which implies that at the order we are working
\begin{align}
\frac{p_1^+}{p_2^-} &= \frac{p_2^+}{p_1^-} = \frac{q^+}{q^-}
\,,&
Y &= \frac{1}{2}\,(y_1 + y_2)
\,,\nn\\
\vec{p}_{1T} &= - \vec{p}_{2T}
\,,&
\vec{p}_{1T}^2 &= \vec{p}_{2T}^2 = \frac{q^2}{2(1+\cosh\Delta y)}
\,.\end{align}
Thus, the leptonic kinematics is described by the four independent variables
$\{q^2, Y, \Delta y, \varphi\}$, with $\{Y, \Delta y\}$ being equivalent to
$\{y_1, y_2\}$. For simplicity, we assume that we do not distinguish the two
leptons, as one would for example by measuring their rapidities $y_i$ or
transverse momenta $p_{iT}$. We can then integrate over $0\leq\varphi\leq 2\pi$
and $-\infty < \Delta y < \infty$ in \eq{leptonPS}, giving an overall factor of
$4\pi$. The leptonic tensor, \eq{Lmunu_def}, now becomes
\begin{align} \label{eq:LmunuDY}
L^{\mu\nu}_{hh'ff'}(q^2, Y)
&= \frac{1}{32\pi^2} \int\!\frac{\df\Delta y\,\df\varphi}{1+\cosh\Delta y}\,
\sum_\mathrm{spins} L^{\dagger\,\mu}_{hf}(p_1, p_2)\, L^\nu_{h'f'}(p_1, p_2)
\nn\\
&= \frac{8\pi \alem^2}{3q^2}
   \Bigl(\frac{q^\mu q^\nu}{q^2} - g^{\mu\nu}\Bigr) L_{hh'ff'}(q^2)
\,,\end{align}
where
\begin{align} \label{eq:LDY}
L_{VVff'}(q^2)
&= Q_f Q_{f'} - \frac{(Q_f v_{f'} + v_f Q_{f'}) v_\ell}{1-m_Z^2/q^2}
+ \frac{v_f v_{f'}(v_\ell^2+a_\ell^2)}{(1-m_Z^2/q^2)^2}
\,,\nn \\
L_{AAff'}(q^2)
&= \frac{a_f a_{f'} (v_\ell^2 + a_\ell^2)}{(1-m_Z^2/q^2)^2}
\,,\nn \\
L_{AVff'}(q^2)
&= \frac{-a_f}{1-m_Z^2/q^2}
\biggl[-Q_{f'} v_\ell + \frac{v_{f'} (v_\ell^2+a_\ell^2)}{1-m_Z^2/q^2}\biggr]
\nn\\
&= L_{VAf'f}(q^2)
\,.\end{align}

\subsection{The Hard Function}

Using parity and charge conjugation invariance of QCD, the matching coefficients
for the vector and axial-vector QCD currents can be written as
\begin{align} \label{eq:CDY}
C_{V f\,q\bar q}^{\mu\,\alpha\beta}(\lb_a, \lb_b)
&= C_{V fq}(q^2)\, (\gamma^\mu_\perp)^{\alpha\beta}
\,,\nn\\
C_{A f\,q\bar q}^{\mu\,\alpha\beta}(\lb_a, \lb_b)
&= C_{A fq}(q^2)\, (\gamma^\mu_\perp\gamma_5)^{\alpha\beta}
\,,\nn\\
C_{A f\,gg}^{\mu\,\rho\sigma}(\lb_a, \lb_b)
&= C_{Ag}(q^2)\, (\lb_a + \lb_b)^\mu\, \img \eps^{\rho\sigma}{}_{\lambda\kappa} \lb_a^\lambda \lb_b^\kappa
\,.\end{align}
By Lorentz invariance (or reparameterization invariance of $n_{a,b}$ and
$\bn_{a,b}$~\cite{Manohar:2002fd}), the scalar coefficients can only depend on
$\lb_a\cdot \lb_b = x_a x_b \ECM^2 = q^2$.  In principle, parity and charge
conjugation would also allow the Dirac structures
$(\lb_a-\lb_b)^\mu\,\delta^{\alpha\beta}$ and
$(\lb_a-\lb_b)^\mu\,(\gamma_5)^{\alpha\beta}$. However, as the vector and
axial-vector currents are chiral even and the matching from QCD conserves
chirality for massless quarks, these cannot be generated. For the gluon
operator, the symmetry of the Wilson coefficient [see \eq{Cgg_symmetry}]
requires it to be proportional to $q^\mu = \lb_a^\mu + \lb_b^\mu$. Current
conservation for the vector current requires $q_\mu C_{Vfq\bar q}^\mu = 0$,
which eliminates this term. Thus, as expected, the only contribution for the
gluon operator is due to the axial anomaly, coming from the diagram in
\fig{DYBglue}. Since we neglect the lepton masses, $q_\mu L^\mu_{Af} = 0$, and
thus $C_{Af gg}$ does not survive the contraction of the leptonic and hadronic
tensors for $L = \ell^+\ell^-$. Hence, the gluon beam functions do not
contribute to Drell-Yan, and the gluon PDF only appears through its contribution
to the quark beam functions. Inserting \eq{CDY} into the general expression for
the hard function in \eq{Hqq_def}, we obtain
\begin{align} \label{eq:HqqDY}
H_{hh' ff'\,q \bar q}^{\mu \nu}(\lb_a, \lb_b)
&= -\frac{1}{2N_c} \Bigl[g^{\mu\nu} \!-\! \frac{1}{2}(n_a^\mu n_b^\nu \!+\! n_a^\nu n_b^\mu) \Bigr]
C^*_{hfq}(q^2) C_{h' f'q}(q^2)
&&\!(\text{for}\ hh'\! = \{VV, AA\})
\,,\nn\\
H_{hh'ff'\,q \bar q}^{\mu\nu}(\lb_a, \lb_b)
&= \frac{1}{4N_c} \,\img \epsilon^{\mu\nu}{}_{\lambda\kappa} n_a^\lambda n_b^\kappa\, C^*_{hfq}(q^2) C_{h'f'q}(q^2)
&&\!(\text{for}\ hh'\! = \{VA, AV\})
\,.\end{align}

At one loop, the vector and axial-vector coefficients are equal and diagonal in
flavor and the SCET matching computation was performed in
Refs.~\cite{Manohar:2003vb, Bauer:2003di}, in agreement with the one-loop form factors
\begin{align} \label{eq:CDY_oneloop}
C_{Vfq}(q^2) &= C_{A fq}(q^2) = \delta_{f q} C(q^2)
\,,\nn \\
C(q^2, \mu) &= 1 + \frac{\alpha_s(\mu)\,C_F}{4\pi} \biggl[-\ln^2 \Bigl(\frac{-q^2-\img 0}{\mu^2}\Bigr) + 3 \ln \Bigl(\frac{-q^2-\img 0}{\mu^2}\Bigr) - 8 + \frac{\pi^2}{6} \biggr]
\,.\end{align}
The vector current coefficient at two loops was obtained in
Refs.~\cite{Idilbi:2006dg, Becher:2006mr} from the known two-loop quark form
factor~\cite{Kramer:1986sg, Matsuura:1987wt, Matsuura:1988sm, Gehrmann:2005pd}.
Starting at three loops, it can have a contribution that is not diagonal in
flavor, i.e., is not proportional to $\delta_{fq}$. The axial-vector coefficient
can also receive additional diagonal and nondiagonal contributions starting at
two loops from the axial anomaly~\cite{Kniehl:1989bb, Kniehl:1989qu,
  Bernreuther:2005rw}. The anomaly contributions cancel in the final result in
the sum over $f$ as long as one sums over massless quark doublets. Therefore,
they will cancel when the hard matching scale is much larger than the top-quark
mass, in which case the top quark can be treated as massless. On the other hand,
they have to be taken into account when the matching scale is below the
top-quark mass, in which case the top quark is integrated out during the
matching step and its mass cannot be neglected.

Combining \eqs{HqqDY}{CDY_oneloop} with \eqs{LmunuDY}{LDY}, the coefficients
$H_{ij}(q^2, Y)$ in \eq{Hij} become
\begin{align}
&\frac{1}{2\ECM^2}\,\frac{8\pi\,\alem^2}{3q^2}\,\frac{1}{N_c}\sum_{ff'} \bigl[L_{VV ff'}(q^2) C^*_{Vfq}(q^2) C_{Vf'q}(q^2) + L_{AA ff'}(q^2) C^*_{Afq}(q^2) C_{Af'q}(q^2)
\bigr]
\nn \\ & \qquad
\equiv\sigma_0 H_{q\bar q}(q^2, \mu)
\,,\end{align}
where at one loop
\begin{align} \label{eq:H_oneloop}
\sigma_0 &= \frac{4\pi\alem^2}{3N_c\ECM^2 q^2}
\,, \\
H_{q\bar q}(q^2, \mu) & = H_{\bar q q}(q^2, \mu) =
\biggl[ Q_q^2 + \frac{(v_q^2 + a_q^2) (v_\ell^2+a_\ell^2) - 2 Q_q v_q v_\ell (1-m_Z^2/q^2)}
{(1-m_Z^2/q^2)^2 + m_Z^2 \Gamma_Z^2/q^4} \biggr] \Abs{C(q^2, \mu)}^2
\,,\nn\end{align}
with $\abs{C(q^2,\mu)}^2$ given by \eq{CDY_oneloop}, and where we also included the
nonzero width of the $Z$. The RGE for the hard function $H_{q\bar q}(q^2, \mu)$
is
\begin{equation} \label{eq:H_RGE}
\mu\,\frac{\df H_{q\bar q}(q^2, \mu)}{\df\mu} = \gamma_H(q^2, \mu)\, H_{q\bar q}(q^2, \mu)
\,,\ \ 
\gamma_H(q^2, \mu) =
2\,\Gamma_\cusp[\alpha_s(\mu)] \ln\frac{q^2}{\mu^2} + \gamma_H[\alpha_s(\mu)]
\,,\end{equation}
where $\Gamma_\cusp$ is the universal cusp anomalous dimension~\cite{Korchemsky:1987wg}, and the one-loop non-cusp term is $\gamma_H[\alpha_s(\mu)] = - 3\alpha_s(\mu)\, C_F/\pi$~\cite{Manohar:2003vb}. The solution of \eq{H_RGE} has the standard form
\begin{equation} \label{eq:Hmatchrun}
H_{q\bar q}(q^2, \mu) = H_{q\bar q}(q^2, \mu_0)\, U_H(q^2, \mu_0, \mu)
\,,\qquad
U_H(q^2, \mu_0, \mu) = e^{K_H(\mu_0, \mu)} \Bigl(\frac{q^2}{\mu_0^2}\Bigr)^{\eta_H(\mu_0, \mu)}
\,,\end{equation}
where $K_H(\mu_0, \mu)$ and $\eta_H(\mu_0, \mu)$ are
\begin{align} \label{eq:KetaH}
K_H(\mu_0, \mu)
&= \int_{\alpha_s(\mu_0)}^{\alpha_s(\mu)}\!\frac{\df\alpha_s}{\beta(\alpha_s)}\,
\biggl[-4\, \Gamma_\cusp(\alpha_s) \int_{\alpha_s(\mu_0)}^{\alpha_s} \frac{\df \alpha_s'}{\beta(\alpha_s')}
   + \gamma_H(\alpha_s) \biggr]
\,,\nn \\
\eta_H(\mu_0, \mu)
&= 2 \int_{\alpha_s(\mu_0)}^{\alpha_s(\mu)}\!\frac{\df\alpha_s}{\beta(\alpha_s)}\, \Gamma_\cusp(\alpha_s)
\,.\end{align}
Together, \eqs{Hmatchrun}{KetaH} sum the large logarithms occurring in isolated Drell-Yan between the scales $\mu_H$ and $\mu_B$. Electroweak corrections to the hard function $H_{q\bar q}(q^2, \mu)$ can be included using the results of Refs.~\cite{Chiu:2007dg, Chiu:2008vv, Chiu:2009mg}.

\subsection{The $q\bar{q}$ Soft Function}
\label{subsec:quarksoftfunction}

The incoming hemisphere soft function contains incoming Wilson lines stretching
from $-\infty$ to $0$ along $n_a$ and $n_b$. Under time reversal, each incoming
Wilson line transforms into a corresponding outgoing Wilson line stretching from
$0$ to $\infty$ along the opposite direction,
\begin{equation}
\mathrm{T}^{-1} Y_{n_a} \mathrm{T}
= \overline{P}\exp\biggl[-\img g\intlim{0}{\infty}{s} n_b\sdt A_\soft(s\,n_b) \biggr]
= \widetilde{Y}_{n_b}
\,,\end{equation}
where $\overline{P}$ denotes anti-path ordering. Since $\mathrm{T}$ itself does
not affect the original ordering of the field operators, time ordering turns
into anti-time ordering and vice versa. In addition $\mathrm{T}\,
n_a\cdot\hp_{a,s}\, \mathrm{T}^{-1} = n_b\cdot\hp_{b,s}$. Therefore, time-reversal
invariance implies
\begin{align} \label{eq:Sqq_T}
S_\hemiin^{q\bar q}(k_a^+, k_b^+)
&\stackrel{\mathrm{T}}{=}
\frac{1}{N_c}\, \tr\,\Mae{0}{ T\bigl[\widetilde{Y}^\dagger_{n_b} \widetilde{Y}_{n_a} \bigr]
 \delta(k_a^+ - n_b\sdt\hp_{b,s})\, \delta(k_b^+ - n_a\sdt\hp_{a,s})\,
  \bT\bigl[\widetilde{Y}^\dagger_{n_a} \widetilde{Y}_{n_b} \bigr]}{0}^*
\nn\\
&= \frac{1}{N_c}\, \tr\,\Mae{0}{ T\bigl[\widetilde{Y}^\dagger_{n_a} \widetilde{Y}_{n_b} \bigr]
 \delta(k_a^+ - n_a\sdt\hp_{a,s})\, \delta(k_b^+ - n_b\sdt\hp_{b,s})\,
  \bT\bigl[\widetilde{Y}^\dagger_{n_b} \widetilde{Y}_{n_a} \bigr]}{0}
\,.\end{align}
In the second step, the complex conjugation has no effect since the matrix
element is real, and we used parity to switch $n_{b,a}$ back to $n_{a,b}$. For
comparison, the hemisphere soft function with outgoing Wilson appearing in the
double-differential hemisphere invariant-mass distribution in $e^+e^- \to 2$
jets~\cite{Korchemsky:1999kt, Korchemsky:2000kp, Fleming:2007qr, Schwartz:2007ib,
Fleming:2007xt, Hoang:2008fs} is
\begin{equation}
S_\hemiout^{q\bar q}(k_a^+, k_b^+)
= \frac{1}{N_c}\, \tr\,\Mae{0}{ \bT\bigl[\widetilde{Y}^\dagger_{n_a} \widetilde{Y}_{n_b} \bigr]
 \delta(k_a^+ - n_a\sdt\hp_{a,s})\, \delta(k_b^+ - n_b\sdt\hp_{b,s})\,
  T\bigl[\widetilde{Y}^\dagger_{n_b} \widetilde{Y}_{n_a} \bigr]}{0}
\,.\end{equation}
This is almost the same as \eq{Sqq_T}, the only difference being the opposite
time ordering. Thus, $S_\hemiin$ and $S_\hemiout$ are equal at one loop, where
the time ordering is still irrelevant. Beyond one loop, $S_\hemiin$ and
$S_\hemiout$ may in general be different.
However, since the beam and jet functions have the same anomalous dimension,
the combined anomalous dimension of the hard and beam functions in isolated
Drell-Yan agrees with that of the hard and jet functions for the $e^+e^-$
hemisphere invariant-mass distribution. The consistency of the RGE in both cases
then requires that $S_\hemiin$ and $S_\hemiout$ have the same anomalous
dimension to all orders in perturbation theory. In addition, the purely virtual
contributions, obtained by inserting the vacuum state, are the same in both
cases,
\begin{equation}
S_\hemiin^{q\bar q,\mathrm{virtual}}(k_a^+, k_b^+)
= \frac{1}{N_c}\, \delta(k_a^+)\, \delta(k_b^+)\,
\tr\,\Abs{\Mae{0}{ T\bigl[\widetilde{Y}^\dagger_{n_a} \widetilde{Y}_{n_b} \bigr] }{0}}^2
= S_\hemiout^{q\bar q,\mathrm{virtual}}(k_a^+, k_b^+)
\,.\end{equation}
Hence, $S_\hemiin$ and $S_\hemiout$ can only differ by finite real-emission
corrections at each order in perturbation theory.

Using the one-loop results for $S_\hemiout^{q\bar q}$ from
Refs.~\cite{Schwartz:2007ib, Fleming:2007xt}, we have
\begin{align} \label{eq:S_oneloop}
S^{q \bar q}_\hemiin(k_a^+, k_b^+) &= \delta(k_a^+)\,\delta(k_b^+)
+ \delta(k_a^+)\,S^\oneloop(k_b^+) + S^\oneloop(k_a^+)\,\delta(k_b^+)
\,,\nn\\
S^\oneloop(k^+) &= \frac{\alpha_s(\mu)\,C_F}{4\pi} \biggl\{
-\frac{8}{\mu} \biggl[\frac{\theta(k^+ / \mu) \ln(k^+ / \mu)}{k^+ / \mu} \biggr]_+ +
\frac{\pi^2}{6}\, \delta(k^+) \biggr\}
\,.\end{align}
The plus distribution is defined in \eq{plus_def}. The one-loop soft function
for beam thrust in \eq{SB} then becomes $S_B(k^+, \mu) = \delta(k^+) + 2\,
S^\oneloop(k^+)$.

\subsection{Final Cross Section for Beam Thrust}

The differential cross section for beam thrust in \eq{dsigma_tauB} including the RGE running is
\begin{align} \label{eq:dsigma_tauB_final}
& \frac{\df\sigma}{\df q^2\, \df Y\, \df \tau_B}
\\ & \qquad
= \sigma_0 \sum_{ij} H_{ij}(q^2, \mu_H)\, U_H(q^2, \mu_H, \mu_S)
\int\!\df t_a\,\df t_b\,Q\,S_B\Bigl(Q\,\tau_B - \frac{t_a + t_b}{Q}, \mu_S \Bigr)
\nn\\ & \qquad \quad \times
\int\!\df t_a'\, B_i(t_a - t_a', x_a, \mu_B)\, U_B(t_a', \mu_B, \mu_S)
\int\!\df t_b'\, B_j(t_b - t_b', x_b, \mu_B)\, U_B(t_b', \mu_B, \mu_S)
\,.\nn\end{align}
We now consider its fixed-order $\alpha_s$ expansion. To our knowledge
$\df\sigma/\df q^2 \df Y \df \tau_B$ has not been considered in perturbation
theory in full QCD even at one loop. To obtain an expression for $\df\sigma/\df
q^2 \df Y \df \tau_B$ at NLO in $\alpha_s$ and leading order in the power
counting, we drop the evolution factors $U_H$ and $U_B$ and expand all functions
to NLO at a common scale $\mu$. From the above NLO results for the hard and soft
functions and the NLO results for the beam functions from \ch{beamf}, we
find
\begin{align} \label{eq:dsigma_tauB_NLO}
\frac{\df\sigma}{\df q^2\, \df Y\, \df \tau_B}
&= \sigma_0 \sum_{i,j}
\biggl[ Q_i^2 + \frac{(v_i^2 + a_i^2) (v_\ell^2+a_\ell^2) - 2 Q_i v_i v_\ell (1-m_Z^2/q^2)}
{(1-m_Z^2/q^2)^2 + m_Z^2 \Gamma_Z^2/q^4} \biggr]
\nn\\ & \quad \times
\int\! \frac{\df \xi_a}{\xi_a}\, \frac{\df \xi_b}{\xi_b}\,
C_{ij}\Bigl(\frac{x_a}{\xi_a},\frac{x_b}{\xi_b}, q^2, \tau_B, \mu \Bigr)\, f_{i/a}(\xi_a,\mu)\, f_{j/b}(\xi_b,\mu)
\,.\end{align}
Here, $f_{i/a}(\xi_a, \mu)$ and $f_{j/b}(\xi_b, \mu)$ are the PDFs for parton $i$ in proton $a$ and parton $j$ in (anti-)proton $b$. At tree level, the nonzero coefficients are
\begin{equation}
C_{q\bar q}^\tree(z_a, z_b, q^2, \tau_B, \mu) = C_{\bar q q}^\tree(z_a, z_b, q^2, \tau_B, \mu)
= \delta(\tau_B)\,\delta(1 - z_a)\,\delta(1 - z_b)
\,.\end{equation}
At one loop, we obtain
\begin{align}
&C^\oneloop_{q \bar q}(z_a,z_b, q^2, \tau_B, \mu)
\nn \\ & \
= \frac{\alpha_s(\mu)\, C_F}{2\pi}\, \delta(1 - z_a)\,\theta(z_b) \biggl\{
\biggl[-2 \biggl[\frac{\theta(\tau_B) \ln\tau_B}{\tau_B}\biggr]_+\!
- \frac{3}{2} \biggl[\frac{\theta(\tau_B)}{\tau_B}\biggr]_+\!
- \delta(\tau_B) \Bigl(4 - \frac{\pi^2}{2}\Bigr) \biggr] \delta(1 - z_b)
\nn\\ &\quad
+ \biggl[\biggl[\frac{\theta(\tau_B)}{\tau_B}\biggr]_+\! + \delta(\tau_B) \ln\frac{q^2}{\mu^2} \biggr] \biggl[\theta(1-z_b)\frac{1 + z_b^2}{1-z_b}\biggr]_+
  + \delta(\tau_B) \biggl[
   \biggl[\frac{\theta(1-z_b) \ln(1-z_b)}{1-z_b}\biggr]_+\!\!  (1+z_b^2)
\nn\\ &\qquad
   + \theta(1-z_b)\biggl(1-z_b - \frac{1 + z_b^2}{1-z_b} \ln z_b\biggr)
   \biggr] \biggr\}  + (z_a\lra z_b)
\,,\nn\\
&C^\oneloop_{\bar qq}(z_a,z_b, q^2, \tau_B, \mu) = C^\oneloop_{q \bar q}(z_a,z_b, q^2, \tau_B, \mu)
\,,\nn\\[1ex]
&C^\oneloop_{qg}(z_a,z_b, q^2, \tau_B, \mu)
\nn \\ & \
= \frac{\alpha_s(\mu)\,T_F}{2\pi}\, \delta(1 - z_a)\,\theta(z_b)\, \theta(1 - z_b) \biggl\{ \biggl[
  \biggl[\frac{\theta(\tau_B)}{\tau_B}\biggr]_+
  + \delta(\tau_B) \ln\frac{q^2}{\mu^2} \biggr]\bigl[z_b^2 + (1-z_b)^2\bigr]
\nn\\ & \ \quad
  + \delta(\tau_B)\biggl[\ln\frac{1 - z_b}{z_b}\bigl[z_b^2 + (1-z_b)^2\bigr] + 2z_b(1-z_b) \biggr] \biggr\}
\,,\nn\\
&C^\oneloop_{\bar q g}(z_a,z_b, q^2, \tau_B, \mu) = C^\oneloop_{q g}(z_a,z_b, q^2, \tau_B, \mu)
\,,\nn\\
&C^\oneloop_{gq}(z_a, z_b, q^2, \tau_B, \mu) = C^\oneloop_{g \bar q}(z_a,z_b, q^2, \tau_B, \mu)
= C^\oneloop_{q g}(z_b,z_a, q^2, \tau_B, \mu)
\,.\end{align}
The single logarithms of $q^2/\mu^2$ are multiplied by the QCD splitting kernels and are
resummed by the PDFs. Thus, in fixed-order perturbation theory the PDFs should
be evaluated at the hard scale $\mu = Q$, such that there are no large
logarithms when integrating over $0 \leq \tau_B \lesssim 1$. However, if the
integration is restricted to $\tau_B \leq \tau_B^\cut \ll 1$, the plus
distributions in $\tau_B$ produce large logarithms $\ln^2\tau_B^\cut$ and
$\ln\tau_B^\cut$, which make a fixed-order expansion unreliable. These are
precisely the logarithms that are resummed by the
combined RGE of hard, jet, and soft functions in \eq{dsigma_tauB_final}.


\chapter{Beam Thrust Cross Section for Drell-Yan}
\label{ch:BT}

In this chapter we present and discuss results for the Drell-Yan beam thrust cross section, to appear in Ref.~\cite{Stewart:2010BT}. Beam thrust was defined in \eq{tauB} and may be written as
\begin{equation} 
\tau_B = \frac{1}{Q} \sum_k \abs{\vec{p}_{kT}} e^{-|\eta_k-Y|}
\,,\end{equation}
where $Q^2$ and $Y$ are the dilepton invariant mass and rapidity. The sum runs over all (pseudo)particles in the final state except the two signal leptons, where $\abs{\vec{p}_{kT}}$ and $\eta_k$ are the measured transverse momenta and rapidities with respect to the beam axis (for simplicity we took all particles to be massless). Beam thrust is the analog of thrust for $e^+e^-\to \text{jets}$ with the thrust axis fixed to the proton beam axis. For $\tau_B\ll 1$, the hadronic final state consists of two back-to-back jets along the beam axis due to the initial state radiation. The measurement of $\tau_B$ can provide a test of our understanding of the initial state at the Tevatron and LHC.

Experimentally, beam thrust is one of the simplest hadronic observables at a hadron collider. It requires no jet algorithms, is boost invariant along the beam axis, and can be directly compared to theory predictions without utilizing parton showering or hadronization from Monte Carlos. 

Beam thrust is also theoretically clean. It is infrared safe, and simple enough to include in theoretical calculations. In \sec{facttheorem} we derived an all-orders factorization theorem for the Drell-Yan beam thrust cross section for small $\tau_B$
\begin{align} \label{eq:DYbeamrun}
&\frac{\df\sigma}{\df Q \df Y \df \tau_B}
\\* &\
= \frac{8\pi\alem^2}{3N_c\Ecm^2}
\sum_{ij} H_{ij}(Q^2, \mu_H)\, U_H(Q^2, \mu_H, \mu)
\int\!\df t_a \df t_a'\, B_i(t_a - t_a', x_a, \mu_B)\, U^i_B(t_a', \mu_B, \mu)
\nn\\ &\ \quad \times\!\!
\int\!\df t_b \df t_b'\, B_j(t_b - t_b', x_b, \mu_B)\, U^j_B(t_b', \mu_B, \mu)
\int\!\df k\,
S_B\Bigl(\tau_B Q - \frac{t_a + t_b}{Q} - k, \mu_S \Bigr) U_S(k, \mu_S, \mu)
\nn\,,\end{align}
where $x_a = (Q/\Ecm) e^{Y}$ and $x_b = (Q/\Ecm) e^{-Y}$ are the partonic momentum fractions transferred to the leptons, and the sum runs over quark flavors $ij = \{u\bar u, \bar u u, d\bar d, \ldots\}$. The hard function $H_{ij}(Q^2, \mu_H)$ contains hard virtual radiation at the hard scale $\mu_H \simeq Q$ (and also includes the leptonic decay). The soft function $S_B(k, \mu_S)$ encodes the effect of soft virtual and real radiation on the measurement of $\tau_B$ and is sensitive to the soft scale $\mu_S \simeq \tau_B Q$. The beam functions $B_i(t_a, x_a, \mu_B)$ and $B_j(t_b, x_b, \mu_B)$ describe extracting the colliding partons out of the proton and the formation of the initial state jets. As we showed in \sec{OPE}, they can be matched onto PDFs $B_i = \sum_k \cI_{ik} \otimes f_k$ at the beam scale $\mu_B \simeq \sqrt{t_{a,b}} \simeq \sqrt{\tau_B} Q$.

The cross section for small $\tau_B\ll 1$ contains large double logarithms $\alpha_s^n \ln^m\tau_B$ with $m \leq 2n$. These are summed to all orders in \eq{DYbeamrun} by evaluating the hard, beam, and soft functions at their natural scale where they contain no large logarithms and using the RG evolution kernels $U_H$, $U_B^{i,j}$, and $U_S$ to run them to a common scale $\mu$. The combination of the evolution kernels is $\mu$ independent and sums the logarithms of $\tau_B$. At NNLL we need the next-to-leading order (NLO) expressions for $H_{ij}$, $\cI_{ik}$, and $S_B$, as well as the NNLL expressions for $U_H$, $U_B^{i,j}$, and $U_S$, given in \sec{DY_final} and \app{pert}. The required convolutions in \eq{DYbeamrun} are carried out analytically following Ref.~\cite{Ligeti:2008ac}. We use the MSTW2008~\cite{Martin:2009bu} parton distribution functions at NLO for $\alpha_s(m_Z) = 0.117$ and with two-loop, five-flavor running for $\alpha_s(\mu)$. Details on our order counting, as well as a technical point related to the running of $\al_s(\mu)$, can be found at the beginning of \ch{beam_plots}.

To estimate the perturbative uncertainties we use the minimum and maximum variation under the following separate scale variations
\begin{align} \label{eq:scales}
\text{a)}&&
\mu_H &= r Q
\,,\
\mu_B = r \sqrt{\tau_B} Q
\,,\
\mu_S = r \tau_B Q
\,,\nn \\*
\text{b)}&&
\mu_H &= Q
\,,\
\mu_B = r^{-(\ln \tau_B)/4} \sqrt{\tau_B}Q
\,,\
\mu_S = \tau_B Q
\,,\nn \\*
\text{c)}&&
\mu_H &= Q
\,,\
\mu_B = \sqrt{\tau_B} Q
\,,\
\mu_S = r^{-(\ln \tau_B)/4} \tau_B Q
\,,\end{align}
with $r = \{1/2, 2\}$, and $r = 1$ corresponds to the central value. 
Only considering the simultaneous variation in a) produces unnaturally small scale uncertainties due to cancelations between the running of the hard, beam, and soft functions.
The exponent of $r$ for cases b) and c) is chosen such that for $\tau_B
= e^{-4}$ the scales $\mu_B$ or $\mu_S$ vary by factors of two,
with smaller variations for increasing $\tau_B$ and no variation for $\tau_B
\to 1$. In this limit, there should only be a single scale $\mu_H = \mu_B = \mu_S$, and thus
the only scale variation should be case a). For the cross section integrated up to
$\tau_B \leq \tau_B^\max$, the scales are chosen by replacing $\tau_B$ with $\tau_B^\max$
in \eq{scales}.

\begin{figure}[t]{%
\centering
\includegraphics[width=0.7\textwidth]{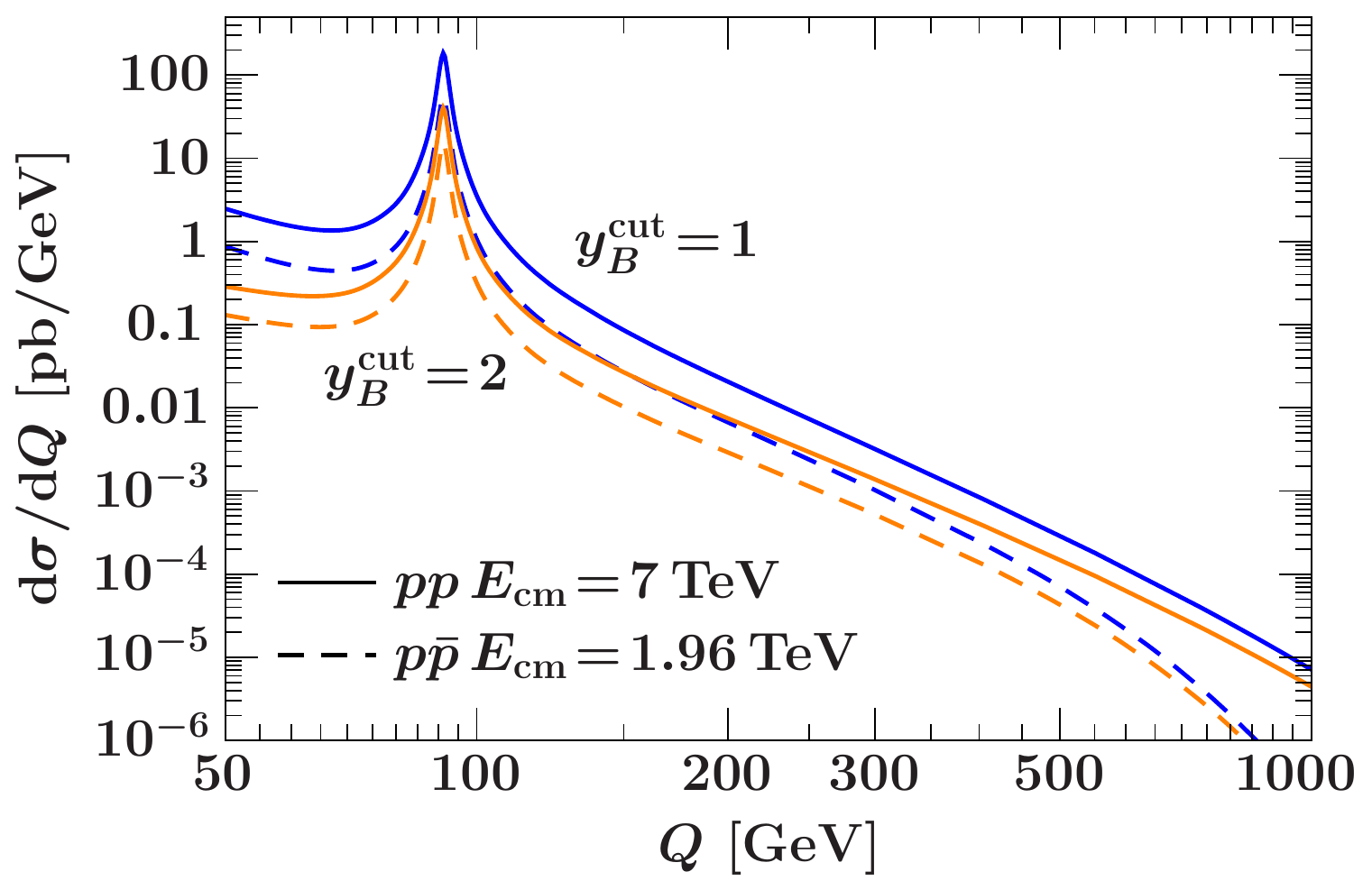}%
\caption[Drell-Yan cross section $\df\sigma/\df Q$ with a cut on beam thrust.]{
  Drell-Yan cross section $\df\sigma/\df Q$ at NNLL with a cut on beam thrust
  $\tau_B \leq \exp(-2y_B^\cut)$ with $y_B^\cut = 1,2$. The solid lines show the cross section
  at the LHC with $\Ecm = 7\TeV$ and the dashed lines the cross section at the Tevatron.}
\label{fig:sigma_q}}
\end{figure}

\begin{figure*}[t]{%
\includegraphics[width=0.49\textwidth]{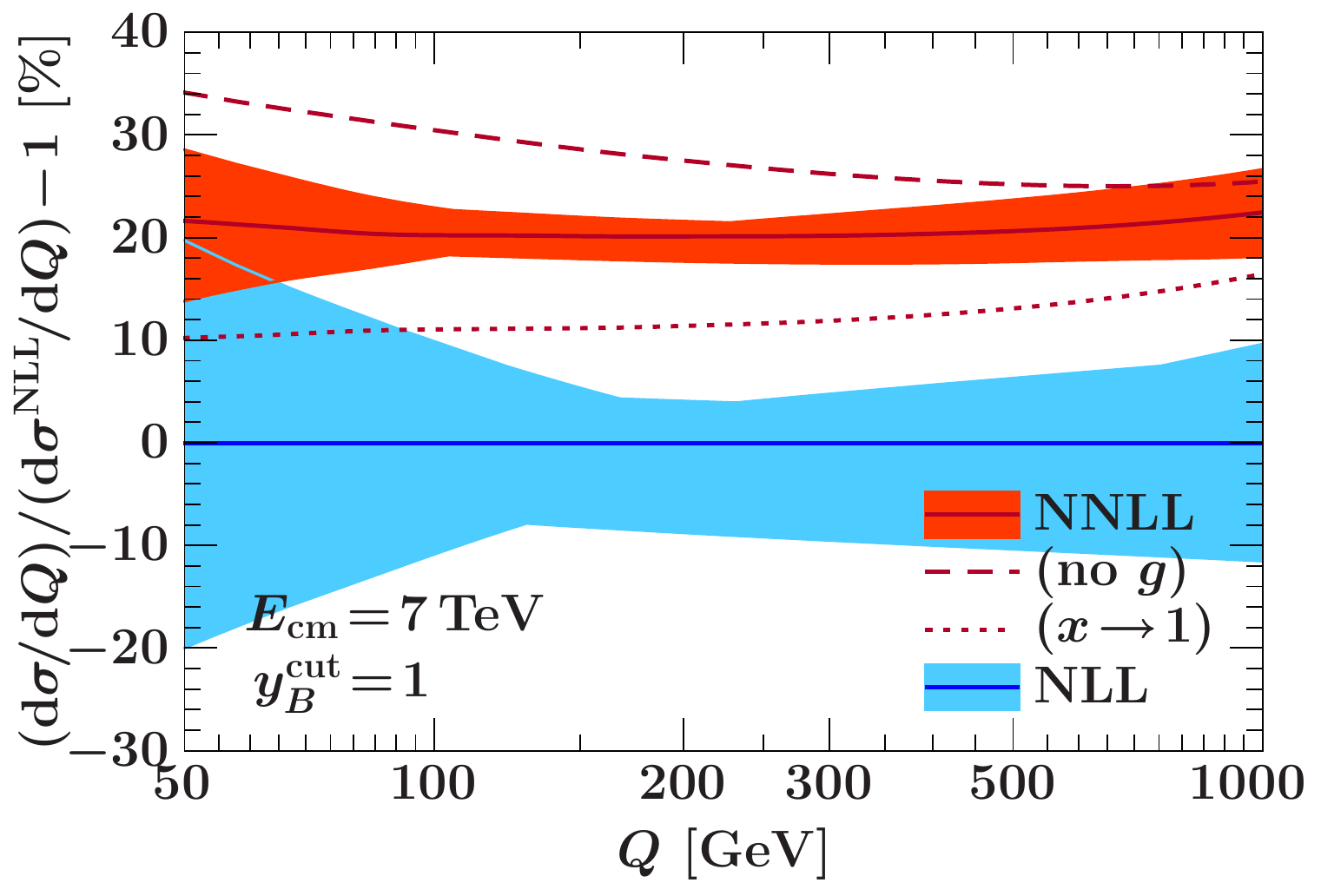}%
\hfill%
\includegraphics[width=0.49\textwidth]{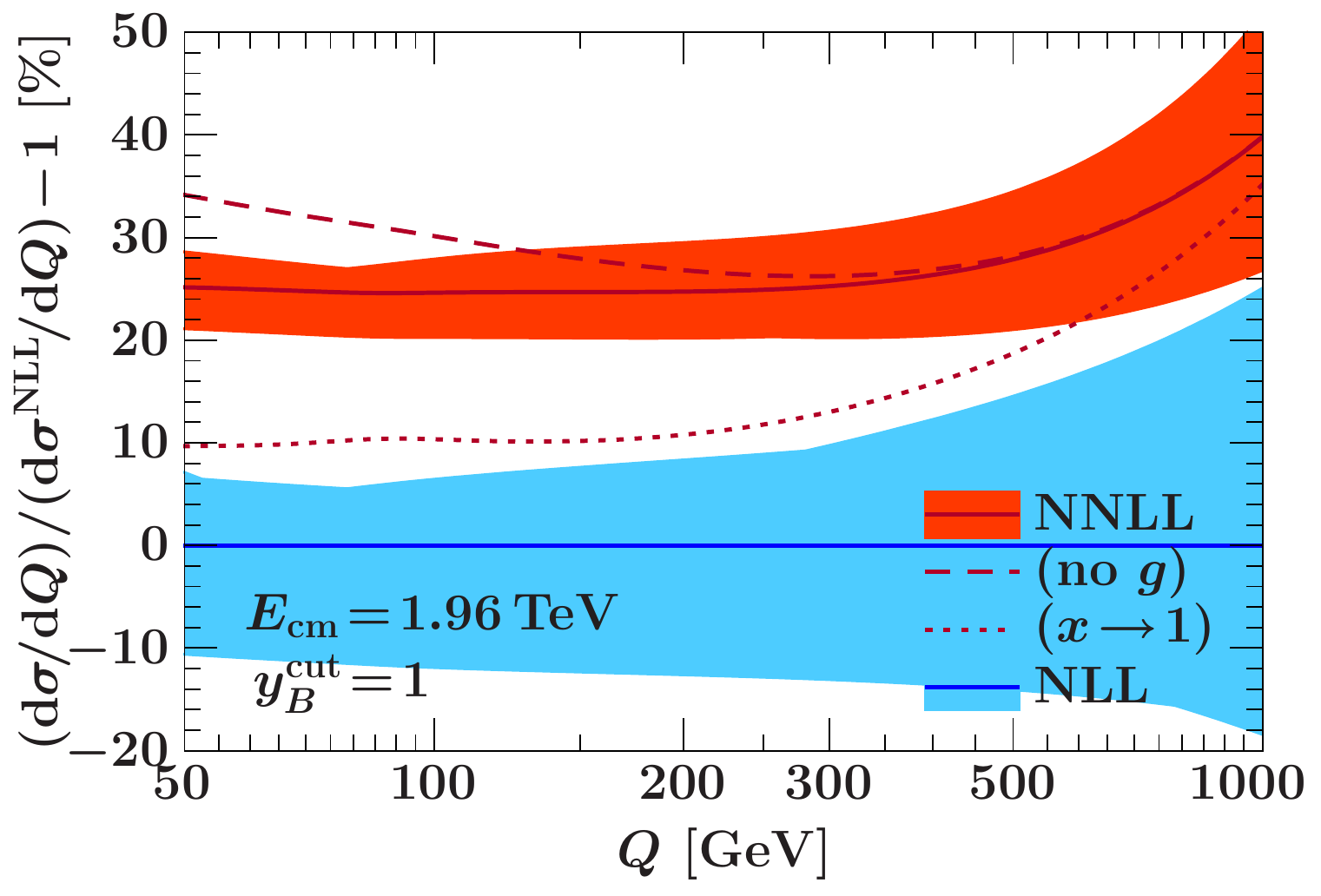}%
\caption[Perturbative corrections and uncertainties for the Drell-Yan cross section with a cut on beam thrust, relative to the NLL result.]{
  The Drell-Yan cross section $\df\sigma/\df Q$ at NNLL with a cut on beam thrust
  $\tau_B \leq \exp(-2y_B^\cut)$ with $y_B^\cut = 1$ at the LHC with $\Ecm =
  7\TeV$ (left panel) and the Tevatron (right panel). All results
  are normalized relative to the NLL result. Shown are the full NNLL result
  (solid), NNLL without the gluon contribution (dashed), and NNLL in the
  threshold limit (dotted). The bands show the scale uncertainties at NLL and NNLL
  as explained in the text.}
\label{fig:sigmarel_q}}
\end{figure*}

As already mentioned, small $\tau_B$ describes events with energetic initial-state
radiation in the forward direction but no hard jets at central rapidities.
More explicitly, a cut on $\tau_B \leq \exp(-2 y_B^\cut)$ vetos energetic radiation with total
energy $\gtrsim Q$ in the rapidity region $\abs{y - Y} \lesssim y_B^\cut$, while allowing
energetic radiation for $\abs{y -Y} \gtrsim y_B^\cut$ with a smooth transition between the two regions.
In \fig{sigma_q} we show the Drell-Yan cross section $\df\sigma/\df Q$ at NNLL
for $y_B^\cut=1$ and $2$ at the LHC with $\Ecm = 7\TeV$ and the Tevatron.
Increasing the restriction on the hadronic final state from $y_B^\cut=1$ to $2$ reduces the
cross section by a factor of few at larger $Q$ which becomes a factor of $10$ at small $Q$.
The $Z$ resonance is visible around $Q = 90\GeV$.

Figure~\ref{fig:sigmarel_q} shows the same cross section for $y_B^\cut = 1$ normalized relative to the
NLL result at the LHC with $\Ecm = 7\TeV$ (left panel) and the Tevatron (right panel).
The bands show the perturbative uncertainties at NLL (medium blue band) and NNLL (dark orange band).
From NLL to NNLL they are reduced by about a factor of two.
As illustration, the dashed line shows the NNLL result without the contribution of the gluon PDF to the quark
beam function, $\cI_{qg}$ in \eq{beam_fact}. The dotted line shows the NNLL result in the threshold limit
$x\to 1$, where in addition to neglecting the gluon contribution we only keep the leading
terms as $x\to 1$ in the quark contribution $\cI_{qq}$, shown in \eq{Ithres}. Hence, the quark non-threshold contributions shift
the result from the dotted to the dashed line, and the (non-threshold) gluon contributions shift it from
the dashed to the solid line. At the LHC, there is a noticable cancelation
between the non-threshold quark and gluon contributions, which is less prominent at the Tevatron. The
reason is that the gluon PDF has a much larger effect on the antiquark beam function than the quark
beam function. The non-threshold quark contributions are as large or larger
than the threshold quark contributions, so there is no indication that the
threshold terms are numerically dominant. The only exception is for the Tevatron, where because
of the lower center-of-mass energy the threshold corrections dominate for large $Q\sim 1\TeV$,
corresponding to $x\sim 0.5$. In this region the absolute cross section in \fig{sigma_q} however
is very small.

\begin{figure*}[t]{%
\includegraphics[width=0.50\textwidth]{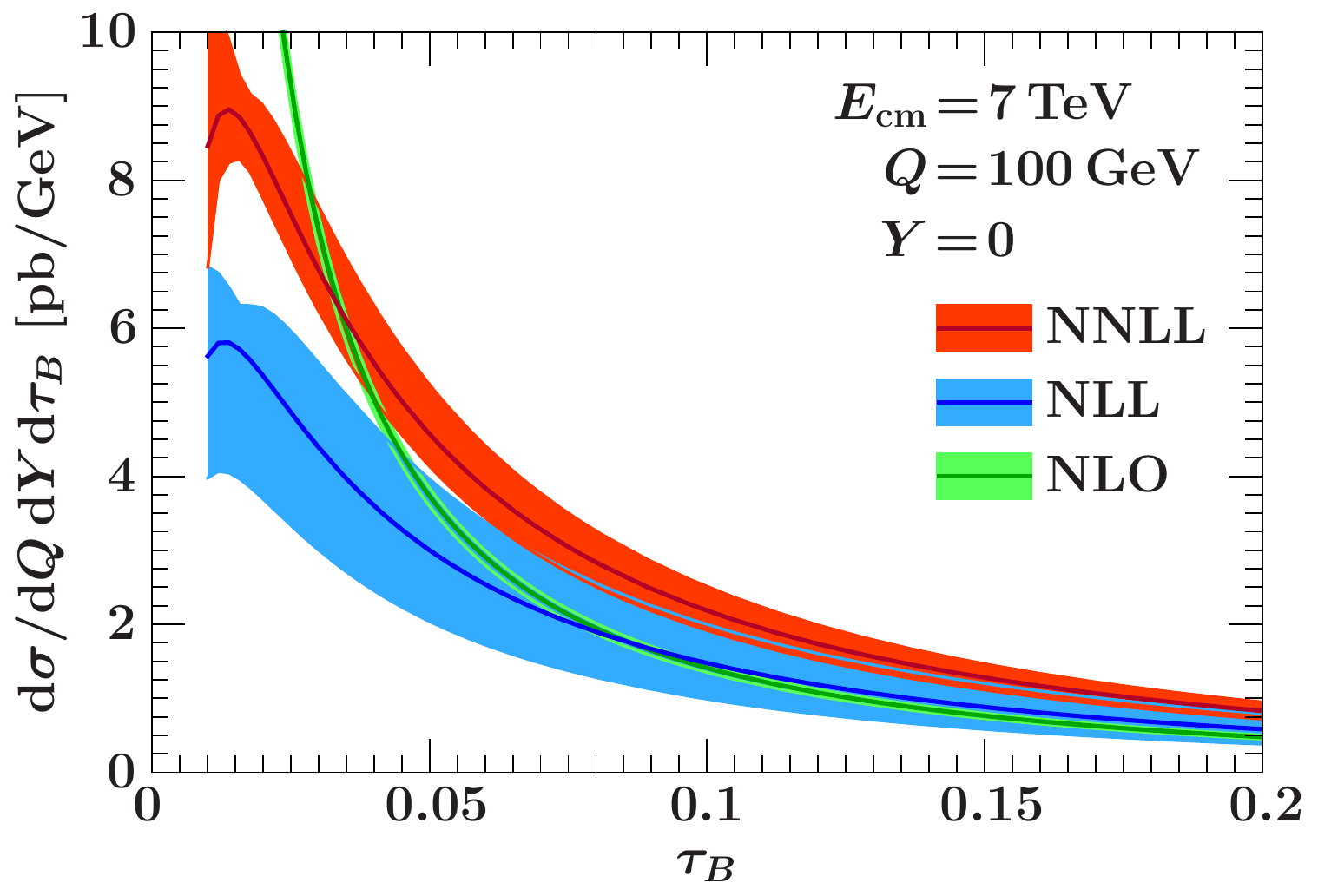}%
\hfill%
\includegraphics[width=0.49\textwidth]{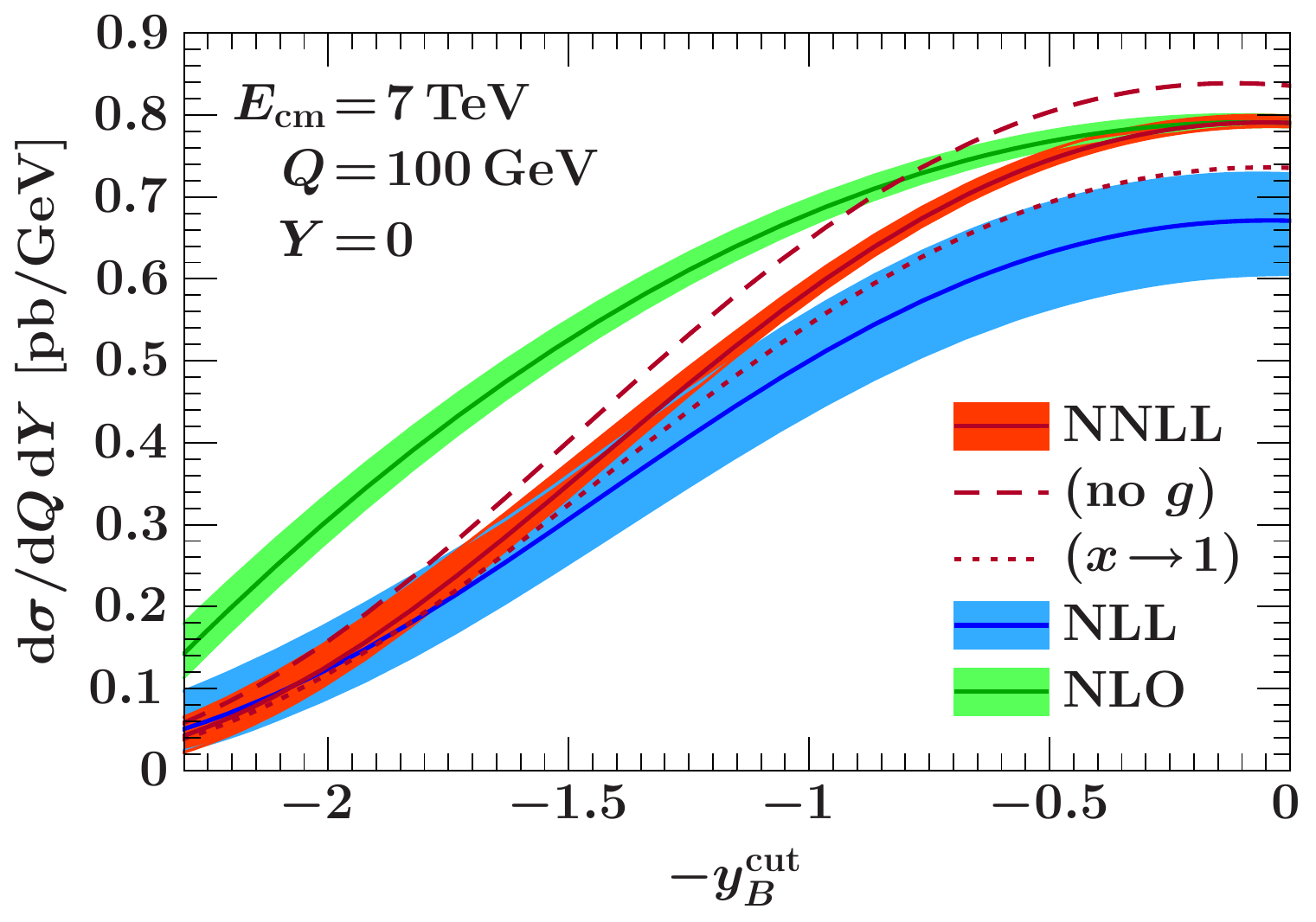}%
\caption[The Drell-Yan cross section as function of beam thrust.]{
  The beam thrust cross section for fixed $Q=100\GeV$ and $Y = 0$ at the LHC
  with $\Ecm = 7\TeV$ at NLO, NLL, and NNLL.
  The bands indicate the perturbative scale uncertainties as explained in the text.
  Left: The differential cross section at small $\tau_B$.
  Right: The cross section  integrated up to $\tau_B \leq \exp(-2y_B^\cut)$
  as function of $y_B^\cut$. Also shown are the NNLL results in the threshold
  limit $x\to 1$ (dotted) and without the gluon contribution (dashed).}
\label{fig:sigtauB}}
\end{figure*}

The left panel of \fig{sigtauB} shows the beam thrust cross section
$\df\sigma/\df Q\df Y\df \tau_B$ at the LHC for small $\tau_B$ and
at fixed $Y=0$ and $Q = 100\GeV$. The bands show the perturbative uncertainties
and results are shown at NNLL (dark orange), NLL (medium blue), and fixed order NLO (light green).
The NLO result is obtained from \eq{DYbeamrun} by using a common fixed scale $\mu_H = \mu_B = \mu_S =
\mu$, with central value $\mu = Q$ and varying $\mu = \{2Q, Q/2\}$. This reproduces the
leading singular terms for $\tau_B\to 0$ shown in \eq{dsigma_tauB_NLO}. The full NLO result, which is not yet known, contains additional nonsingular
terms that are relevant for $\tau_B \sim 1$, but for $\tau_B\ll 1$ constitute power corrections and
are neglected in \eq{DYbeamrun}. The summation has a large effect at small
$\tau_B$ and effectively regularizes the IR singularity in the fixed-order result. The
curves are not plotted for $\tau_B\leq 0.01$, because at this point the soft scale drops
below $1\GeV$. Near this cutoff, the soft function becomes nonperturbative and so we expect
large corrections to our purely perturbative results.

In the right panel of \fig{sigtauB} we plot the corresponding cross section integrated up to $\tau_B
\leq \exp(-2 y_B^\cut)$ as function of $y_B^\cut$. For nonzero $y_B^\cut$ (small $\tau_B^\max$)
the summation is again important, while for $y_B^\cut \to 0$ ($\tau_B^\max\to 1$)
no summation is necessary. The full NNLL result provides a smooth
interpolation between these two regimes, approaching the singular NLO result for
$y_B^\cut \to 0$ and the NLL result for $y_B^\cut \gtrsim 1.5$.

In conclusion, beam thrust in Drell-Yan provides an experimentally and theoretically
clean measure of initial-state radiation in $q\bar{q} \to e^+e^-$, similar to how usual thrust measures final-state radiation in $e^+e^-\to q\bar{q}$. The experimental measurement of beam thrust will contribute very valuable information to our understanding of initial-state radiation at hadron colliders. It should be a priority in the early LHC runs when backgrounds from pile-up effects are still small. We have presented theoretical results for the Drell-Yan beam thrust cross section at NNLL, showing that the effects of initial-state radiation in quark initiated processes are important but also under good theoretical control. We will see an example of a gluon initiated process in the next chapter.


\chapter{Beam Thrust Cross Section for Higgs Production}
\label{ch:higgs}

In this chapter we apply the factorization theorem of \ch{fact} to Higgs production through gluon fusion. In \sec{higgs_intro} we discuss the experimental relevance of a central jet veto for Higgs searches in the $H\to W^+W^-\to \ell^+\nu \ell^- \bar\nu$ and $H\to \ga\ga$ channel and present the corresponding factorization theorems. Plots for the cross section are shown and discussed in \sec{higgs_plots}. These results will appear in Ref. \cite{Berger:2010BgNLO}.

\section{Introduction}
\label{sec:higgs_intro}

The discovery of the Higgs boson is a major goal of the Large Hadron Collider
(LHC) and current analyses at the Tevatron. The $H\to W^+W^-\to \ell^+\nu
\ell^- \bar\nu$ channel has strong discovery potential and plays a very important
role for early searches that are statistics limited.  It is the dominant channel
in the current Tevatron exclusion limit~\cite{Aaltonen:2010yv}.
The presence of the final-state neutrinos does not allow the reconstruction of the
Higgs invariant mass, and hence for this channel sideband methods cannot be used to determine the backgrounds
directly from data. At the LHC, $t\bar t \to W^+W^-b\bar b$ events constitute by far the largest
background, dominating the signal by a factor of more than $30:1$ after
applying lepton selection criteria~\cite{Aad:2009wy, CMSnoteWW}
(see Ref.~\cite{Berger:2010nc} for a sensitivity update at $7\TeV$). Requiring
a minimum missing energy is not effective against this background since it
also contains two neutrinos. To eliminate the huge background from top-quark decays one
imposes a stringent jet veto to define a $0$-jet sample for the search. For example,
ATLAS rejects all events that contain any hard  jet with $p_T> 20\,{\rm GeV}$ and
pseudorapidity $\abs{\eta}<4.8$, which reduces the $t\bar t$
background by a factor of more than $300$~\cite{Aad:2009wy}. After the jet veto, the main
background is from the direct production $pp\to W^+W^-$, which at this point
still dominates the signal by a factor of $4:1$, and the final background
discrimination is achieved by a fit to several kinematic variables.

Theoretically, the inclusive Higgs production cross section has been studied
extensively in the literature and is known to NNLO~\cite{Dawson:1990zj, Djouadi:1991tka, Spira:1995rr, Harlander:2002wh, Anastasiou:2002yz, Ravindran:2003um, Pak:2009dg, Harlander:2009my} 
(for reviews and additional references see e.g.\ Refs.~\cite{Djouadi:2005gi, Boughezal:2009fw}).
However, Higgs production in a
$0$-jet sample differs substantially from inclusive Higgs production. In particular, the
jet veto induces large double logarithms that are not present in the inclusive cross
section, and also induces dependence on the choice of jet algorithm used to
define the veto. The current method employed by experiments to study the effect of the jet veto
and the accompanying large logarithms is to use parton-shower Monte Carlos, such as
MC@NLO~\cite{Frixione:2002ik, Frixione:2003ei} and Pythia~\cite{Sjostrand:2007gs}.
This allows one to take into account
the dependence of the $0$-jet sample on the choice of jet algorithm, but
for the large logarithms it limits the accuracy to the leading-logarithmic
summation inherent to parton showers.

Theoretical studies of the jet veto at fixed NNLO~\cite{Catani:2001cr, Anastasiou:2004xq, Davatz:2006ut}
and including additional kinematic selection cuts~\cite{Anastasiou:2007mz, Anastasiou:2008ik, Grazzini:2008tf, Anastasiou:2009bt}
are available (see also Ref.~\cite{Berger:2010nc}).
The comparison~\cite{Davatz:2006ut, Anastasiou:2008ik, Anastasiou:2009bt} of the fixed NLO results with
MC@NLO, which combines NLO with a LL parton-shower resummation, shows indeed discrepancies that indicate the importance
of resumming the phase-space logarithms caused by the jet veto.
The disagreements are reduced but not eliminated when the fixed NNLO prediction is compared
with those from MC@NLO, Herwig~\cite{Corcella:2000bw, Corcella:2002jc}, and Pythia (reweighted to the same inclusive
NNLO cross section). There are also notable differences between the Herwig and Pythia parton-level results, which is another
indication that subleading phase-space logarithms are important.

Theoretically, one can also study the Higgs production as a
function of the Higgs transverse momentum, $p_T^H$, both in fixed-order perturbation
theory for large $p_T^H$~\cite{deFlorian:1999zd, Ravindran:2002dc, Glosser:2002gm, Anastasiou:2005qj} and with a
resummation of logarithms of $p_T^H$ at small $p_T^H$~\cite{Collins:1984kg, Balazs:2000wv, Berger:2002ut, Bozzi:2003jy,
Kulesza:2003wn, Bozzi:2005wk, Mantry:2009qz}. The latter is motivated by the fact that the jet veto
automatically forces $p_T^H$ to be small, see e.g.\ Refs.~\cite{Davatz:2004zg, Anastasiou:2008ik, Anastasiou:2009bt}.
However, a restriction to small $p_T^H$ by itself does not provide a jet veto as
it still allows for configurations with back-to-back hard final-state jets, so a study of the small-$p_T^H$ spectrum can only
provide a qualitative template for the effect of the jet veto.

Here, we use beam thrust to impose the jet veto in $pp\to H X$ and $p\bar p\to HX$. This allows us to directly predict a $0$-jet Higgs production cross section, including higher-order logarithmic resummation and without relying on parton showers or hadronization models from Monte Carlo.
The corresponding cross section for the $pp\to WW + 0j$ background can be computed in
a similar manner, which is left for future work.

While $H\to WW$ provides the most obvious motivation for studying
the effect of jet vetos, we will also consider the case of $H\to \gamma\gamma$.
Here, sideband methods are available to experimentally control the overwhelming
QCD background (which is up to six orders of magnitude larger than the signal at the
LHC), but it may still be interesting to study sidebands in combination
with an explicit jet veto using an inclusive variable such as beam thrust.
By suppressing the backgrounds from $pp\to jj$ and $pp\to j\gamma$ this
could reduce the sensitivity of the measurement to the details of the photon isolation cuts.
For our purposes, we use this channel to study the effect
of the jet veto for the case where the Higgs invariant mass distribution
and rapidity can be measured.  With the ability to measure these
variables the optimal beam thrust variable for $H\to \gamma\gamma$ differs
slightly from that for $H\to WW$, as we discuss below.

It should be emphasized that Tevatron Higgs
searches, which exclude the range $m_H= 162-166\GeV$ at $95\%$ confidence
level~\cite{Aaltonen:2010yv, Aaltonen:2010cm, Abazov:2010ct},
already analyse their data using a jet algorithm and Monte Carlo to implement a jet veto
on jets with $p_T \geq 15 \GeV$ and $\abs{\eta_j} \leq 2.4-2.5$.
The motivation is again to eliminate backgrounds with additional jets.
The CDF analysis~\cite{Aaltonen:2010cm} explicitly uses separate $0$-jet, $1$-jet, and $\geq 2$-jet signal samples,
where the latter are included to maximize the discovery reach due to the presence of $ZH$, $WH$,
and vector boson fusion production channels. The D0 search previously used a $(0+1)$-jet sample,
where in the latest update~\cite{Abazov:2010ct} the number of jets is used as input to a neural network.
Even though the different jet samples are combined in the
final exclusion limit, they are likely to have different efficiencies and systematics, with
most of the sensitivity coming from the $0$-jet sample. Hence,
it is important to have a theoretical handle on the $0$-jet production
cross section also at the Tevatron. Furthermore, it will be important to use the
available Tevatron data to test our method in preparation for the LHC searches.

At the LHC, the Higgs plus $2$-jet channel with $H\to WW$ is also considered.
In this thesis we restrict ourselves to studying an
inclusive $0$-jet veto using beam thrust. An analogous inclusive variable, $N$-jettiness, that
works in the presence of $N$-signal jets will be discussed in \ch{njet} and can be used to study the exclusive $H+1j$ and $H+2j$ cross section.

To implement a jet veto in $H\to WW$ where missing energy plays an important
role, the appropriate version of beam thrust is defined in the hadronic center-of-mass
frame
\begin{equation}
\tau_B^\cm = \frac{\Tau_B^\cm}{m_H}
\,,\qquad
\Tau_B^\cm = \sum_k \abs{\vec p_{kT}}\, e^{-|\eta_k|}
\,.\end{equation}
For the case studied here where the mass of the Higgs is unknown, the dimension-one variable
$\Tau_B^\cm$ is more convenient than the dimensionless $\tau_B^\cm$.
The sum over $k$ runs over all particles in the final state, with $\vec{p}_{kT}$ and $\eta_k$
the measured transverse momentum and rapidity of each particle, excluding the signal leptons
from the $W$ decays. For simplicity we take all particles to be massless.

The production cross section from gluon fusion, $gg\to H$, for
$\Tau_B \ll m_H$ (i.e. $\tau_B^\cm \ll 1$) is given by the factorization
theorem [see \eq{fact_bhat}]
\begin{align}\label{eq:higgsprodW}
\frac{\df\sigma}{\df \Tau_B^\cm}
&= \sigma_0\, H_{gg}(m_H, m_t, \mu)  \int\!\df Y \int\!\df t_a\, \df t_b\,
B_i(t_a, x_a, \mu)\, B_j(t_b, x_b, \mu)
\nn\\ &\quad \times
S_B^{gg}\Bigl(\Tau_B^\cm - \frac{e^{-Y} t_a + e^Y t_b}{m_H}, \mu\Bigr)
\biggl[1 + \ORD{\frac{\lqcd}{m_H},\tau_B^\cm} \biggr]
\,,
\end{align}
where
\begin{equation}
x_a = \frac{m_H}{\Ecm}\,e^{Y}
\,,\qquad
x_b = \frac{m_H}{\Ecm}\,e^{-Y}
\,,\qquad
\sigma_0 = \frac{\sqrt{2} G_F\, m_H^2}{72 \pi (N_c^2-1) \Ecm^2}
\,,\end{equation}
and the limits on the $Y$ integration are $\ln(m_H/\Ecm) \leq Y \leq - \ln(m_H/\Ecm)$.

For $H\to \gamma\gamma$ the rapidity $Y$ of the Higgs is measurable and provides
an estimate of the boost of the partonic hard collision relative to the hadronic
center-of-mass frame. In this case it makes sense to account for the
boost of the hard collision when imposing the jet veto. Therefore we use the usual beam thrust
\begin{align}
\tau_B = \frac{\Tau_B}{m_H}
\,,\qquad
\Tau_B = \sum_k \abs{\vec p_{kT}} \, e^{-|\eta_k-Y|}
\,.\end{align}
In this case, the factorization theorem for the production cross section is
\begin{align} \label{eq:higgsprod}
\frac{\df\sigma}{ \df Y\, \df \Tau_B}
& = \sigma_0\, H_{gg}(m_t, m_H, \mu)
\int\!\df t_a\,\df t_b\, B_g(t_a, x_a, \mu)\, B_g(t_b, x_b, \mu)
\nn\\ &\quad \times
S_B^{gg}\Bigl(\Tau_B - \frac{t_a + t_b}{m_H}, \mu \Bigr)
 \biggl[1 + \ord{\frac{\lqcd}{m_H},\tau_B} \biggr]
\,.\end{align}

We will focus on the Higgs production cross section. The leptonic decay of the Higgs, which
is of course important in practical applications, does not alter the structure of the factorization theorem.
It can be included straightforwardly as was done in Ref.~\cite{Stewart:2009yx} for the simpler case of
$pp\to Z/\gamma\to \ell^+\ell^-$. In the narrow width approximation one can simply multiply the production cross section
by the relevant differential branching fraction.

\section{Plots}
\label{sec:higgs_plots}

We will now show results for the 0-jet Higgs production cross section from gluon fusion. We restrict ourselves to the cross section for $\Tau_B$ in \eq{higgsprod}, because there is no distinction between $\Tau_B$ and $\Tau_B^\cm$ at NLL and the difference at NNLL is very small for the cross section integrated over $Y$ (the distinction is important for large $Y$ but this only gives a small contribution to the total cross section). 
We evaluate the hard, beam and soft functions at their natural scale, where they contain no large logarithms, and run them to a common scale $\mu$ using their RGE, which effectively sums the large logarithms. All the necessary perturbative results are collected in \app{pert}, and we evaluate the convolutions of plus distributions using the identities from App.~B of Ref.~\cite{Ligeti:2008ac}. We use the MSTW2008~\cite{Martin:2009bu} parton distributions at NLO for $\alpha_s(m_Z) = 0.12$ and with two-loop, five-flavor running for $\alpha_s(\mu)$. 

The perturbative uncertainty bands in the plots are estimated by taking the envelope of the following three separate scale variations 
\begin{align}\label{eq:scales2}
\text{a)}&&
\mu_H &= r m_H
\,,\
\mu_B = r \sqrt{\Tau_B m_H}
\,,\
\mu_S = r \Tau_B
\,,\nn \\*
\text{b)}&&
\mu_H &= m_H
\,,\
\mu_B = r^{-(\ln \Tau_B/m_H)/4} \sqrt{\Tau_B m_H}
\,,\
\mu_S = \Tau_B
\,,\nn \\*
\text{c)}&&
\mu_H &= m_H
\,,\
\mu_B = \sqrt{\Tau_B m_H}
\,,\
\mu_S = r^{-(\ln \Tau_B/m_H)/4}\, \Tau_B
\,,\end{align}
with $r = \{1/2, 2\}$, and $r = 1$ corresponds to the central value. This procedure for estimating the uncertainties is identical to Drell-Yan and is discussed further around \eq{scales}.
For the cross section integrated up to $\Tau_B \leq \Tau_B^\cut$, the scales are chosen by replacing $\Tau_B$ with $\Tau_B^\cut$ in \eq{scales2}.

\begin{figure}[t]{%
\centering
\includegraphics[width=0.49\textwidth]{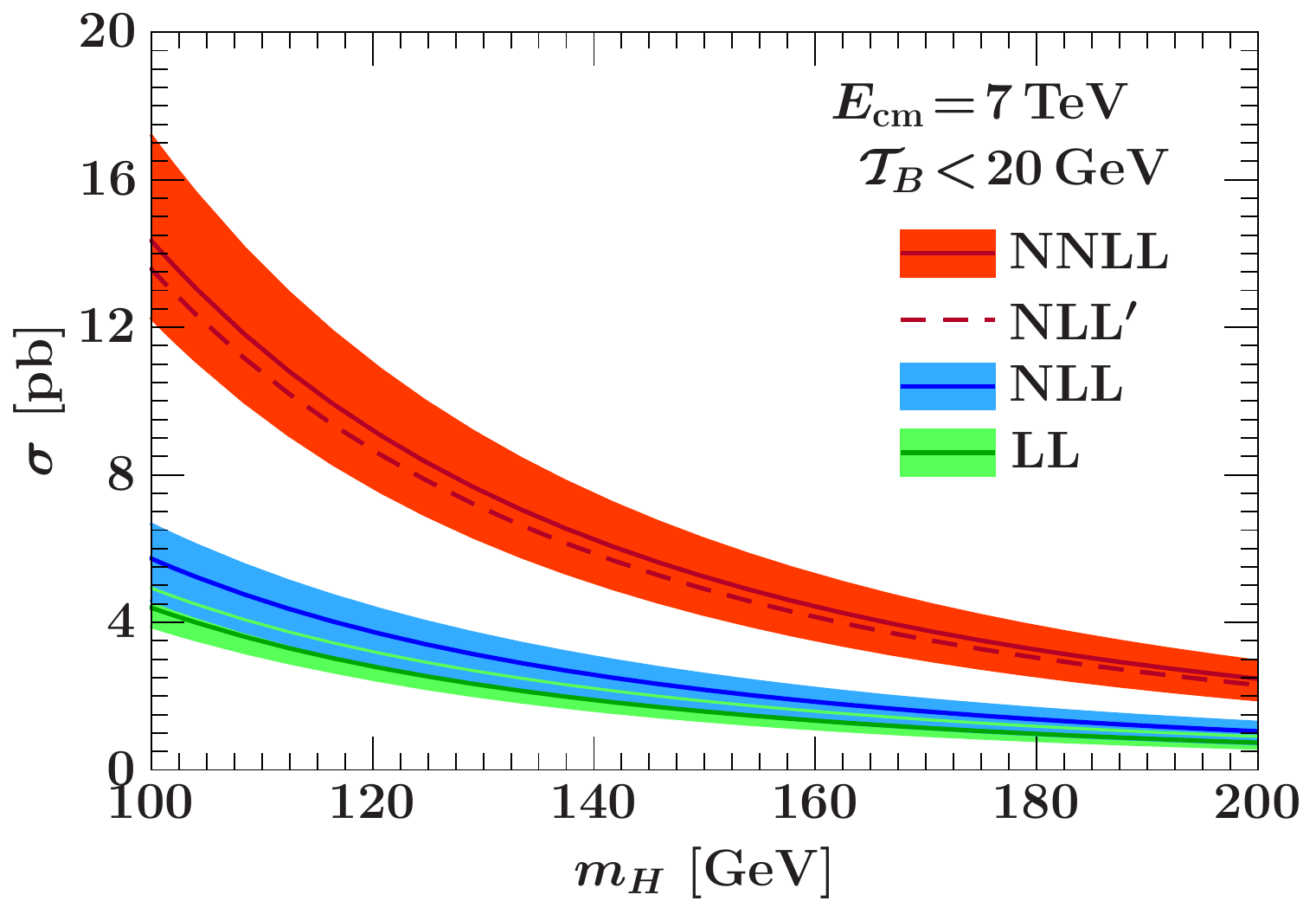}%
\includegraphics[width=0.5\textwidth]{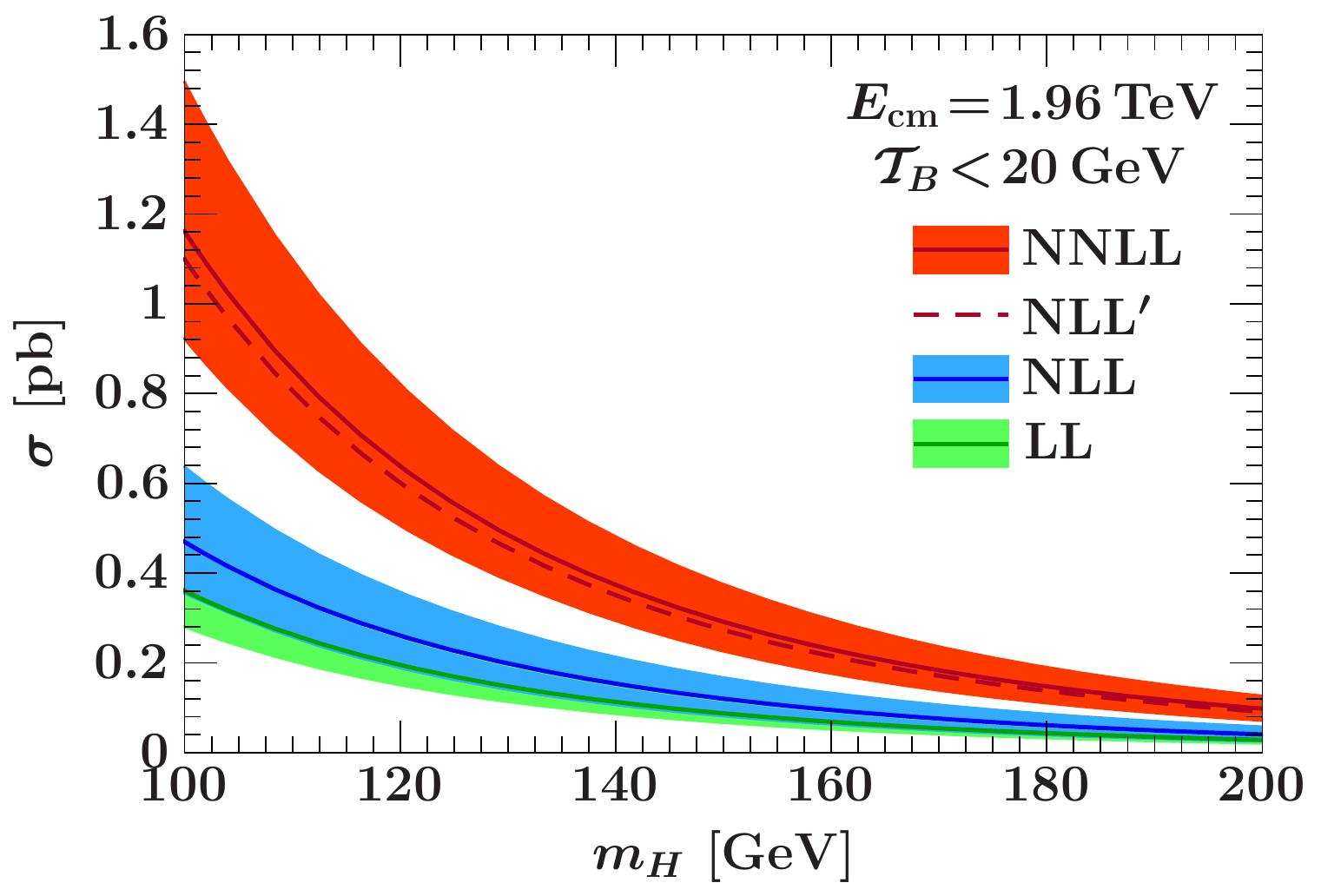}%
\caption[Higgs production cross section from gluon fusion with a cut on beam thrust.]{
The Higgs production cross section from gluon fusion at the LHC with $\Ecm = 7\TeV$ (left panel) and the Tevatron (right panel) with a cut on beam thrust $\Tau_B < 20 \GeV$. Shown are the LL, NLL and NNLL results with perturbative uncertainties as well as the NLL$'$ results, which combine NLO fixed order corrections with NLL resummation.}
\label{fig:sigma_mh}}
\end{figure}

\begin{figure*}[t]{%
\includegraphics[width=0.49\textwidth]{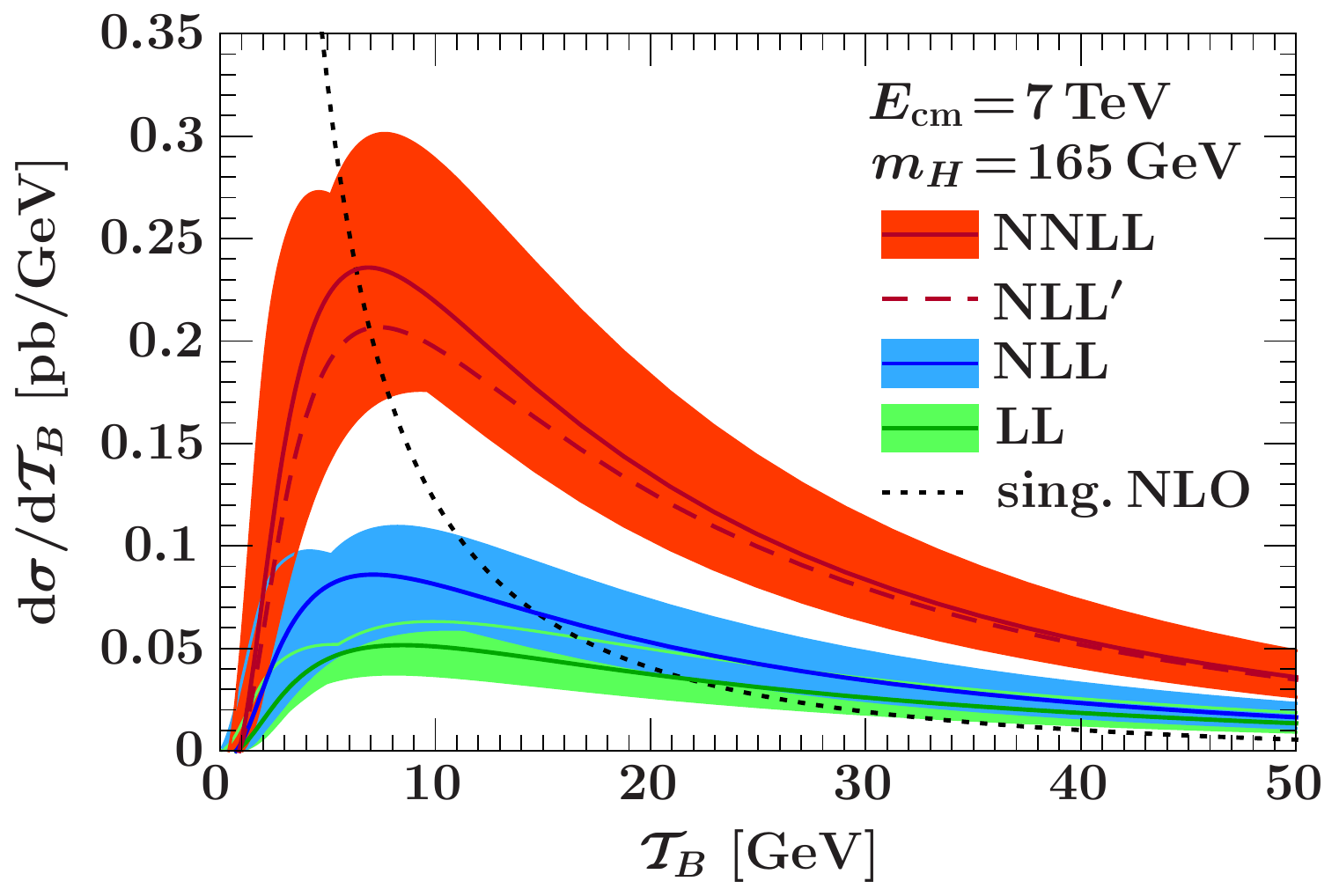}%
\hfill%
\includegraphics[width=0.49\textwidth]{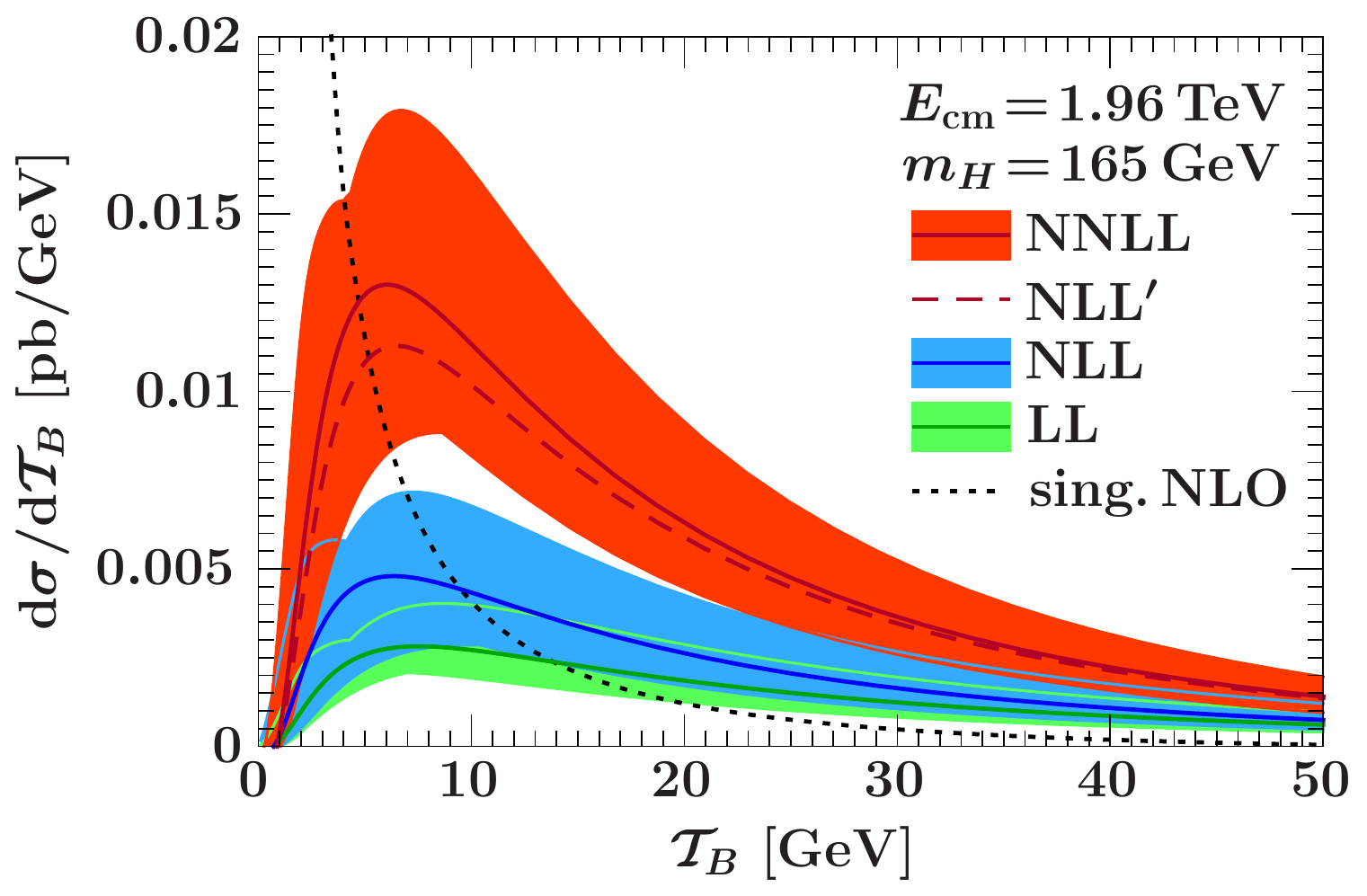}%
\\
\phantom{bla}
\includegraphics[width=0.46\textwidth]{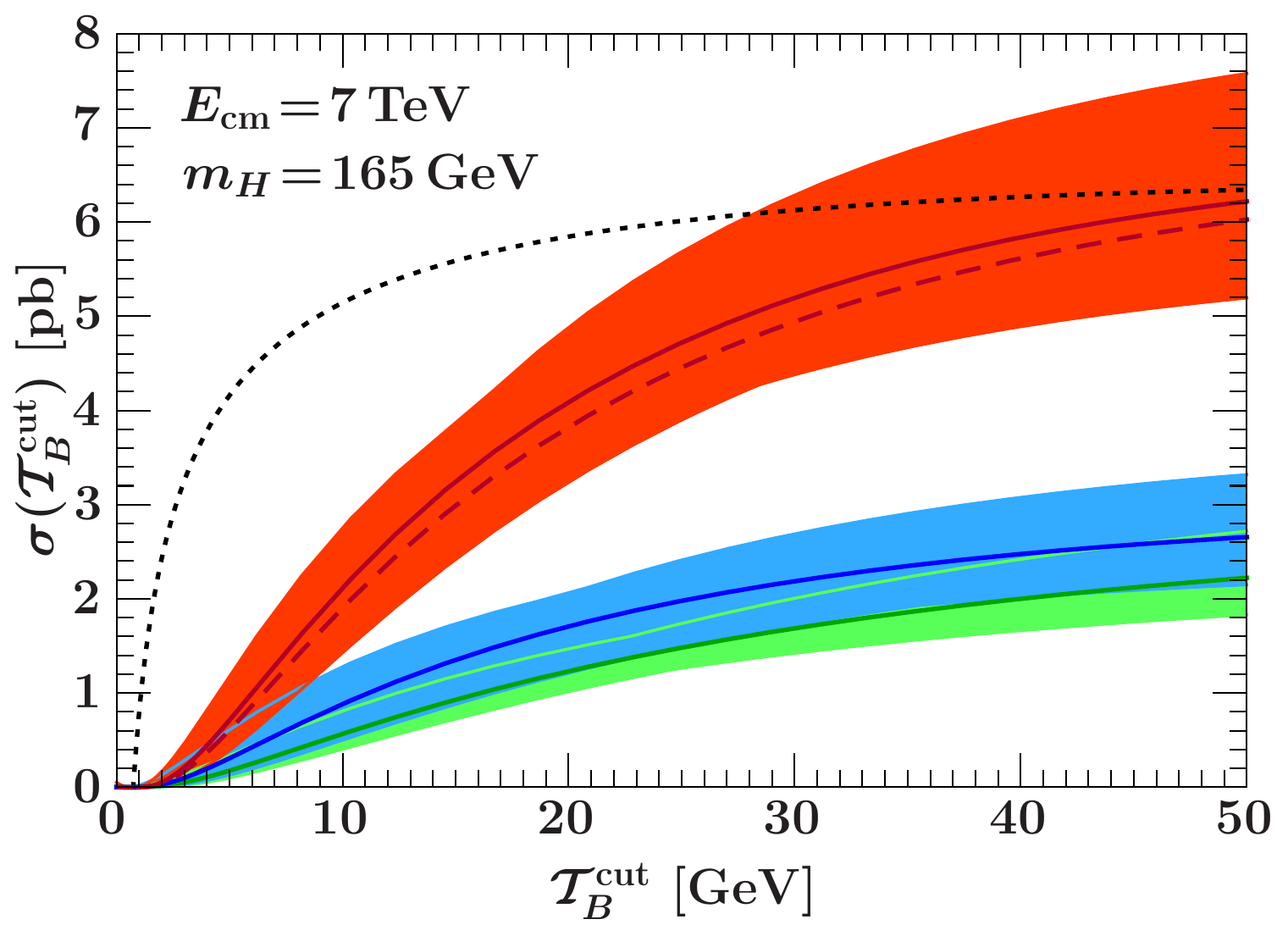}%
\phantom{a}
\includegraphics[width=0.48\textwidth]{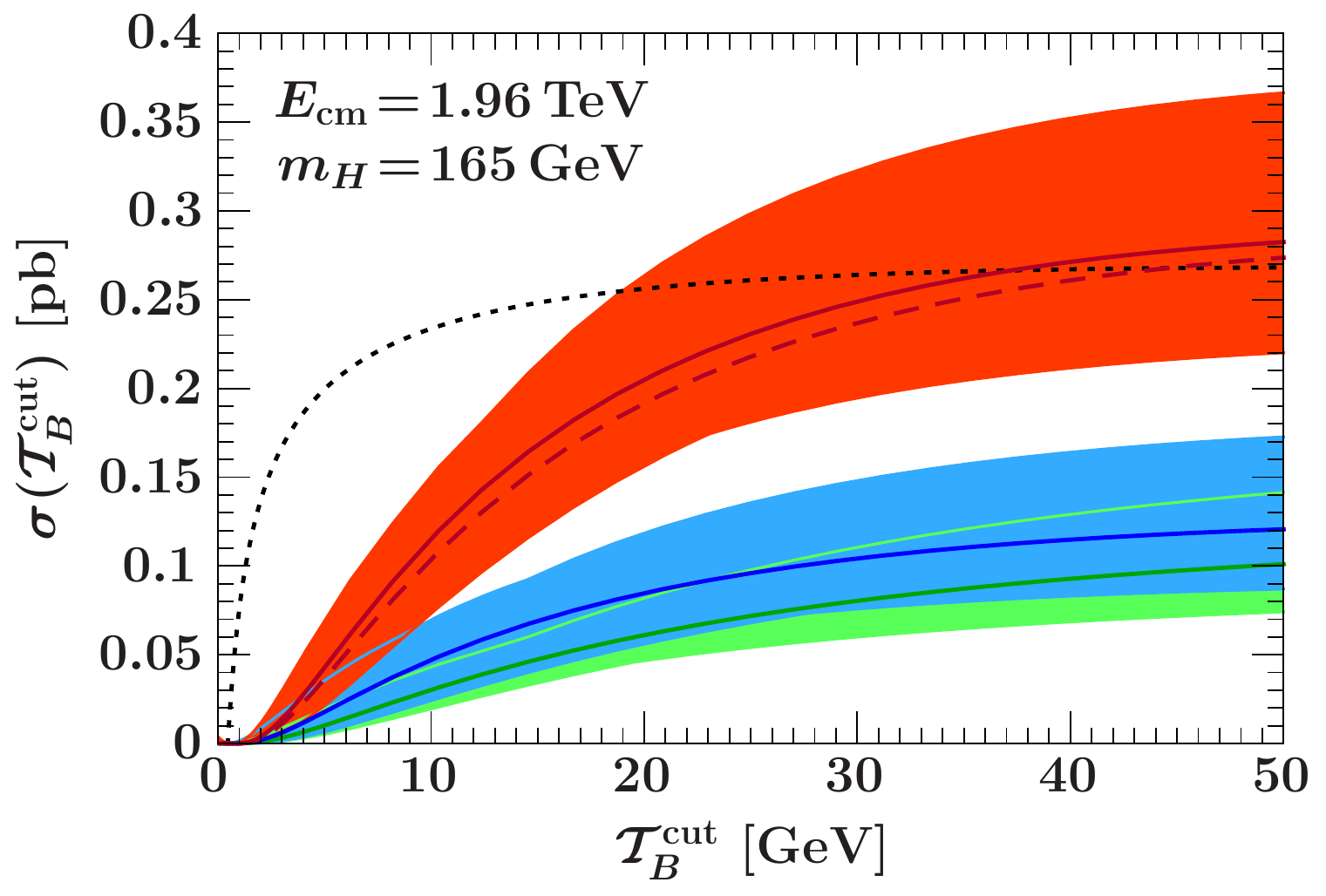}%
\caption[Higgs production cross section from gluon fusion as function of beam thrust.]{
The Higgs production cross section though gluon fusion for $m_H=165 \GeV$ at the LHC with $\Ecm=7\TeV$ (left column) and the Tevatron (right column). Top row: The cross section differential in $\Tau_B$. Bottom row: The cross section integrated up to $\Tau_B<\Tau_B^\cut$ as function of $\Tau_B^\cut$. Shown are the LL, NLL and NNLL results with perturbative uncertainties and the NLL$'$ and singular NLO result.}
\label{fig:sigma_mh_taub}}
\end{figure*}

In \fig{sigma_mh} we show the Higgs production cross section from gluon fusion at the LHC with $\Ecm = 7\TeV$ (left panel) and the Tevatron (right panel), where we imposed the jet veto by requiring $\Tau_B < 20 \GeV$. The large shift from NLL to NNLL is due to the large NLO corrections rather than NNLL resummation. This is clear from the NLL$'$ result (dashed), which combines the fixed order NLO corrections with NLL resummation and accounts for almost the entire shift. These large perturbative corrections are well-known from fixed order calculations. Although it is not very visible, the relative uncertainties decrease slightly when going from NLL to NNLL. The larger relative uncertainties compared to Drell-Yan are not surprising since $C_A/C_F = 9/4$.

Figure~\ref{fig:sigma_mh_taub} shows the beam thrust dependence of the production cross section for a Higgs with mass $m_H = 165 \GeV$ at the LHC with $\Ecm = 7\TeV$ (left column) and the Tevatron (right column). In these plots we include the singular NLO result, which corresponds to evaluating the factorization theorem in \eq{higgsprod} at $\mu = Q$ without any running, i.e.~without any resummation. The top row of \fig{sigma_mh_taub} shows the cross section as function of $\Tau_B$. We observe that the resummation of logarithms regularizes the IR singularity for $\Tau_B \to 0$ in the fixed-order NLO result, as was the case for Drell-Yan. Compared to the Drell-Yan result in the left panel of \fig{sigtauB}, the peak occurs at a larger value  of $\Tau_B$ and the tail is larger as well. This implies that there is more initial state radiation, gluons radiate more than quarks. We did not plot the curves for $\Tau_B\leq 1 \GeV$, because $\mu_S = \Tau_B$ and we therefore expect large corrections to our perturbative soft function in that region.

The bottom row in \fig{sigma_mh_taub} shows the corresponding cross sections integrated up to $\Tau_B < \Tau_B^\cut$ as a function of $\Tau_B^\cut$. For $\Tau_B^\cut \to 0$ the resummation is most important, while for large $\Tau_B^\cut$ the perturbative corrections are most important. We see that the NNLL result interpolates between these regions, approaching the singular NLO result as $\Tau_B^\cut$ becomes large and the NLL and LL result for $\Tau_B^\cut \to 0$. Compared to the results for Drell-Yan shown in the right panel of \fig{sigtauB}, the size of the $\al_s$ corrections as well as the effect of resummation is much larger. 

\chapter{\boldmath$N$-Jettiness: An Inclusive Event Shape to Veto Jets}
\label{ch:njet}

In this chapter we will extend our work to events with $N$ signal jets and consider the veto on additional unwanted jets. We introduce a global event shape ``$N$-jettiness'' $\tau_N$ to impose this veto, which for $N=0$ reduces to the beam thrust. We discuss the experimental relevance of jet vetos and the benefits of using $\tau_N$ in theory calculations. In particular, $\tau_N$ allows us to sum large logarithms due to phase space restrictions and leads to a factorization formula with inclusive jet and beam functions. The work presented in this chapter was first reported in Ref.~\cite{Stewart:2010tn}.

\section{Introduction}

At the LHC or Tevatron, hard interactions involving Higgs or new-physics particles
are identified by looking for signals with
a characteristic number of energetic jets, leptons, or photons~\cite{Ball:2007zza,:1999fr}.
The backgrounds come from Standard Model processes producing
the same signature of hard objects possibly with
additional jets. An example are top quarks decaying into $W$ plus
$b$-jet, which is a major background for $H\to WW$~\cite{Aaltonen:2010yv}.
When reconstructing masses and decay chains of new-physics particles
additional jets can cause large combinatorial backgrounds.
Standard Model processes can also fake a signal when a jet is
misidentified as lepton or photon, a typical example being $H \to \gamma\gamma$.

Thus, a veto on additional undesired jets is an effective and sometimes necessary method to
clean up the events and discriminate signal and the various backgrounds.
More generally, one would like to measure an ``exclusive''
$N$-jet cross section, $pp\to X L(Nj)$, to produce $N$ signal jets $j$
where the remaining $X$ contains no hard (central) jets.
Here, $N \geq 0$ and $L$ denotes the hard leptons or photons required as part of the signal.

We introduce an inclusive event shape ``$N$-jettiness'', denoted $\tau_N$ and
defined below in \eq{tauN}. For an event with at least $N$ energetic jets,
$\tau_N$ provides an inclusive measure of how $N$-jet-like the event looks. In
the limit $\tau_N \to 0$ the event contains exactly $N$ infinitely narrow jets.
For $\tau_N \sim 1$ the event has hard radiation between the $N$ signal jets.
Requiring $\tau_N \ll 1$ constrains the radiation outside the signal and beam jets,
providing an inclusive way to veto additional central jets. It yields an inclusive
definition of an exclusive $N$-jet cross section with a smooth transition
between the case of no jet veto, $\tau_N \sim 1$, and the extremely exclusive
case $\tau_N\to 0$.

Vetoing additional jets imposes a phase-space restriction on the underlying
inclusive $N$-jet cross section to produce $N$ or more jets with the same $L$.
Irrespectively of its precise definition, the jet veto introduces a jet
resolution scale $\mu_J$ that characterizes this restriction, i.e.\ the
distinction between $N$ and $N\!+\!1$ jets. Hence, the exclusive $N$-jet
cross section contains phase-space logarithms $\alpha_s^n
\ln^m(\mu_J^2/\mu_H^2)$, where $m\leq 2n$ and $\mu_H$ is the scale of the hard
interaction. For $\tau_N$, $\mu_J^2/\mu_H^2 \simeq \tau_N \ll 1$. Generically
there is always a hierarchy $\mu_J \ll \mu_H$, which becomes larger the stronger
the restrictions are. These large logarithms must be summed to obtain reliable predictions.

Jet vetoes are typically implemented by using a jet
algorithm to find all jets in the event and vetoing events with too many energetic jets.
Jet algorithms are good tools to identify the signal
jets. However, they are not necessarily well-suited to veto unwanted jets, because the corresponding
phase-space restrictions are complicated and depend in detail on the algorithm.
This makes it difficult to incorporate the jet veto into explicit theoretical calculations and
in particular inhibits a systematic summation of the resulting large logarithms.
In this case, usually the only way to predict the corresponding exclusive $N$-jet
cross section is to rely on parton shower Monte Carlos to sum the
leading logarithmic (LL) series.

In contrast, vetoing jets by cutting on an inclusive variable like $\tau_N$ has
several advantages. First, we can go beyond LL order, because the logarithms
from the phase-space restriction, $\alpha_s^n \ln^m\!\tau_N$, are simple enough
to allow their systematic summation to higher orders. Moreover, the theory
predictions with factorization can be directly compared with experiment without
having to utilize Monte Carlos for parton showering or hadronization.
Experimentally, $\tau_N$ reduces the dependence on jet algorithms and might help
improve the background rejection.

\begin{figure*}[t!]
\subfigure[\hspace{1ex}$e^+e^- \to 2$ jets.]{%
\includegraphics[width=0.48\textwidth]{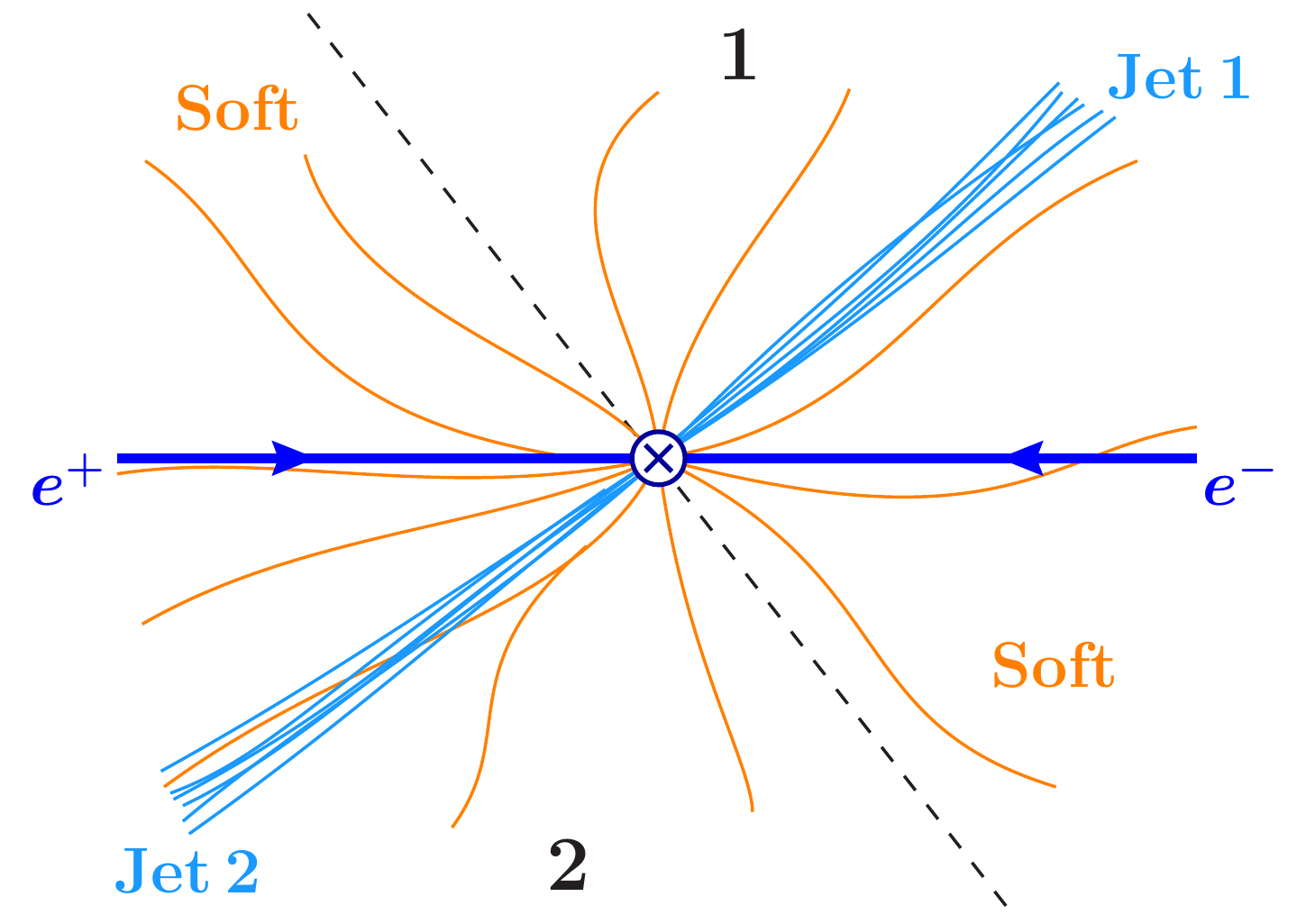}%
\label{fig:ee2jets}%
}\hfill%
\subfigure[\hspace{1ex}Isolated Drell-Yan.]{%
\includegraphics[width=0.48\textwidth]{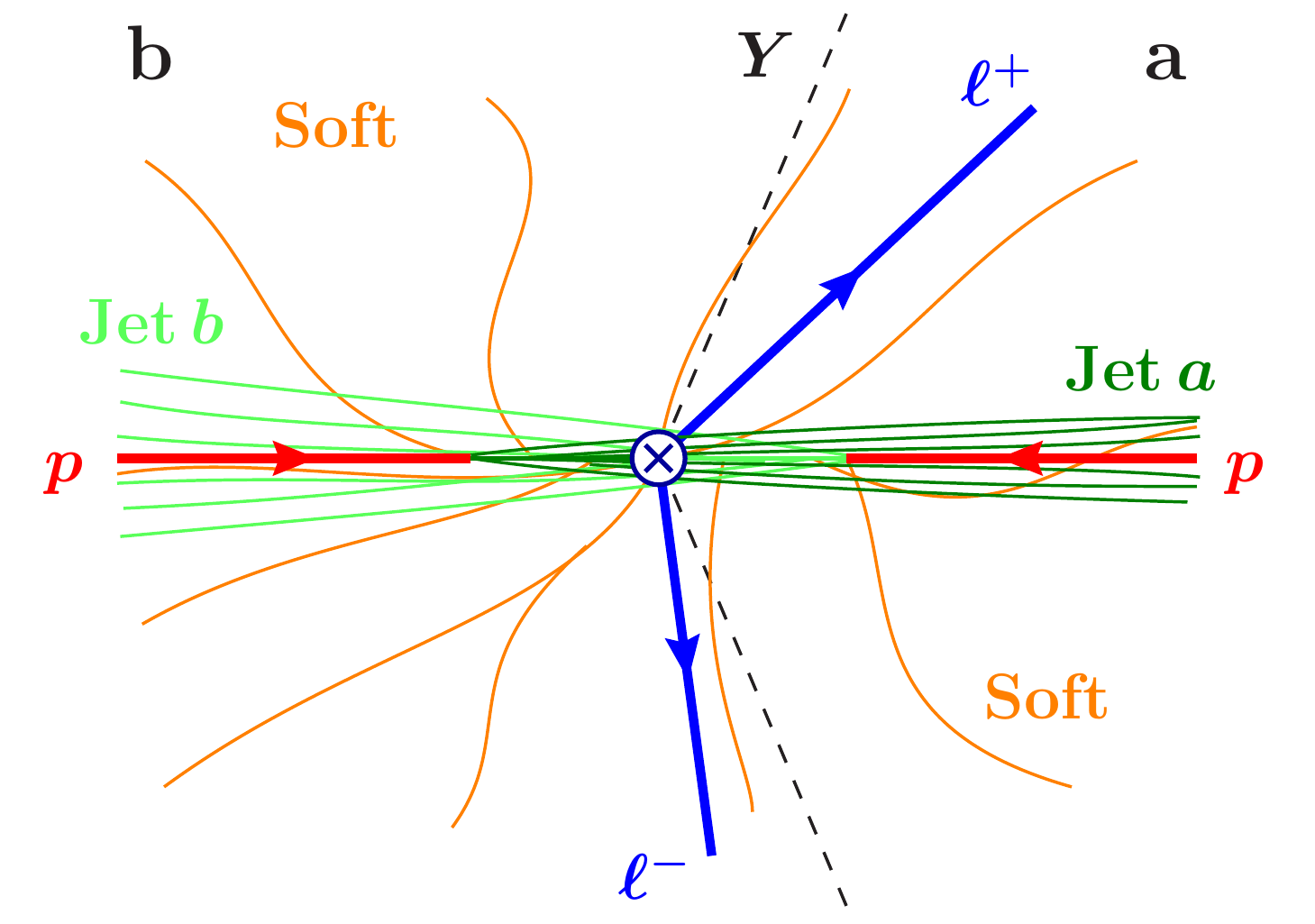}%
\label{fig:drellyan}%
}\\[1ex]
\subfigure[\hspace{1ex}$pp\to $ leptons plus jets.]{%
\includegraphics[width=0.48\textwidth]{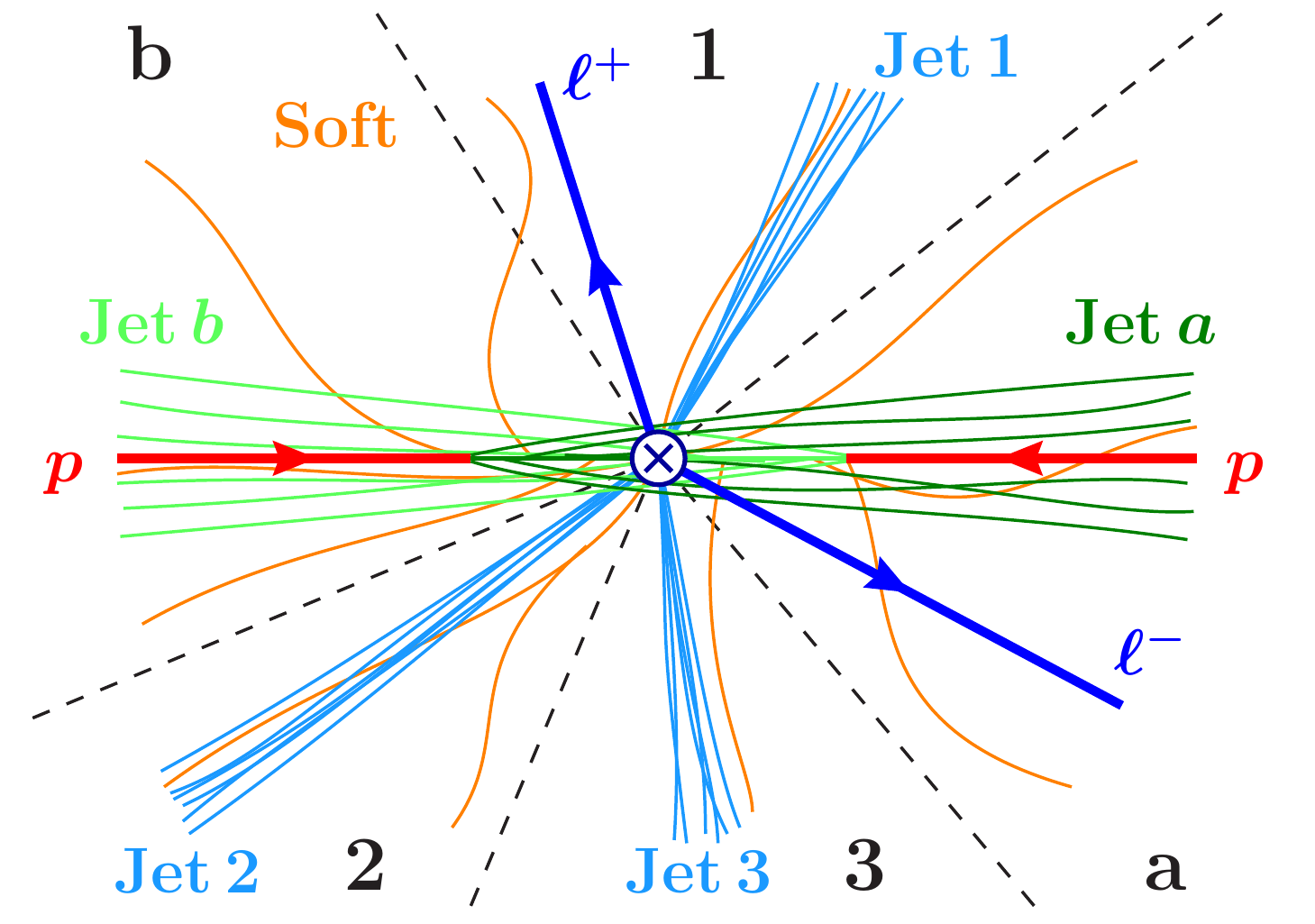}%
\label{fig:3jet}%
}%
\caption{Different situations for the application of $N$-jettiness.}
\end{figure*}

\section{Definition}

$N$-jettiness is defined as
\begin{equation}\label{eq:tauN}
\tau_N = \frac{2}{Q^2}\sum_k
\min \bigl\{ q_a\cdot p_k,\, q_b\cdot p_k,\, q_1\cdot p_k,\, \ldots,\, q_N\cdot p_k \bigr\}
\,.\end{equation}
As we discuss below, this definition of $\tau_N$ yields a factorization formula
with inclusive jet and beam functions and allows the summation of logarithms to
next-to-next-to-leading logarithmic (NNLL) order.  The sum over $k$ in \eq{tauN}
runs over the momenta $p_k$ of all measured (pseudo-)particles in the final
state excluding the signal leptons or photons in $L$. (Any other leptons or
photons, e.g.\ from hadronic decays, are included in the sum.) For simplicity we
take all $p_k$ to be massless.  The $q_a$, $q_b$, and $q_1$, ..., $q_N$ are a
fixed set of massless reference momenta for the two beams and the $N$ signal
jets,
\begin{align}\label{eq:mom}
q_{a,b}^\mu &= \tfrac{1}{2}x_{a,b} \Ecm\, n_{a,b}^\mu
\,,& n_a^\mu &= (1, 0,0,1)\,,\quad n_b^\mu = (1, 0,0,-1)
\,,\nn\\
q_J^\mu &= E_J (1, \hat{n}_J)
\,,& J &= \{1,\ldots,N\}
\,.\end{align}
The $E_J$ and $\hat{n}_J$ correspond to the energies and directions of the $N$ signal jets. Their choice is discussed below. The beam reference momenta $q_a$ and $q_b$
are the large momentum components of the colliding partons along the beam axis (taken to be the $z$ axis). They are defined by
\begin{equation}\label{eq:xab2}
x_a \Ecm = n_b \cdot(q_1 + \dotsb + q_N + q)
\,,\end{equation}
and analogously for $x_b$ with $a\leftrightarrow b$. Here, $q$ is the total momentum of the non-hadronic signal $L$.
In \eq{tauN}, $Q^2 = x_a x_b\Ecm^2$ is the hard interaction scale, and
the distance of a particle with momentum $p_k$ from the jets or beams is measured by $q_m\cdot p_k$.
If $L$ contains missing energy, so $q$ and $x_{a,b}$ are not known,
one can use a modified distance measure as we discuss below \eq{geo}.

The minimum for each $k$ in \eq{tauN} associates the particle with the closest beam or jet,
appropriately dividing the hadronic initial-state radiation (ISR) and
final-state radiation (FSR). Soft particles and energetic particles near any jet
or beam only give small contributions to the sum. For $2\to N$ scattering of
massless partons, $\tau_N = 0$. Energetic particles far away from all jets and
beams give large contributions. Hence, for $\tau_N \ll 1$ the final state has
$N$ jets, two forward beam jets, and only soft radiation between them. In this
limit $x_{a,b}$ are the momentum fractions of the annihilated partons, and $Y =
\ln(x_a/x_b)/2$ is the boost of the partonic center-of-mass frame.

\subsection[$N = 2$ for $e^+e^-\to $ Jets.]{\boldmath $N = 2$ for $e^+e^-\to $ Jets.}

In $e^+e^-$ collisions there is no hadronic ISR, so we drop the $q_{a,b}\cdot p_k$ entries in \eq{tauN}. Now $Q^2$ is the total invariant mass of the leptons and $Y = 0$. In the two-jet limit, the jet directions are close to the thrust axis $\hat{t}$, so we can choose
\begin{equation}
q_1^\mu = \frac{1}{2}\, Q\,(1,\hat{t}\,)
\,,\qquad
q_2^\mu = \frac{1}{2}\, Q\,(1,-\hat{t}\,)
\end{equation}
as reference momenta, and \eq{tauN} becomes
\begin{equation}\label{eq:tau2ee}
\tau^{ee}_2
= \frac{1}{Q} \sum_k E_k \min \bigl\{ 1-\cos\theta_k,\, 1 + \cos\theta_k \bigr\}
\,,\end{equation}
where $\theta_k$ is the angle between $\vec{p}_k$ and $\hat{t}$.
The minimum divides all particles into the two hemispheres perpendicular to $\hat{t}$ as shown in \fig{ee2jets}. For $\tau_2^{ee} \ll 1$,
the total invariant mass in each hemisphere is much smaller than $Q$, so the final state contains two narrow jets. In this limit, $\tau_2^{ee} = 1 - T$, where $T$ is thrust, and a factorization theorem exists for $\df\sigma/\df\tau_2^{ee}$, which can be used to sum logarithms of $\tau_2^{ee}$~\cite{Korchemsky:1999kt,Catani:1992ua,Becher:2008cf}.

\subsection[$N = 0$ for Drell-Yan]{\boldmath $N = 0$ for Drell-Yan}

Next, consider isolated Drell-Yan, $pp \to X\ell^+ \ell^-$ with no hard central jets, shown in \fig{drellyan}.
We now have ISR from the incoming partons, but no FSR from jets. From \eq{xab2} we have
\begin{equation} \label{eq:tau0_xab}
x_a \Ecm = e^{+Y}\! \sqrt{q^2 + \vec{q}_T^{\,2}}
\,,\quad
x_b \Ecm = e^{-Y}\! \sqrt{q^2 + \vec{q}_T^{\,2}}
\,,\end{equation}
where $q^2$ and $\vec{q}_T$ are the dilepton invariant mass and transverse momentum, and $Y$ equals the dilepton rapidity. Now, $Q^2 = q^2 + \vec{q}^{\,2}_T$ and \eq{tauN} becomes
\begin{equation} \label{eq:tau0}
\tau_0 = \frac{1}{Q} \sum_k \abs{\vec{p}_{kT}} \min\bigl\{e^{Y-\eta_k},\, e^{-Y+\eta_k} \bigr\}
\,.\end{equation}
Thus, the $q_a$ and $q_b$ dependence in \eq{tauN} explicitly accounts for the boost of the partonic center-of-mass frame. For $Y = 0$, the minimum in \eq{tau0} divides all particles into two hemispheres perpendicular to the beam axis (analogous to the thrust axis in $e^+e^-$). For $Y \neq 0$, the hemispheres are boosted with their boundary now at $Y$, and the beam jet in the direction of the boost is narrower than the other, as depicted in \fig{drellyan}. Contributions to $\tau_0$ from large rapidities are exponentially suppressed by $\abs{\vec{p}_{kT}} e^{-\abs{\eta_k}} \approx 2E_k e^{-2\abs{\eta_k}}$, so particles beyond the detector's rapidity reach give negligible contributions.

Beam thrust is given by $\tau_B = \sqrt{1 + \vec{q}_T^{\,2}/q^2}\, \tau_0$. It is obtained by choosing $x_{a,b}\Ecm = \sqrt{q^2} e^{\pm Y}$ in case $q^2$ and $Y$ are measured rather than the longitudinal components $n_{a,b}\cdot q$ in \eq{tau0_xab}. For $\tau_0 \ll 1$ the hadronic final state can only contain soft radiation plus energetic radiation in the forward directions. In this limit the leptons become back-to-back in the transverse plane, $\vec{q}_T \ll Q$, so $\tau_B = \tau_0$. In \ch{fact} we derived a factorization theorem for $\df\sigma/\df\tau_B$ at small $\tau_B$, which we used to sum logarithms of $\tau_B$ to NNLL.

\subsection{General Case}

For $pp\to XL (Nj)$ we have both ISR and FSR. We select candidate signal events by measuring $L$ and running a jet algorithm to find the $N$ signal jets and their momenta $p_J$. The conditions on the jets and $L$ that define the signal are encoded in the cross section by a measurement function $F_N(\{p_J\}, L)$. Generically, $F_N$ will enforce that there are at least $N$ energetic jets that are sufficiently separated from each other and the beams. We now use the measured jet energies and directions to define the jet reference momenta $q_J$ in \eq{mom},
\begin{equation}
E_J = p_J^0
\,,\qquad
\hat{n}_J = \vec{p}_J/\abs{\vec{p}_J}
\,,\end{equation}
while $q_a$ and $q_b$ are given by \eqs{mom}{xab}.

Taking the minimum in \eq{tauN} combines the previous cases in \eqs{tau2ee}{tau0}. It divides all particles into jet and beam regions that are unique for a given set of reference momenta and whose union covers all of phase space, as illustrated in \fig{3jet}. The boundary between any two neighboring regions is part of a cone and is such that the sum of the total invariant masses in the regions is minimized
(or in case of a beam region the virtuality of the incoming colliding parton).

For events with small $\tau_N$ all jet algorithms should agree how energetic radiation is split up between the jets and beams, and only differ in their treatment of softer particles. Thus, they all give
the same $\hat{n}_J$ and $E_J$ up to power corrections, while the split up of the soft radiation is determined by $\tau_N$ itself. Hence, the dependence of $\tau_N$ on the jet algorithm is formally power suppressed, $\tau_N^\mathrm{alg. 1} = \tau_N^\mathrm{alg. 2} + \ord{\tau_N^2}$,
as seen in \eq{sigma} below.

To measure $\tau_N$, we still rely on having a suitable jet algorithm to find the $N$ signal jets but not more so than if we were not measuring $\tau_N$. Imagine the jet size in the algorithm is chosen too small such that the algorithm divides what should be a single signal jet into several narrow jets~\footnote{This can be tested by comparing the total energy in each region defined by $\tau_N$ with the energy from the jet algorithm. If these are very different, but at the same time $\tau_N$ is small, then there are additional energetic particles near the signal jets that the algorithm should have included.}.
In this case, the jet algorithm yields a poorly reconstructed signal irrespective of measuring $\tau_N$.

Since the jet veto is now provided by $\tau_N$, this situation can be avoided because we do not have to rely on the jet algorithm to identify additional jets and so can use an algorithm that can be forced to always yield at most $N$ jets. This is in fact the most natural thing to do when one is looking for $N$ jets. Therefore, using $\tau_N$ as jet veto could also help improve the signal reconstruction.

\subsection{Generalizations}

We can generalize $\tau_N$ to
\begin{equation}
\tau_N^d = \sum_k \min \bigl\{ d_a(p_k), d_b(p_k), d_1(p_k),\ldots, d_N(p_k) \bigr\}
\,,\end{equation}
where $d_m(p_k)$ can be any infrared-safe distance measure. In \eq{tauN},
$d_m(p_k) = 2q_m\cdot p_k/Q^2$ with
\begin{align} \label{eq:qipk}
2q_a \cdot p_k &= \abs{\vec p_{kT}}\, Q\, e^{Y-\eta_k}
\,,\nn\\
2q_J \cdot p_k &= \abs{\vec p_{kT}}\, \abs{\vec q_{JT}}\, (2\cosh \Delta \eta_{Jk} - 2\cos \Delta \phi_{Jk})
\,.\end{align}
Here, $\Delta\eta_{Jk}$ and $\Delta\phi_{Jk}$ are the rapidity and azimuthal distances between $q_J$ and $p_k$. If these are small, the factor in brackets reduces to the familiar $R^2 = (\Delta \eta)^2 + (\Delta \phi)^2$.

Raising a given measure to a power, $d_m(p_k)^{\alpha_m}$ changes the relative weight between the region's center and periphery in the sum over $k$, while the powers $\alpha_m$ can be used to change the division between beams and jets.
Different measures that are boost-invariant along the beam axis can be obtained by raising each factor in \eq{qipk} to a different power.
A geometric measure, which is independent of $\abs{\vec{q}_{JT}}$, is
\begin{align} \label{eq:geo}
d_a(p_k) &= \frac{\abs{\vec p_{kT}}}{Q}\, e^{Y -\eta_k}
\,,\qquad
d_b(p_k) = \frac{\abs{\vec p_{kT}}}{Q}\, e^{-Y +\eta_k}
\,,\nn\\
d_J(p_k) &= \frac{\abs{\vec p_{kT}}}{Q}\, (2\cosh \Delta \eta_{Jk} - 2\cos \Delta \phi_{Jk})
\,.\end{align}
It evenly divides the area rather than invariant mass between neighboring regions, such that more energetic jets also get more invariant mass.

If $L$ contains missing energy, $x_{a,b}$ in \eq{xab2} and thus $Q$ and $Y$ are not known. For $Q$, one can use any hard scale, like the $\abs{\vec{q}_{JT}}$ of the hardest jet, since it only serves as an overall normalization. In the beam measures $d_{a,b}(p_k)$ in \eq{geo} we can simply set $Y = 0$, which defines them in the hadronic center-of-mass frame and effectively averages over all possible boosts $Y$.

$N$-jettiness does not split events into $N$, $N+1$, $N+2$, etc. jets like a traditional jet algorithm.
But we can consider using $\tau_N$ to define an ``exclusive $N$-jet
algorithm'' as follows: First, we use a geometric measure and find the
directions $\hat{n}_J$ and boost $Y$ that minimize $\tau_N$.  This is analogous
to determining the thrust axis in $e^+ e^- \to$ jets.  It might actually allow
one to get an estimate of $Y$ even in the case of missing energy by exploiting
the asymmetry in the beam jets.  Second, we determine the jet energies by
summing over the particles in each jet region. (To reduce the sensitivity to the
underlying event and pile-up, one can weigh the sum over energies by the
distance from $\hat{n}_J$.)

\section{Factorization formula}

For simplicity, we now use $\tau_N$ again as defined in \eq{tauN}.
For $\tau_N \ll 1$, QCD ISR and FSR can be described in soft-collinear effective theory~\cite{Bauer:2000ew,Bauer:2000yr,Bauer:2001ct,Bauer:2001yt,Bauer:2002nz}
at leading power by $N+2$ independent sectors for collinear particles close to each $q_m$ with $m = \{a,b,J\}$ and a separate sector for soft particles. By power counting, $J$-collinear particles are closest to $q_J$, so the sum over $k$ for the $J$-collinear sector is
\begin{equation}
\sum_{k\in \mathrm{coll}_J} \min_m \bigl\{2q_m\cdot p_k \bigr\}
= \sum_{k\in \mathrm{coll}_J}  2q_J\cdot p_k = s_J
\,,\end{equation}
where (up to power corrections) $s_J$ is the total invariant mass in the $J$-collinear sector. Similarly, the sum over the beam collinear sectors yields the total (transverse) virtuality of the colliding partons, $t_a$ and $t_b$. Therefore,
\begin{equation} \label{eq:tauN_SCET}
\tau_N Q^2 = t_a + t_b + \sum_J s_J + \sum_{k\in\mathrm{soft}} \min_m \bigl\{2 q_m \cdot p_k \bigr\}
\,.\end{equation}
The sum in the last term is now restricted to the soft sector.  Combining
\eq{tauN_SCET} with the analyses in Ref.~\cite{Bauer:2008jx} and \ch{fact}
yields the factorization formula for $N$-jettiness \footnote{Here, $F_N$
  enforces distinct collinear sectors with $1-\hat n_l\cdot\hat n_m \gg \tau_N$
  and $E_m/Q \gg \tau_N$. We assume $F_N$ only depends on the large components
  $q_J$ of the jet momenta, $p_J = q_J[1 + \ord{\tau_N}]$, and that $L$ only
  couples to the QCD subprocess via a hard interaction. We also assume that Glauber gluons
do not spoil this factorization.}
\begin{align} \label{eq:sigma}
\frac{\df\sigma}{\df \tau_N}
&=
\int\!\df x_a \df x_b \int\!\df^4 q\,\df\Phi_L(q) \int\! \df \Phi_N(\{q_J\})
F_N(\{q_m\}, L)\, (2\pi)^4 \delta^4\Bigl(q_a + q_b - \sum_J q_J - q\Bigr)
\nn\\ &\quad \times
\sum_{ij,\kappa} \tr\, \hH_{ij\to\kappa}(\{q_m\}, L, \mu) 
\int\!\df t_a\, B_i(t_a, x_a, \mu)
\int\!\df t_b\, B_j(t_b, x_b, \mu)
\prod_J \int\!\df s_J\, J_{\kappa_J}(s_J, \mu)
\nn\\ &\quad \times
\hS_N^{ij\to\kappa} \Bigl(\tau_N - \frac{t_a + t_b + \sum_J s_J}{Q^2}, \{q_m\}, \mu\Bigr)
\,.\end{align}
$\hH_{ij\to\kappa}(\{q_m\}, L)$ contains the underlying hard process $i(q_a)j(q_b) \to L(q) \kappa_1(q_1) \dotsb \kappa_N(q_N)$, where $i,j,\kappa_J$ denote the parton types, and the sum over $ij,\kappa$ is over all relevant partonic channels. It is a matrix in color space given by the IR-finite parts (in pure dim. reg.) of the squared partonic matrix elements in each channel. The $N$-body phase space for the massless momenta $q_J$ is denoted $\df\Phi_n(\{q_J\})$, and that for $L$ by $\df\Phi_L(q)$.

The distributions in $s_J$ and $t_{a,b}$ are described by inclusive jet and beam functions, $J_{\kappa_J}(s_J)$ and $B_{i,j}(t_{a,b}, x_{a,b})$. 
The last term in \eq{tauN_SCET} is the contribution to $\tau_N$ from soft particles in the ``underlying event''. It is described by the soft function $\hS_N^{ij\to\kappa}(\tau_N^\mathrm{soft}, \{q_m\})$, which depends on the jet's angles $\hat{n}_l\cdot \hat{n}_m$ and energy fractions $E_l/E_m$. Like $\hH$, it is a color matrix, and the trace in \eq{sigma} is over $\tr(\hH \hS)$.

In \eq{sigma}, all functions are evaluated at the same renormalization scale $\mu$. Large logarithms of $\tau_N$ in $\df\sigma/\df\tau_N$ are summed by first computing $\hH(\mu_H)$, $J(\mu_J)$, $B(\mu_B)$, $\hS(\mu_S)$ at the scales $\mu_H \simeq Q$, $\mu_J\simeq\mu_B\simeq \sqrt{\tau_N}Q$, $\mu_S\simeq \tau_N Q$, where the functions contain no large logarithms, and then evolving them to the scale $\mu$.
This evolution is known analytically~\cite{Neubert:2004dd,Fleming:2007xt,Ligeti:2008ac} and the required anomalous dimensions are already known to NNLL~\cite{Korchemsky:1987wg,Moch:2004pa,Kramer:1986sg,Harlander:2000mg,MertAybat:2006mz,Ferroglia:2009ii}, because we have inclusive jet and beam functions. NNLL also requires the $\ord{\alpha_s}$ corrections for each function, which are known for $J$ and $B$. The $\ord{\alpha_s}$ hard function is determined by the one-loop QCD matrix elements. For $\tau_N \gg \Lambda_\mathrm{QCD}/Q$, $\hS(\mu_S)$ can be computed perturbatively and will be given in a future publication.


\chapter{Conclusions and Outlook}
\label{ch:concl}

Typical Higgs and new physics searches at the LHC or Tevatron require a certain number of hard jets, leptons and photons. Jet algorithms are good for identifying jets, but, as we have seen, a veto on additional unwanted jets is most easily incorporated in theory calculations by using an event shape. This approach allowed us to sum the corresponding large phase space logarithms and make predictions without relying on Monte Carlo programs for parton showers or hadronization models. 

A lot of our work has focussed on the 0-jet case, for which we implemented the jet veto by restricting the beam thrust variable $\tau_B \ll 1$. A central jet veto is needed to reduce the large background in $H\to WW \to \ell \nu \ell \bar\nu$ from top quarks decaying into a $W$ plus $b$-jet. We derived a factorization theorem that allowed us to sum the large phase space logarithms $\al_s^n \ln^m \tau_B$. The central jet veto also probes the initial-state radiation, which requires beam function to describe the initial state. Beam functions describe extracting a parton out of the proton as well as the formation of an initial-state jet. They can be related to PDFs through a perturbative matching calculation, which we carried out at one loop. Our calculations allow us to study the effect of initial-state radiation in perturbation theory and the next step is to compare our analytical results with Monte Carlo programs. Another interesting extension of our work would be to include the measurement of the recoil of the initial-state jets, which requires $p_T$ dependent beam functions. 

We concluded this thesis by considering the extension to $N$ signal jets, for which we introduced the event shape $N$-jettiness to veto unwanted additional jets. Many of the features of the 0-jet case carry over, including the resummation of large logarithms and the presence of beam functions. In the 0-jet case there was no jet algorithm dependence, in this case a jet algorithm is used to identify the jets but the dependence on the choice of jet algorithm is power suppressed. We still need to show the cancellation of Glauber gluons. Once the one-loop soft function has been calculated we can use our factorization formula to determine cross sections for processes with final state jets at NNLL. This will allow us to study the jet energies and directions, which are of interest to new physics searches. To study other properties of the jets, such as their invariant masses, requires an extension of our factorization formula.

\appendix

\chapter{Notation Cheat Sheet}
\label{app:notation}

\section{Lightcone Coordinates}

We use lightcone vectors $n^\mu$ and $\bn^\mu$ with $n^2 = \bn^2 = 0$ and $\bn \cdot n =2$ to decompose four-vectors $p^\mu = (p^+,p^-,p_\perp^\mu)$
\begin{equation}
  p^\mu = p^- \frac{n^\mu}{2} + p^+ \frac{\bn^\mu}{2} + p_\perp^\mu\,, \qquad
  p^+ =  n \sdt p\,, \qquad
  p^- = \bn \sdt p\,.
\end{equation}
We have two collinear directions $n_a^\mu = (1,0,0,1)$ and $n_b^\mu = (1,0,0,-1)$ corresponding to the directions of the incoming protons, and take $n_b = \bn_a$ and $n_a = \bn_b$. Lightcone coordinates of a momentum with subscript $a$ or $b$ will be with respect to $n_a$ or $n_b$, e.g. $B_a^+ = n_a \sdt B_a$, $B_b^+ = n_b \sdt B_b$. We will sometimes write $p_T = p_\perp$.

For an energetic particle collinear to the $n^\mu$ direction we separate its momentum $p^\mu$ into a (discrete) large $\tilde p^\mu$ and (continuous) small momentum $p_r^\mu$ in SCET, $p^\mu = \tilde p^\mu + p_r^\mu$. The large momentum $\tilde p^\mu = (0,\tilde p^-,\tilde p_\perp^\mu)$ has components $\tilde p^- = \bn \sdt p$ and $p_\perp^\mu \sim \la p^-$, where $\la$ is the power counting parameter. The continuous small momentum has scaling $p_r^\mu = \la^2 \tilde p^-$. The large momentum appears as a label on the field and is picked out by the label operator $\cP$. The small momentum is picked out by $\hat p$ and is often written in the conjugate position space coordinate $y$. We absorb the residual $p_r^-$ and $p_{r\perp}^\mu$ into the label $\tilde p$, to give continuous labels. This leaves us with just one residual coordinate $y^-$. More details on SCET can be found in Secs.~\ref{sec:SCET_intro}, \ref{sec:definition} and \ref{sec:SCET}.

\section{Symbols}

Below we list the most common symbols and their definition. Whenever there is a subscript $a$ there will be a corresponding symbol with a subscript $b$, corresponding to the other proton, parton or hemisphere. The subscripts $i$ and $j$ run over flavors (quark or gluon). \ \\

\begin{table}[h!]
  \begin{tabular}{l | l l}
  & Symbol & Definition \\
  \hline
  General
  &$\varepsilon$ & polarization vector \\
  &$L$ & non-hadronic final state ($\ell^+ \ell^-$ for Drell-Yan) \\
  &$n_a^\mu$ & lightcone vectors in the direction of the incoming protons \\
  &$p_n$ & n-collinear proton \\
  &$u_n(p)$ & n-collinear spinor \\  
  &$X$ & total hadronic final state \\
  \hline
  Momenta
  &$b_a^\mu$ & collinear radiation from the incoming parton $a$ \\
  &$B_a^\mu$ & hadronic momentum in hemisphere $a$, $B_a^\mu = b_a^\mu + r_a^\mu + k_a^\mu$ \\
  &$k_a^\mu$ & soft momentum in the hemisphere $a$ \\
  &$P_a^\mu$ & momentum of proton $a$, $P_a^\mu = \ECM n_a^\mu/2$ \\
  &$p_a^\mu$ & momenta of incoming parton $a$ \\
  &$p_{1,2}^\mu$ & momenta of the leptons for Drell-Yan \\
  &$r_a^\mu$ & momentum of the proton remnant for proton $a$ \\
  &$q^\mu$ & total non-hadronic momentum ($q^\mu = p_1^\mu + p_2^\mu$ for Drell-Yan) \\
  \hline
  Kinematics
  &$\w_{a,b}$ & large momentum component of incoming parton, $\w_a = p_a^-$ \\
  &$Q$ & inv. mass of non-hadronic particles, hard scale, $Q = \sqrt{q^2}$ \\
  &$t$ & transverse virtuality of incoming parton, $t_a = \w_a b_a^+ $ \\
  &$\tau$ & $\tau = q^2/\ECM^2$ \\
  &$\tau_B$ & beam thrust \\
  &$x_a$ & momentum fraction of parton, $x_a = \w_a/P_a^-$ \\
  &$\xi_a$ & momentum fraction of parton in the PDF \\
  &$y_B^\cut$ & rapidity ``cut" corresponding to beam thrust, $\tau_B \leq \exp(-2y_B^\cut)$ \\
  &$Y$ & total rapidity of non-hadronic particles \\
  &$z$ & parton momentum fraction of in partonic calculation \\
  \hline
  Scales
  &$\mu_B$ & beam scale, $\mu_B \sim \sqrt{\tau_B}Q$ for beam thrust \\
  &$\mu_H$ & hard scale, $\mu_H \sim Q$ \\
  &$\mu_\Lambda$ & low scale at which the PDFs are defined \\
  &$\mu_S$ & soft scale, $\mu_S \sim \tau_B Q$ for beam thrust \\     
  \hline
  \end{tabular}
\end{table}
\begin{table}
  \begin{tabular}{l | l l}
  & Symbol & Definition \\
  \hline  
  Factor-
  &$B_i$ & beam function \\
  ization
  &$\tB_i$ & beam function integrated up to $t\leq t_\max$  \\  
  &$f_i$ & parton distribution function \\ 
  &$H_{ij}$ & hard function (same for threshold and isolated processes)\\
  &$\cI_{ij}$ & Wilson coefficient for initial state jet, $\cB = \cI \otimes f$ \\
  &$J_i$ & (final state) jet function \\
  &$\la$ & power counting parameter, $\la \sim B^+/Q \sim \tau_B$ \\
  &$\op_i$ & beam function operator, $B_i = \mae{P^-}{\op_i}{P^-}$ \\ 
  &$\oq_i$ & PDF operator,  $f_i = \mae{P^-}{\oq_i}{P^-}$ \\
  &$S^{ij}$ & soft function (differs between threshold and isolated) \\
  &$\sigma_0$ & Born cross section \\
  \hline    
  Renorm-
  &$\eta_\Gamma$ & shows up in RGE solution for cusp piece \\
  alization
  &$\gamma(\dots,\mu)$ & anomalous dimension \\
  &$\gamma(\alpha_s)$ & non-cusp anomalous dimension \\
  &$\Gamma_\cusp$ & cusp anomalous dimension \\
  &$K_\Gamma$ & exponent in RGE solution for cusp piece \\
  &$K_\gamma$ & exponent in RGE solution for non-cusp piece \\
  &$P_{jk}(z)$ & splitting functions in PDF evolution \\
  &$U$ & evolution function \\
  &$Z$ & renormalization factor \\  
  \hline
  SCET
  &$A_n$ & n-colinear gluon field \\
  &$A_s$ & soft gluon field \\
  &$\cB_{n\perp}$ & gauge invariant collinear gluon field \\
  &$\chi_n$ & gauge invariant n-collinear quark field, $\chi_n(y) = W_n^\dagger(y) \xi_n(y)$ \\
  &$\la$ & power counting parameter, $\la \sim B^+/Q \sim \tau_B$ \\ 
  &$\hat p^\mu$ & residual momentum operator \\
  &$\cP^\mu$ & label momentum operator, $\cP^\mu \xi_{n,\tilde p}(y) = \tilde p^\mu \xi_{n,\tilde p}(y)$ \\
  &$\bnP$ & minus label momentum operator, $\bnP = \cP^-$ \\  
  &$W_n$ & n-collinear Wilson line \\
  &$\xi_n$ & n-collinear quark field \\
  &$y^-$ & residual minus position coordinate \\
  &$Y_n$ & soft Wilson line along the $n^\mu$ direction (fundamental repr.) \\
  &$\cY_n$ & soft Wilson line along the $n^\mu$ direction (adjoint repr.) \\
  \hline
  \end{tabular}
\end{table}

\chapter{Plus Distributions and Discontinuities}
\label{app:plusdisc}

The standard plus distribution for some function $g(x)$ can be defined as
\begin{equation} \label{eq:plus_def}
\bigl[\theta(x) g(x)\bigr]_+
= \lim_{\beta \to 0} \frac{\df}{\df x} \bigl[\theta(x-\beta)\, G(x) \bigr]
\qquad\text{with}\qquad
G(x) = \int_1^x\!\df x'\, g(x')
\,,\end{equation}
satisfying the boundary condition $\int_0^1 \df x\, [\theta(x) g(x)]_+ = 0$. Two special cases we need are
\begin{align} \label{eq:plusdef}
\cL_n(x)
&\equiv \biggl[ \frac{\theta(x) \ln^n x}{x}\biggr]_+
 = \lim_{\beta \to 0} \biggl[
  \frac{\theta(x- \beta)\ln^n x}{x} +
  \delta(x- \beta) \, \frac{\ln^{n+1}\!\beta}{n+1} \biggr]
\,,\nn\\
\cL^\eta(x)
&\equiv \biggl[ \frac{\theta(x)}{x^{1-\eta}}\biggr]_+
 = \lim_{\beta \to 0} \biggl[
  \frac{\theta(x - \beta)}{x^{1-\eta}} +
  \delta(x- \beta) \, \frac{x^\eta - 1}{\eta} \biggr]
\,.\end{align}
In addition, we need the identity
\begin{equation} \label{eq:distr_id}
 \frac{\theta(x)}{x^{1+\eps}} = - \frac{1}{\eps}\,\delta(x) + \cL_0(x)
  - \eps \cL_1(x) + \ord{\eps^2}
\,,\end{equation}
the Fourier transform
\begin{equation} \label{eq:distr_FT}
\cL_0(x) = - \int\!\frac{\df y}{2\pi}\,e^{\img x y}\,\ln\bigl[\img(y - \img 0) e^{\gamma_E}\bigr]
\,,\end{equation}
and the two limits
\begin{align} \label{eq:limits}
\lim_{\bt\to 0}\biggl[\frac{\theta(x-\bt)\ln(x-\bt)}{x} + \delta(x-\bt)\,\frac{1}{2}\ln^2 \bt\biggr]
&= \cL_1(x) - \frac{\pi^2}{6}\,\delta(x)
\,,\nn\\
\lim_{\bt\to 0} \frac{\theta(x-\bt)\,\bt}{x^2}
&= \delta(x)
\,.\end{align}
Away from $x=0$ these relations are straightforward, while the behavior at $x = 0$ is obtained by taking the integral of both sides. General relations for the rescaling and convolutions of $\cL_n(x)$ and $\cL^\eta(x)$ can be found in App.~B of Ref.~\cite{Ligeti:2008ac}.

The discontinuity of a function $g(x)$ is defined as
\begin{align} \label{eq:Disc_def2}
\Disc_x\, g(x) = \lim_{\beta\to 0} \bigl[ g(x + \img\beta) - g(x - \img\beta) \bigr]
\,.\end{align}
If we are only interested in the discontinuity in some interval in $x$, we simply multiply the right-hand side with the appropriate $\theta$ functions, as in \eq{Disc_def}. If $g(x)$ is real then $\Disc_x g(x) = 2\img\, \Im\, g(x+\img0)$. Two useful identities are
\begin{align} \label{eq:disc_os}
\frac{\img}{2\pi} \Disc_x\, \frac{1}{x^{n+1}}
= \frac{(-1)^n}{n!}\, \delta^{(n)}(x)
\,,\qquad
\frac{\img}{2\pi}\, \Disc_x\, (-x)^{n-\eps}
= (-1)^{n-1} \frac{\sin\pi\eps}{\pi}\, \theta(x) x^{n-\eps}
\,.\end{align}
To derive the last identity, note that
\begin{equation}
(-x-\img 0)^{n-\eps} = \exp[ (n-\eps)\ln(-x-\img 0)] \\
= |x|^{n-\eps} \exp[-\img\pi (n-\eps)\theta(x)]
\,,
\end{equation}
so taking the imaginary part gives
$\Im (-x-\img0)^{n-\eps} =  (-1)^n \sin(\pi\eps)\, \theta(x)\, x^{n-\eps}$.
To calculate the discontinuities of the various graphs in \sec{gluoncalc} we need the relations,
\begin{align} \label{eq:disc}
  & -\theta(z) \frac{\img}{2\pi}\, \Disc_{t>0} \int_0^1 \df\al\, \De^{-1-\eps}
  = \theta(z)\, \frac{\sin \pi\eps}{\pi\eps}\,  \frac{\theta(t)}{t^{1+\eps}}\,
  \theta(1-z) z^{1+\eps} (1-z)^{-\eps}
  \,, \\
  & -\theta(z) \frac{\img}{2\pi}\, \Disc_{t>0} \int_0^1 \df\al\, (1-\al) \De^{-1-\eps}
  = \theta(z)\, \frac{\sin \pi\eps}{\pi\eps(1-\eps)}\,  \frac{\theta(t)}{t^{1+\eps}}\,
  \theta(1-z) z^{1+\eps} (1-z)^{1-\eps}
  \,, \nn \\
  & -\theta(z) \frac{\img}{2\pi}\, \Disc_{t>0} \int_0^1 \df\al\, (1-\al)\, t \De^{-2-\eps}
  = \theta(z)\, \frac{\sin \pi\eps}{\pi\eps(1+\eps)}\,  \frac{\theta(t)}{t^{1+\eps}}\,
  \theta(1-z) z^{2+\eps} (1-z)^{-\eps}
  \,, \nn
\end{align}
where $\De=t(1-\al/z)$.


\chapter{Renormalization Structure of the Beam Function}
\label{app:renorm}

In this appendix we derive the general structure of the beam function RGE in \eq{B_RGE} to all orders in perturbation theory, which was presented in Ref.~\cite{Stewart:2010qs}. The two essential ingredients will be the known all-order renormalization properties of lightlike Wilson lines~\cite{Brandt:1981kf, Korchemsky:1987wg, Korchemskaya:1992je, Korchemsky:1992xv} and the factorization theorem for the isolated $pp\to X L$ cross section, where $X$ is the hadronic and $L$ the non-hadronic final state. In \ch{fact} we proved that to all orders in perturbation theory and leading order in the power counting this cross section factorizes as
\begin{align} \label{eq:sigBab}
\frac{\df\sigma}{\df q^2 \df Y \df B_a^+\df B_b^+}
&= \sum_{ij} H_{ij}(q^2, Y, \mu) \int\!\df k_a^+\, \df k_b^+\, S^{ij}_\hemiin(k_a^+,k_b^+, \mu)
\nn\\ & \quad\times
 q^2 B_i[\w_a(B_a^+ - k_a^+), x_a, \mu] B_j[\w_b(B_b^+ -k_b^+),x_b, \mu]
\,.\end{align}
The sum over $ij$ runs over parton species $ij = \{gg, u\bar u, \bar u u, d\bar d, \bar d d, \ldots\}$. The soft function does not depend on the quark flavor, and its superscript only refers to the color representation. 

The three ingredients in \eq{sigBab} are the renormalized hard, beam, and soft functions, $H_{ij}(q^2, Y, \mu)$, $B_i(t, x, \mu)$, $S^{ij}_\hemiin(k_a^+, k_b^+, \mu)$. Their dependence on the renormalization scale $\mu$ must cancel in \eq{sigBab}, because the cross section must be $\mu$ independent. The structure of the RGE for the hard and soft functions thus uniquely determines the allowed structure of the beam function RGE.

The hard function is a contraction between the relevant leptonic matrix element squared and the square of the Wilson coefficients of the color-singlet $q\bar q$ and $gg$ local SCET currents (see \subsec{matchQCD})
\begin{equation}
O_{q\bar{q}}^{\alpha\beta}
= \bar\chi_{n_a,-\w_a}^\alpha\,\chi_{n_b,\w_b}^\beta
\,,\qquad
O_{gg}^{\mu\nu}
= \sqrt{\w_a\,\w_b}\,\cB_{n_a,-\w_a\perp }^{\mu c}\, \cB_{\bn_b,-\w_b\perp}^{\nu c}
\,,\end{equation}
where $\alpha$ and $\beta$ are spin indices. In each collinear sector, total label momentum and fermion number for each quark flavor are conserved. Thus, the currents cannot mix with each other and are multiplicatively renormalized. Furthermore, RPI-III invariance implies that the RGE for the currents can only depend on $q^2 = \w_a \w_b$. The renormalization of these SCET currents also does not depend on their spin structure, so the RGE for the hard function must have the same structure as for the currents. Therefore, to all orders in perturbation theory we have (with no sum on $ij$)
\begin{equation} 
\mu \frac{\df}{\df\mu} H_{ij}(q^2, Y, \mu) = \gamma^{ij}_H(q^2, \mu)\, H_{ij}(q^2, Y, \mu)
\,.\end{equation}

Next, the incoming hemisphere soft function, $S^{ij}_\hemiin(k_a^+, k_b^+, \mu)$, is given by the vacuum matrix element of incoming soft lightlike Wilson lines along the $n_a$ and $n_b$ directions. In position space,
\begin{equation}
\tS^{ij}_\hemiin(y_a^-, y_b^-, \mu)
= \int\! \df k_a^+ \df k_b^+\, e^{-\img (k_a^+ y_a^- + k_b^+ y_b^-)/2}\,S^{ij}_\hemiin(k_a^+, k_b^+, \mu)
\end{equation}
has two cusps, one at spacetime position $0$ and one at $y = y_a^- n_a/2 + y_b^- n_b/2$. The renormalization properties of lightlike Wilson lines with cusps~\cite{Brandt:1981kf, Korchemsky:1987wg, Korchemskaya:1992je, Korchemsky:1992xv} then imply that to all orders in perturbation theory,
\begin{align} \label{eq:tS_RGE}
\mu \frac{\df}{\df\mu} \tS^{ij}_\hemiin(y_a^-, y_b^-, \mu)
&= \tga_S^{ij}(y_a^-, y_b^-, \mu)\, \tS^{ij}_\hemiin(y_a^-, y_b^-, \mu)
\,,\\\nn
\tga_S^{ij}(y_a^-, y_b^-, \mu)
&= 2\Gamma^i_\cusp(\alpha_s) \Bigl[
- \ln\Bigl(\img \frac{y_a^-\! - \img 0}{2} \mu e^{\gamma_E}\Bigr)
- \ln\Bigl(\img \frac{y_b^-\! - \img 0}{2} \mu e^{\gamma_E}\Bigr) \Bigr]
\!+ \gamma_S^{ij}(\alpha_s)
\,,\end{align}
where $\Gamma^i_\cusp$ is the cusp anomalous dimension for quarks/antiquarks or gluons, and $\gamma_S^{ij}[\alpha_s(\mu)]$ and $\Gamma_\cusp^i[\alpha_s(\mu)]$ depend only indirectly on $\mu$ via $\alpha_s(\mu)$. Dimensional analysis and RPI-III invariance imply that the single logarithm multiplying $2\Gamma^i_\cusp$ scales like $\ln(y_a^- y_b^- \mu^2)$. (The additional dimensionless factors are chosen for convenience. Any change in them can be absorbed into $\gamma_S^{ij}(\alpha_s)$.) The correct overall sign and $\img 0$ prescription for the logarithms can be deduced from the explicitly known one-loop result see \eq{S_oneloop} and ~\cite{Schwartz:2007ib, Fleming:2007xt}.

Taking the Fourier transform of the cross section in \eq{sigBab} with respect to $B_a^+$ and $B_b^+$ and differentiating the result with respect to $\mu$ yields
\begin{align}
0 &= \mu\frac{\df}{\df\mu} \biggl[
\sum_{ij} H_{ij}(q^2, Y, \mu)
\tB_i\Bigl(\frac{y_a^-}{2\w_a}, x_a, \mu \Bigr) \tB_j\Bigl(\frac{y_b^-}{2\w_b}, x_b, \mu \Bigr)
\tS^{ij}_\hemiin(y_a^-, y_b^-, \mu) \biggr]
\nn\\
&= \sum_{ij} H_{ij}(q^2, Y, \mu) \tS^{ij}_\hemiin(y_a^-, y_b^-, \mu)
\nn\\ & \quad\times
\Bigl[ \gamma_H^{ij}(\w_a \w_b, \mu) + \tga_S^{ij}(y_a^-, y_b^-, \mu) + \mu\frac{\df}{\df\mu} \Bigr]
\tB_i\Bigl(\frac{y_a^-}{2\w_a}, x_a, \mu\Bigr) \tB_j\Bigl(\frac{y_b^-}{2\w_b}, x_b, \mu\Bigr)
\,.\end{align}
The factorization theorem for the cross section neither depends on the choice of $L$, which affects the form of $H_{ij}$ for different $ij$, nor the type of the colliding hadrons. This implies that each term in the sum over $ij$ must vanish separately. (For example, choosing Drell-Yan, $L = \ell^+\ell^-$, there is no contribution from $ij = gg$, so the quark and gluon contributions are separately zero. Then, by assigning arbitrary electroweak quark charges, the contribution from each quark flavor must vanish separately. Finally, the $ij = q\bar{q}$ and $ij = \bar{q}q$ contributions for a single quark flavor $q$ must vanish separately by choosing various different incoming hadrons.) Therefore, the RGE for the product of the two beam functions is
\begin{equation} \label{eq:tBB_RGE}
\Bigl[ \gamma_H^{ij}(\w_a \w_b, \mu) + \tga_S^{ij}(y_a^-, y_b^-, \mu) + \mu\frac{\df}{\df\mu} \Bigr]
\tB_i\Bigl(\frac{y_a^-}{2\w_a}, x_a, \mu\Bigr) \tB_j\Bigl(\frac{y_b^-}{2\w_b}, x_b, \mu\Bigr) = 0
\,,\end{equation}
which shows that the beam functions in position space renormalize multiplicatively and independently of $x_{a,b}$. The RGE for each individual beam function can only depend on the RPI-III invariant $y^-/2\w$ and obviously cannot depend on the variables of the other beam function. Hence, we find that to all orders in perturbation theory
\begin{equation} \label{eq:tB_RGE}
\mu\frac{\df}{\df\mu} \tB_i\Bigl(\frac{y-}{2\w}, x, \mu\Bigr)
= \tga_B^i\Bigl(\frac{y^-}{2\w}, \mu \Bigr) \tB_i\Bigl(\frac{y-}{2\w}, x, \mu\Bigr)
\,,\end{equation}
which is the result we set out to prove in this Appendix. Using \eq{tB_RGE} together with \eq{tBB_RGE}, the anomalous dimensions must satisfy the consistency condition
\begin{equation}
0 =
\gamma_H^{ij}(\w_a \w_b, \mu) + \tga_S^{ij}(y_a^-, y_b^-, \mu)
+ \tga_B^i\Bigl(\frac{y^-_a}{2\w_a}, \mu \Bigr) + \tga_B^j\Bigl(\frac{y^-_b}{2\w_b}, \mu \Bigr)
\,.\end{equation}
Given the form of $\tga_S^{ij}$ in \eq{tS_RGE}, it follows that the anomalous dimensions are given to all orders by
\begin{align} \label{eq:gammas}
\gamma_H^{ij}(\w_a \w_b, \mu)
&= 2\Gamma^i_\cusp(\alpha_s) \ln\frac{\w_a \w_b}{\mu^2} + \gamma_H^{ij}(\alpha_s)
\,,\nn\\
\tga_B^i\Bigl(\frac{y^-}{2\w}, \mu \Bigr)
&= 2\Gamma^i_\cusp(\alpha_s) \ln\Bigl(\img \frac{y^-\!- \img 0}{2\w}\mu^2 e^{\gamma_E}\Bigr)
+ \gamma_B^i(\alpha_s)
\,,\nn\\
\gamma_S^{ij}(\alpha_s) &= -\gamma_H^{ij}(\alpha_s) - \gamma_B^i(\alpha_s) - \gamma_B^j(\alpha_s)
\,.\end{align}
Taking the Fourier transform using \eq{distr_FT}, the momentum-space anomalous dimensions become
\begin{align} \label{eq:gammas2}
\gamma_S^{ij}(k_a^+, k_b^+, \mu)
&= 2\Gamma^i_\cusp(\alpha_s)
\biggl[\frac{1}{\mu}\cL_0\Bigl(\frac{k_a^+}{\mu}\Bigr) \delta(k_b^+) + \delta(k_a^+) \frac{1}{\mu}\cL_0\Bigl(\frac{k_b^+}{\mu}\Bigr) \biggr] + \gamma_S^{ij}(\alpha_s)\, \delta(k_a^+) \delta(k_b^+)
\,,\nn\\
\gamma_B^i(t, \mu)
&= - 2\Gamma^i_\cusp(\alpha_s)\, \frac{1}{\mu^2}\cL_0\Bigl(\frac{t}{\mu^2}\Bigr)
+ \gamma_B^i(\alpha_s)\, \delta(t)
\,.\end{align}

The same all-order structure of the soft anomalous dimension as in \eq{gammas2} was obtained in Ref.~\cite{Fleming:2007xt} for the hemisphere soft function with outgoing Wilson lines in $e^+e^-\to 2$ jets using analogous consistency conditions. In fact, the hard SCET currents here and there are the same and in \sec{B_RGE} we proved that the anomalous dimensions for the beam and jet function are the same, $\gamma_B^i = \gamma_J^i$. Hence, the hemisphere soft functions with incoming and outgoing Wilson lines have in fact identical anomalous dimensions to all orders.


\chapter{Quark Beam Function Matching in Pure Dimensional Regularization}
\label{app:dimreg}

Here we repeat the NLO matching calculation from \ch{quark} using dimensional regularization for both the UV and IR, which was reported in Ref.~\cite{Stewart:2010qs}. Since we only change the IR regulator, the final results for the matching coefficients $\cI_{ij}(t, z, \mu)$ should not be affected.

In pure dimensional regularization all the loop diagrams contributing to the bare matrix elements of $\oq_q$ vanish, since by dimensional analysis there is no Lorentz invariant quantity they can depend on. Hence, including the counter terms in \eq{Zpdf} to subtract the UV divergences, the renormalized matrix elements consist of pure IR divergences with opposite signs to the UV divergences,
\begin{align} \label{eq:fDR}
\Mae{q_n}{\oq_q(\w,\mu)}{q_n}^\one
&= - \frac{1}{\eps}\,\frac{\alpha_s(\mu) C_F}{2\pi}\, \theta(z) P_{qq}(z)
\,,\nn\\
\Mae{g_n}{\oq_q(\w,\mu)}{g_n}^\one
&=  - \frac{1}{\eps}\,\frac{\alpha_s(\mu) T_F}{2\pi}\, \theta(z) P_{qg}(z)
\,.\end{align}
This shows explicitly that the conventional $\overline{\rm MS}$ definition of the PDFs in QCD, which also yields \eq{fDR}, is indeed identical to the SCET definition used in our OPE for the beam function.

Considering the beam function matrix elements, the bare results for \figs{Bone_c}{Bone_d} now vanish, because their loop integrals are again scaleless. For the remaining diagrams we can reuse the intermediate results from \sec{NLO_B} before carrying out the Feynman parameter integrals and taking the discontinuity. Setting $t' = 0$ the denominator in the Feynman parameter integrals in \eq{I12} becomes
$(1-\alpha)A - \alpha B = t(1 - \alpha/z)$. In this case it easier to carry out the integral after taking the discontinuity. The discontinuity we need is
\begin{equation}
\frac{\img}{2\pi}\text{Disc}_{t>0} \Bigl[\Bigl(1-\frac{\alpha}{z}\Bigr)t\Bigr]^{-1-\eps}
=  \frac{\sin \pi\eps}{\pi}\, \frac{\theta(t)}{t^{1+\eps}}\, \theta\Bigl(\frac{\alpha}{z}-1\Bigr) \Bigl(\frac{\alpha}{z}-1\Bigr)^{-1-\eps}
\,,\end{equation}
where we used \eq{disc_os}. Since we require $z > 0$, the first $\theta$ function becomes $\theta(\alpha - z)$, and so we have
\begin{align} \label{eq:discI12}
-\theta(z)\, \frac{\img}{2\pi}\Disc_{t>0}\, I_1(A, B, \eps)
&= -\theta(z)\,\frac{\sin \pi\eps}{\pi}\, \frac{\theta(t)}{t^{1+\eps}}
\int_0^1\!\df\alpha\, \theta(\alpha - z) \Bigl(\frac{\alpha}{z}-1\Bigr)^{-1-\eps}
\nn\\
&= \theta(z)\,\frac{\sin\pi\eps}{\pi\eps}\, \frac{\theta(t)}{t^{1+\eps}}\, \theta(1-z) z^{1+\eps}(1-z)^{-\eps}
\,,\nn\\
-\theta(z)\, \frac{\img}{2\pi}\Disc_{t>0}\, I_2(A, B, \eps)
&= \theta(z)\,\frac{\sin\pi\eps}{\pi\eps(1-\eps)}\, \frac{\theta(t)}{t^{1+\eps}}\, \theta(1-z)z^{1+\eps}(1-z)^{1-\eps}
\,.\end{align}
For \fig{Bone_a}, using \eqs{Ba}{discI12} we obtain
\begin{align} \label{eq:BaDR}
& \Mae{q_n}{\theta(\w)\op^\bare_q(t,\w)}{q_n}^{(a)}
\nn\\ & \qquad
= \frac{\alpha_s(\mu)C_F}{2\pi}\,\frac{\theta(z)}{z}\, \Gamma(1+\eps) (e^{\gamma_E} \mu^2)^\eps (1-\eps)^2
  \Bigl[-\frac{\img}{2\pi}\Disc_{t>0}\, I_2(A, B, \eps) \Bigr]
\nn\\ & \qquad
= \frac{\alpha_s(\mu)C_F}{2\pi}\,\theta(z)\theta(1-z)(1-z) \Gamma(1+\eps) (e^{\gamma_E} \mu^2)^\eps (1-\eps)
\frac{\sin\pi\eps}{\pi\eps}\, \frac{\theta(t)}{t^{1+\eps}}\, \Bigl(\frac{z}{1-z}\Bigr)^{\eps}
\nn \\ & \qquad
=  \frac{\alpha_s(\mu)C_F}{2\pi}\, \theta(z) \theta(1-z) (1-z) \biggl\{
\frac{1}{\mu^2} \cL_0\Bigl(\frac{t}{\mu^2}\Bigr)
+ \delta(t) \Bigl(-\frac{1}{\eps} + \ln \frac{1-z}{z} + 1\Bigr)
\biggr\}
\,,\end{align}
where in the last step we used \eq{distr_id} to expand in $\eps$. For \fig{Bone_b}, we start from the third line in \eq{Bb1} and using \eqs{discI12}{distr_id} we get
\begin{align} \label{eq:BbDR}
 & \Mae{q_n}{\theta(\w)\op_q^\bare(t,\w)}{q_n}^{(b)}
\nn\\ & \qquad
= \frac{\alpha_s(\mu)C_F}{\pi}\, \frac{\theta(z)}{1 -z}\, \Gamma(1+\eps)(e^{\gamma_E} \mu^2)^\eps
  \Bigl[-\frac{\img}{2\pi}\Disc_{t>0}\, I_1(A, B, \eps) \Bigr]
\nn\\ & \qquad
= \frac{\alpha_s(\mu)C_F}{\pi}\, \theta(z) \Gamma(1+\eps)(e^{\gamma_E} \mu^2)^\eps
\frac{\sin\pi\eps}{\pi\eps}\, \frac{\theta(t)}{t^{1+\eps}}\, \frac{\theta(1-z)z^{1+\eps} }{(1-z)^{1+\eps}}
\nn\\ & \qquad
= \frac{\alpha_s(\mu) C_F}{\pi}\, \theta(z) \biggl\{
\biggl[- \frac{1}{\eps}\,\delta(t) + \frac{1}{\mu^2} \cL_0\Bigl(\frac{t}{\mu^2}\Bigr) \biggr]
\Bigl[- \frac{1}{\eps}\,\delta(1-z) + \cL_0(1-z)z \Bigr]
\nn \\ & \qquad\quad
+ \frac{1}{\mu^2} \cL_1\Bigl(\frac{t}{\mu^2}\Bigr) \delta(1 - z)
+  \delta(t) \Bigl[\cL_1(1 - z)z  - \cL_0(1 - z) z \ln z -  \frac{\pi^2}{12} \delta(1 - z) \Bigr]
 \biggr\}
\,.\end{align}
Adding up \eqs{BaDR}{BbDR}, the bare quark matrix element in pure dimensional regularization becomes
\begin{align} \label{eq:BqbareDR}
&\Mae{q_n}{\theta(\w)\op_q^\bare(t,\w)}{q_n}^\one
\nn\\ & \qquad
= \frac{\alpha_s(\mu) C_F}{2\pi}\,\theta(z) \biggl\{
   \biggl[ \delta(t) \Bigl(\frac{2}{\eps^2} +
    \frac{3}{2\eps} \Bigr) -
    \frac{2}{\eps}\, \frac{1}{\mu^2} \cL_0\Bigl(\frac{t}{\mu^2}\Bigr)
  \biggr]\delta(1-z)
  - \frac{1}{\eps}\,\delta(t) P_{qq}(z)
\nn \\ & \qquad\quad
  +\frac{2}{\mu^2} \cL_1\Bigl(\frac{t}{\mu^2}\Bigr)\delta(1-z) +
  \frac{1}{\mu^2} \cL_0\Bigl(\frac{t}{\mu^2}\Bigr)\cL_0(1-z)(1 + z^2)
\nn\\ & \qquad\quad
  + \delta(t) \biggl[
  \cL_1(1 - z)(1 + z^2)
  -\frac{\pi^2}{6}\, \delta(1 - z)
  + \theta(1 - z)\Bigl(1 - z - \frac{1 + z^2}{1 - z}\ln z \Bigr) \biggr]
  \biggr\}
\,.\end{align}
We can now proceed in two ways to obtain the matching coefficient $\cI_{qq}(t, z, \mu)$.

First, we can subtract $\delta(t)$ times \eq{fDR} from \eq{BqbareDR} to obtain the bare matching coefficient. This simply removes the $(1/\eps)\delta(t) P_{qq}(z)$ in the first line of \eq{BqbareDR}. Assuming that the IR divergences between the PDF and beam function cancel (and including the vanishing zero-bin) the remaining poles in the first line are of UV origin and determine the necessary $\overline{\mathrm{MS}}$ counter term, reproducing our previous result for $Z_B^q(t, \mu)$ in \eq{ZB}.

Alternatively, we can use our general result that the beam function has the same renormalization as the jet function. In this case, we subtract the one-loop counter term for $\op_q^\bare$ in \eq{ZB} (which is already known from the jet function's renormalization) from \eq{BqbareDR} to obtain the renormalized quark matrix element, which equals \eq{BqbareDR} without the $[...]\delta(1-z)$ term in the first line. The remaining $1/\eps$ pole must then be of IR origin, so we again have an explicit check that the IR divergences in the beam function match those of the PDF in \eq{fDR}. Either way, the finite terms in the last two lines of \eq{BqbareDR} determine the renormalized matching coefficient $\cI_{qq}(t, z, \mu)$, which agrees with our previous result in \eq{Iresult}.

For the gluon matrix element, \fig{Bone_f} again does not contribute. For \fig{Bone_e}, starting from the third line of \eq{Be}, we find
\begin{align}  \label{eq:BgbareDR}
 & \Mae{g_n}{\theta(\w)\op_q^\bare(t,\w)}{g_n}^\one
\\ & \qquad
= \frac{\alpha_s(\mu)T_F}{2\pi}\,\frac{\theta(z)}{z}\,
 \Gamma(1+\eps) (e^{\gamma_E} \mu^2)^\eps \Bigl(\frac{1-\eps}{1-z} - 2z\Bigr)
  \Bigl[-\frac{\img}{2\pi}\Disc_{t>0}\, I_2(A, B, \eps) \Bigr]
\nn\\ & \qquad
= \frac{\alpha_s(\mu)T_F}{2\pi}\,\theta(z)\theta(1-z)
 \Gamma(1+\eps) (e^{\gamma_E} \mu^2)^\eps (1 - 2z + 2z^2 -\eps)
\frac{\sin\pi\eps}{\pi\eps(1-\eps)}\, \frac{\theta(t)}{t^{1+\eps}}\, \Bigl(\frac{z}{1-z}\Bigr)^\eps
\nn\\\nn & \qquad
 = \frac{ \alpha_s(\mu) T_F }{2\pi}\, \theta(z) \biggl\{
\frac{1}{\mu^2} \cL_0\Bigl(\frac{t}{\mu^2}\Bigr) P_{qg}(z)
 + \delta(t) \biggl[P_{qg}(z)\Bigl(-\frac{1}{\eps} + \ln\frac{1-z}{z} - 1\Bigr) +  \theta(1-z) \biggr]
\biggr\}
\,.\end{align}
The same discussion as for the quark matrix element above can be repeated for the gluon matrix element. The $(1/\eps)\delta(t)P_{qg}(z)$ term matches the IR divergence in the PDF in \eq{fDR}. Since there are no further poles, no UV renormalization is required and the quark and gluon operators do not mix. The finite terms in \eq{BgbareDR} then determine the matching coefficient $\cI_{qg}(t, z, \mu)$, reproducing our previous result in \eq{Iresult}.

\chapter{Perturbative Results}
\label{app:pert}

In this appendix we collect perturbative results relevant for the Drell-Yan beam thrust cross section
in \eq{DYbeamrun} and the cross section for isolated Higgs production through gluon fusion in \eq{higgsprod}. These were included as appendices in Ref.~\cite{Stewart:2010qs,Berger:2010BgNLO}.

\section{Fixed-Order Results for Drell-Yan}
\label{app:HandS}

We summarize the results of \sec{DY_final}.
The one-loop hard function for Drell-Yan is given by the square of Wilson coefficients in SCET for which the relevant one-loop matching from QCD onto SCET was carried out in Refs.~\cite{Manohar:2003vb, Bauer:2003di} 
\begin{align}
H_{q\bar q}(q^2, \mu) = H_{\bar q q}(q^2, \mu)
&=
\biggl[ Q_q^2 + \frac{(v_q^2 + a_q^2) (v_\ell^2+a_\ell^2) - 2 Q_q v_q v_\ell (1-m_Z^2/q^2)}
{(1-m_Z^2/q^2)^2 + m_Z^2 \Gamma_Z^2/q^4} \biggr]
\nn \\ & \quad\times
\biggl[1 + \frac{\alpha_s(\mu)\,C_F}{2\pi}
\biggl(-\ln^2 \frac{q^2}{\mu^2} +
3 \ln \frac{q^2}{\mu^2} - 8 + \frac{7\pi^2}{6} \biggr) \biggr]
\,.\end{align}
We included the prefactor from the leptonic matrix element, where $Q_q$ is the quark charge in units of $\abs{e}$, $v_{\ell,q}$ and $a_{\ell,q}$ are the standard vector and axial couplings of the leptons and quarks, and $m_Z$ and $\Gamma_Z$ are the mass and width of the $Z$ boson.
The beam thrust soft function can be extracted form the one-loop incoming hemisphere soft function~\cite{Schwartz:2007ib, Fleming:2007xt} and is given by
\begin{equation} \label{eq:SBquark}
S_B^{q\bar q}(k^+,\mu) = \delta(k^+) + \frac{\alpha_s(\mu)\,C_F}{2\pi} \biggl[
-\frac{8}{\mu} \cL_1\Bigl(\frac{k^+}{\mu}\Bigr) + \frac{\pi^2}{6}\, \delta(k^+) \biggr]
\,.\end{equation}
Our one-loop results for the matching coefficients in the beam function OPE in \eq{beam_fact} are for the quark beam function given in \eq{Iresult}.

\section{Fixed-Order Results for Higgs Production}

For Higgs production through gluon fusion we will integrate out the top quark and hard off-shell modes in one step. This avoids the often used expansion $m_H \ll m_t$, but does not allow us to sum logarithms of $m_H^2/m_t^2$. In the narrow width approximation the virtuality of the Higgs is simply $q^2 = m_H^2$, but we will keep $q^2$ general in the below discussion. In pure dimensional regularization the matching coefficient $C_{ggH}(m_t, q^2)$ is given by the infrared-finite part of the on-shell $ggH$ form factor \cite{Harlander:2005rq,Anastasiou:2006hc,Harlander:2009bw,Pak:2009bx}. At NLO this yields
\begin{align}
H_{ggH}(m_t, q^2, \mu)
= \alpha_s^2(\mu) \Abs{F^\zero(\rho/4)}^2
\biggl\{1 + \frac{\alpha_s(\mu)}{2\pi}
\biggl[ C_A \Bigl(-\ln^2 \frac{q^2}{\mu^2} + \frac{7\pi^2}{6} \Bigr) + F^\one(\rho) \biggr]
\biggr\}
\,,\end{align}
where $\rho = m_H^2/m_t^2$ and
\begin{align}
F^\zero(x) &= \frac{3}{2x} - \frac{3}{2x}\Bigl|1 - \frac{1}{x}\Bigr|
\begin{cases}
\arcsin^2(\sqrt{x})\, , & 0 < x \leq 1 \,,\\
\ln^2[-\img(\sqrt{x} + \sqrt{x-1})] \,, \quad & x > 1
\,,\end{cases}
\\
F^\one(\rho) &=
\Bigl(5 - \frac{19}{90} \rho - \frac{1289}{75600} \rho^2 - \frac{155}{72576} \rho^3 - \frac{5385047}{16765056000} \rho^4\Bigr) C_A
\nn\\ & \quad
+ \Bigl(-3 + \frac{307}{360} \rho + \frac{25813}{302400} \rho^2 + \frac{3055907}{254016000} \rho^3 +
   \frac{659504801}{335301120000} \rho^4 \Bigr) C_F + \ord{\rho^5}
\,.\nn\end{align}
For $\rho\to 0$ (corresponding to $m_t\to\infty$) we have $F^\zero(0) = 1$ and $F^\one(0) = 5C_A - 3C_F$.

The gluon beam thrust soft function has Wilson lines in the adjoint rather than fundamental representation. At one loop we can simply replace $C_F$ by $C_A$ (Casimir scaling) in the quark result in \eq{SBquark}
\begin{equation}
S_B^{gg}(k^+,\mu) = \delta(k^+) + \frac{\alpha_s(\mu)\,C_A}{2\pi} \biggl\{
-\frac{8}{\mu} \biggl[\frac{\theta(k^+ / \mu) \ln(k^+ / \mu)}{k^+ / \mu} \biggr]_+ +
\frac{\pi^2}{6}\, \delta(k^+) \biggr\}
\,. \nn \\
\end{equation}
The one-loop coefficients for matching the gluon beam function onto PDFs are given in \eq{I_results}.

\section{Renormalization Group Evolution}
\label{app:rge}

The RGE and anomalous dimension for the hard function are [see \eqs{H_RGE}{gammas}]
\begin{align}
\mu \frac{\df}{\df\mu} H_{ij}(q^2, \mu) = \gamma_H^{ij}(q^2, \mu) H_{ij}(q^2, \mu)
\,, \nn\\
\gamma_H^{ij}(q^2, \mu) =
2 \Gamma_\cusp^i(\alpha_s) \ln\frac{q^2}{\mu^2} + \gamma_H^{ij}(\alpha_s)
\,.\end{align}
The expansion coefficients of $\Gamma_\cusp^i(\alpha_s)$ and $\gamma_H^{ij}(\alpha_s)$ are given below in \eqs{Gacuspexp}{gaHexp}. Note that $\Gamma_\cusp^q = \Gamma_\cusp^{\bar q}$ so $\Ga_\cusp^i = \Ga_\cusp^j$. The RGE in \eq{H_RGE} has the standard solution
\begin{align} \label{eq:Hrun}
H_{ij}(q^2, \mu) &= H_{ij}(q^2, \mu_0)\, U_H^{ij}(q^2, \mu_0, \mu)
\,, \nn\\
U_H^{ij}(q^2, \mu_0, \mu) &= e^{K_H^{ij}(\mu_0, \mu)} \Bigl(\frac{q^2}{\mu_0^2}\Bigr)^{\eta_H^{ij}(\mu_0, \mu)}
\,,\nn \\
K_H^{ij}(\mu_0,\mu) &= -4 K^i_\Gamma(\mu_0,\mu) + K_{\gamma_H^{ij}}(\mu_0,\mu)
\,, \qquad
\eta_H^{ij}(\mu_0,\mu) = 2\eta_\Gamma^i(\mu_0,\mu)
\,,\end{align}
where the functions $K_\Gamma^i(\mu_0, \mu)$, $\eta_\Gamma^i(\mu_0, \mu)$ and $K_\gamma$ are given below in \eq{Keta_def}.

The beam function RGE is [see \eqs{B_RGE}{gaB_gen}]
\begin{align}
\mu \frac{\df}{\df \mu} B_i(t, x, \mu) &= \int\! \df t'\, \gamma_B^i(t-t',\mu)\, B_i(t', x, \mu)
\,,\nn\\
\gamma_B^i(t, \mu)
&= -2 \Gamma^i_{\cusp}(\alpha_s)\,\frac{1}{\mu^2}\cL_0\Bigl(\frac{t}{\mu^2}\Bigr) + \gamma_B^i(\alpha_s)\,\delta(t)
\,,\end{align}
and its solution is~\cite{Balzereit:1998yf, Neubert:2004dd, Fleming:2007xt, Ligeti:2008ac} [see \eq{Brun}]
\begin{align} \label{eq:Brun_full}
B_i(t,x,\mu) & =  \int\! \df t'\, B_i(t - t',x,\mu_0)\, U_B^i(t',\mu_0, \mu)
\,, \nn \\
U_B^i(t, \mu_0, \mu) &= \frac{e^{K_B^i -\gamma_E\, \eta_B^i}}{\Gamma(1+\eta_B^i)}\,
\biggl[\frac{\eta_B^i}{\mu_0^2} \cL^{\eta_B^i} \Bigl( \frac{t}{\mu_0^2} \Bigr) + \delta(t) \biggr]
\,, \nn \\
K_B^i(\mu_0,\mu) &= 4 K^i_\Gamma(\mu_0,\mu) + K_{\gamma_B^i}(\mu_0,\mu)
\,, \qquad
\eta_B^i(\mu_0,\mu) = -2\eta^i_{\Gamma}(\mu_0,\mu)
\,.\end{align}

The beam thrust soft function is given in terms of $S_{\hemiin}^{ij}$ by
\begin{equation}
S_B^{ij}(k^+, \mu)
= \!\int\!\df k_a^+ \df k_b^+\, S_\hemiin^{ij}(k_a^+, k_b^+, \mu )\,\delta(k^+\! - k_a^+ - k_b^+)
\,.\end{equation}
Its RGE is easily obtained by integrating \eqs{tS_RGE}{gammas2},
\begin{align} \label{eq:SB_RGE}
\mu\frac{\df}{\df\mu} S_B^{ij}(k^+, \mu)
&= \int\! \df \ell^+\, \gamma_S^{ij}(k^+\! - \ell^+, \mu)\, S_B^{ij}(\ell^+, \mu)
\,, \\
\gamma_S^{ij}(k^+, \mu)
&= 4\,\Gamma_\cusp^i(\alpha_s)\, \frac{1}{\mu} \cL_0\Big(\frac{k^+}{\mu}\Big) +
\gamma_S^{ij}(\alpha_s)\, \delta(k^+)
\,, \quad
\gamma_S^{ij}(\alpha_s) 
= - \gamma_H^{ij}(\alpha_s) - 2\gamma_B^i(\alpha_s)
\,, \nn \end{align}
whose solution is completely analogous to \eq{Brun_full},
\begin{align} \label{eq:SBrun}
S_B^{ij}(k^+,\mu) & =  \int\! \df \ell^+\, S_B^{ij}(k^+\! - \ell^+,\mu_0)\, U_S^{ij}(\ell^+,\mu_0,\mu)
\,, \nn \\
U_S^{ij}(k^+, \mu_0, \mu) & = \frac{e^{K_S^{ij} -\gamma_E\, \eta_S^{ij}}}{\Gamma(1+\eta_S^{ij})}\,
\biggl[\frac{\eta_S^{ij}}{\mu_0} \cL^{\eta_S^{ij}} \Big( \frac{k^+}{\mu_0} \Big) + \delta(k^+) \biggr]
\,, \nn \\
K_S^{ij}(\mu_0,\mu) &= -4K_\Gamma^i(\mu_0,\mu) + K_{\gamma_S^{ij}}(\mu_0,\mu)
\,, \quad
\eta_S^{ij}(\mu_0,\mu) = 4\eta_\Gamma^i(\mu_0,\mu)
\,.\end{align}
The functions $K_\Gamma^i(\mu_0, \mu)$, $\eta_\Gamma^i(\mu_0, \mu)$, $K_\gamma(\mu_0, \mu)$ in the above RGE solutions are defined as
\begin{align} \label{eq:Keta_def}
K_\Gamma^i(\mu_0, \mu)
& = \int_{\alpha_s(\mu_0)}^{\alpha_s(\mu)}\!\frac{\df\alpha_s}{\beta(\alpha_s)}\,
\Gamma_\cusp^i(\alpha_s) \int_{\alpha_s(\mu_0)}^{\alpha_s} \frac{\df \alpha_s'}{\beta(\alpha_s')}
\,,\quad
\eta_\Gamma^i(\mu_0, \mu)
= \int_{\alpha_s(\mu_0)}^{\alpha_s(\mu)}\!\frac{\df\alpha_s}{\beta(\alpha_s)}\, \Gamma_\cusp^i(\alpha_s)
\,,\nn \\
K_\gamma(\mu_0, \mu)
& = \int_{\alpha_s(\mu_0)}^{\alpha_s(\mu)}\!\frac{\df\alpha_s}{\beta(\alpha_s)}\, \gamma(\alpha_s)
\,.\end{align}
Expanding the beta function and anomalous dimensions in powers of $\alpha_s$,
\begin{align}
&\beta(\alpha_s) =
- 2 \alpha_s \sum_{n=0}^\infty \beta_n\Bigl(\frac{\alpha_s}{4\pi}\Bigr)^{n+1}
\,, \
&\Gamma^i_\cusp(\alpha_s) = 
\sum_{n=0}^\infty \Gamma^i_n \Bigl(\frac{\alpha_s}{4\pi}\Bigr)^{n+1}
\,, \
\gamma(\alpha_s) = \sum_{n=0}^\infty \gamma_n \Bigl(\frac{\alpha_s}{4\pi}\Bigr)^{n+1}
\,,\end{align}
their explicit expressions at NNLL are, 
\begin{align} \label{eq:Keta}
K_\Gamma(\mu_0, \mu) &= -\frac{\Gamma_0}{4\beta_0^2}\,
\biggl\{ \frac{4\pi}{\alpha_s(\mu_0)}\, \Bigl(1 - \frac{1}{r} - \ln r\Bigr)
   + \biggl(\frac{\Gamma_1 }{\Gamma_0 } - \frac{\beta_1}{\beta_0}\biggr) (1-r+\ln r)
   + \frac{\beta_1}{2\beta_0} \ln^2 r
\nn\\ & \hspace{10ex}
+ \frac{\alpha_s(\mu_0)}{4\pi}\, \biggl[
  \biggl(\frac{\beta_1^2}{\beta_0^2} - \frac{\beta_2}{\beta_0} \biggr) \Bigl(\frac{1 - r^2}{2} + \ln r\Bigr)
  + \biggl(\frac{\beta_1\Gamma_1 }{\beta_0 \Gamma_0 } - \frac{\beta_1^2}{\beta_0^2} \biggr) (1- r+ r\ln r)
\nn\\ & \hspace{10ex}
  - \biggl(\frac{\Gamma_2 }{\Gamma_0} - \frac{\beta_1\Gamma_1}{\beta_0\Gamma_0} \biggr) \frac{(1- r)^2}{2}
     \biggr] \biggr\}
\,, \nn\\
\eta_\Gamma(\mu_0, \mu) &=
 - \frac{\Gamma_0}{2\beta_0}\, \biggl[ \ln r
 + \frac{\alpha_s(\mu_0)}{4\pi}\, \biggl(\frac{\Gamma_1 }{\Gamma_0 }
 - \frac{\beta_1}{\beta_0}\biggr)(r-1)
\nn \\ & \hspace{10ex}
 + \frac{\alpha_s^2(\mu_0)}{16\pi^2} \biggl(
    \frac{\Gamma_2 }{\Gamma_0 } - \frac{\beta_1\Gamma_1 }{\beta_0 \Gamma_0 }
      + \frac{\beta_1^2}{\beta_0^2} -\frac{\beta_2}{\beta_0} \biggr) \frac{r^2-1}{2}
    \biggr]
\,, \nn\\
K_\gamma(\mu_0, \mu) &=
 - \frac{\gamma_0}{2\beta_0}\, \biggl[ \ln r
 + \frac{\alpha_s(\mu_0)}{4\pi}\, \biggl(\frac{\gamma_1 }{\gamma_0 }
 - \frac{\beta_1}{\beta_0}\biggr)(r-1) \biggr]
\,.\end{align}
Here $r = \alpha_s(\mu)/\alpha_s(\mu_0)$ and we have suppressed the superscript $i$ on $K_\Ga^i$, $\eta_\Ga^i$ and $\Ga^i_n$.

Up to three loops, the coefficients of the beta function~\cite{Tarasov:1980au, Larin:1993tp} and cusp anomalous dimension~\cite{Korchemsky:1987wg, Moch:2004pa} in $\overline{\mathrm{MS}}$ are
\begin{align} \label{eq:Gacuspexp}
\beta_0 &= \frac{11}{3}\,C_A -\frac{4}{3}\,T_F\,n_f
\,,\\
\beta_1 &= \frac{34}{3}\,C_A^2  - \Bigl(\frac{20}{3}\,C_A\, + 4 C_F\Bigr)\, T_F\,n_f
\,, \nn\\
\beta_2 &=
\frac{2857}{54}\,C_A^3 + \Bigl(C_F^2 - \frac{205}{18}\,C_F C_A
 - \frac{1415}{54}\,C_A^2 \Bigr)\, 2T_F\,n_f
 + \Bigl(\frac{11}{9}\, C_F + \frac{79}{54}\, C_A \Bigr)\, 4T_F^2\,n_f^2
\,,\nn\\[2ex]
\Gamma^q_0 &= 4C_F
\,,\nn\\
\Gamma^q_1 &= 4C_F \Bigl[\Bigl( \frac{67}{9} -\frac{\pi^2}{3} \Bigr)\,C_A  -
   \frac{20}{9}\,T_F\, n_f \Bigr]
\,,\nn\\
\Gamma^q_2 &= 4C_F \Bigl[
\Bigl(\frac{245}{6} -\frac{134 \pi^2}{27} + \frac{11 \pi ^4}{45}
  + \frac{22 \zeta_3}{3}\Bigr)C_A^2
  + \Bigl(- \frac{418}{27} + \frac{40 \pi^2}{27}  - \frac{56 \zeta_3}{3} \Bigr)C_A\, T_F\,n_f
\nn\\* & \hspace{8ex}
  + \Bigl(- \frac{55}{3} + 16 \zeta_3 \Bigr) C_F\, T_F\,n_f
  - \frac{16}{27}\,T_F^2\, n_f^2 \Bigr]
\,,\\[2ex]
\Gamma^g_n &= \frac{C_A}{C_F}\, \Gamma^q_n \quad \text{for }n\leq 2.
\end{align}

The $\overline{\mathrm{MS}}$ non-cusp anomalous dimension for the hard function $H^{q\bar q}$ can be obtained~\cite{Idilbi:2006dg, Becher:2006mr} from the IR divergences of the on-shell massless quark form factor which is known to three loops~\cite{Moch:2005id}. Similarly, the anomalous dimension for $H^{gg}$ can be extracted~\cite{Becher:2009qa} from the gluon form factor, which has also been calculated to three loops~\cite{Moch:2005tm}.
\begin{align} \label{eq:gaHexp}
\gamma_{H\,0}^{q\bar q} &= -12 C_F
\,,\nn\\
\gamma_{H\,1}^{q\bar q}
&= -2 C_F \Bigl[
  \Bigl(\frac{82}{9} - 52 \zeta_3\Bigr) C_A
+ (3 - 4 \pi^2 + 48 \zeta_3) C_F
+ \Bigl(\frac{65}{9} + \pi^2 \Bigr) \beta_0 \Bigr]
\,,\nn\\
\gamma_{H\,2}^{q\bar q}
&= -4C_F \Bigl[
  \Bigl(\frac{66167}{324} - \frac{686 \pi^2}{81} - \frac{302 \pi^4}{135} - \frac{782 \zeta_3}{9} + \frac{44\pi^2 \zeta_3}{9} + 136 \zeta_5\Bigr) C_A^2
\nn\\ & \qquad\hspace{6ex}
+ \Bigl(\frac{151}{4} - \frac{205 \pi^2}{9} - \frac{247 \pi^4}{135} + \frac{844 \zeta_3}{3} + \frac{8 \pi^2 \zeta_3}{3} + 120 \zeta_5\Bigr) C_F C_A
\nn\\ & \qquad\hspace{6ex}
+ \Bigl(\frac{29}{2} + 3 \pi^2 + \frac{8\pi^4}{5} + 68 \zeta_3 - \frac{16\pi^2 \zeta_3}{3} - 240 \zeta_5\Bigr) C_F^2
\nn\\ & \qquad\hspace{6ex}
+ \Bigl(-\frac{10781}{108} + \frac{446 \pi^2}{81} + \frac{449 \pi^4}{270} - \frac{1166 \zeta_3}{9} \Bigr) C_A \beta_0
\nn\\ & \qquad\hspace{6ex}
+ \Bigl(\frac{2953}{108} - \frac{13 \pi^2}{18} - \frac{7 \pi^4 }{27} + \frac{128 \zeta_3}{9}\Bigr)\beta_1
+ \Bigl(-\frac{2417}{324} + \frac{5 \pi^2}{6} + \frac{2 \zeta_3}{3}\Bigr)\beta_0^2
\Bigr]
\,, \\[2ex]
\ga_{H\,0}^{gg} &= -4 \bt_0
\,,\nn\\
\ga_{H\,1}^{gg}
&= \Big(-\frac{236}{9} + 8\zeta_3\Big)C_A^2 +
\Big(-\frac{76}{9}+\frac{2\pi^2}{3}\Big) C_A\, \bt_0 - 4 \bt_1
\,,\nn\\
\ga_{H\,2}^{gg}
&= \Big(-\frac{60875}{81} + \frac{1268 \pi^2}{81} +\frac{16\pi^4}{5}+\frac{3944\zeta_3}{9}
- \frac{80\pi^2 \zeta_3}{9} - 64 \zeta_5 \Big)C_A^3
\nn \\ & \quad +
\Big(\frac{7649}{27}+\frac{268 \pi^2}{81} - \frac{122\pi^4}{45}
 - \frac{1000\zeta_3}{9}\Big) C_A^2\, \bt_0 +
\Big(\frac{932}{81}+\frac{10\pi^2}{9}-\frac{56 \zeta_3}{3}\Big) C_A\, \bt_0^2
\nn \\ & \quad +
\Big(-\frac{1819}{27} + \frac{2\pi^2}{3} + \frac{8\pi^4}{45} + \frac{304\zeta_3}{9}\Big) C_A\, \bt_1
-4 \bt_2
\,.\end{align}
Denoting $\gamma_f^q$ the coefficient of the $\delta(1-z)$ in the quark PDF anomalous dimension, \eq{PDF_RGE} (which gives the non-cusp part of the anomalous dimension in the threshold limit $z\to 1$), the factorization theorem for DIS at threshold implies that $\gamma_H + \gamma_J^q + \gamma_f^q = 0$, which was used in Ref.~\cite{Becher:2006mr} to obtain $\gamma_J^q$ at three loops from the known three-loop result for $\gamma_f^q$~\cite{Moch:2004pa}. This argument can repeated for a gluons. As we showed in \sec{B_RGE}, the anomalous dimension for the beam function equals that of the jet function, $\gamma_B^i = \gamma_J^i$, so the three-loop result for $\gamma_f^i$ together with \eq{gaHexp} yield the non-cusp three-loop anomalous dimension for the beam functions
\begin{align}
\gamma_{B\,0}^q &= 6 C_F
\,,\nn\\
\gamma_{B\,1}^q
&= C_F \Bigl[
  \Bigl(\frac{146}{9} - 80 \zeta_3\Bigr) C_A
+ (3 - 4 \pi^2 + 48 \zeta_3) C_F
+ \Bigl(\frac{121}{9} + \frac{2\pi^2}{3} \Bigr) \beta_0 \Bigr]
\,,\nn\\
\gamma_{B\,2}^q
&= 2 C_F \Bigl[
  \Bigl(\frac{52019}{162} - \frac{841\pi^2}{81} - \frac{82\pi^4}{27} -\frac{2056\zeta_3}{9} + \frac{88\pi^2 \zeta_3}{9} + 232 \zeta_5\Bigr)C_A^2
\nn\\ & \quad\hspace{6ex}
+ \Bigl(\frac{151}{4} - \frac{205\pi^2}{9} - \frac{247\pi^4}{135} + \frac{844\zeta_3}{3} + \frac{8\pi^2 \zeta_3}{3} + 120 \zeta_5\Bigr) C_A C_F
\nn\\ & \quad\hspace{6ex}
+ \Bigl(\frac{29}{2} + 3 \pi^2 + \frac{8\pi^4}{5} + 68 \zeta_3 - \frac{16\pi^2 \zeta_3}{3} - 240 \zeta_5\Bigr) C_F^2
\nn\\ & \quad\hspace{6ex}
+ \Bigl(-\frac{7739}{54} + \frac{325}{81} \pi^2 + \frac{617 \pi^4}{270} - \frac{1276\zeta_3}{9} \Bigr) C_A\beta_0
\nn\\ & \quad\hspace{6ex}
+ \Bigl(-\frac{3457}{324} + \frac{5\pi^2}{9} + \frac{16 \zeta_3}{3} \Bigr) \beta_0^2
+ \Bigl(\frac{1166}{27} - \frac{8 \pi^2}{9} - \frac{41 \pi^4}{135} + \frac{52 \zeta_3}{9}\Bigr) \beta_1
\Bigr]
\,,\\
\ga_{B\,0}^g &= 2 \bt_0
\,,\nn\\
\ga_{B\,1}^g
&= \Big(\frac{182}{9} - 32\zeta_3\Big)C_A^2 +
\Big(\frac{94}{9}-\frac{2\pi^2}{3}\Big) C_A\, \bt_0 + 2\bt_1
\,,\nn\\
\ga_{B\,2}^g
&= \Big(\frac{49373}{81} - \frac{944 \pi^2}{81} - \frac{16\pi^4}{5} - \frac{4520 \zeta_3}{9}
+ \frac{128\pi^2 \zeta_3}{9} + 224 \zeta_5 \Big)C_A^3
\nn \\ & \quad +
\Big(-\frac{6173}{27} - \frac{376 \pi^2}{81} + \frac{13\pi^4}{5}
+ \frac{280\zeta_3}{9}\Big) C_A^2\, \bt_0 +
\Big(-\frac{986}{81}-\frac{10\pi^2}{9}+\frac{56 \zeta_3}{3}\Big) C_A\, \bt_0^2
\nn \\ & \quad +
\Big(\frac{1765}{27} - \frac{2\pi^2}{3} - \frac{8\pi^4}{45} - \frac{304\zeta_3}{9}\Big) C_A\, \bt_1
+ 2 \bt_2
\,. \end{align}
At NNLL, we only need the one- and two-loop coefficients of $\gamma_B^i$ and $\gamma_H^{ij}$. The three-loop coefficients, $\gamma^{ij}_{H\,2}$ and $\gamma_{B\,2}^i$, are given here for completeness. They are required for the resummation at N$^3$LL, where one would also need the four-loop beta function and cusp anomalous dimension, the latter of which is has not been calculated so far. In addition, the full N$^3$LL would also require the two-loop fixed-order corrections, which are known for the hard function, but not yet for the beam and soft functions.


\begin{singlespace}
\bibliography{main.bib}
\bibliographystyle{plain}
\end{singlespace}

\end{document}